\documentclass[
aps,prd,twocolumn,preprintnumbers,
nofootinbib,floatfix,
amsmath,amssymb,longbibliography
]{revtex4-2}
\usepackage[utf8]{inputenc}
\usepackage{siunitx}
\usepackage{amsfonts}
\usepackage{braket}
\usepackage{mathtools}
\usepackage{cancel}
\usepackage{slashed}
\usepackage{pifont}
\usepackage{soul}
\usepackage{comment}

\usepackage{enumerate}

\usepackage{yhmath} % widehat

\usepackage{booktabs,array}
\usepackage{hhline}
\usepackage{dcolumn}
    \newcolumntype{d}[1]{D{.}{.}{#1}}
\usepackage{graphicx}
\usepackage[caption=false]{subfig}

\usepackage{multirow}

\usepackage{dcolumn}
\usepackage{mathtools}
\usepackage{upgreek}
\newcolumntype{T}{D{.}{.}{10}}
\newcolumntype{E}{D{.}{.}{11}}
\newcolumntype{F}{D{.}{.}{5}}

\usepackage[normalem]{ulem}

\usepackage[pdftex]{color}
\usepackage[dvipsnames]{xcolor}
\usepackage{hyperref}% add hypertext capabilities
\hypersetup{
    colorlinks=true,     % false: boxed links; true: colored links
    linkcolor=blue,      % color of internal links
    citecolor=blue,      % color of links to bibliography
    filecolor=blue,      % color of file links
    urlcolor=blue        % color of external links
}

\def\mybot{{\!\bot\!}}
\newcommand{\vev}[1]{\ensuremath{\langle #1 \rangle}}
\renewcommand{\vec}[1]{{\mathbf{#1}}}

\makeatletter 
\renewcommand\onecolumngrid{% <<<<<<
\do@columngrid{one}{\@ne}%
\def\set@footnotewidth{\onecolumngrid}% <<<<<<<<<<<<<<<<
\def\footnoterule{\kern-6pt\hrule width 1.5in\kern6pt}%
}
\renewcommand\twocolumngrid{% <<<<<<
        \def\footnoterule{% restore rule
        \dimen@\skip\footins\divide\dimen@\thr@@
        \kern-\dimen@\hrule width.5in\kern\dimen@}
        \do@columngrid{mlt}{\tw@}
}%
\makeatother

\begin{document}
\title{Gluon gravitational structure of hadrons of different spin}

\newcommand{\getMITAffiliation}{\affiliation{Center for Theoretical Physics, Massachusetts Institute of Technology, Cambridge, MA 02139, U.S.A.}}

\author{Dimitra A. Pefkou}
\author{Daniel C. Hackett}
\author{Phiala E. Shanahan}

\getMITAffiliation

\begin{abstract}
The gravitational form factors (GFFs) of hadrons encode the matrix elements of the energy-momentum tensor of QCD. These quantities describe how energy, spin, and various mechanical properties of hadrons are carried by their quark and gluon constituents. We present the gluon GFFs of the
pion, nucleon, $\rho$ meson, and
$\Delta$ baryon as functions of the squared momentum transfer $t$
in the region
$0 \leq -t < 2 \; \text{GeV}^2$, 
as determined in a lattice QCD study with pion mass $m_{\pi} = 450(5) \; \text{MeV}$.
By fitting the extracted GFFs using multipole
and z-parameter expansion functional forms, we extract various gluon contributions to the
energy, pressure, and shear force distributions of
the hadrons in the 3D
and 2D Breit frames as well as in the infinite
momentum frame. We also obtain estimates for
the corresponding gluon mechanical and mass radii, as well as the forward-limit gluon contributions to the momentum fraction and angular momentum of the hadrons.
\end{abstract}

\preprint{MIT-CTP/5318}

\maketitle

\section{INTRODUCTION}

Understanding the internal dynamics of hadrons
in terms of their fundamental quark and gluon constituents has
been a goal of particle and nuclear physics
since the first experimental probe of proton
substructure at SLAC~\cite{Hofstadter:1956qs}
and the subsequent development of the
theory of quantum chromodynamics (QCD)~\cite{GellMann:1964nj,Zweig:1964jf,Zweig:1964ruk}.
However, many aspects of hadron structure have not yet been fully constrained from theory or experiment,
including the gravitational form factors
(GFFs)~\cite{Pagels:1966zza} of hadrons, defined from matrix elements of the QCD energy-momentum tensor (EMT).
These form factors
describe how energy, spin, 
pressure, and shear forces are distributed
within hadrons~\cite{Polyakov:2002yz};
therefore,
their determination is of fundamental
significance.

The off-forward hadron matrix elements of the symmetric%
\footnote{In general the QCD EMT does not need to be symmetric, and its matrix elements for hadrons of spin $> 0$ include additional GFFs associated with antisymmetric Lorentz structures~\cite{Lorce:2017wkb}. Here we only consider the symmetric part.} EMT $T_i^{\mu\nu}$, with $i\in\{q,g\}$ indexing the gluon or quark component, can generically be decomposed into $N_h$ terms with distinct Lorentz structures as
\begin{equation}
\bra{h(p,s)} T_i^{\mu\nu} \ket{h(p',s')} = \sum_{j=1}^{N_h} K^{\{\mu\nu\},h,j}_{ss'}(P, \Delta) ~ G_{i}^{h,j}(t) ,
\label{eqn:general-gff-decomp}
\end{equation}
where $\ket{h(p,s)}$ denotes a hadronic state with four-momentum $p$ and polarization $s$, and $K_{ss'}^{\{\mu\nu\},h,j}$ are kinematic coefficients symmetrized over their Lorentz indices as ${a_{\{\mu}b_{\nu\}} = (a_{\mu}b_{\nu}+
b_{\mu}a_{\nu})/2}$, written in terms of ${P = (p + p')/2}$ and ${\Delta = p' - p}$. The GFFs $G_{i}^{h,j}(t)$ are functions of the Mandelstam variable $t=\Delta^2$, and $j$ indexes the different GFFs in the decomposition for hadron $h$. An analogous decomposition of the total conserved EMT ${T^{\mu\nu}=\sum_q T^{\mu\nu}_q + T^{\mu\nu}_g}$ yields the total GFFs ${G^{h,j}(t) = \sum_q G^{h,j}_{q}(t) + G^{h,j}_{g}(t)}$.

The GFFs associated with the symmetric 
traceless part of the EMT correspond to the second Mellin moments of the corresponding generalized parton distributions (GPDs)
\cite{Ji:1996nm,Mueller:1998fv,Radyushkin:1996nd}, 
which allows them to be constrained by experimental data
from deeply virtual Compton scattering (DVCS)~\cite{Ji:1996ek,dHose:2016mda,Kumericki:2016ehc} and meson production~\cite{Collins:1996fb,Mankiewicz:1997bk}. For example, data from the Belle experiment at KEKB \cite{Belle:2015oin,Savinov:2013hda} has been used to constrain the pion quark GFFs \cite{Kumano:2017lhr}, while the nucleon quark GFFs have been studied from DVCS with the CLAS detector \cite{Burkert:2018bqq,Pasquini:2014vua,CLAS:2007clm,CLAS:2015uuo} at the Thomas Jefferson National Accelerator Facility (JLab).
The PANDA experiment at the Facility for Antiproton and Ion Research (FAIR) \cite{PANDA:2009yku}, as well as future experiments at SuperKEKB, the International Linear Collider (ILC), the Japan proton accelerator complex (J-PARC) \cite{LoI} and the nuclotron-based ion collider facility (NICA) \cite{NICA} will further
constrain the quark GPDs and thus GFFs of various hadrons.

There has also been significant progress in the theoretical determination of quark and total GFFs, in particular through lattice QCD, phenomenology, and models.
For example, the total and quark GFFs of the pion \cite{Hudson:2017xug} and the nucleon \cite{Chen:2001pva,Belitsky:2002jp,Ando:2006sk,Diehl:2006ya,Dorati:2007bk} have been studied via chiral perturbation theory, chiral quark models like the spectral quark model and 
the Nambu-Jona-Lasinio (NJL) model have been used to constrain the pion~\cite{Broniowski:2008hx,Freese:2019bhb} and the $\rho$ GFFs~\cite{Freese:2019bhb},
the bag model to investigate those of the nucleon, $\rho$ meson, and $\Delta$ baryon \cite{Neubelt:2019sou}, the Skyrme model those of the nucleon \cite{Cebulla:2007ei,Kim:2012ts} and $\Delta$ baryon \cite{Kim:2020lrs}, the chiral quark-soliton model \cite{Petrov:1998kf,Schweitzer:2002nm,Ossmann:2004bp,Wakamatsu:2005vk,Wakamatsu:2006dy,Wakamatsu:2007uc,Goeke:2007fp,Goeke:2007fq} instanton model \cite{Polyakov:2018exb} and light-cone QCD sum rule formalism \cite{Azizi:2019ytx} those of the nucleon, and the light-cone constituent quark model \cite{Sun:2020wfo} and AdS/QCD model
\cite{Abidin:2008ku} those of the $\rho$ meson.  
Lattice QCD has also been used to study
the quark GFFs of the pion \cite{Brommel:2005jC,Brommel:2007zz} and
nucleon \cite{Alexandrou:2018ii,Alexandrou:2019ali,Hagler:2007xi,Bali:2016wqg,Alexandrou:2020sml}, and these quantities have been studied within the large-$N_c$ approach~\cite{Masjuan:2012sk}. 

\begin{table*}[!t]
\begin{center}
\begin{tabular}{cccccccccccccc}
% \hline \hline
\toprule
$L/a$   & $T/a$ & $\beta$ & $a m_l$ & $a m_s$ & $a$ [fm] & $L$ [fm] & $T$ [fm] & $m_{\pi}$ [MeV] & $m_K$ [MeV] &
$m_{\pi}L$ & $m_{\pi}T$ & $N_{\text{cfgs}}$ & $\bar{N}_{\text{meas}}$ \\ \midrule
$32$ & $96$ & $6.1$ & $-0.2800$ &
$-0.2450$ & $0.1167(16)$ &
$3.7$ & $11.2$ & $450(5)$ & 
$596(6)$ & $8.5$ & $25.6$ &
$2820$ & $235$ \\
\bottomrule
\end{tabular}
\end{center}
\caption{\label{tab:ensemble}Specifics of the
ensemble used for the lattice QCD
calculation. An average of $\bar{N}_{\text{meas}}$ sources are measured on each of $N_{\text{cfgs}}$ configurations. For more information see Ref.~\cite{Orginos:2015aya}.}
\end{table*}

The gluon contributions to the GFFs, on the other hand, are far less well constrained and have so far only
been studied in an extended holographic light-front QCD framework for the pion and nucleon~\cite{deTeramond:2021lxc}, in lattice QCD calculations for the pion~\cite{Shanahan:2018pib}, nucleon \cite{Yang:2018bft,Alexandrou:2018ii,Yang:2018nqn,Shanahan:2018pib,Alexandrou:2017oeh,Alexandrou:2020sml}, and $\phi$ meson~\cite{Detmold:2017oqb}, in almost all cases at larger-than-physical values of the quark masses. 
While no experimental constraints on the gluon GFFs of any hadron have been achieved to date, the gluon GFFs of the nucleon are accessible via photo- or leptoproduction of $J/\psi$ and $\Upsilon$~
\cite{Mamo:2019mka,Hatta:2018ina,Boussarie:2020vmu}; $J/\psi$ production is studied in experiments that are ongoing at JLab~\cite{GlueX:2019mkq}, while $\Upsilon$ production studies are planned at the electron-ion collider (EIC)
\cite{AbdulKhalek:2021gbh}. Improved QCD constraints on the gluon GFFs of the nucleon and other hadrons are particularly valuable at the current time as they can inform the target kinematics for these experiments and provide theory predictions to test against future experimental results.

In this work, we present a lattice QCD calculation of gluon GFFs of the pion and nucleon at unphysically heavy quark masses, incorporating additional data corresponding to spin-nonconserving channels and an improved statistical analysis compared with the previous study of Ref.~\cite{Shanahan:2018pib}. We further undertake a first study
of the complete set of gluon GFFs of the $\rho$ meson
and $\Delta$ baryon, which are stable at these quark masses, to investigate the gluon radii and gluon energy,
pressure, and shear force
distributions of hadrons of higher spin. In Sec.~\ref{sec:gffs-from-lqcd} we outline the lattice QCD calculation and analysis, discussed more extensively in Appendix~\ref{sec:lattice-details}, and show the extracted
renormalized GFFs for all hadrons considered.
In Sec.~\ref{sec:densities}, we present our results for
the radii and densities in
different frames.
In Sec.~\ref{sec:conclusion}, we provide a summary and outlook.

\section{GRAVITATIONAL FORM FACTORS FROM LATTICE QCD}
\label{sec:gffs-from-lqcd}

In this section we discuss the decompositions of hadronic matrix elements of the
gluon EMT into gluon GFFs for the pion, nucleon, $\rho$ meson, and $\Delta$ baryon, and
present the results of our lattice QCD extraction of these quantities.
We use a single ensemble of 2820 configurations of lattice volume $32^3 \times 96$
with $N_f=2+1$ quark flavors, with a heavier-than-physical pion mass of ${m_\pi = 450(5)~\text{MeV}}$
and lattice spacing ${a = 0.1167(16)~\text{fm}}$~\cite{s_Meinel}.
The ensemble was generated using the L\"uscher-Weisz gauge action \cite{Luscher:1984xn} and clover-improved Wilson quarks \cite{Sheikholeslami:1985ij} with clover coefficient set to the tree-level tadpole-improved value and constructed using stout-smeared links \cite{Morningstar:2003gk}.
The specifics of the ensemble are summarized in Table \ref{tab:ensemble} \cite{Orginos:2015aya,s_stefan}.
Our results for the nucleon and pion GFFs are consistent with but more precise than those of Ref.~\cite{Shanahan:2018pib}, which studied those states on a subset of the data used in this work, including only spin-conserving channels.

Our methods are similar to those of Ref.~\cite{Shanahan:2018pib}, but with
an improved statistical analysis and with the necessary extensions to treat hadrons
of higher spin.
The calculation proceeds independently but analogously for each hadron, in several stages detailed in Appendix~\ref{sec:lattice-details}:
\begin{enumerate}[(i)]
    \item Compute (hadron-independent) measurements of the symmetric traceless gluon EMT,
    discretized and projected to irreducible representations 
    (irreps) of the hypercubic group that are protected from mixing with lower-dimensional operators (Appendix~\ref{sec:lattice-gluon-EMT}).
    \item Compute hadronic two-point correlation functions (Appendix~\ref{sec:two-point-fns}) and three-point correlation functions (Appendix~\ref{sec:three-pt-fns}) including insertions of the gluon EMT.
    \item Extract hadronic matrix elements of the gluon EMT by fitting ratios of three- and two-point correlation functions (Appendix~\ref{sec:coeffs-binning-and-ratio-fits}).
    \item Extract the renormalized gluon GFFs by fitting the constraints defined by the measured matrix elements and Eq.~\eqref{eqn:general-gff-decomp} (Appendixes~\ref{sec:constraint-fitting}, \ref{sec:dagostini}), incorporating the renormalization factors for the different irreps computed using a nonperturbative RI-MOM prescription and a one-loop perturbative matching to $\overline{\mathrm{MS}}$ at $\mu = 2~\mathrm{GeV}$. We neglect effects due to mixing with the quark GFFs under renormalization, 
    which are expected to be $\mathcal{O}(10\%)$ \cite{Alexandrou:2016ekb,Alexandrou:2020sml} and thus small compared with the statistical uncertainty of the calculation.
\end{enumerate}

\begin{figure*}[tp]
\centering
\subfloat[\centering  ]
{{\includegraphics[height=5.8cm,width=7.8cm]{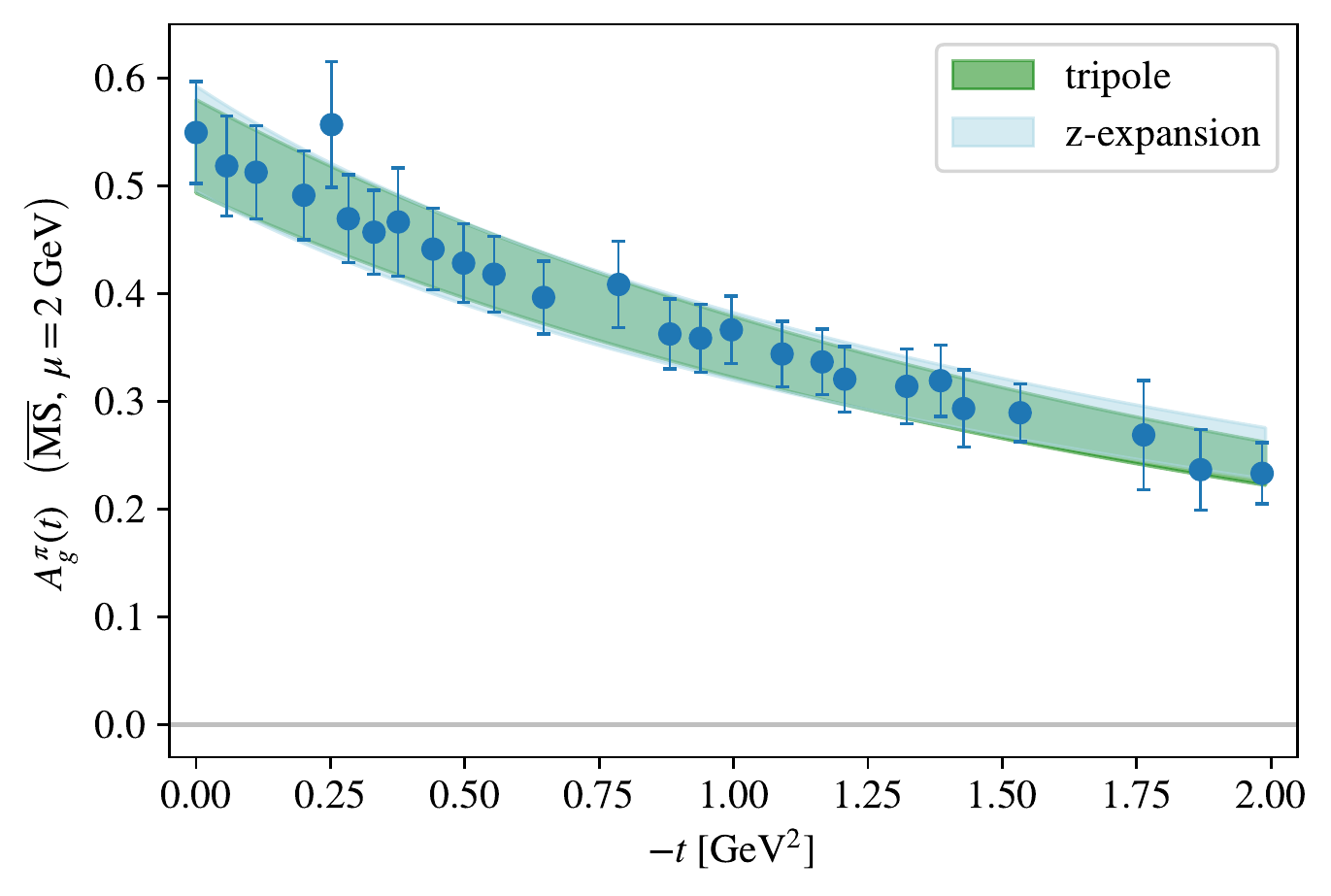} }}
\!
\subfloat[\centering  ]
{{\includegraphics[height=5.8cm,width=7.8cm]{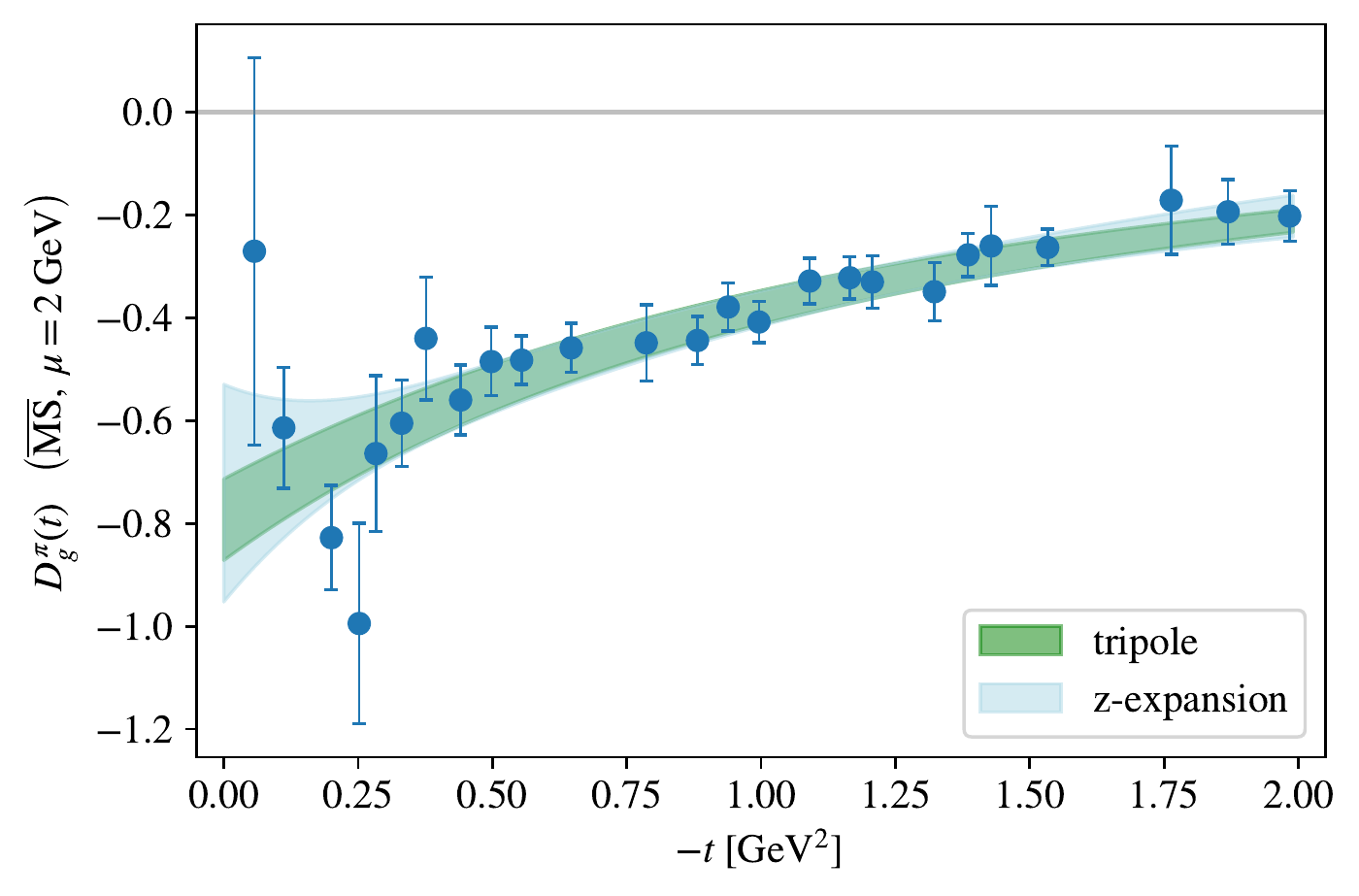} }}
\caption{$A^{\pi}_g(t)$ and $D^{\pi}_g(t)$ renormalized
at $\mu = 2~\text{GeV}$ in the $\overline{\text{MS}}$ scheme.
The bands correspond to the multipole form (Eq.~\eqref{eq:multipole}) with $n=3$ and the modified z-expansion (Eq.~\eqref{eq:z-expansion}) with $k_{\text{max}} = 2$, with fit parameters shown in Tab.~\ref{tab:pion}.
}
\label{fig:pionGFF}
\end{figure*}

Our lattice calculation yields the GFFs at a set of discrete values of the squared momentum transfer $t$ but, as discussed in Sec.~\ref{sec:densities}, subsequent extrapolations to the forward limit and derivations of densities and radii require  models of the $t$ dependence of the GFFs.
We consider two different ans{\"a}tze, a multipole and a modification of the z-parameter expansion or ``z-expansion'' \cite{Hill:2010yb,Shanahan:2018nnv}.
The multipole form is defined as
\begin{equation}\label{eq:multipole}
\text{G}_n(t) = \frac{\alpha}{(1-t/\Lambda^2)^n},
\end{equation}
where $\alpha$ and $\Lambda$ are free parameters and we set $n=3$ (tripole) in order for all the integrals that define the energy, pressure, and shear force densities discussed in Sec.~\ref{sec:densities} to converge.
As introduced in Ref.~\cite{Shanahan:2018nnv}, the modified z-expansion we consider is
\begin{equation} \label{eq:z-expansion}
\text{G}_{z,n}(t) = \frac{1}{(1-t/\Lambda^2)^n}\sum_{k=0}^{k_{\text{max}}}
\alpha_k[z(t)]^k \;,
\end{equation}
where
\begin{equation}
z(t) = \frac{\sqrt{t_{\text{cut}}-t}-\sqrt{t_{\text{cut}}-t_0}}{\sqrt{t_{\text{cut}}
-t}+\sqrt{t_{\text{cut}}-t_0}},
\end{equation}
$\alpha_k$ are free parameters, and $\Lambda$ is constrained as described below.
This functional form can be interpreted as a series of corrections to the envelope defined by the multipole form, which coincides with Eq.~\eqref{eq:z-expansion} when $k_{\text{max}} = 0$.
The multipole envelope is necessary for convergence of the density integrals discussed in Sec.~\ref{sec:densities}.
Following Ref.~\cite{Shanahan:2018nnv}, we set $k_{\text{max}} = 2$, 
$t_0 = t_{\text{cut}}( 1 - \sqrt{1 + (2\; \text{GeV})^2 /t_{\text{cut}}} )$, and 
$t_{\text{cut}} = 4 m_{\pi}^2$, using $m_\pi = 450~\mathrm{MeV}$.
For each GFF, we use $\Lambda$ obtained from the multipole fit to the same GFF as a prior for the parameter $\Lambda$ in the z-expansion, retaining correlations between the prior and the data,\footnote{Compare with the discussion of ``chained fitting'' in Ref.~\cite{Bouchard:2014ypa}.}
thereby reducing the number of free parameters to prevent overfitting and explicitly enforcing the notion of the modified z-expansion as a correction to the multipole envelope.
As detailed in Appendix~\ref{sec:lattice-details}, we fit the models to bare GFFs and renormalize afterwards to circumvent the d'Agostini bias~\cite{DAgostini:1993arp}; in this section we present the resulting renormalized parameters $\alpha$ and $\alpha_k$.

\subsection{Pion}

The pion matrix element of the symmetric gluon or quark contribution to the energy-momentum tensor can be decomposed as
\begin{equation} \label{eq:pionME}
\begin{aligned}
\bra{\pi(p')} T_i^{\mu\nu} \ket{\pi(p)} 
&= 2P^{\mu}P^{\nu}A^{\pi}_i(t) + 
2m_{\pi}^2\bar{c}^{\pi}_i(t)g^{\mu\nu} & \\
& ~~ +\frac{1}{2}
(\Delta^{\mu}\Delta^{\nu}-g^{\mu\nu}\Delta^2)D^{\pi}_i(t)
\\ &\equiv 
\mathcal{O}^{\mu\nu}_{(\pi)}[A^{\pi}_i(t),D^{\pi}_i(t)] +
\text{trace} \;,
\end{aligned}
\end{equation}
where $i\in\{q,g\}$,
$g_{\mu\nu}$ is the Minkowski
space-time metric, and we have defined the traceless piece $\mathcal{O}^{\mu\nu}_{(\pi)}$ for later convenience.
$A^{\pi}_i(0)$ is the traceless contribution to the
momentum fraction carried by the quarks or gluons and must satisfy
$A^{\pi}(0) = \sum_i A^{\pi}_i(0) = 1$ because of Poincar\'e symmetry. 
$D^{\pi}_i(t)$ is related to the mechanical properties of the pion. 
In the forward and chiral limits, the total
$D^{\pi}(0)$, also called the {$D$-term} or Druck term, is predicted to be
$-1$ up to chiral-symmetry breaking effects \cite{Hudson:2017xug,Polyakov:1999gs,Donoghue:1991qv}. 
$\bar{c}^{\pi}_i(t)$ appears due
to the nonconservation of the separate quark and gluon contributions and
vanishes for the total EMT, i.e.~${\bar{c}^{\pi}(t) = \bar{c}^{\pi}_g(t) + \sum_q \bar{c}^{\pi}_q(t) = 0}$.

\begin{table}[t]
\begin{center}
\begin{tabular}{SD{:}{}{2.7}D{:}{}{2.7}D{:}{}{2.7}D{:}{}{2.7}}
\toprule
\multicolumn{1}{c}{tripole} & \multicolumn{1}{c}{$\alpha$} & \multicolumn{1}{c}{$\Lambda$~[GeV]} && \multicolumn{1}{c}{$\chi^2/\text{d.o.f.}$} \\ \midrule
{$A_g^{\pi}(t)$} & 0:.537(45) & 2:.561(43)  && 0:.9 \\[2pt]
{$D_g^{\pi}(t)$} &  -0:.793(84) & 1:.90(11) && 1:.3 \\
\midrule\midrule
\multicolumn{1}{c}{z-expansion} & \multicolumn{1}{c}{$\alpha_0$} & \multicolumn{1}{c}{$\alpha_1$} & \multicolumn{1}{c}{$\alpha_2$} & \multicolumn{1}{c}{$\chi^2/$d.o.f.}\\ \midrule
{$A_g^{\pi}(t)$} & 0:.540(45) & 0:.14(10) & 0:.70(52) & 0:.8\\[2pt]
{$D_g^{\pi}(t)$} & -0:.793(70) & 0:.21(75) & 2:.0(7.3) & 1:.1\\
\bottomrule
\end{tabular}
\end{center}
\caption{\label{tab:pion}
Fit parameters of the multipole [Eq.~\eqref{eq:multipole}] with $n=3$
and the modified z-expansion [Eq.~\eqref{eq:z-expansion}] with $k_{\text{max}} = 2$ and $n=3$ models for the $t$ dependence of the renormalized pion GFFs. For all GFFs, the parameter $\Lambda$ of the z-expansion fit is consistent with the prior provided by the tripole fit and is thus not shown.
The parameters $\alpha$ and $\alpha_k$ are renormalized at $\mu = 2\;\text{GeV}$ after fitting the bare GFFs as described in Appendix~\ref{sec:constraint-fitting}.
}
\end{table}

\begin{table}[t] 
\begin{center}
\begin{tabular}{S@{\hskip 0.2in}D{:}{}{2.7}@{\hskip 0.2in}D{:}{}{2.7}}
\toprule
& \multicolumn{1}{c}{tripole} & \multicolumn{1}{c}{z-expansion}  \\ \midrule
{$A_g^{\pi}(0)$} & 0:.537(45) & 0:.544(46)  \\[2pt]
{$D_g^{\pi}(0)$} & -0:.793(84) & -0:.74(21) \\
\bottomrule
\end{tabular}
\end{center}
\caption{\label{tab:pionQuantities}The forward-limit values of the
gluon momentum fraction and the gluon $D$-term, obtained from
the tripole and modified z-expansion fits to the pion GFFs, renormalized at $\mu = 2\;\text{GeV}$ in the $\overline{\text{MS}}$ scheme, 
with parameters shown in Table~\ref{tab:pion}.
}
\end{table}

Our results for the two renormalized traceless gluon GFFs of the pion, $A^{\pi}_g(t)$
and $D^{\pi}_g(t)$, are shown in Fig.~\ref{fig:pionGFF}.
The fit parameters for the two ans{\"a}tze, Eqs.~\eqref{eq:multipole} and \eqref{eq:z-expansion}, are shown in Table~\ref{tab:pion}, and
the predicted forward-limit gluon momentum fraction and $D$-term are shown
in Table~\ref{tab:pionQuantities}.
We note that the sum of our gluon $D$-term with the
value
$D_{u+d}^{\pi}(0) = -0.264(32)$ (extrapolated to the physical pion mass) from Ref.~\cite{Brommel:2007zz}
is statistically consistent
with the chiral prediction, although this may be a coincidence that does not survive a chiral and continuum limit extrapolation.

\subsection{Nucleon}
\label{sec:nucGFFs}

For the nucleon, the GFFs of the symmetric gluon or quark parts of the EMT are defined by
\begin{align}\label{eq:protonME}
\bra{N(p',\sigma')} T_i^{\mu\nu} \ket{N(p,\sigma)} &= \nonumber \\
\bar{u}(\vec{p}',\sigma')
\bigg[ \gamma_{\{\mu}P_{\nu\}}A^N_i(t)
+& \frac{iP_{\{\mu}\sigma_{\nu\}\rho}\Delta^{\rho}}{2m_N}
B^N_i(t)   \nonumber \\
 + m_N g_{\mu\nu}\bar{c}^N_i(t)
 +& \frac{\Delta_{\mu}
\Delta_{\nu} - g_{\mu\nu}\Delta^2}{4m_N}D_i^N(t) \bigg] u(\vec{p},\sigma) \nonumber\\
 \equiv \bar{u}(\vec{p}',\sigma')\mathcal{O}^{\mu\nu}_{(N)}[&A^N_i(t),...]
u(\vec{p},\sigma)
+\text{trace} \; ,
\end{align} 
where $m_N$ denotes the nucleon mass, $\sigma_{\mu\nu} = \frac{i}{2}[\gamma_{\mu},\gamma_{\nu}]$, 
and $u(\vec{p},\sigma)$ is the Dirac spinor, which satisfies
\begin{equation}
    \sum_{\sigma} u(\vec{p},\sigma) \bar{u}(\vec{p}, \sigma) = \cancel{p} + m_N\;,
\end{equation}
where $\sigma \in \{-1/2, +1/2\}$.
Equation~\ref{eq:protonME} is often expressed in
terms of the form factor $J^N_i(t) = (A^N_i(t) + B^N_i(t))/2$ instead
of $B^N_i(t)$, where the total $J^N(0) = 1/2$ is the spin of the
nucleon.
As for the total GFFs for spin-0 states, $A^N(0) = 1$ and $\bar{c}^N(t) = 0$ for the GFFs of spin-1/2 states. The new form factor $B^N(t)$ that appears for hadrons of spin $>0$ must
obey $B^N(0) = 0$ due to the vanishing of the anomalous gravitomagnetic
moment of spin-1/2 fermions~\cite{Kobzarev:1962wt,Pagels:1966zza,Ji:1996ek,Teryaev:1999su,Brodsky:2000ii,Silenko:2006er,Teryaev2016,
Lowdon:2017idv,Polyakov:2018zvc}.
The are no \textit{a priori} restrictions on the $D$-term of
a spin-1/2 hadron, but for a free
fermion $D^N(0) = 0$ \cite{Hudson:2017oul}.

\begin{figure*}[bt]
\captionsetup[subfloat]{captionskip=-3pt}
\centering
\subfloat[\centering  ]
{{\includegraphics[height=5.6cm,width=7.8cm]{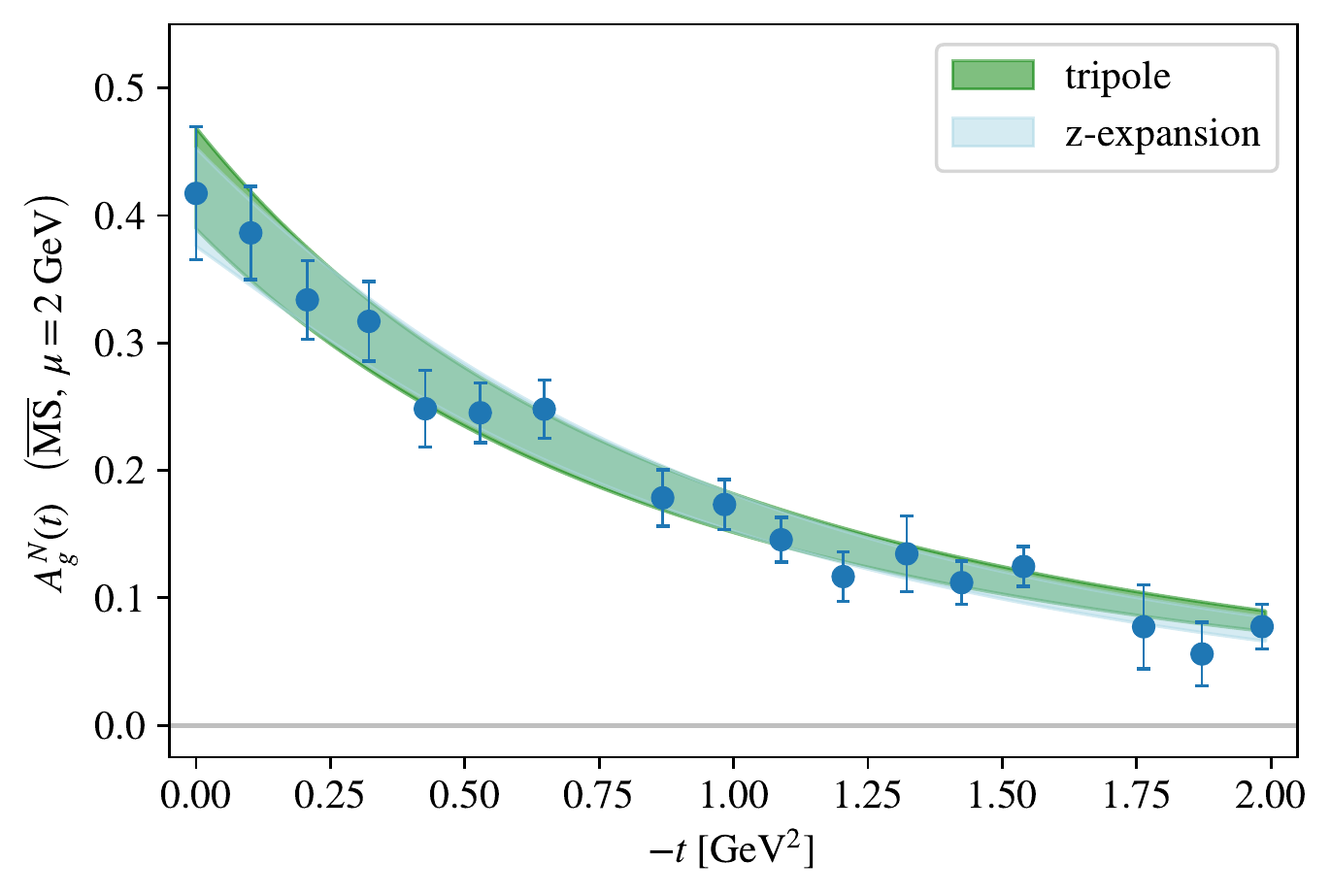} }}
\!
\subfloat[\centering  ]
{{\includegraphics[height=5.6cm,width=7.8cm]{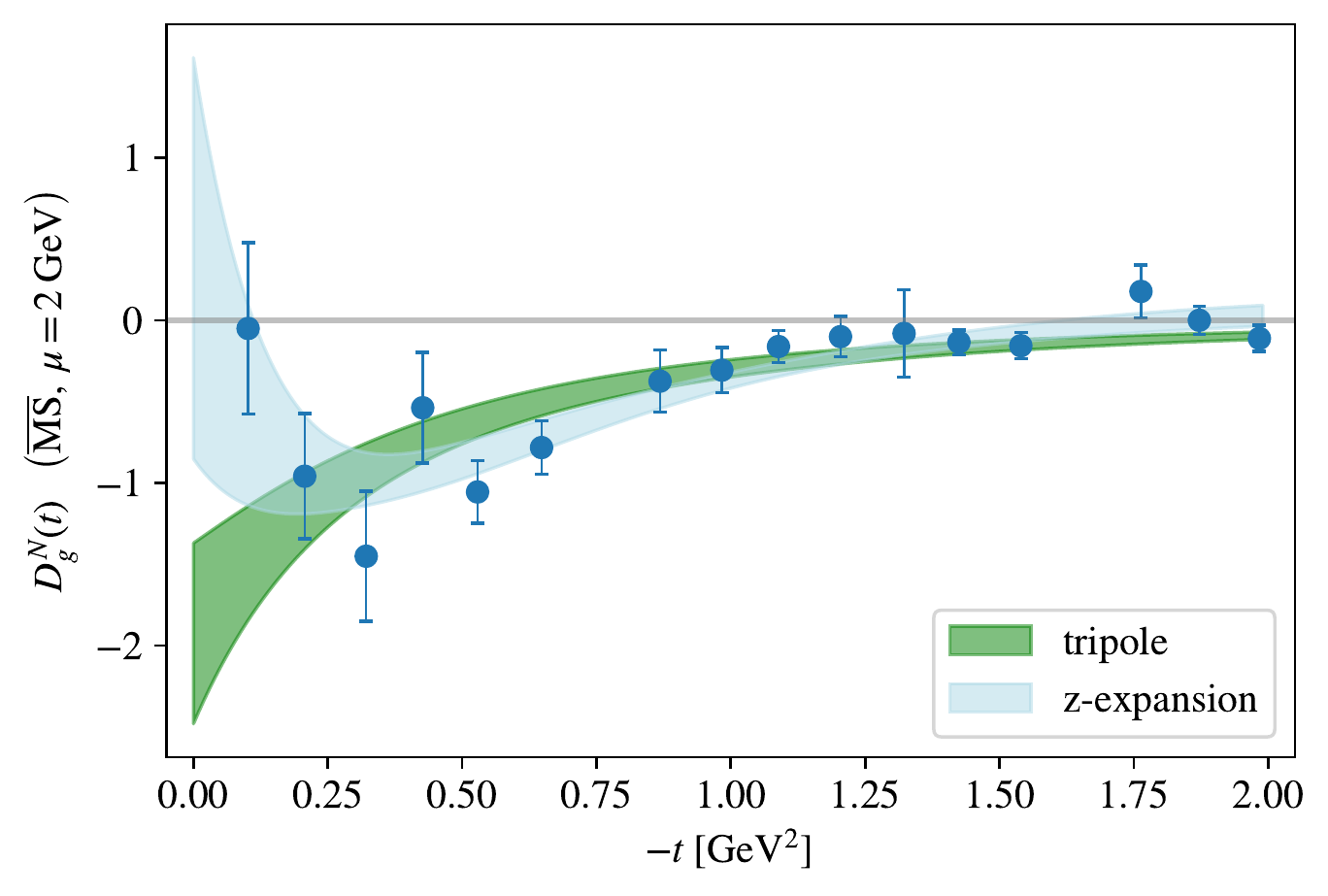} }}
\\[-0.5ex]
\subfloat[\centering  ]
{{\includegraphics[height=5.6cm,width=7.8cm]{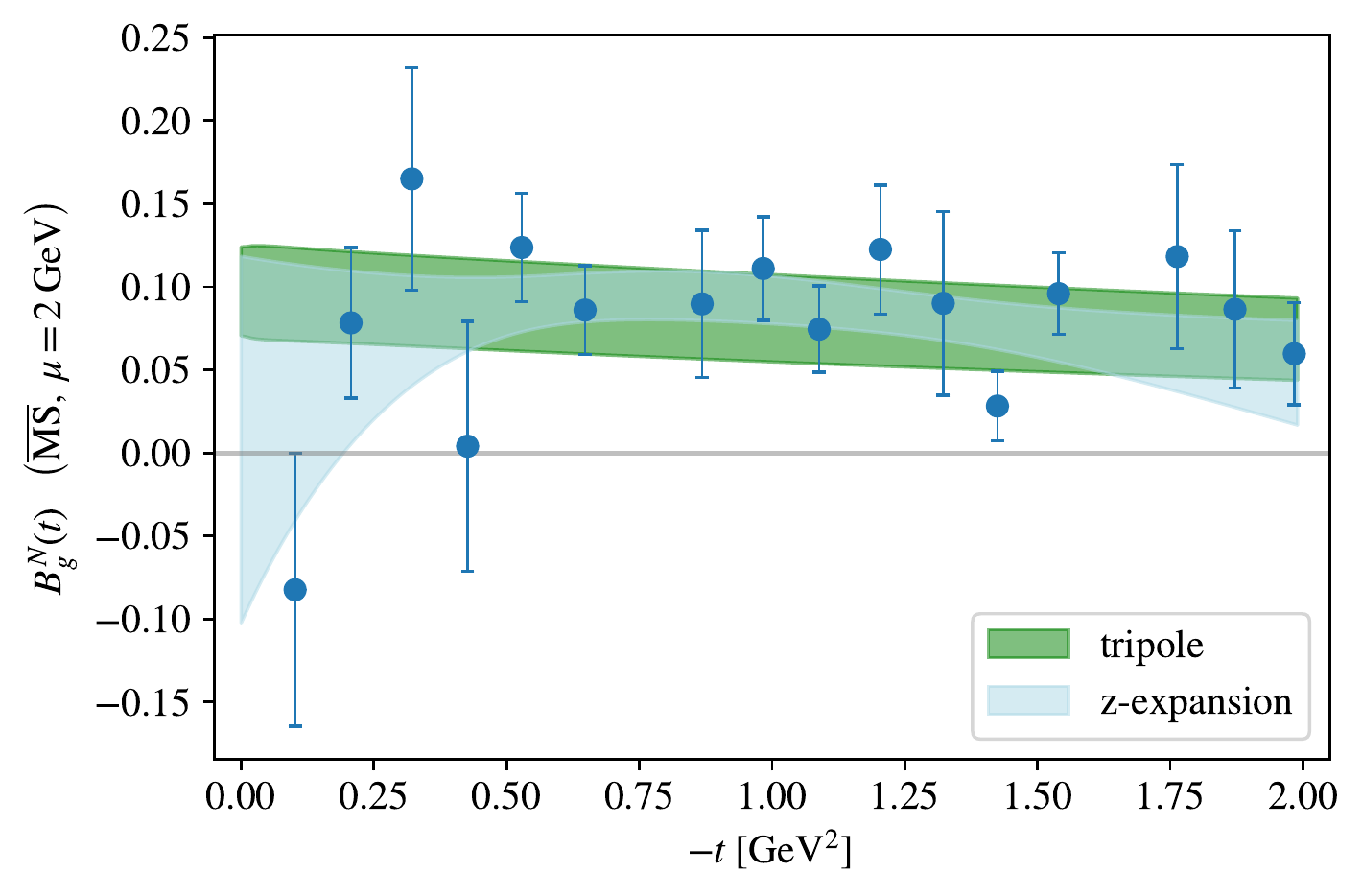} }}
\caption{$A^{N}_g(t)$ (a), $D^{N}_g(t)$ (b), and $B^{N}_g(t)$ (c), renormalized
at $\mu = 2~\text{GeV}$ in the $\overline{\text{MS}}$ scheme.
The bands correspond to the multipole form [Eq.~\eqref{eq:multipole}] with $n=3$ and the modified z-expansion [Eq.~\eqref{eq:z-expansion}] with $k_{\text{max}} = 2$, with fit parameters shown in Table~\ref{tab:nuc}.}
\label{fig:nucGFF}
\end{figure*}

\begin{table}[tp]
\begin{center}
\begin{tabular}{SD{:}{}{2.7}@{\hskip 0.1in}D{:}{}{2.7}@{\hskip 0.1in}D{:}{}{2.7}D{:}{}{2.7}}
\toprule
\multicolumn{1}{c}{tripole}  & \multicolumn{1}{c}{$\alpha$} & \multicolumn{1}{c}{$\Lambda$~[GeV]} &&\multicolumn{1}{c}{$\chi^2/$d.o.f.}  \\ \midrule
{$A_g^{N}(t)$} & 0:.429(39) & 1:.641(43) && 1:.1 \\[2pt]
{$B_g^{N}(t)$} & 0:.097(26) & 4:.0(2.5) && 1:.4 \\[2pt]
{$D_g^{N}(t)$} & -1:.93(53) & 1:.07(12) && 2:.0 \\
\midrule \midrule
\multicolumn{1}{c}{z-expansion}  & \multicolumn{1}{c}{$\alpha_0$} & \multicolumn{1}{c}{$\alpha_1$} & \multicolumn{1}{c}{$\alpha_2$} & \multicolumn{1}{c}{$\chi^2/$d.o.f.} \\ \midrule
{$A_g^{N}(t)$} & 0:.426(38) & -0:.22(22) & -1:.2(1.2)  & 0:.9\\[2pt]
{$B_g^{N}(t)$} & 0:.108(16) & -0:.19(20) & -2:.9(3.2) & 1:.1\\[2pt]
{$D_g^{N}(t)$} & -2:.09(35) & 18:.0(7.9) & 132:(58) & 1:.0\\ \bottomrule
\end{tabular}
\end{center}
\caption{
Fit parameters of the multipole [Eq.~\eqref{eq:multipole}] with $n=3$
and the modified z-expansion [Eq.~\eqref{eq:z-expansion}] with $k_{\text{max}} = 2$ and $n=3$ models for the $t$ dependence of the renormalized nucleon GFFs. 
For all GFFs, the parameter $\Lambda$ of the z-expansion fit is consistent with the prior provided by the tripole fit and is thus not shown.
The parameters $\alpha$ and $\alpha_k$ are renormalized at $\mu = 2\;\text{GeV}$ after fitting the bare GFFs as described in Appendix~\ref{sec:constraint-fitting}.
}
\label{tab:nuc}
\end{table}

\begin{table}[!t]
\begin{center}
\begin{tabular}{S@{\hskip 0.2in}D{:}{}{2.7}@{\hskip 0.2in}D{:}{}{2.7}}
\toprule
& \multicolumn{1}{c}{tripole} & \multicolumn{1}{c}{z-expansion} \\ \midrule
{$A_g^{N}(0)$} & 0:.429(39) & 0:.414(40)\\[2pt]
{$D_g^{N}(0)$} & -1:.93(53) & 0:.4(1.2)\\[2pt]
{$J_g^{N}(0)$} & 0:.263(26) & 0:.211(57)\\ \bottomrule
\end{tabular}
\end{center}
\caption{The forward-limit values of the
gluon contributions to the momentum fraction, $D$-term, and 
angular momentum of the nucleon, obtained from
the tripole and modified z-expansion fits of the nucleon GFFs renormalized at ${\mu=2\;\text{GeV}}$ in the $\overline{\text{MS}}$ scheme,
with parameters shown in Table~\ref{tab:nuc}.}
\label{tab:nucQuantities}
\end{table}

Our results for the three renormalized traceless gluon GFFs of the nucleon are shown in Fig.~\ref{fig:nucGFF}, the tripole and modified z-expansion fit parameters are shown in Table~\ref{tab:nuc}, and the predictions for the forward-limit gluon contributions to the momentum fraction, $D$-term, and angular momentum are shown in  Table~\ref{tab:nucQuantities}. Note that the increase order-by-order of the parameters in the modified z-expansion fit for the $D$-term are not of concern, as we consider here a modification of the standard z-expansion for which the guarantees of convergence for the standard form do not apply.

These results are based on a superset of the data presented in Ref.~\cite{Shanahan:2018pib}, including a larger number of sources per configuration and all four nucleon polarization channels, rather than only the two spin-conserving ones. The additional data allows a nonzero functional fit for $B_g^N(t)$ to be resolved. Additionally, the behavior of $A_g^N(t)$ and $D_g^N(t)$ in the lower half of the $-t$ region studied is shifted slightly compared to what was found in Ref.~\cite{Shanahan:2018pib}. These updated results show a qualitative difference between the low-$|t|$ behavior of the model obtained for $D^N_g(t)$ with the tripole functional form (that is by definition monotonic), and the modified z-expansion fit (that is allowed to be nonmonotonic). In general, monotonically increasing behavior is universally expected for the total $D$-term of any stable mechanical system  \cite{Gegelia:2021wnj}; however, no such prediction exists for the individual quark and gluon contributions. It is not clear whether the suppression of the lowest-$|t|$ point of $D^N_g(t)$ is physical or due to a statistical fluctuation or unquantified systematic uncertainty. If physical, it would imply a qualitatively different $t$ dependence of the gluon GFFs compared with that typically assumed for the quark GFFs, and suggest that the multipole functional form often used by default for these quantities is not a good model at low $|t|$. If the first data point is considered an outlier and excluded from the z-expansion fit, the resulting error bands do not exclude nonmonotonicity but encompass both monotonic and nonmonotonic forms.

Our predictions for $A_g^N(0)$ and $J_g^N(0)$ are statistically consistent with the equivalent quantities
found in a lattice QCD calculation
at quark masses corresponding to the physical value of the pion mass \cite{Alexandrou:2020sml}.

\subsection{$\rho$ Meson}
\label{sec:rhoGFFs}

Following the conventions of Ref.~\cite{Polyakov:2019lbq}, the matrix elements of the gluon or quark contribution to the symmetric EMT 
for the $\rho$ meson can be decomposed as
\begin{widetext}
\begin{equation}\label{eq:rhoME}
\scalebox{0.95}{$
\begin{split}
\bra{\rho(p',\lambda')} T_i^{\mu\nu} \ket{\rho(p,\lambda)}
&=\epsilon_{\alpha'}^*(\vec{p}',\lambda')
\epsilon_{\alpha}(\vec{p},\lambda)
\bigg[
2P^{\mu}P^{\nu}
\left(-g^{\alpha\alpha'}
A^{\rho}_{0,i}(t)
 + \frac{P^{\alpha}P^{\alpha'}}{m_{\rho}^2}A^{\rho}_{1,i}(t)\right) \\ 
&\quad + \frac{1}{2}(\Delta^{\mu}\Delta^{\nu}-g^{\mu\nu}\Delta^2)
\left(g^{\alpha\alpha'}D^{\rho}_{0,i}(t)+
\frac{P^{\alpha}P^{\alpha'}}
{m_{\rho}^2}D^{\rho}_{1,i}(t)\right)  
+4[P^{\{\mu}g^{\nu\}\alpha'}P^{\alpha}+P^{\{\mu}g^{\nu\}\alpha}
P^{\alpha'}]J^{\rho}_i(t) \\
&\quad +[g^{\alpha\{\mu}g^{\nu\}\alpha'}\Delta^2 - 2g^{\alpha'\{\mu}\Delta^{\nu\}}P^{\alpha}
+2g^{\alpha\{\mu}\Delta^{\nu\}}P^{\alpha'}
-4g^{\mu\nu}P^{\alpha}P^{\alpha'}]E_i^{\rho}(t)  \\
&\quad +[2g^{\alpha\{\mu}g^{\nu\}\alpha'}-\frac{1}{2}g^{\alpha\alpha'}
g^{\mu\nu}]m_{\rho}^2 \bar{f}_i^{\rho}(t)
+g^{\mu\nu}[g^{\alpha\alpha'}m_{\rho}^2\bar{c}_i^{0,\rho}(t)+P^{\alpha}P^{\alpha'}
\bar{c}_i^{1,\rho}(t)] 
\bigg] \\
&\equiv \epsilon_{\alpha'}^*(\vec{p}',\lambda')
\epsilon_{\alpha}(\vec{p},\lambda)
\mathcal{O}^{\alpha\mu\nu\alpha'}_{(\rho)}[A_{0,i}^{\rho}(t),
A_{1,i}^{\rho}(t),D_{0,i}^{\rho}(t),D_{1,i}^{\rho}(t),
J_i^{\rho}(t),E_i^{\rho}(t),\bar{f}_i^{\rho}(t)] + \text{trace},
\end{split}
$}
\end{equation}
\end{widetext}
where $m_{\rho}$ denotes the mass of the $\rho$ meson, and $\epsilon_{\alpha}(\vec{p},\lambda)$ is the 
polarization 4-vector for a massive spin-1 particle, which 
satisfies
\begin{equation}
\sum_{\lambda} \epsilon_{\alpha}(\vec{p},\lambda)
\epsilon_{\beta}(\vec{p},\lambda)=
-g_{\alpha\beta} + \frac{p_{\alpha}p_{\beta}}{m_{\rho}^2}
\equiv \Lambda_{\alpha \beta}^{(\rho)}(\vec{p})
\label{eqn:lambda-rho-def}
\end{equation}
% \end{widetext}
with $\lambda \in \{1,0,-1\}$. Note that the subscript $\rho$ in 
$m_{\rho}$ and the superscript $\rho$ in the GFFs such as $A_{0,i}^{\rho}(t)$ is a label for the $\rho$ meson and not a Lorentz
index.

The momentum sum rule constrains
the total momentum fraction to be $A^{\rho}(0) \equiv A_0^{\rho}(0) = 1$, and 
the forward-limit angular momentum must be equal
to the spin of the hadron, i.e.,
$J^{\rho}(0) = 1$.
The interpretation of the $D$-term of
the $\rho$ meson
is more complicated than those of the
pion and the nucleon, since there are
three such terms \cite{Polyakov:2019lbq,Sun:2020wfo}, one of monopole and two of quadrupole order, corresponding to
frame-dependent linear
combinations of $D_0^{\rho}(0)$, $D_1^{\rho}(0)$, and $E^{\rho}(0)$.
Here we focus on the forward
limit of the form factor that is the coefficient of the same Lorentz structure corresponding to the $D$-term GFFs
of the nucleon
and pion, namely $D^{\rho}(0) \equiv -D_0^{\rho}(0)$,
which is unconstrained from theory.
There are two GFFs that arise from the trace of the EMT, $\bar{c}_i^{1,\rho}(t)$ and $\bar{c}_i^{2,\rho}(t)$, and their contribution to the total EMT must be equal to zero.
In contrast to the pion and the nucleon GFF decompositions, the traceless piece of the EMT matrix element for
hadrons of spin-1
gives rise to a nonconserved GFF, $\bar{f}_i^{\rho}(t)$, which we can access in our
calculation and vanishes
when summed over the quark and gluon contributions.

\begin{figure*}[!p]
\captionsetup[subfloat]{captionskip=-3pt}
\centering
\subfloat[\centering  ]
{{\includegraphics[height=5.25cm,width=7.8cm]{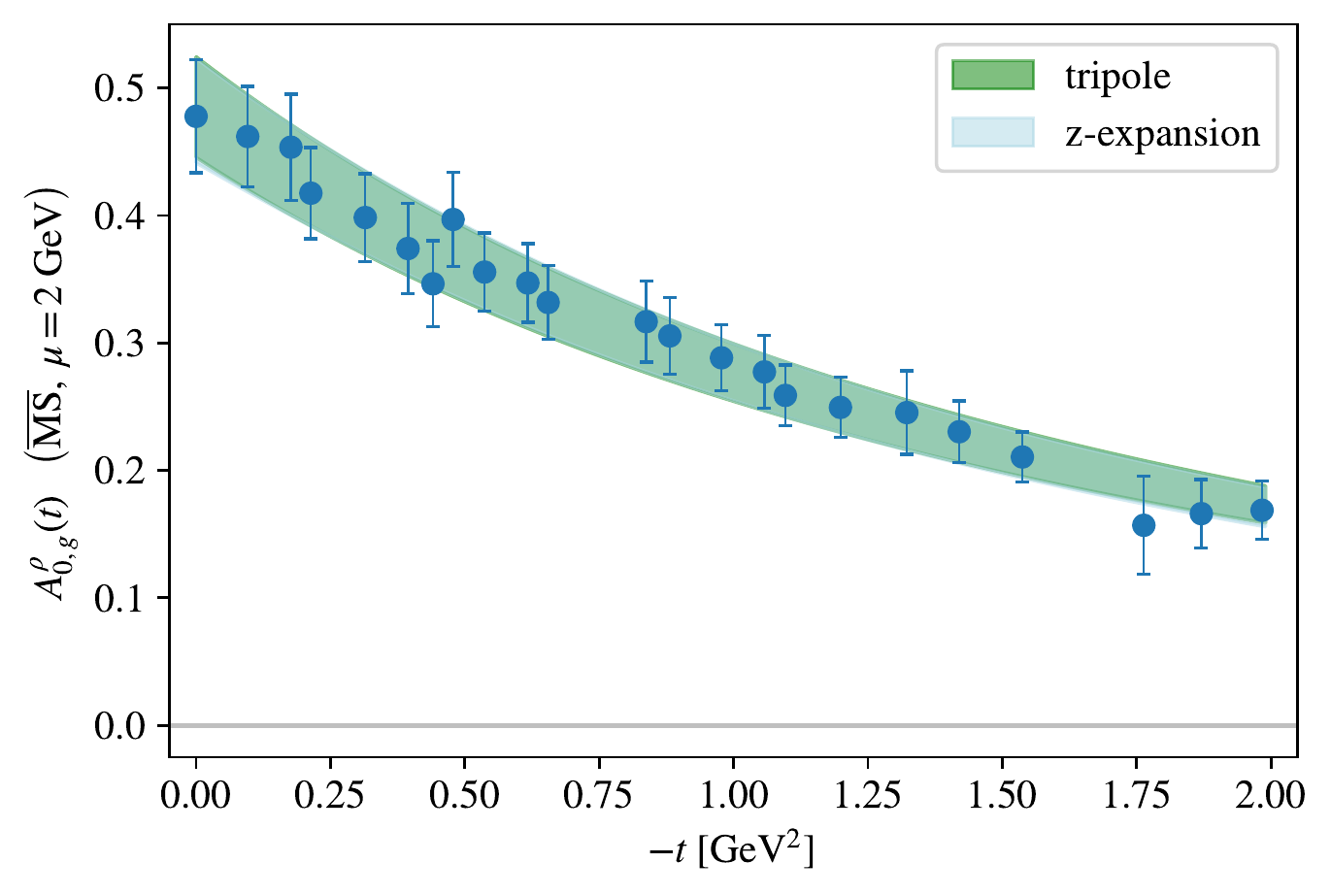} }}
\!
\subfloat[\centering  ]
{{\includegraphics[height=5.25cm,width=7.8cm]{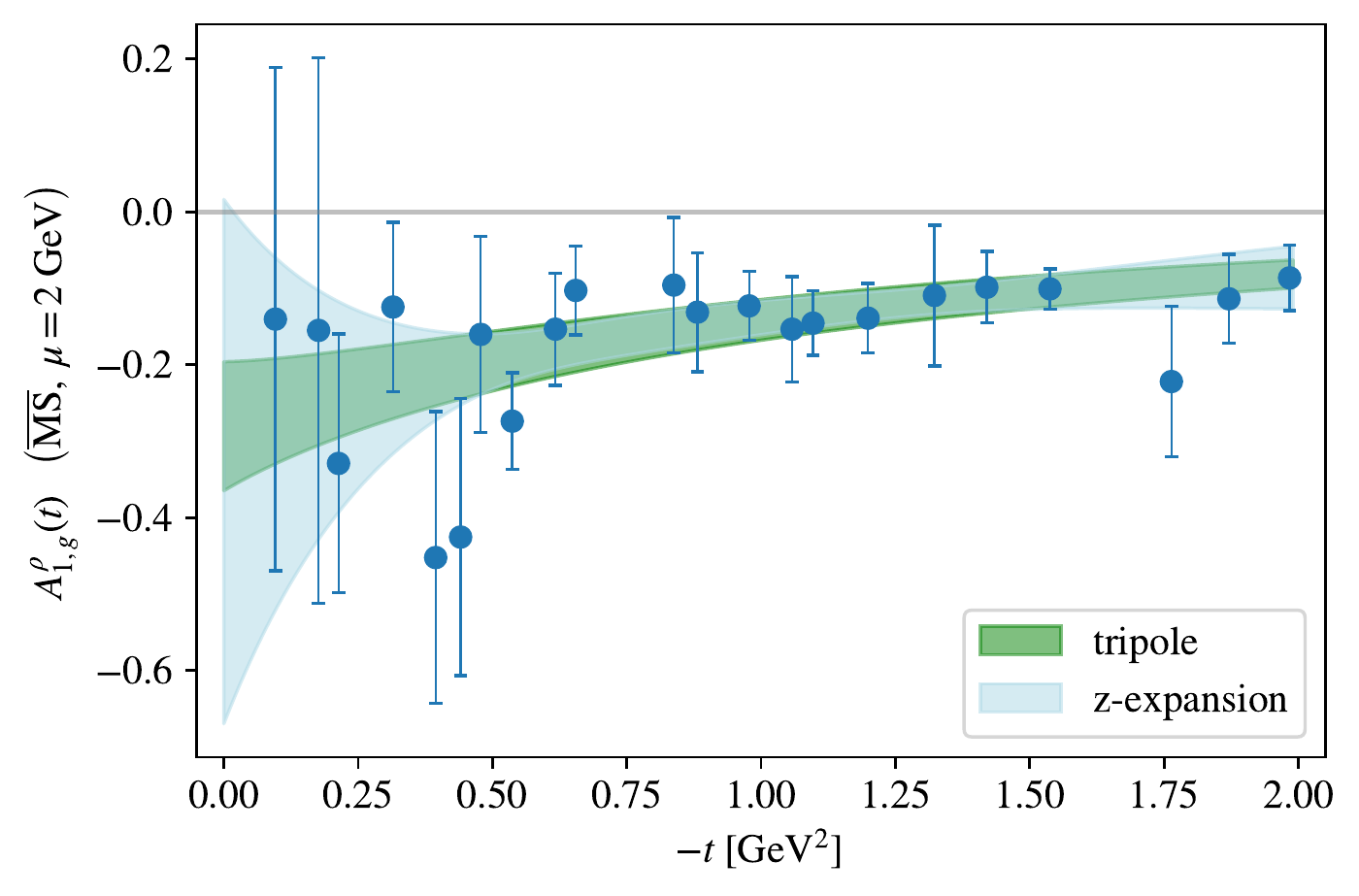} }}
\\[-3ex]
\subfloat[\centering  ]
{{\includegraphics[height=5.25cm,width=7.8cm]{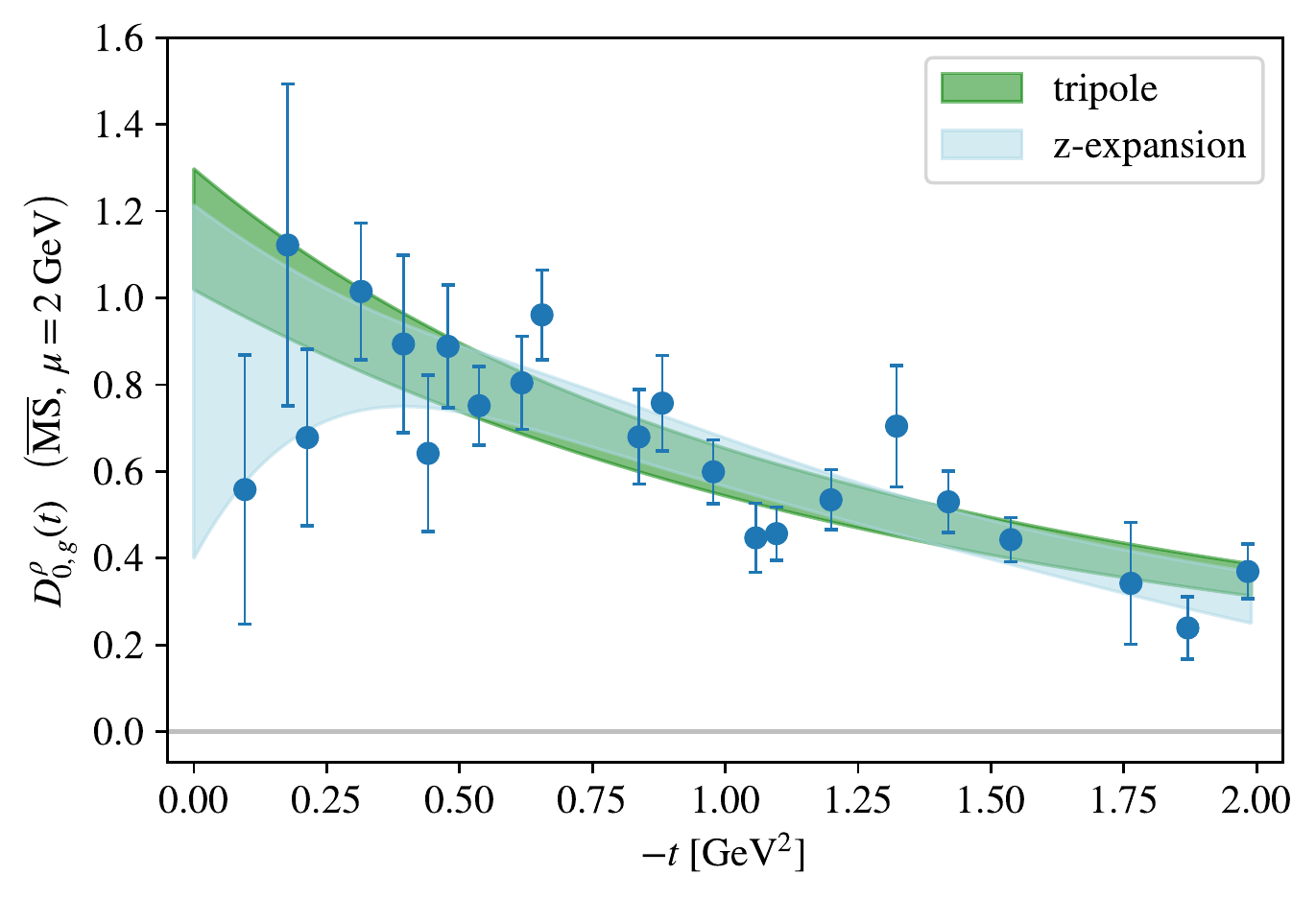} }}
\!
\subfloat[\centering  ]
{{\includegraphics[height=5.25cm,width=7.8cm]{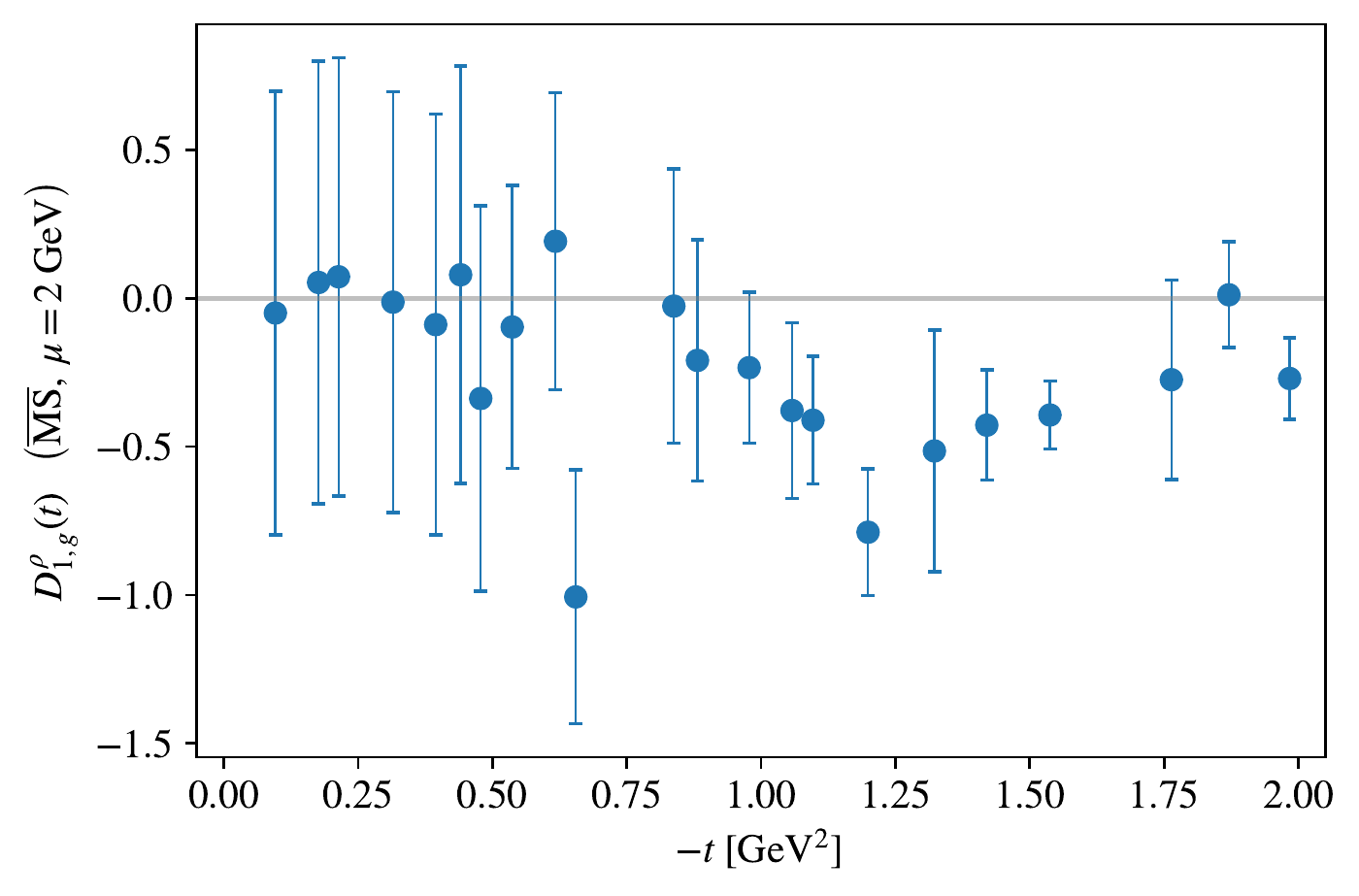} }}
\\[-3ex]
\subfloat[\centering  ]
{{\includegraphics[height=5.25cm,width=7.8cm]{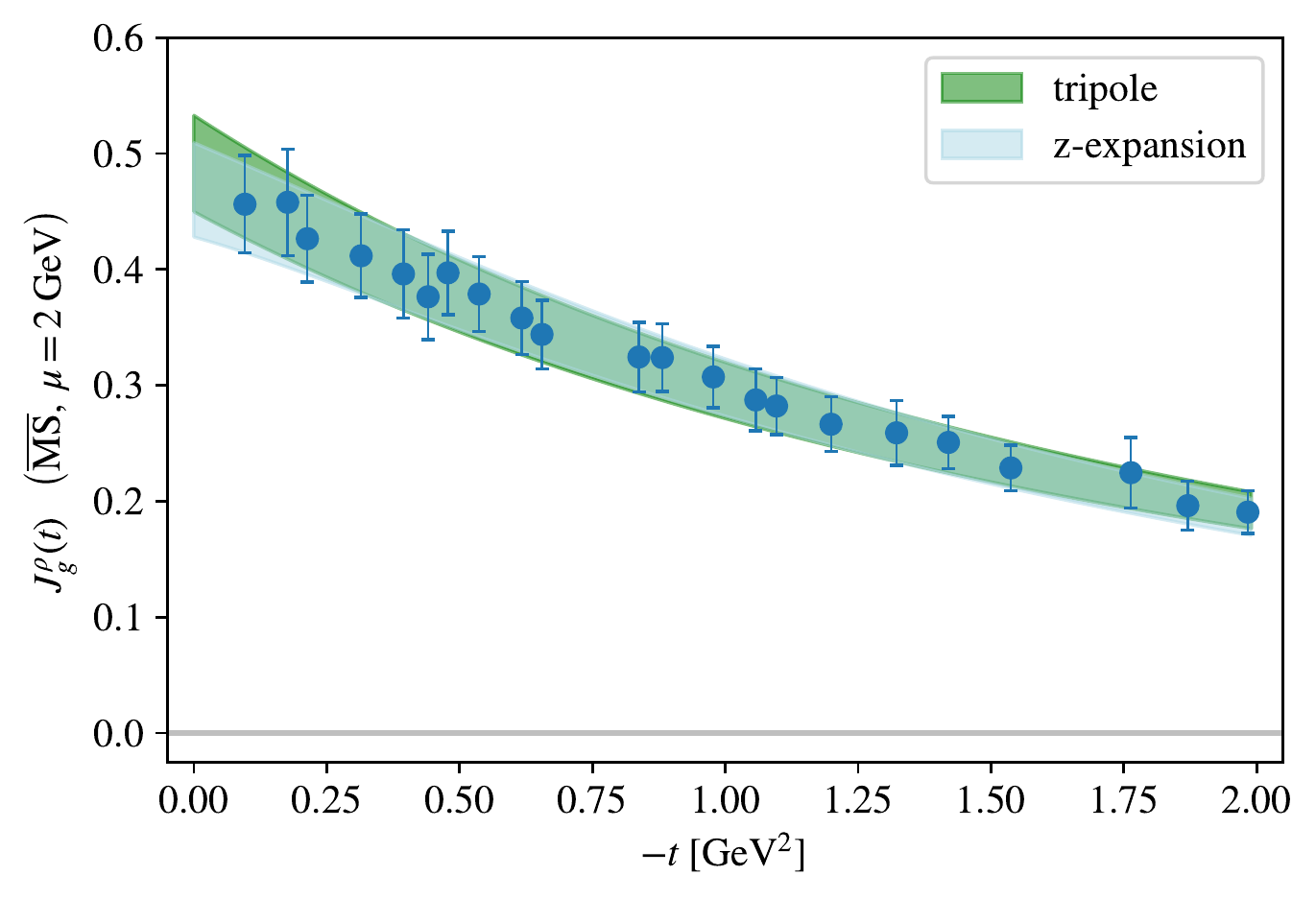} }}
\!
\subfloat[\centering  ]
{{\includegraphics[height=5.25cm,width=7.8cm]{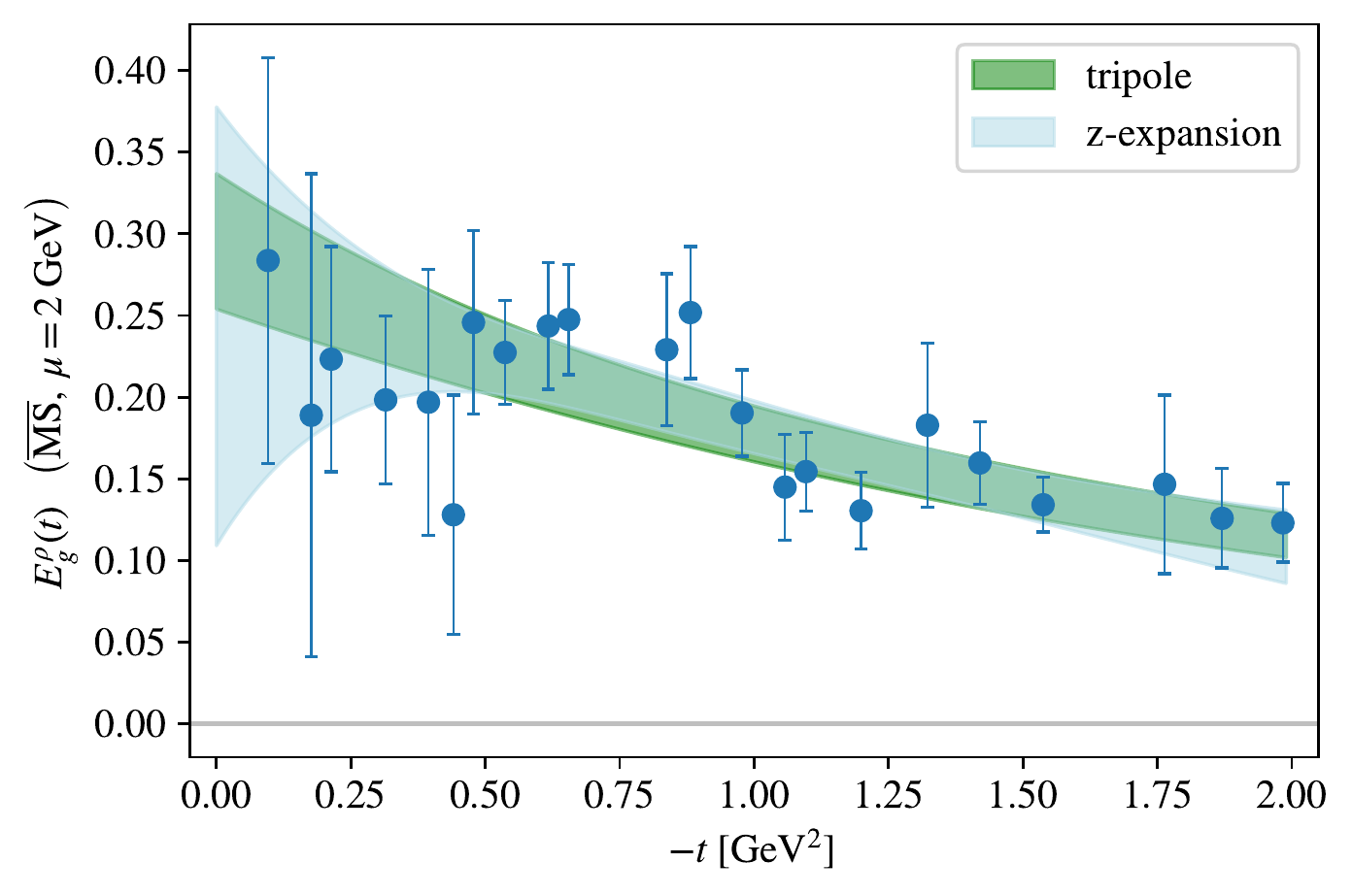} }}
\\[-0.5ex]
\subfloat[\centering  ]
{{\includegraphics[height=5.25cm,width=7.8cm]{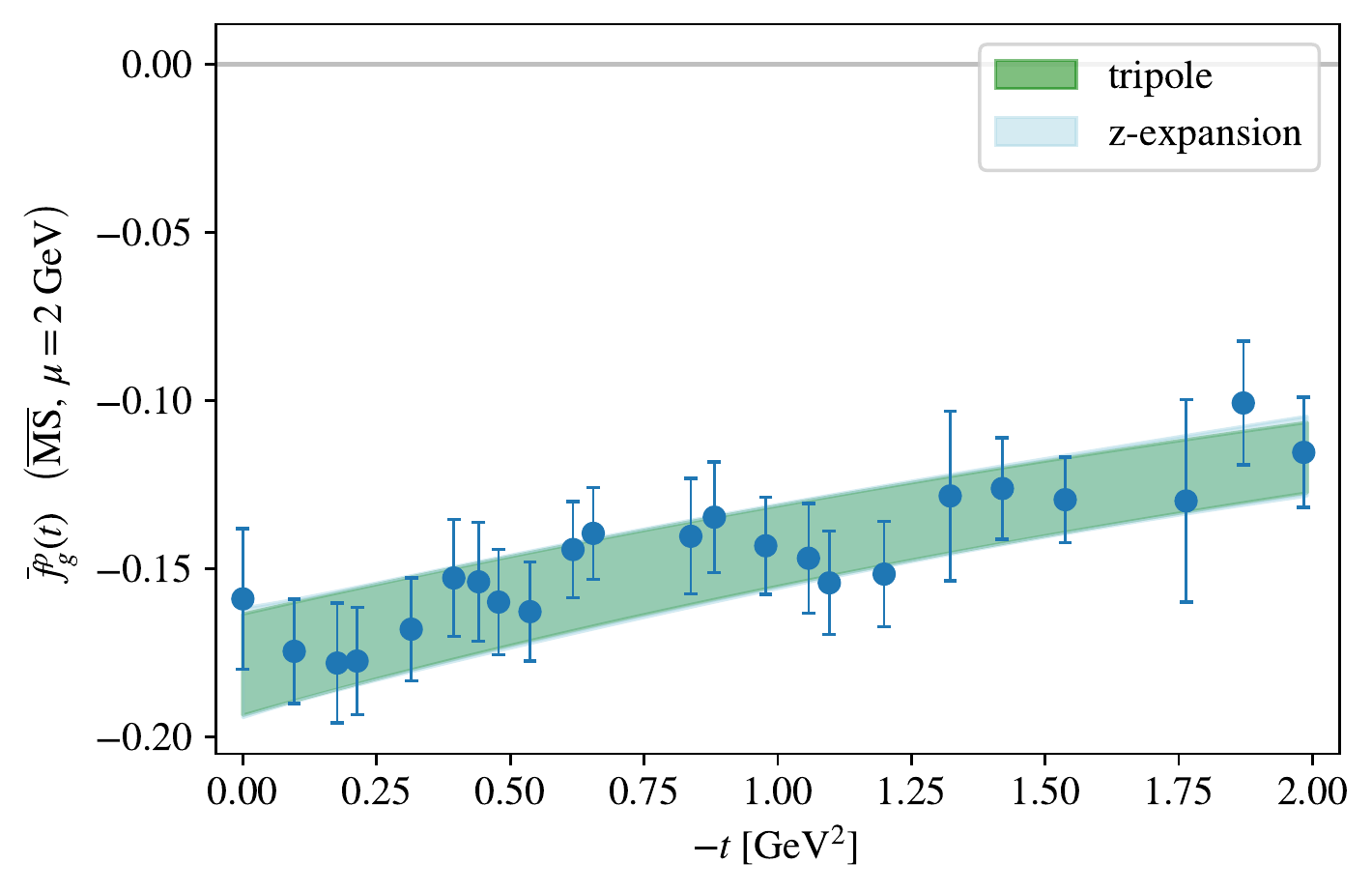} }}
\caption{$A^{\rho}_{0,g}(t)$ (a), $A^{\rho}_{1,g}(t)$ (b), $D^{\rho}_{0,g}(t)$ (c), $D^{\rho}_{1,g}(t)$ (d), $J^{\rho}_{g}(t)$ (e), $E^{\rho}_{g}(t)$ (f), and $\overline{f}^{\rho}_{g}(t)$ (g), renormalized
at $\mu = 2\;\text{GeV}$ in the $\overline{\text{MS}}$ scheme.
The bands correspond to the multipole form [Eq.~\eqref{eq:multipole}] with $n=3$ and the modified z-expansion [Eq.~\eqref{eq:z-expansion}] with $k_{\text{max}} = 2$, with fit parameters shown in Table~\ref{tab:rho}. 
No fit is shown for $D_1^g(t)$ because the corresponding data is not well described by either functional form.}
\label{fig:rhoGFF}
\end{figure*}

Our results for the seven renormalized traceless gluon GFFs are shown in Fig.~\ref{fig:rhoGFF}.
Model fit parameters are tabulated in Table~\ref{tab:rho}, excluding for the GFF $D_{1,g}^{\rho}(t)$, which is not well described by either model ansatz.
The conserved gluon predictions of these quantities are presented in Table~\ref{tab:rhoQuantities}.

We find that approximately half of the angular momentum of the $\rho$ meson is carried by gluons. Interestingly, the NJL model \cite{Freese:2019bhb} predicts that half is carried by the quark spin. Just as in the nucleon case, we find
a significant difference between the forward
limit of the $D$ form factor
resulting from the tripole fit and the z-expansion (see Table~\ref{tab:rho})
which can be traced to the difference of the two
fits in the low momentum region of $D_{0,g}^{\rho}(t)$
as seen in Fig. \ref{fig:rhoGFF}(c). Here the
first data point again suggests nonmonotonic 
behavior that cannot be captured by a multipole form, but it is unclear whether this is a physical effect or due to a statistical fluctuation or underestimated systematic uncertainty.
In Ref.~\cite{Sun:2020wfo}, the total $D_0^{\rho}(t)$ from the light-front 
constituent quark model was found to be nonmonotonic, in contrast
 with the prediction from the NJL model \cite{Freese:2019bhb}.

\begin{table}[!tp] 
\begin{center}
\begin{tabular}{SD{:}{}{2.7}@{\hskip 0.12in}D{:}{}{2.7}@{\hskip 0.12in}D{:}{}{2.7}D{:}{}{2.7}D{:}{}{2.7}}
\toprule
\multicolumn{1}{c}{tripole}  & \multicolumn{1}{c}{$\alpha$} & \multicolumn{1}{c}{$\Lambda$~[GeV]} && \multicolumn{1}{c}{$\chi^2/$d.o.f.}  \\ \midrule
{$A_{0,g}^{\rho}(t)$} & 0:.485(41) & 2:.205(41) && 0:.8 \\[2pt]
{$A_{1,g}^{\rho}(t)$} & -0:.281(84) & 1:.97(43) && 0:.5 \\[2pt]
{$D_{0,g}^{\rho}(t)$} & 1:.16(14) & 2:.01(14) && 1:.9 \\ [2pt]
{$J_{g}^{\rho}(t)$} & 0:.491(42) & 2:.327(44) && 0:.6 \\ [2pt]
{$E_{g}^{\rho}(t)$} & 0:.295(41) & 2:.33(25) && 0:.7 \\ [2pt]
{$\bar{f}_{g}^{\rho}(t)$} & -0:.178(15) & 3:.63(22) && 0:.8 \\
\midrule \midrule
\multicolumn{1}{c}{z-expansion} & \multicolumn{1}{c}{$\alpha_0$} & \multicolumn{1}{c}{$\alpha_1$} & \multicolumn{1}{c}{$\alpha_2$} & \multicolumn{1}{c}{$\chi^2/$d.o.f.} \\ \midrule
{$A_{0,g}^{\rho}(t)$} & 0:.485(41) & -0:.04(11) & -0:.23(66)  & 0:.7 \\[2pt]
{$A_{1,g}^{\rho}(t)$} & -0:.276(42) & -0:.10(71) & -2:(11)  & 0:.5 \\[2pt]
{$D_{0,g}^{\rho}(t)$} & 1:.19(11) & -0:.86(85) & -12:(12) & 1:.6 \\[2pt]
{$J_{g}^{\rho}(t)$} & 0:.494(42) & -0:.105(69) & -1:.01(66)  & 0:.4 \\[2pt]
{$E_{g}^{\rho}(t)$} & 0:.301(29) & -0:.11(24) & -1:.7(3.8)  & 0:.6 \\[2pt]
{$\bar{f}_{g}^{\rho}(t)$} & -0:.179(15) & 0:.005(54) & 0:.04(37)  & 0:.7 \\
\bottomrule
\end{tabular}
\end{center}
\caption{
Fit parameters of the multipole [Eq.~\eqref{eq:multipole}] with $n=3$
and the modified z-expansion [Eq.~\eqref{eq:z-expansion}] with $k_{\text{max}} = 2$ and $n=3$ models for the $t$ dependence of the renormalized $\rho$ GFFs. 
The signal for $D_{1,g}^{\rho}(t)$ is not well described by either functional form and therefore no fit is shown.
For all GFFs that have been fit, the parameter $\Lambda$ of the z-expansion fit is consistent with the prior provided by the tripole fit and is thus not shown.
The parameters $\alpha$ and $\alpha_k$ are renormalized at $\mu = 2\;\text{GeV}$ after fitting the bare GFFs as described in Appendix~\ref{sec:constraint-fitting}.
}
\label{tab:rho}
\end{table}

\begin{table}[!tp]
\begin{center}
\begin{tabular}{S@{\hskip 0.2in}D{:}{}{2.7}@{\hskip 0.2in}D{:}{}{2.7}}
\toprule
& \multicolumn{1}{c}{tripole} & \multicolumn{1}{c}{z-expansion} \\ \midrule
{$A_{g}^{\rho}(0)$} & 0:.485(41) & 0:.482(42)  \\[2pt]
{$D_{g}^{\rho}(0)$} & -1:.16(14) & -0:.81(38) \\[2pt]
{$J_g^{\rho}(0)$} & 0:.491(42) & 0:.469(42) \\
\bottomrule
\end{tabular}
\end{center}
\caption{The forward-limit values of the
conserved gluon contribution to the momentum fraction, $D$ form factor, and
angular momentum of the $\rho$ meson, with definitions provided in the text. The values are obtained from
the tripole and modified z-expansion fits of the $\rho$ GFFs, renormalized at ${\mu=2\;\text{GeV}}$ in the $\overline{\text{MS}}$ scheme,
with parameters shown in Table~\ref{tab:rho}.}
\label{tab:rhoQuantities}
\end{table}

\subsection{$\Delta$ Baryon}
\label{sec:DeltaGFFs}

\begin{table}[tp] 
\begin{center}
\begin{tabular}{SD{:}{}{2.7}@{\hskip 0.2in}D{:}{}{2.7}D{:}{}{2.7}D{:}{}{2.7}D{:}{}{2.7}}
\toprule
\multicolumn{1}{c}{tripole}  & \multicolumn{1}{c}{$\alpha$} & \multicolumn{1}{c}{$\Lambda$~[GeV]} && \multicolumn{1}{c}{$\chi^2/$d.o.f.}  \\ \midrule
{$F_{10}^{\Delta,g}(t)$} & 0:.393(36) & 1:.788(66) && 1:.2 \\ [2pt]
{$F_{20}^{\Delta,g}(t)$} & -1:.80(69) & 1:.10(17) && 2:.3 \\ [2pt]
{$F_{40}^{\Delta,g}(t)$} & 0:.587(78) & 1:.421(78) && 2:.0 \\ [2pt]
{$F_{60}^{\Delta,g}(t)$} & 0:.0492(89) & 2:.26(38) && 2:.1 \\
\midrule \midrule
\multicolumn{1}{c}{z-expansion} & \multicolumn{1}{c}{$\alpha_0$} & \multicolumn{1}{c}{$\alpha_1$} & \multicolumn{1}{c}{$\alpha_2$} & \multicolumn{1}{c}{$\chi^2/$d.o.f.} \\ \midrule
{$F_{10}^{\Delta,g}(t)$} & 0:.395(35) & -0:.19(21) & -1:.2(1.4) & 1:.0 \\ [2pt]
{$F_{20}^{\Delta,g}(t)$} & -2:.36(42) & 18:.8(8.4) & 153:(68)  & 1:.3 \\ [2pt]
{$F_{40}^{\Delta,g}(t)$} & 0:.623(62) & -0:.56(49) & -7:.0(6.0)  & 1:.6\\ [2pt]
{$F_{60}^{\Delta,g}(t)$} & 0:.0554(69) & -0:.154(97) & -1:.49(93)  & 1:.5 \\ \bottomrule
\end{tabular}
\end{center}
\caption{
Fit parameters of the multipole [Eq.~\eqref{eq:multipole}] with $n=3$
and the modified z-expansion [Eq.~\eqref{eq:z-expansion}] with $k_{\text{max}} = 2$ and $n=3$ models for the $t$ dependence of the renormalized $\Delta$ GFFs. 
Only four of the eight GFFs are fit, as the others are not resolved from zero as seen in Fig.~\ref{fig:deltaGFF}.
For all GFFs that have been fit, the parameter $\Lambda$ of the z-expansion fit is consistent with the prior provided by the tripole fit and is thus not shown.
The parameters $\alpha$ and $\alpha_k$ are renormalized at $\mu = 2\;\text{GeV}$ after fitting the bare GFFs as described in Appendix~\ref{sec:constraint-fitting}.
}
\label{tab:delta}
\end{table}

\begin{table}[tp] 
\begin{center}
\begin{tabular}{S@{\hskip 0.2in}D{:}{}{2.7}@{\hskip 0.2in}D{:}{}{2.7}}
\toprule
& \multicolumn{1}{c}{tripole} & \multicolumn{1}{c}{z-expansion} \\ \midrule
{$A^{\Delta}_g(0)$} & 0:.393(36) & 0:.378(38) \\[2pt]
{$D^{\Delta}_g(0)$} & -1:.80(69) & 0:.9(1.5) \\[2pt]
{$J^{\Delta}_g(0)$} & 0:.588(78) & 0:.41(18) \\
\bottomrule
\end{tabular}
\end{center}
\caption{The forward-limit values of the conserved
gluon contribution to the momentum fraction, $D$ form factor, and
angular momentum of the $\Delta$ baryon, obtained from
the tripole and modified z-expansion fits of the $\Delta$ GFFs, renormalized at $\mu=2\;\text{GeV}$ in the $\overline{\text{MS}}$ scheme,
with parameters shown in Table~\ref{tab:delta}.}
\label{tab:deltaQuantities}
\end{table}

\begin{figure*}[p]
\captionsetup[subfloat]{captionskip=-3pt}
\centering
\subfloat[\centering  ]
{{\includegraphics[height=5.25cm,width=7.8cm]{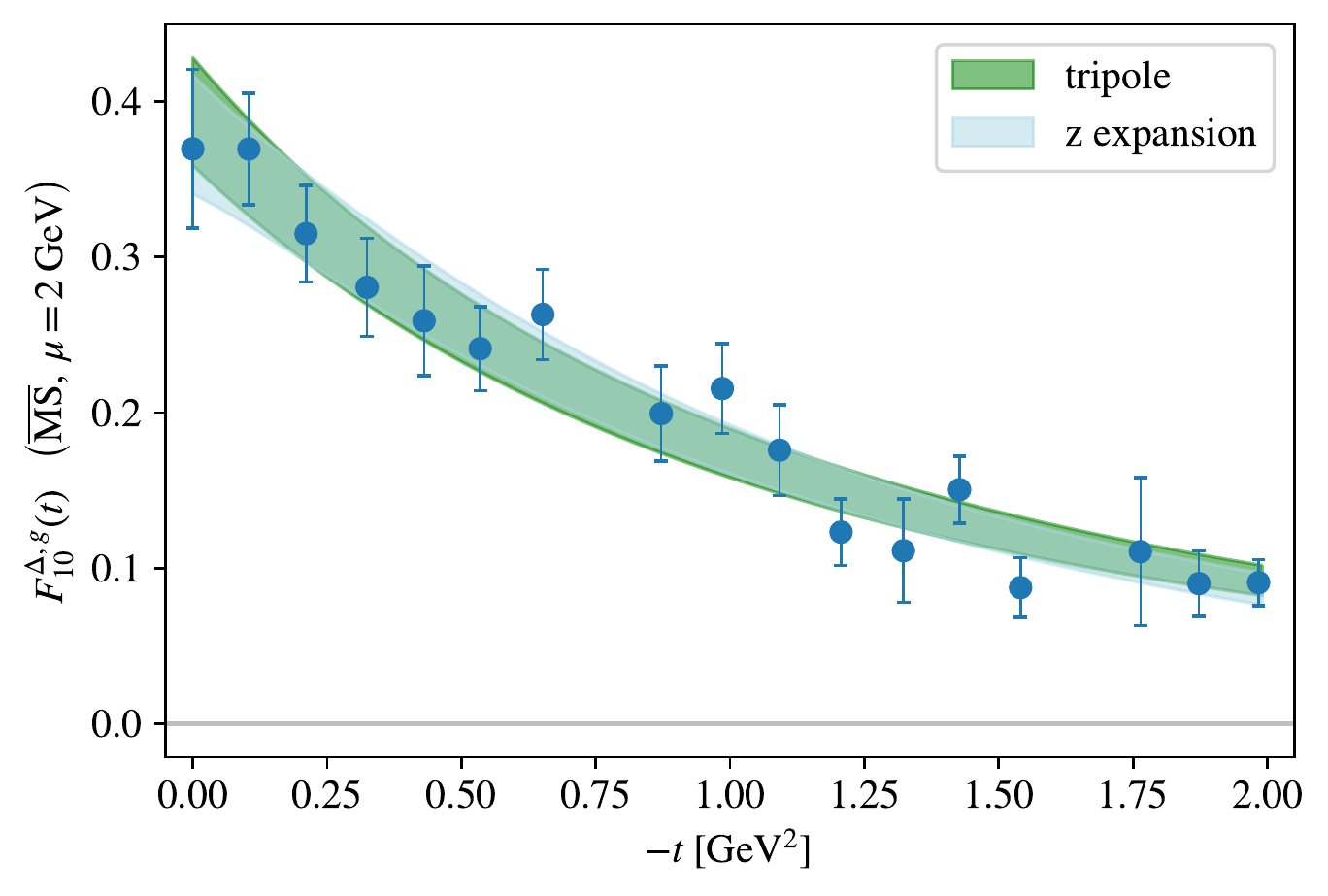} }}
\!
\subfloat[\centering  ]
{{\includegraphics[height=5.25cm,width=7.8cm]{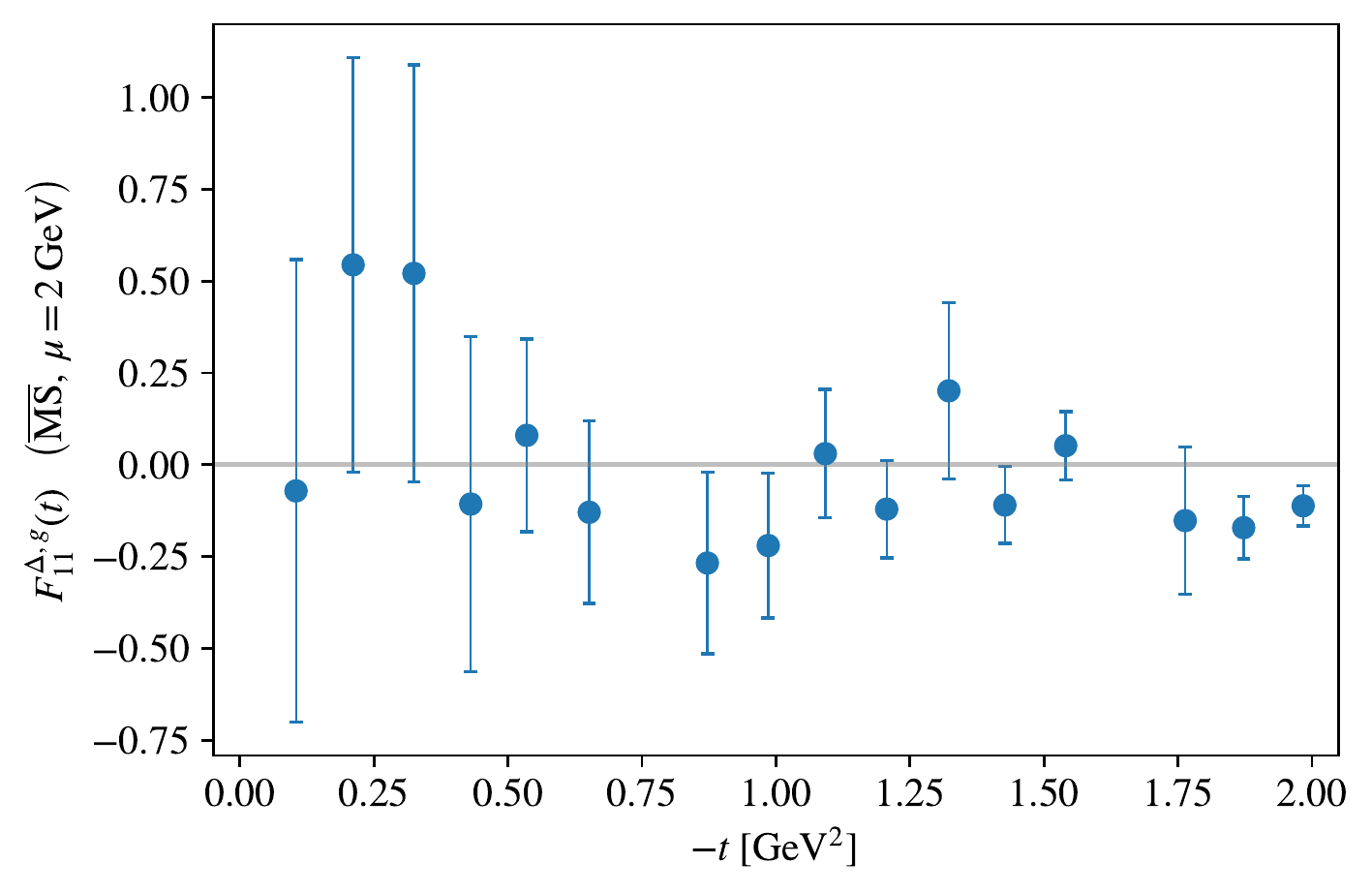} }}
\\[-2.5ex]
\subfloat[\centering  ]
{{\includegraphics[height=5.25cm,width=7.8cm]{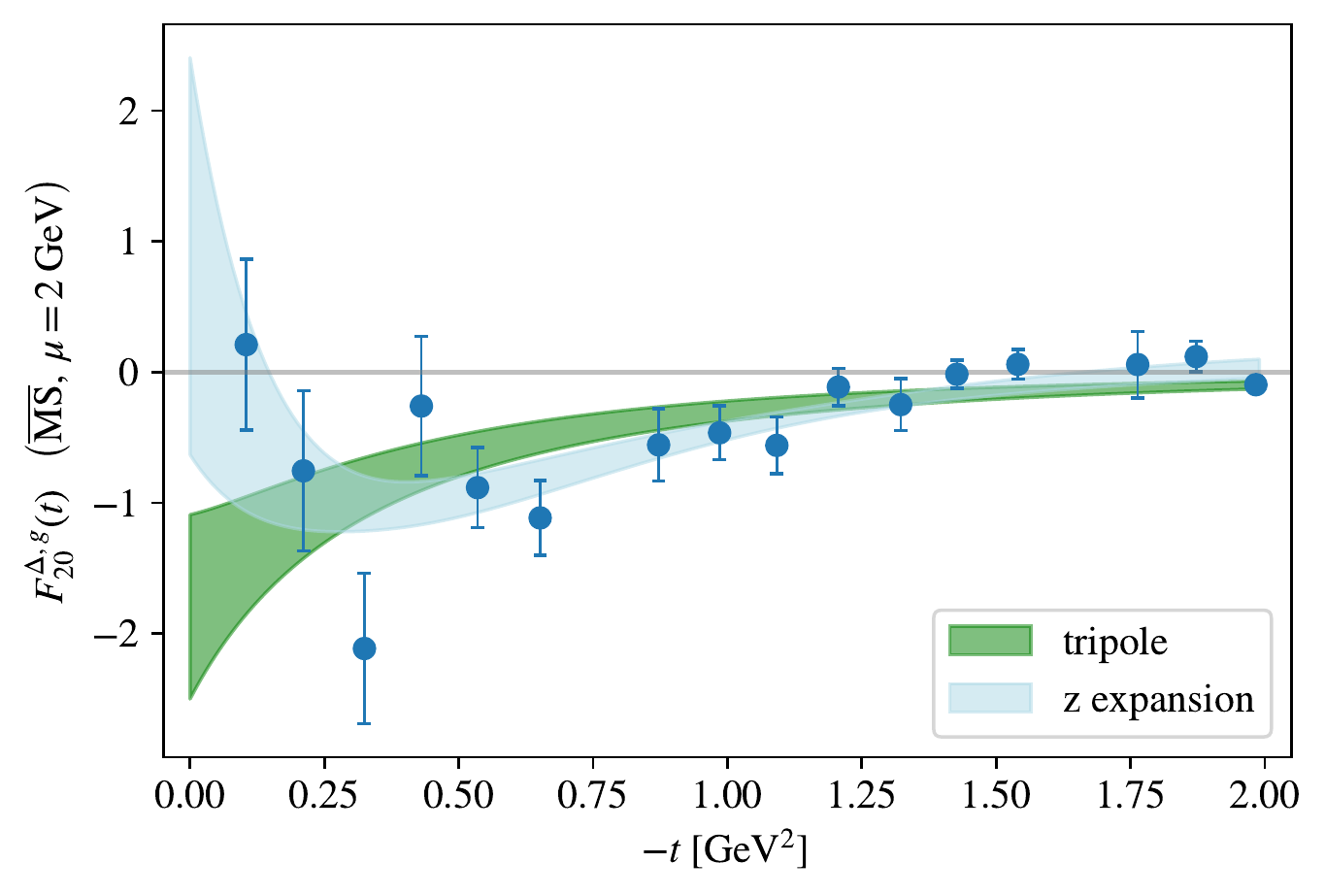} }}
\!
\subfloat[\centering  ]
{{\includegraphics[height=5.25cm,width=7.8cm]{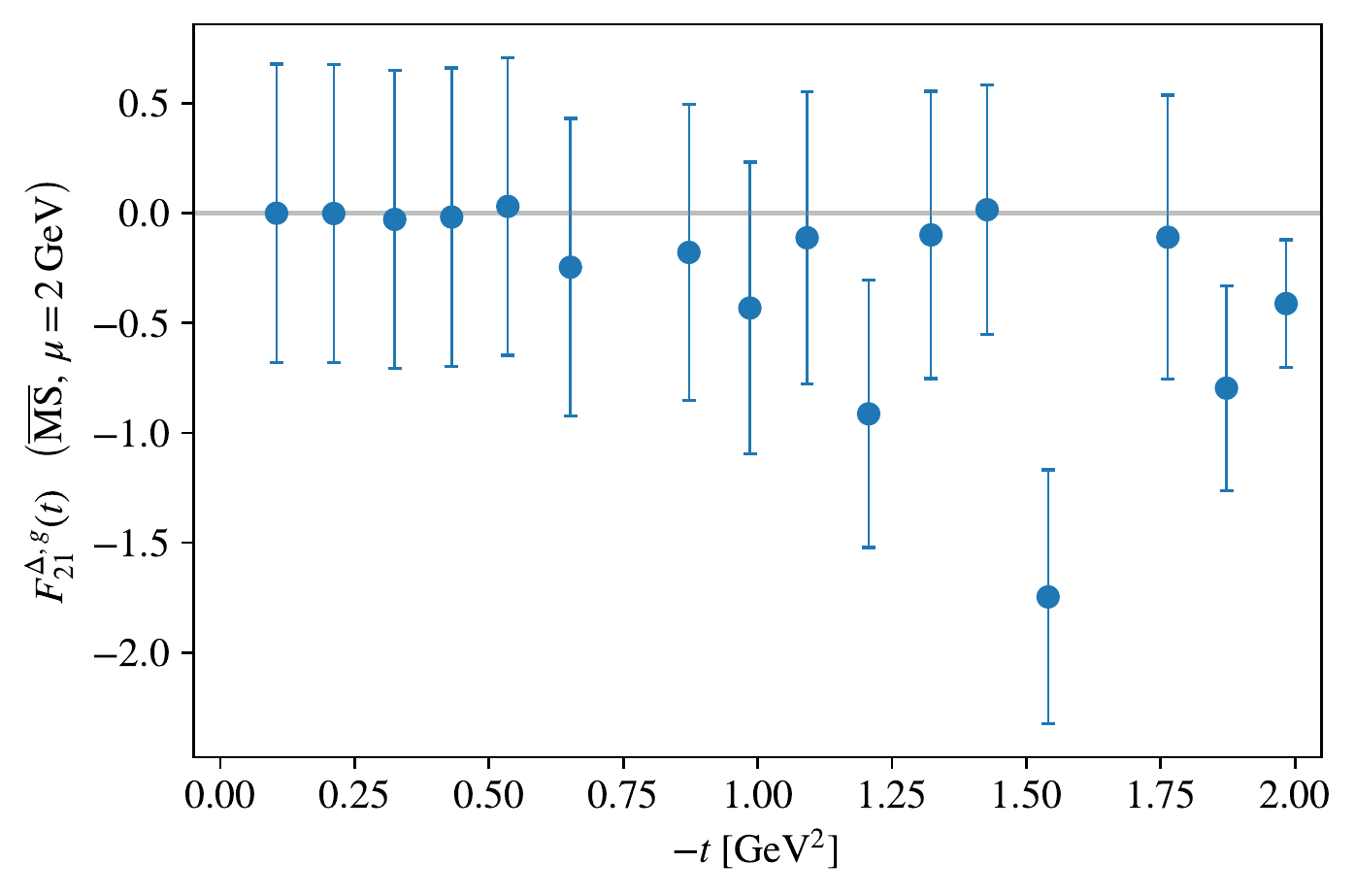} }}
\\[-2.5ex]
\subfloat[\centering  ]
{{\includegraphics[height=5.25cm,width=7.8cm]{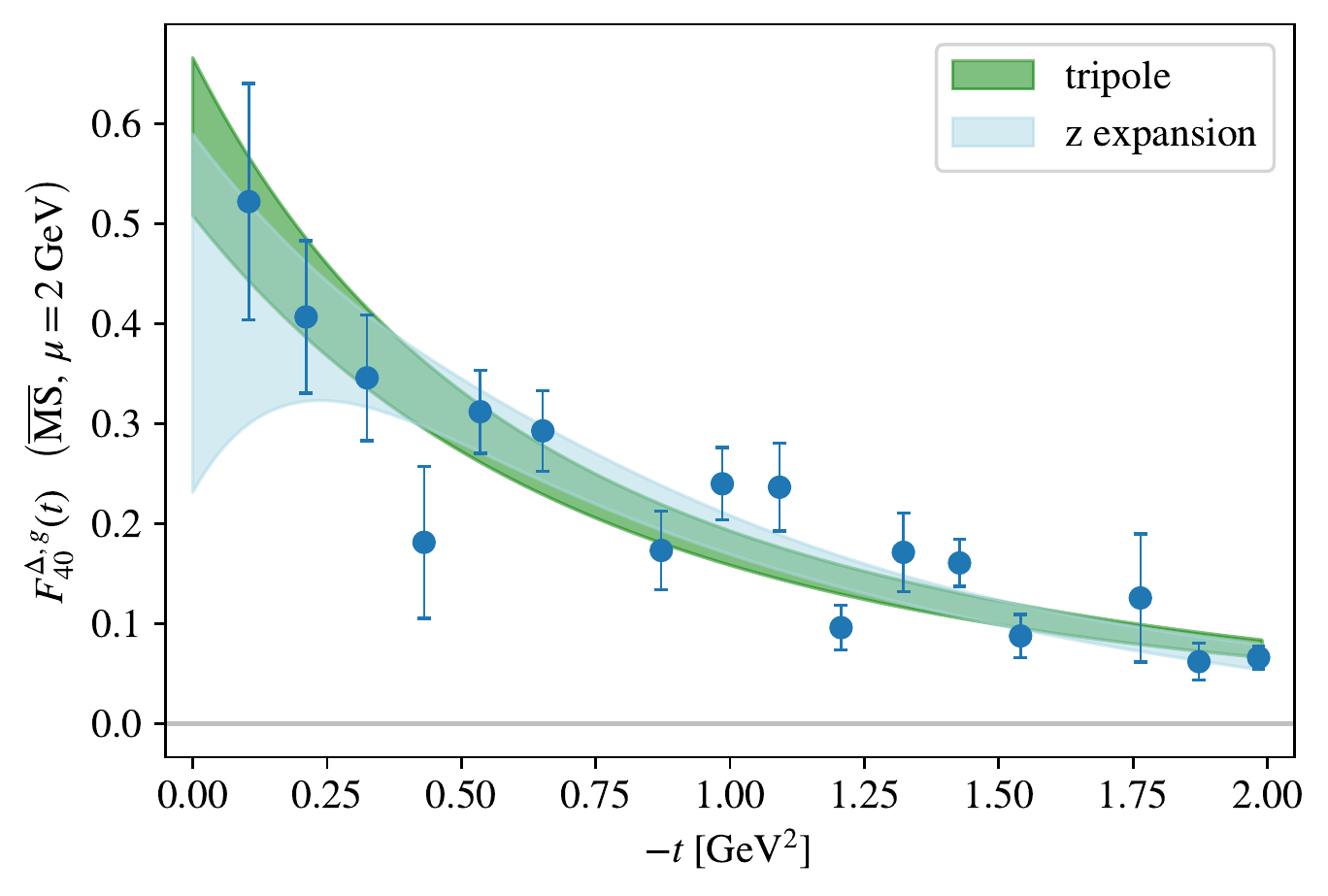} }}
\!
\subfloat[\centering  ]
{{\includegraphics[height=5.25cm,width=7.8cm]{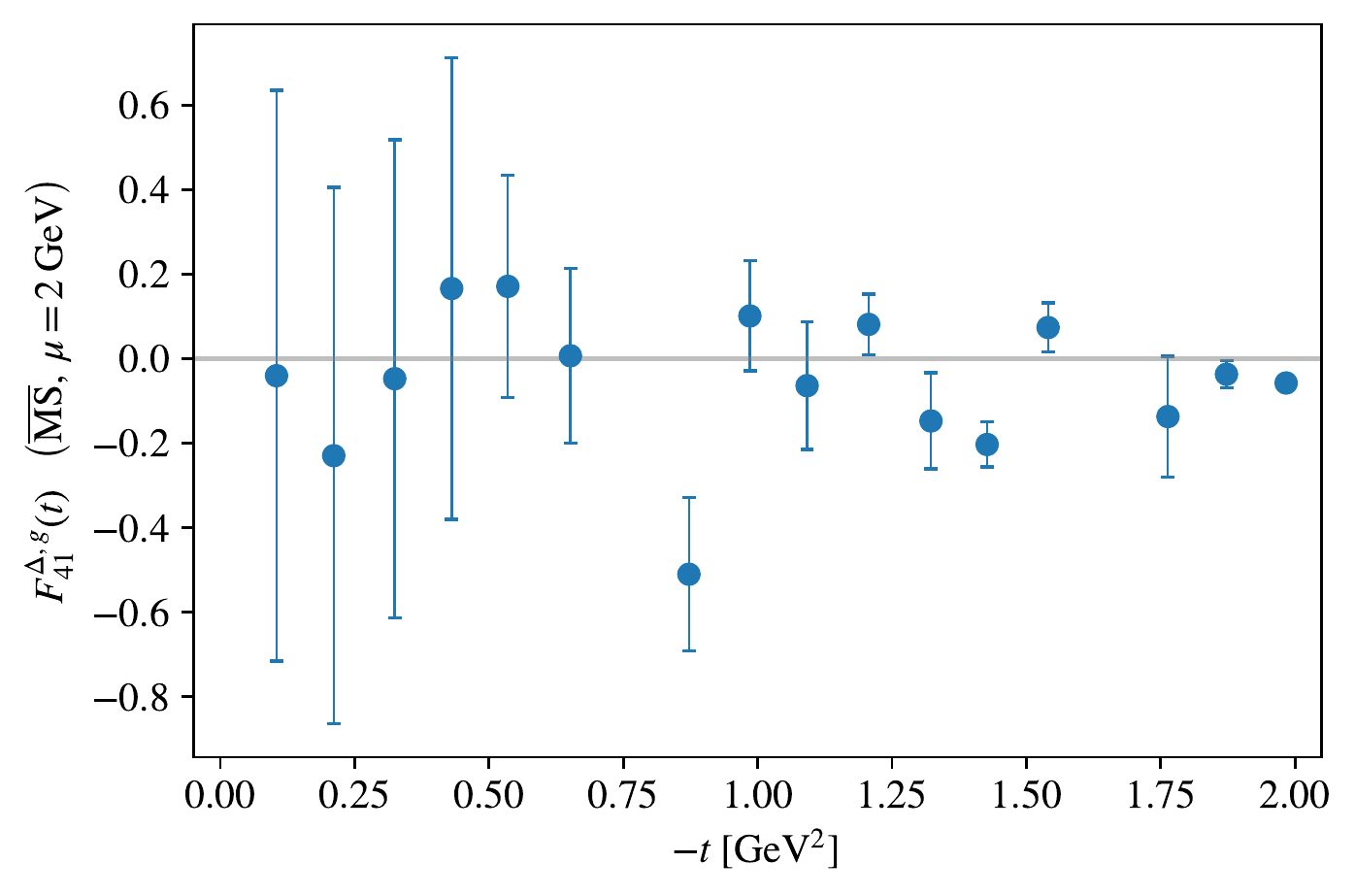} }}
\\[-2.5ex]
\subfloat[\centering  ]
{{\includegraphics[height=5.25cm,width=7.8cm]{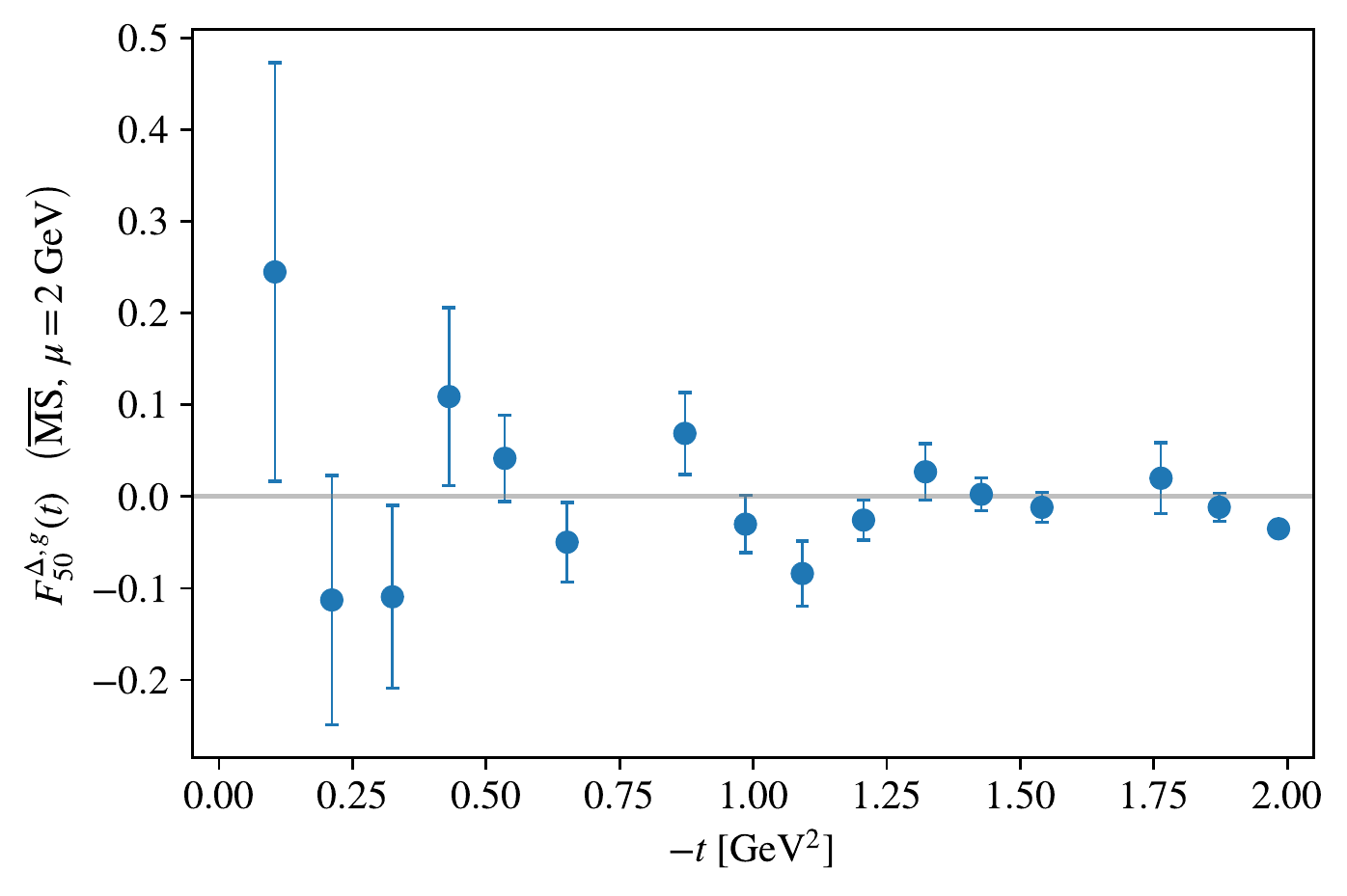} }}
\!
\subfloat[\centering  ]
{{\includegraphics[height=5.25cm,width=7.8cm]{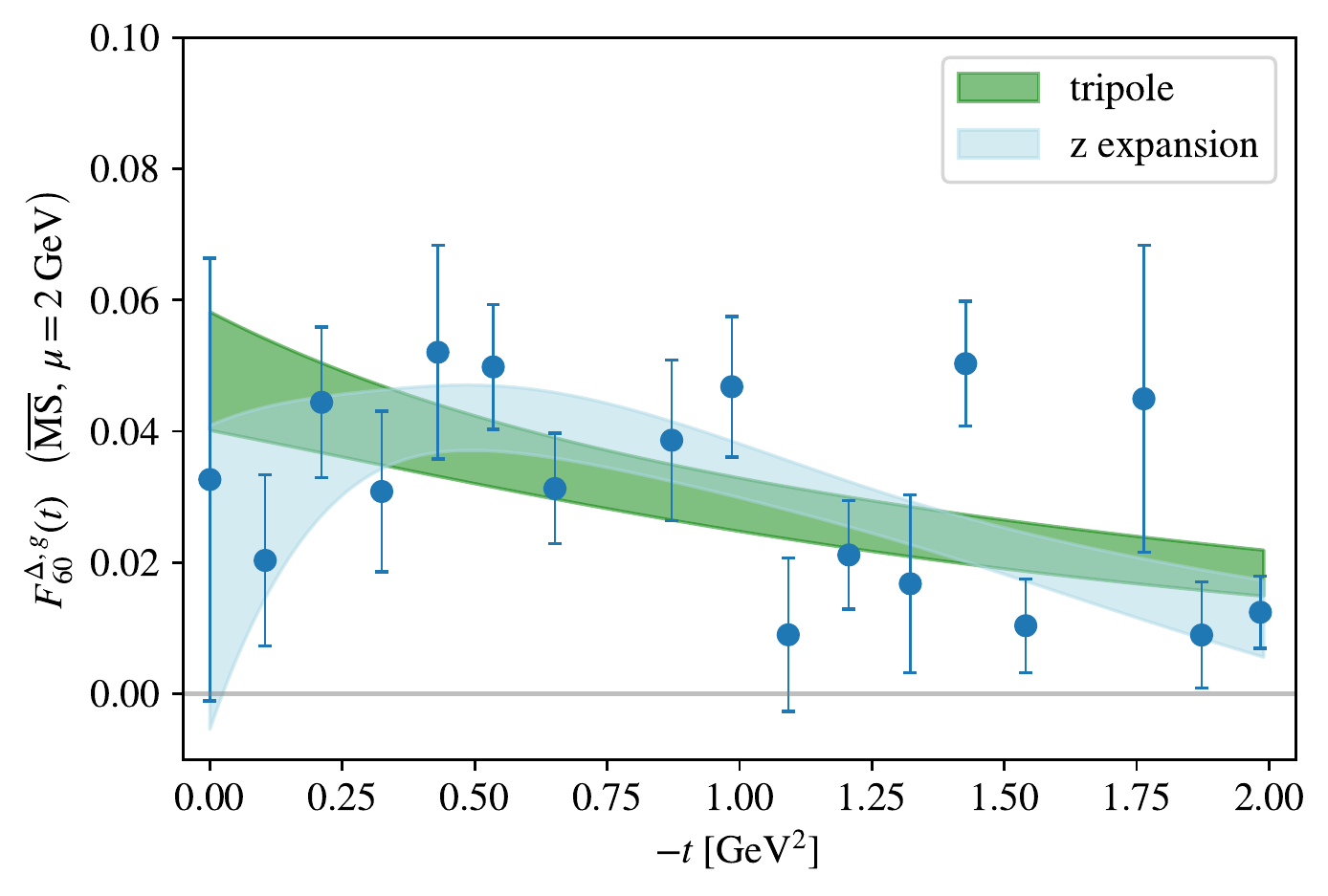} }}
\caption{$F^{\Delta,g}_{10}(t)$ (a), $F^{\Delta,g}_{11}(t)$ (b), $F^{\Delta,g}_{20}(t)$ (c), $F^{\Delta,g}_{21}(t)$ (d), $F^{\Delta,g}_{40}(t)$ (e), $F^{\Delta,g}_{41}(t)$ (f), $F^{\Delta,g}_{50}(t)$ (g), $F^{\Delta,g}_{60}(t)$ (h), renormalized
at $\mu = 2~\text{GeV}$ in the $\overline{\text{MS}}$ scheme.
The bands correspond to the multipole form [Eq.~\eqref{eq:multipole}] with $n=3$ and the modified z-expansion [Eq.~\eqref{eq:z-expansion}] with $k_{\text{max}} = 2$, with fit parameters shown in Table~\ref{tab:delta}. 
No fit is shown for the four GFFs that are not resolved from zero.
}
\label{fig:deltaGFF}
\end{figure*}

For the $\Delta$ baryon, the matrix element of 
the quark or gluon symmetric EMT
can be decomposed as \cite{Kim:2020lrs}
\begin{widetext}
\begin{equation} \label{eq:deltaME}
\scalebox{0.95}{$
\begin{split}
&\bra{\Delta(p',\xi')} T_i^{\mu\nu} \ket{\Delta(p,\xi)} \\
&\qquad = \bar{u}_{\alpha'}(\vec{p}',\xi')\bigg[
 \frac{P^{\mu}P^{\nu}}{m_{\Delta}}\left(-g^{\alpha\alpha'}  F^{\Delta,i}_{10}(t) + \frac{\Delta^{\alpha}
\Delta^{\alpha'}}{2m_{\Delta}^2}F^{\Delta,i}_{11}(t)\right) 
+ \frac{\Delta^{\mu}\Delta^{\nu}-g^{\mu\nu}\Delta^2}{4m_{\Delta}}\left(-g^{\alpha\alpha'}F^{\Delta,i}_{20}(t)+\frac{\Delta^{\alpha}\Delta^{\alpha'}}
{2m_{\Delta}^2}F^{\Delta,i}_{21}(t)\right)  \\
&\qquad \qquad \quad +m_{\Delta}g^{\mu\nu}\left(-g^{\alpha\alpha'}F^{\Delta,i}_{30}(t) +\frac{\Delta^{\alpha}\Delta^{\alpha'}}{2m_{\Delta}^2}F^{\Delta,i}_{31}(t)\right) 
 +\frac{i P^{\{\mu}\sigma^{\nu\}\rho}\Delta_{\rho}}{m_{\Delta}}\left(-g^{\alpha'\alpha}F^{\Delta,i}_{40}(t)+
\frac{\Delta^{\alpha'}\Delta^{\alpha}}{2m_{\Delta}^2}F^{\Delta,i}_{41}(t)\right) \\
&\qquad \qquad \quad +\frac{2}{m_{\Delta}}(\Delta^{\{\mu}g^{\nu\}\{\alpha'}\Delta^{\alpha\}}
-g^{\mu\nu}\Delta^{\alpha}\Delta^{\alpha'}-g^{\alpha'\{\mu}
g^{\nu\}\alpha}\Delta^2 )
F^{\Delta,i}_{50}(t)
 -2 g^{\alpha'\{\mu}g^{\nu\}\alpha}
m_{\Delta}F^{\Delta,i}_{60}(t)  \bigg] u_{\alpha}(\vec{p},\xi) \\
&\qquad \equiv \bar{u}_{\alpha'}(\vec{p}',\xi')
\mathcal{O}_{(\Delta)}^{\alpha\mu\nu\alpha'}[F^{\Delta,i}_{10}(t),F^{\Delta,i}_{11}(t),F^{\Delta,i}_{20}(t),F^{\Delta,i}_{21}(t),F^{\Delta,i}_{40}(t),F^{\Delta,i}_{41}(t),F^{\Delta,i}_{50}(t),F^{\Delta,i}_{60}(t)]u_{\alpha}(\vec{p},\xi) + \text{trace} \;,
\end{split}
$}
\end{equation}
\end{widetext}
where $m_{\Delta}$ denotes the mass of the $\Delta$ baryon, and $u_{\alpha}(\vec{p},\xi)$ is the Rarita-Schwinger spin-vector satisfying
\begin{equation}
\begin{split}
\sum_{\xi} u_{\sigma}(\vec{p},\xi)\bar{u}_{\tau}(\vec{p},\xi) &\\=
-\frac{\cancel{p} +
m_{\Delta}}{2m_{\Delta}}\bigg(g_{\sigma\tau}-&\frac{1}{2}\gamma_{\sigma}
\gamma_{\tau} -\frac{2p_{\sigma}p_{\tau}}{3m_{\Delta}^2}+
\frac{p_{\sigma}\gamma_{\tau}-p_{\tau}\gamma_{\sigma}}{3m_{\Delta}}\bigg) \\
 \equiv \Lambda^{(\Delta)}_{\sigma \tau}(p) \qquad\qquad &
\end{split}
\label{eqn:lambda-delta-def}
\end{equation}
with $\xi \in \{3/2, 1/2, -1/2, -3/2 \}$.

\newpage

The total momentum fraction $A^{\Delta}(0) \equiv F^{\Delta}_{10}(0)$ is constrained to equal $1$. As with the $\rho$, there are three total $D$-terms of
different order and we again focus our discussion on the form factor that is the coefficient of the same Lorentz structure as the nucleon and pion $D$-terms, namely $D^{\Delta}(0) \equiv F^{\Delta}_{20}(0)$. The total forward-limit angular momentum, ${J^{\Delta}(0) = F^{\Delta}_{40}(0)}$, is constrained to be equal to $3/2$ \cite{Kim:2020lrs}. Just as in the case of the $\rho$ meson,
there are nonconserved GFFs related to
the trace piece, $F_{30}^{\Delta,i}(t)$ and
$F_{31}^{\Delta,i}(t)$, that we do not have
access to in this calculation, and one
nonconserved GFF $F_{60}^{\Delta,i}(t)$, that arises from the traceless EMT and that we are able to
constrain.

\newpage

Our results for the eight renormalized traceless gluon GFFs of the $\Delta$ baryon are shown in Fig.~\ref{fig:deltaGFF}. Only four of them are resolved from zero, and their fit parameters are shown in Table~\ref{tab:delta}. 
The conserved gluon contributions to the forward limit quantities obtained from the tripole and $z$-expansion fits are shown in Table~\ref{tab:deltaQuantities}.

\section{DENSITIES AND RADII FROM GFFs}
\label{sec:densities}

In the decomposition of the
matrix element
$\bra{h(p',s')} T_i^{\mu\nu} \ket{h(p,s)}$, the GFFs
are Lorentz scalars but their coefficients depend on the 
reference frame.
The spatial energy, pressure, and shear force densities of hadrons are related to Fourier transforms of the momentum space
matrix elements, and therefore are also 
frame dependent. In this work we consider these densities in two frames, namely the Breit frame and the infinite momentum frame.

\textit{Breit frame (BF)}.---The ``brick-wall'' frame in which there is no energy transfer to
the system, $\Delta^0 = 0$, and additionally $\vec{P} = 0$. This is
the frame traditionally used to define spatial distributions, such as the charge distribution in terms of the electromagnetic form factors~\cite{Sachs:1962zzc}.
The equivalent 3D density for the EMT in the BF (the BF3 density) is
\begin{equation} \label{eq:BFEMT3D}
T^{\mu\nu}_{i,\text{BF3}}(r) = \int \frac{d^3\Delta e^{-i \vec{\Delta}\cdot\vec{r}}}{{2P^0}(2\pi)^3}
\bra{h(p',s')} T_i^{\mu\nu} \ket{h(p,s)}
\bigr\rvert_{\vec{P}=0} \;,
\end{equation}
where $P^0=\sqrt{m_{h}^2+\vec{\Delta}^2/4}$,
while in
a 2D plane, the (BF2) density is equal to
\begin{equation} \label{eq:BFEMT2D}
T^{\mu\nu}_{i,\text{BF2}}(r) = \int
\frac{d^2\Delta_{\mybot}e^{-i \vec{\Delta}_{\mybot}\cdot\vec{r}}}{{2P^0 (2\pi)^2}}
\bra{h(p',s')}T_i^{\mu\nu} \ket{h(p,s}
\bigr\rvert_{\vec{P}=0} \;.
\end{equation}

It is known that the
 Fourier transform of a form factor in the BF
is not a relativistically correct 
way to define the corresponding spatial distributions
\cite{Miller:2018ybm,Jaffe:2020ebz} since one is free to multiply the distribution by a boost factor that cannot be uniquely defined in 
relativistic quantum field theory.
However, following the phase-space approach  introduced in 
Refs.~\cite{Lorce:2018egm,Lorce:2020onh},
Eqs.~\eqref{eq:BFEMT3D} and \eqref{eq:BFEMT2D} can be interpreted as quasidistributions instead of densities, and there is no ambiguity
with respect to the boost factor.

\vspace{5mm}

\textit{Infinite momentum frame (IMF)}.---The elastic frame in which
$\vec{\Delta}\cdot \vec{P} = 0$ and $P_z \rightarrow \infty$.
In this frame there is Galilean symmetry in the transverse plane and
thus 2D Fourier transforms of the momentum tensor matrix elements can be interpreted as spatial
densities.\footnote{Another frame that can be considered
is the front-form Drell-Yan frame, in which $\Delta_- = 0$ and
$\Delta_+ = 0$. 2D Fourier
transforms in the Drell-Yan frame can be correctly interpreted
as spatial distributions \cite{Burkardt:2002hr,Lorce:2018egm,Freese:2019bhb,Freese:2021czn,Freese:2021qtb},
since in the 
light-cone transverse boosts are Galilean \cite{Brodsky:1997de}. We choose not
to discuss this frame here, since the energy
density corresponds to a different component
of the energy momentum tensor and is thus
not directly comparable with the instant-form
energy density. The pressure and shear forces
for the pion and the nucleon are identical
in the infinite momentum frame
and the Drell-Yan frame.} The expression for
the EMT density in this frame is
\begin{equation} \label{eq:IMFEMT}
T^{\mu\nu}_{i,\text{IMF}}(r) \!=\!\int\!
\frac{d^2\Delta_{\mybot}e^{-i \vec{\Delta}_{\mybot}\cdot\vec{r}}}{{2P^0 (2\pi)^2}}\!
\bra{h(p',s')}T_i^{\mu\nu} \ket{h(p,s)}
\bigr\rvert_{\vec{P}\cdot\vec{\Delta}=0}^{P_z\rightarrow \infty}\;,
\end{equation}
where $P_0 = \sqrt{m_h^2+\vec{\Delta}^2/4+P_z^2}$.
We note that for the case of 
the pressure and shear
forces of spherically symmetric hadrons, it was recently shown
that the densities in the two frames are related by Abel 
transformations~\cite{Panteleeva:2021iip}.

\subsection{Pion}

The expressions for the various EMT distributions of the pion in terms of the corresponding GFFs are
listed in Appendix~\ref{sec:piondensities}. In Fig.~\ref{fig:pionall}, we present our results
for the gluon contribution to the
energy density $\varepsilon(r)$, the pressure $p(r)$, and the
shear forces $s(r)$ in the 3D and 2D BF,
and in the IMF. The
definitions of the energy and pressure densities 
of the individual constituents include
the nonconserved GFF $\bar{c}(t)$, which cancels between the quark and gluon contributions. Since we
cannot constrain this term from our calculations, the results
shown in Fig.~\ref{fig:pionall}, and for the rest of the hadrons in the
sections below, include only the
traceless gluon contribution to the
densities.

\begin{figure*} 
\centering
\subfloat[\centering  ]
{{\includegraphics[height=5.8cm,width=7.8cm]{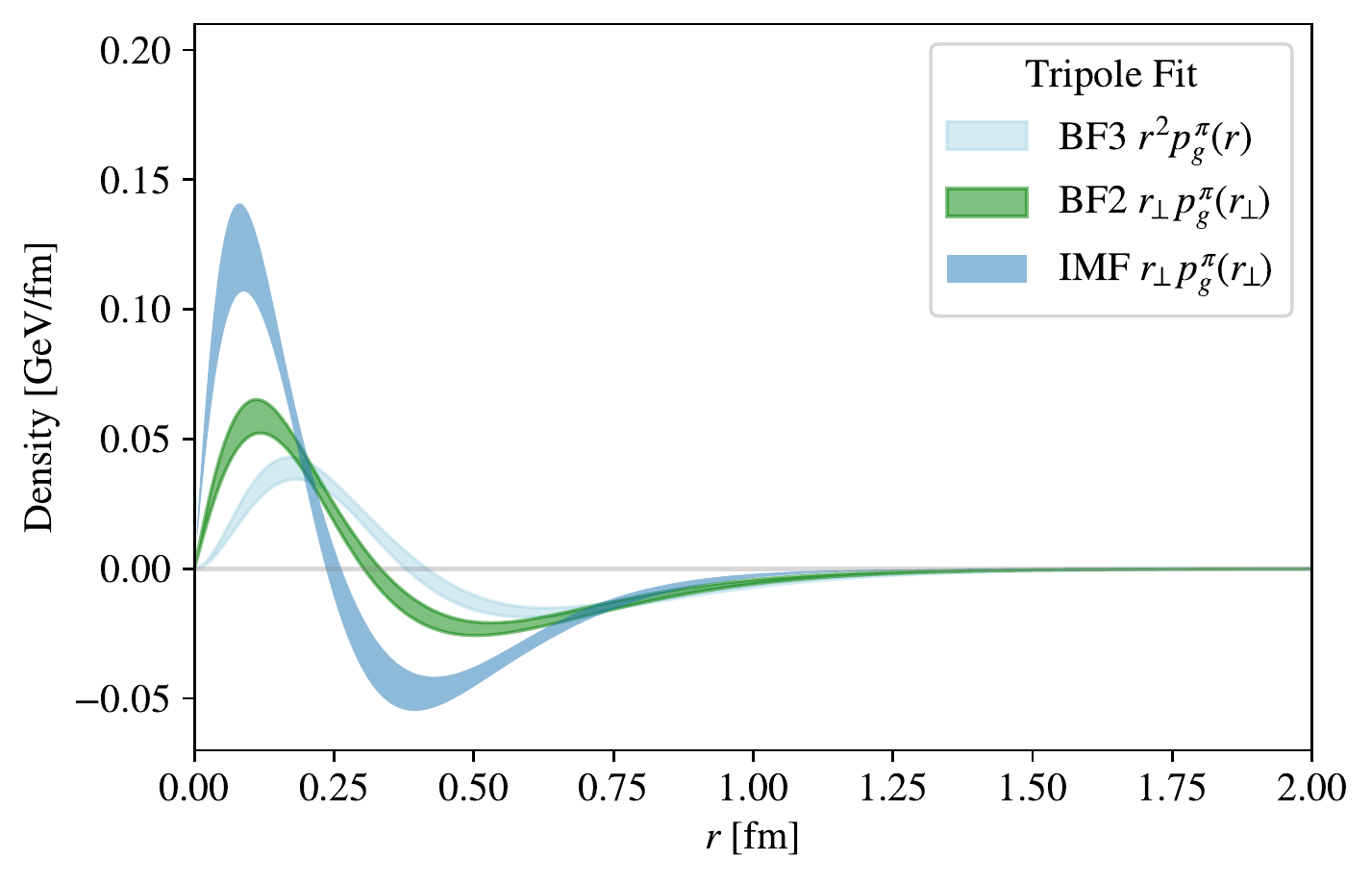} }}
\!
\subfloat[\centering  ]
{{\includegraphics[height=5.8cm,width=7.8cm]{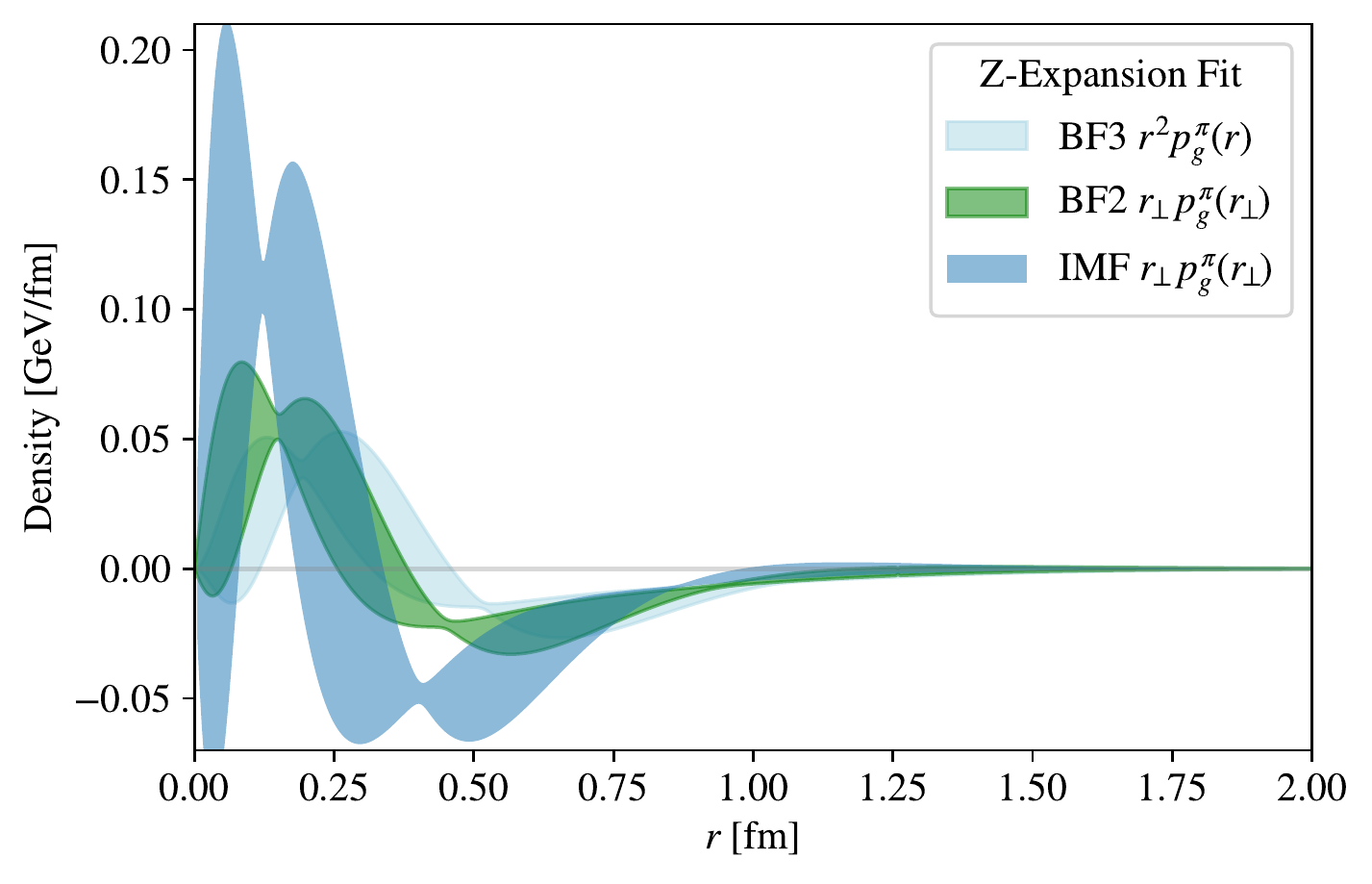} }}
\\
\subfloat[\centering  ]
{{\includegraphics[height=5.8cm,width=7.8cm]{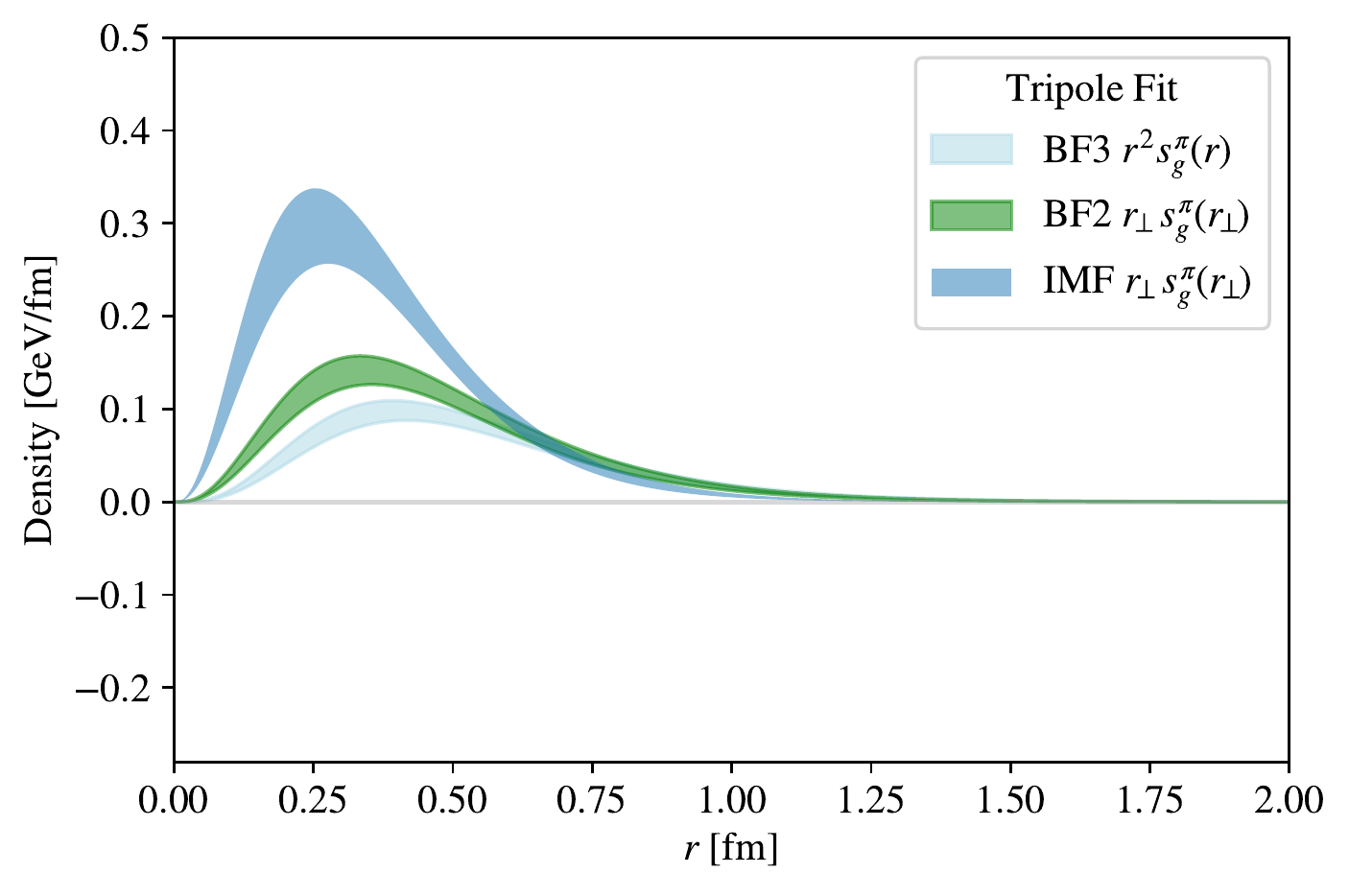} }}
\!
\subfloat[\centering  ]
{{\includegraphics[height=5.8cm,width=7.8cm]{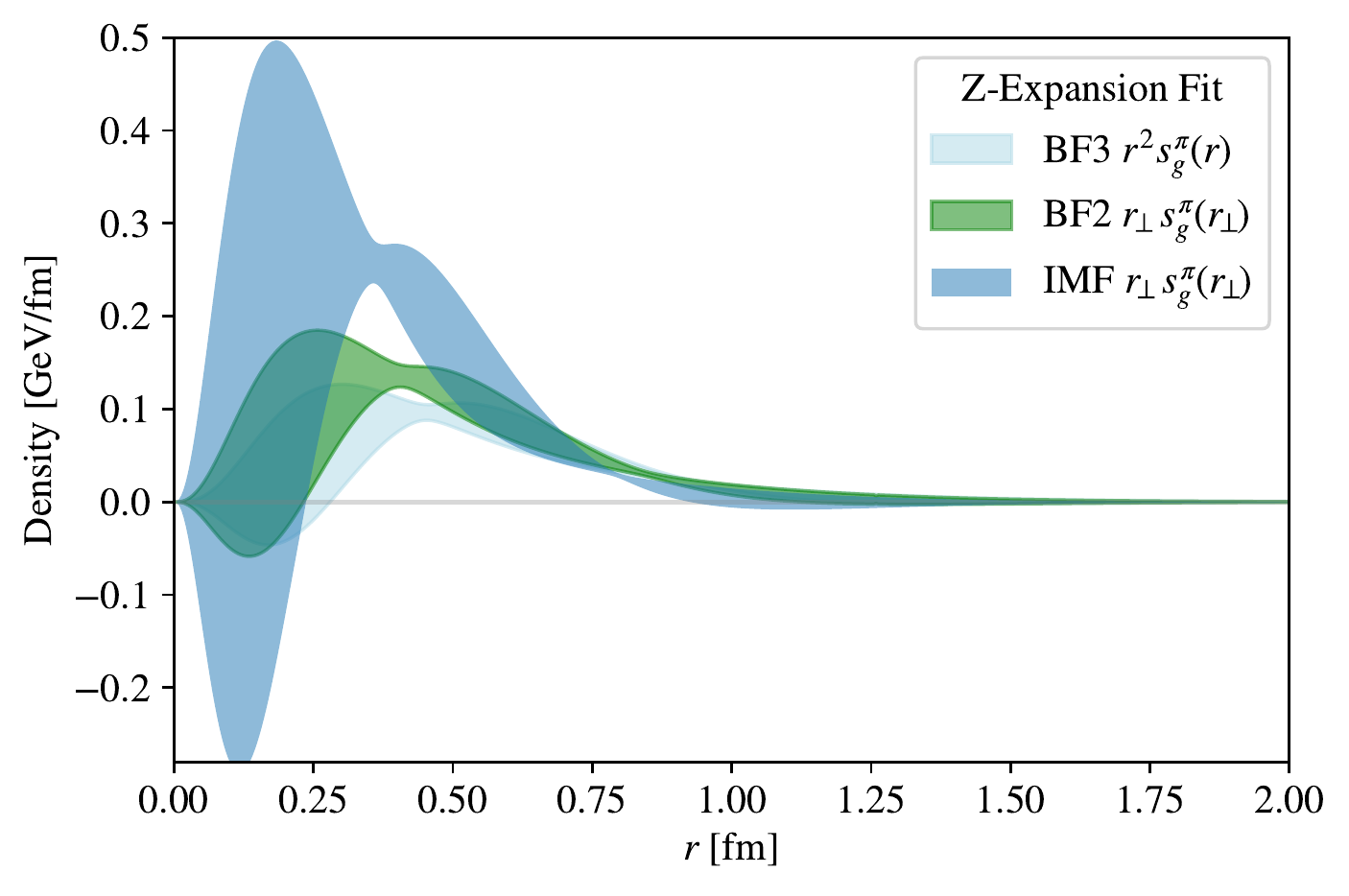} }}
\\
\subfloat[\centering  ]
{{\includegraphics[height=5.8cm,width=7.8cm]{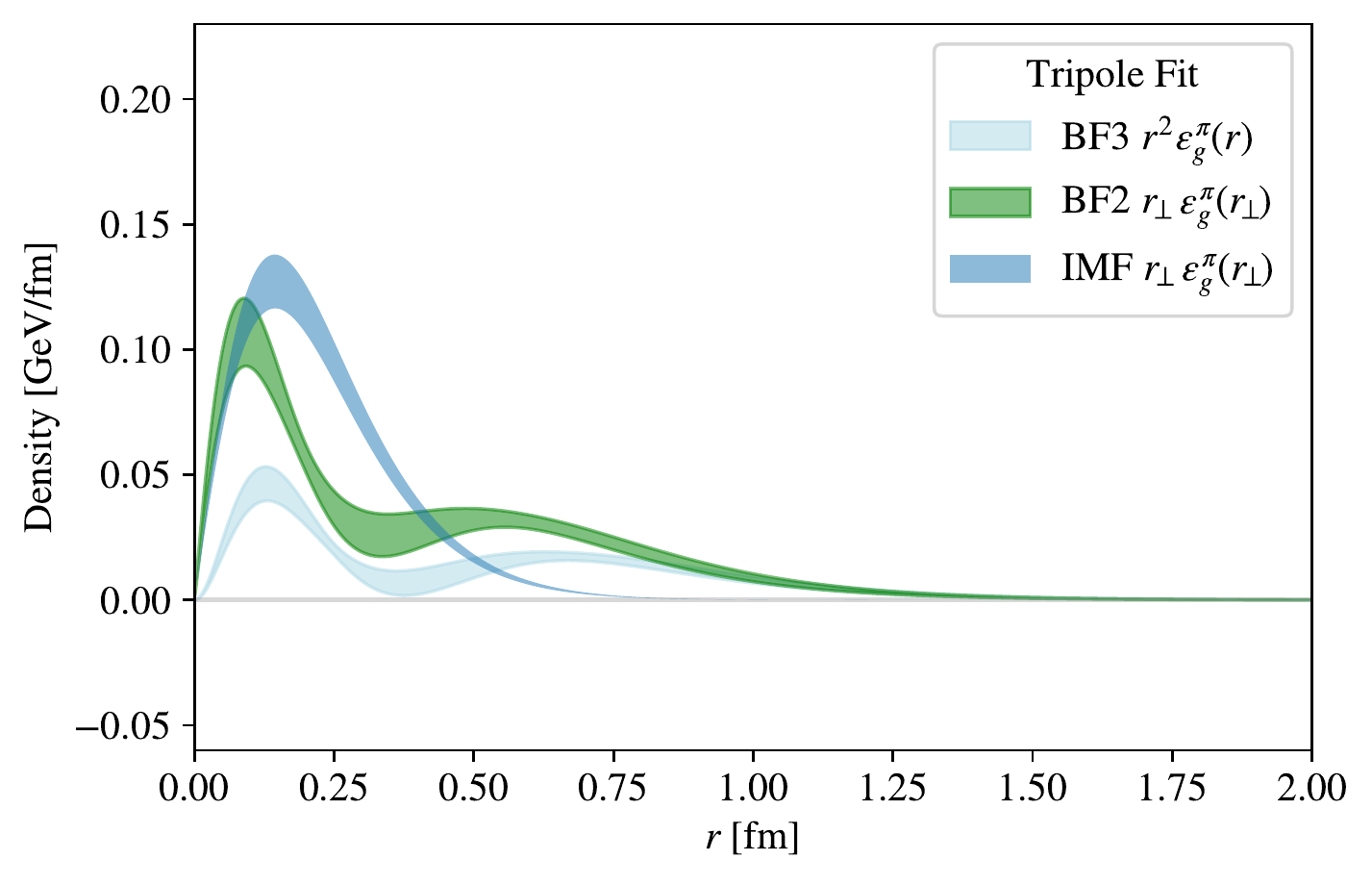} }}
\!
\subfloat[\centering  ]
{{\includegraphics[height=5.8cm,width=7.8cm]{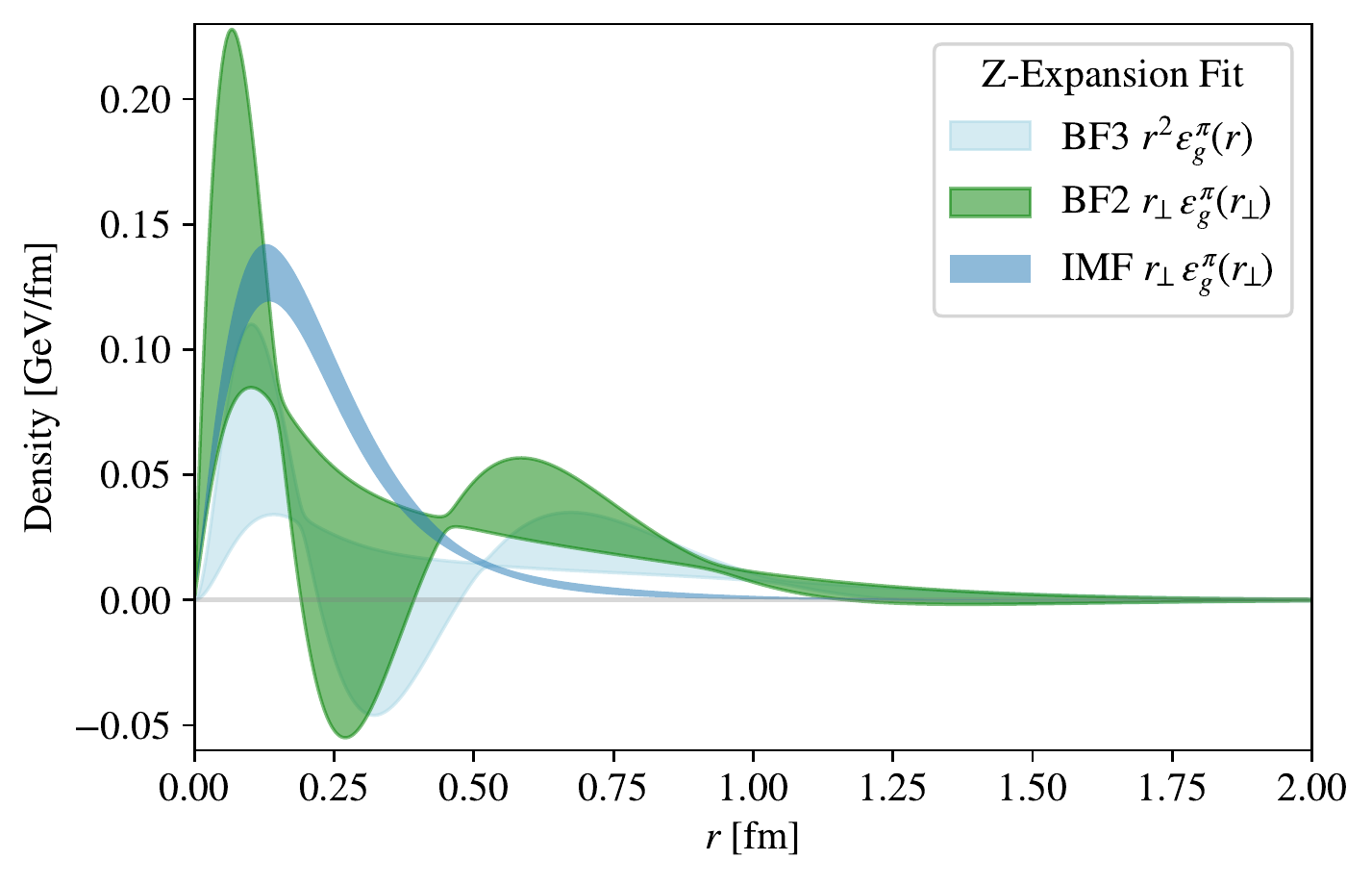} }}
\caption{Traceless gluon contributions to 
the pressure (a-b), shear force (c-d), and
energy (e-f) distributions of the pion
in the 3D BF, 2D BF, and in the IMF. 
The figures in the left (right) column
correspond to densities computed from the tripole (modified z-expansion) fits, with fit
parameters given in Table~\ref{tab:pion}.}
\label{fig:pionall}
\end{figure*}

From the pressure density, it is interesting 
to test whether the 3D and 2D von Laue stability conditions \cite{VonLaue} 
for the
total pressure of a composite particle
\begin{equation} \label{eq:vonLaue}
\int_0^{\infty} dr ~ r^2 ~ p_{\text{BF3}}(r) = 0, \quad  \int_0^{\infty}
dr_{\mybot} ~ r_{\mybot} ~ p_{\text{BF2/IMF}}(r_{\mybot}) = 0
\end{equation}
hold for the traceless gluon piece alone. 
Indeed, by numerical integration we find that the pressures are consistent with the
von Laue condition in all frames
and using both multipole and z-expansion functional forms to model the $t$ dependence of the GFFs. Another important
stability condition first shown in Ref.~\cite{Polyakov:2018zvc}
and recently extended in Ref.~\cite{Freese:2021czn} is
that for the total $D$-term
\begin{equation}
D(0) \leq 0 \;,
\end{equation}
which is satisfied by the gluon contribution to the pion $D$-term
in Table~\ref{tab:pion}.
We also find that the hadron stability conditions
\cite{Polyakov:2018zvc,Freese:2021qtb}
\begin{equation} \label{eq:mechstab}
\begin{split}
p_{\text{BF3}}(r)+\frac{2}{3}
s_{\text{BF3}}(r) > 0 \\
p_{\text{BF2/IMF}}(r_{\mybot})
+\frac{1}{2}s_{\text{BF2/IMF}}(r_{\mybot}) > 0
\end{split}
\end{equation}
hold for the traceless gluon piece of the pion pressure and shear force, which allows us to define
a partial gluon
mechanical mean square radius for the pion  $\braket{r^2_{\pi,g}}$ [see 
Eq.~\eqref{eq:mechrad}].
Our results for the mechanical radii as
well as the mass radii, defined in Appendix~\ref{sec:piondensities} 
as appropriate averages of the distance from
the center of the hadron weighted by the energy density, are
presented in Table~\ref{tab:allradii}.

\begin{table}[] 
\begin{center}
\begin{tabular}{SD{:}{}{2.7}D{:}{}{2.7}D{:}{}{2.7}D{:}{}{2.7}}
\toprule
\multicolumn{1}{c}{$\sqrt{\braket{r^2_{\pi,g}}}$ [fm]} & \multicolumn{1}{c}{BF3} & \multicolumn{1}{c}{BF2} & \multicolumn{1}{c}{$\qquad$ IMF $\qquad$} \\ \midrule
{Mech. tripole} & 0:.465(19) & 0:.380(16) & 0:.294(16) \\
{Mech. z-expansion} & 0:.42(36) & 0:.34(29) & 0:.305(30) \\
{Mass tripole} & 0:.435(25) & 0:.355(20) & 0:.1594(74) \\
{Mass z-expansion} & 0:.452(68) & 0:.369(56) & 0:.215(17) \\ \midrule \midrule
\multicolumn{1}{c}{$\sqrt{\braket{r^2_{N,g}}}$ [fm]} & \multicolumn{1}{c}{} & \multicolumn{1}{c}{} & \multicolumn{1}{c}{} \\ \midrule
{Mech. tripole} & 0:.631(68) & 0:.517(57) & 0:.517(57) \\
{Mass tripole} & 0:.382(33) & 0:.312(27) & 0:.213(14) \\
{Mass z-expansion} & 0:.27(11) & 0:.217(91) & 0:.238(38) \\
\midrule \midrule
\multicolumn{1}{c}{$\sqrt{\braket{r^2_{\rho,g}}}$ [fm]} & \multicolumn{1}{c}{} & \multicolumn{1}{c}{} & \multicolumn{1}{c}{} \\ \midrule
{Mech. tripole} & 0:.278(54) & 0:.227(35) &  \\
{Mass tripole} & 0:.371(19) & 0:.285(15) & 0:.216(10) \\
{Mass z-expansion} & 0:.329(37) & 0:.248(32) & 0:.215(10) \\
\midrule \midrule
\multicolumn{1}{c}{$\sqrt{\braket{r^2_{\Delta,g}}}$ [fm]} & \multicolumn{1}{c}{} & \multicolumn{1}{c}{} & \multicolumn{1}{c}{} \\ \midrule
{Mech. tripole} & 0:.588(96) & 0:.471(81) & 0:.503(77) \\
{Mass tripole} & 0:.387(33) & 0:.289(28) & 0:.239(16) \\
{Mass z-expansion} & 0:.20(15) & 0:.16(23) & 0:.235(12) \\
\bottomrule
\end{tabular}
\end{center}
\caption{Conserved gluon contributions to the
mechanical and mass radii of the four hadrons, as defined in Eq.~\eqref{eq:mechrad} and
Appendix~\ref{sec:density-defs}. For the $\Delta$ and $\rho$, the contributions
from GFFs that were not fit in Sec.~\ref{sec:gffs-from-lqcd} are neglected.
Only the mechanical radii corresponding to frames and models for which
the mechanical stability requirement of
Eq.~\ref{eq:mechstab} is satisfied are shown.}
\label{tab:allradii}
\end{table}

\subsection{Nucleon}

\begin{figure*}[h!]
\centering
\subfloat[\centering  ]
{{\includegraphics[height=5.8cm,width=7.8cm]{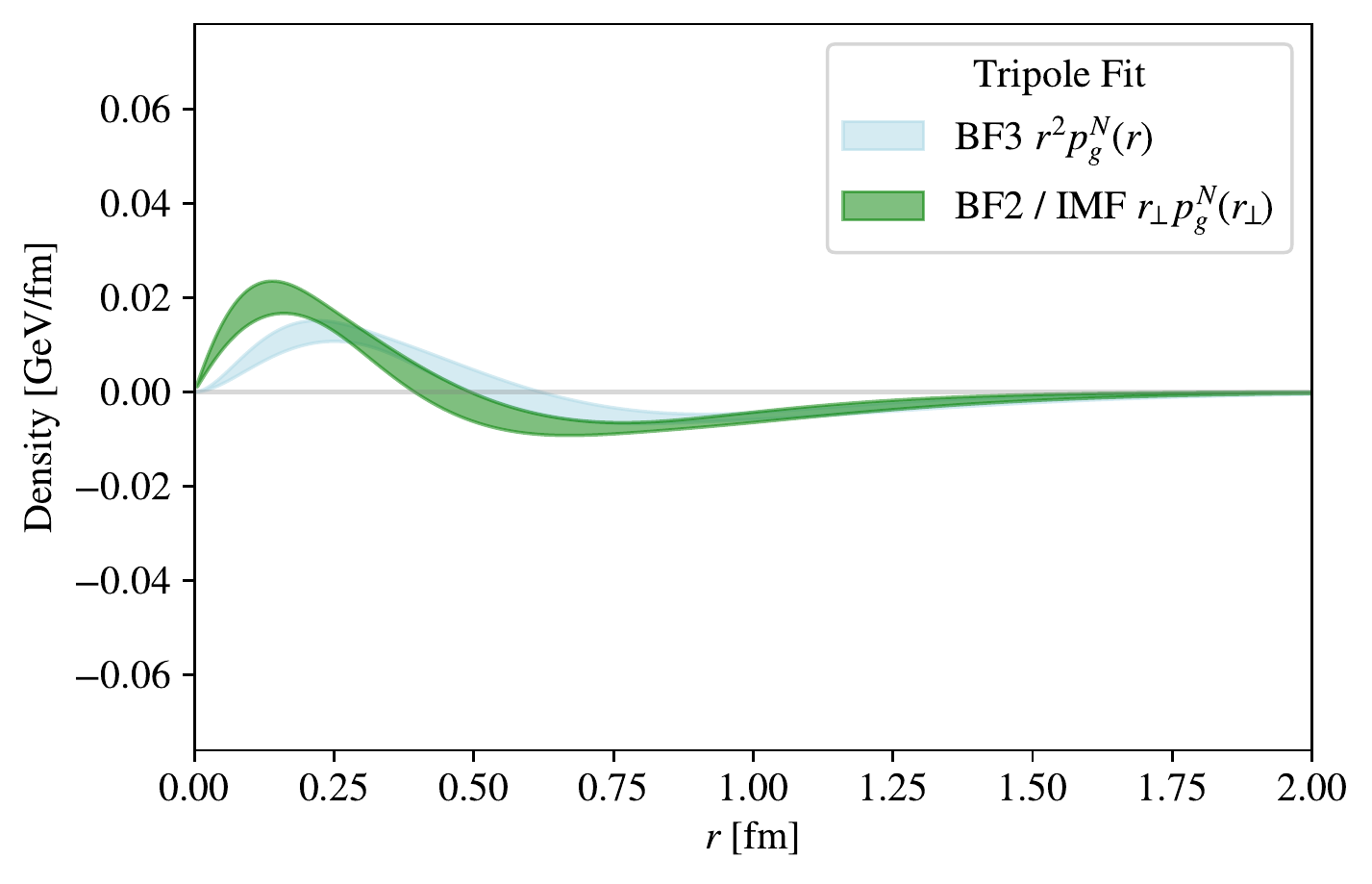} }}
\!
\subfloat[\centering  ]
{{\includegraphics[height=5.8cm,width=7.8cm]{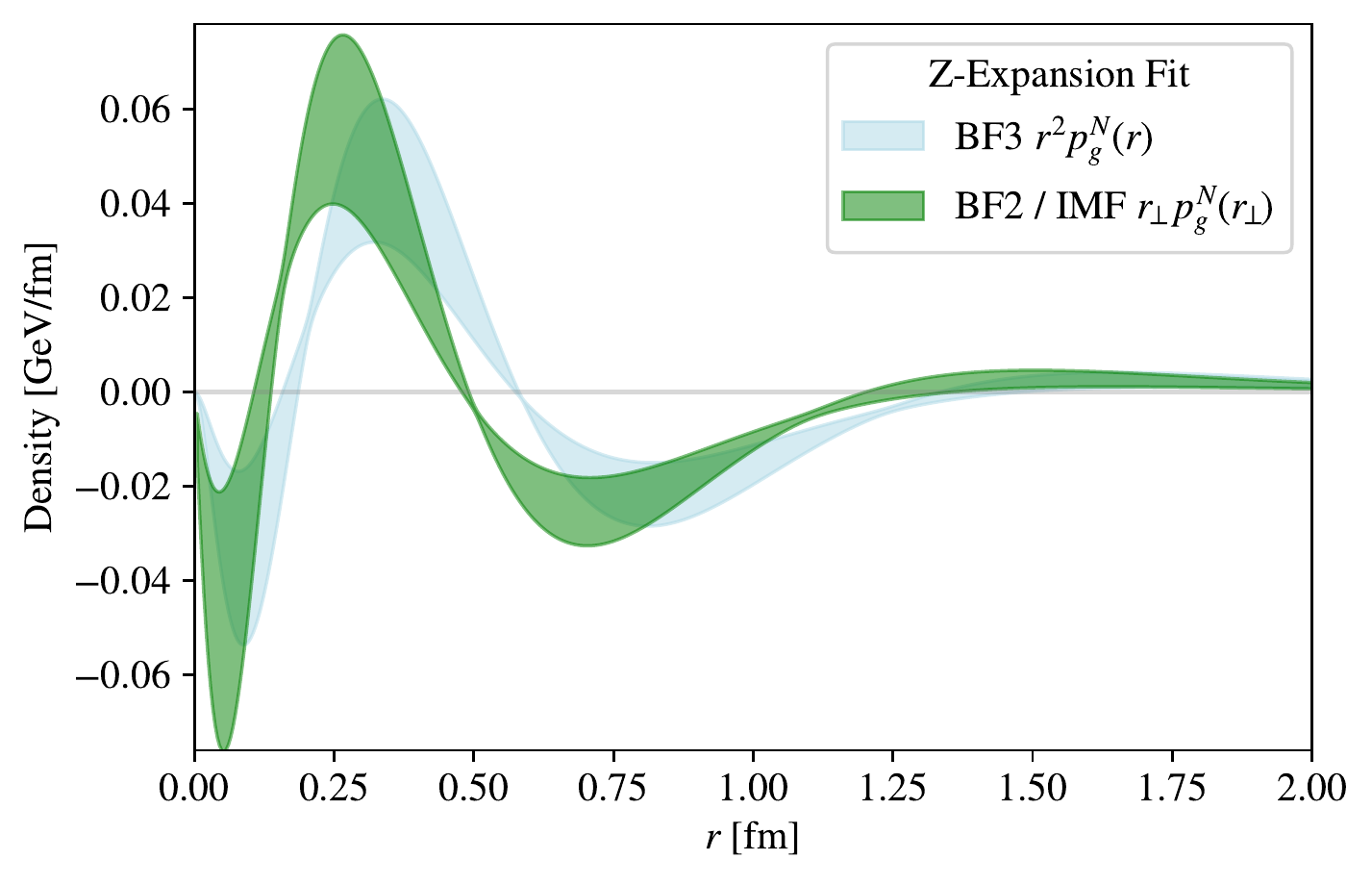} }}
\\
\subfloat[\centering  ]
{{\includegraphics[height=5.8cm,width=7.8cm]{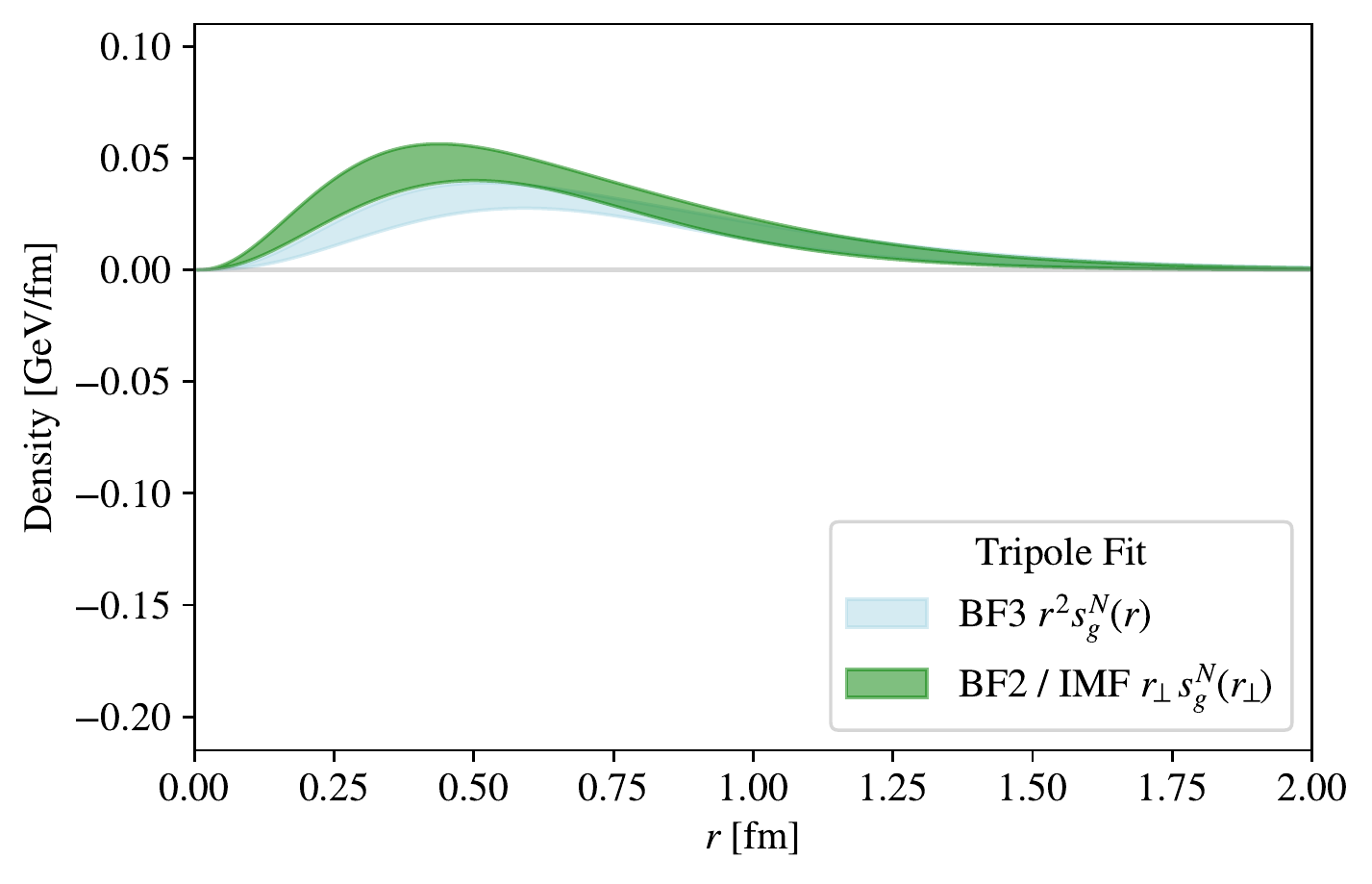} }}
\!
\subfloat[\centering  ]
{{\includegraphics[height=5.8cm,width=7.8cm]{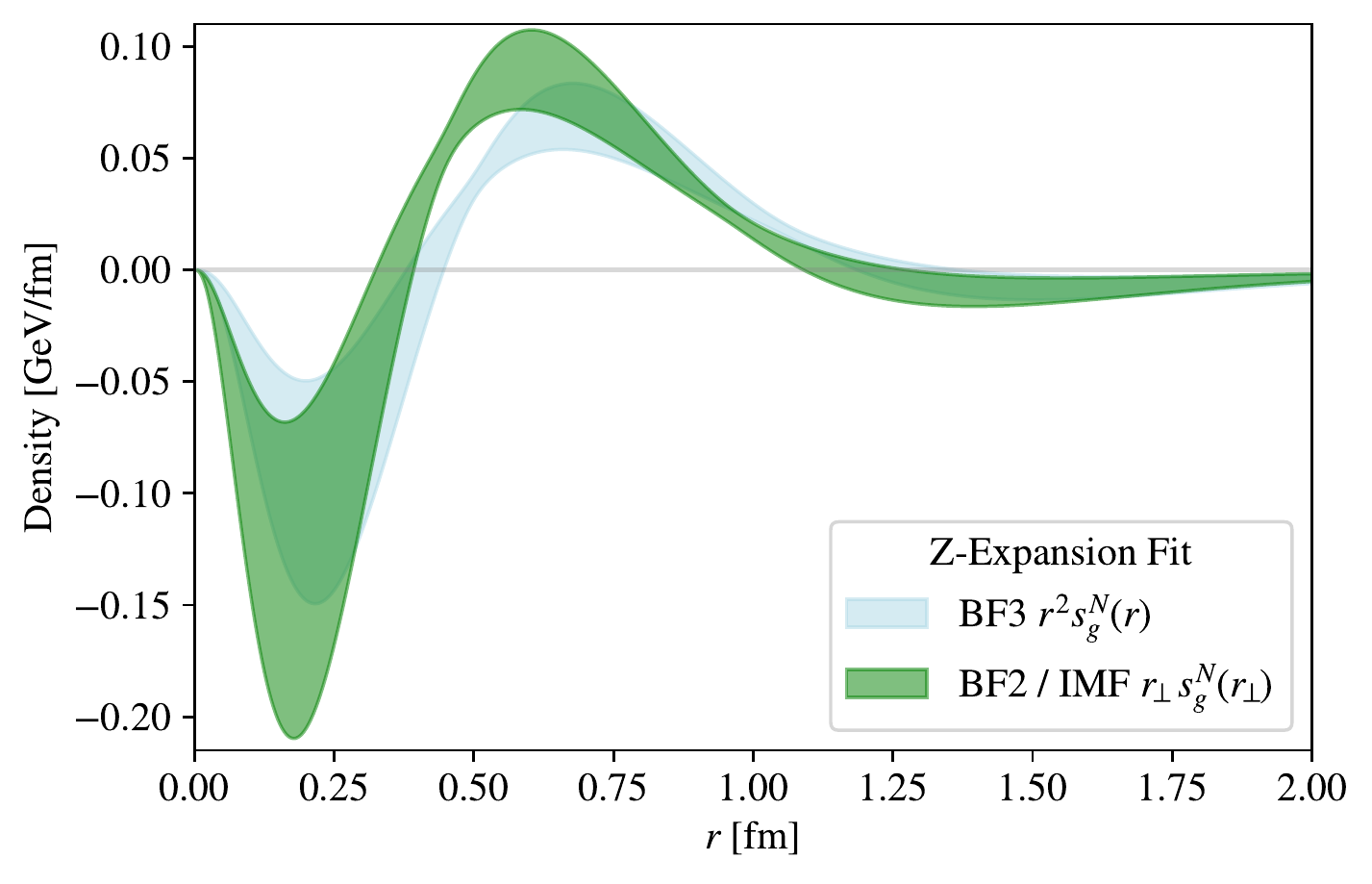} }}
\\
\subfloat[\centering  ]
{{\includegraphics[height=5.8cm,width=7.8cm]{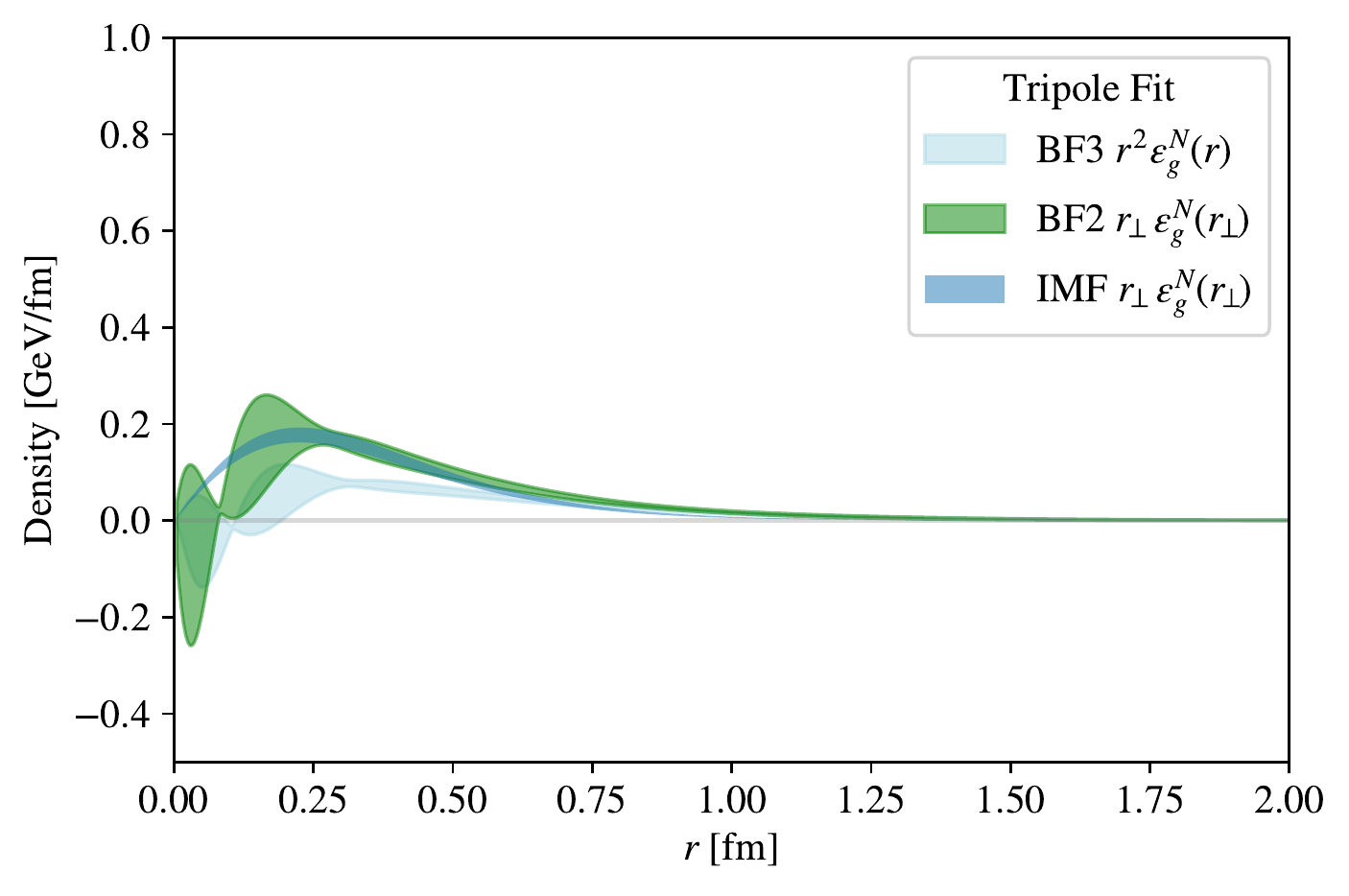} }}
\!
\subfloat[\centering  ]
{{\includegraphics[height=5.8cm,width=7.8cm]{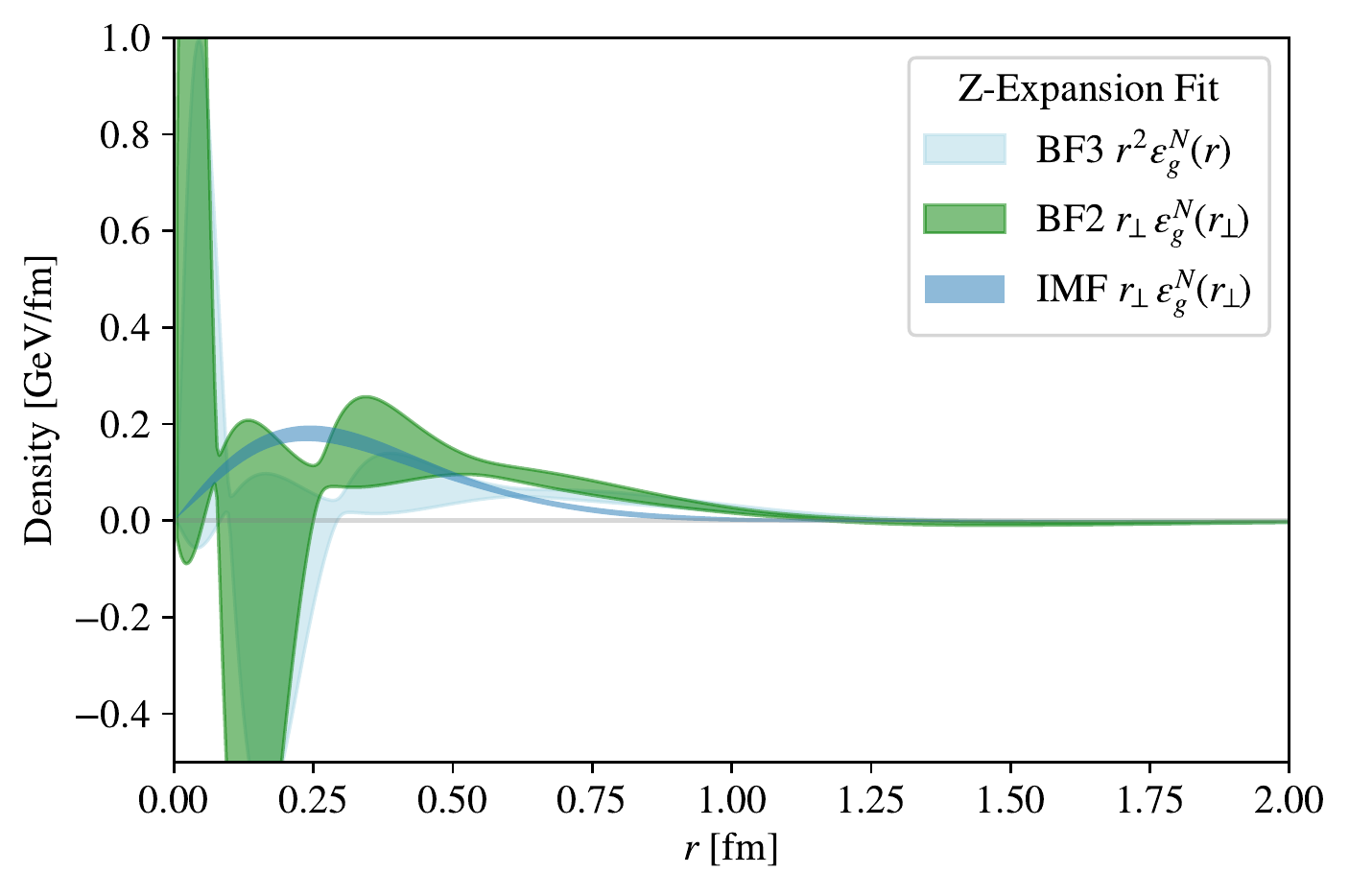} }}
\caption{Traceless gluon contributions to 
the pressure (a-b), shear force (c-d), and
energy (e-f) distributions of the nucleon
in the 3D BF, 2D BF, and in the IMF. 
The figures in the left (right) column
correspond to the tripole (z-expansion) fits, with fit
parameters in Table~\ref{tab:nuc}. The
expressions for the pressure and shear force
are identical in the 2D BF
and the IMF.}
\label{fig:nucall}
\end{figure*}

The BF and IMF densities of the nucleon EMT have been studied previously
\cite{Lorce:2018egm,Shanahan:2018nnv,Burkert:2018bqq,Polyakov:2018zvc}
and are defined in Appendix~\ref{sec:nucdensities}.
We note that the 2D Breit frame pressure and shear force coincide
with their IMF equivalents.
Our results for the symmetric traceless gluon contributions to the densities 
are shown in Fig.~\ref{fig:nucall}. The
difference between the results based on tripole and z-expansion fits to the GFFs is
due to the difference between the two fits
of $D_g^N(t)$ in the low $-t$ region, as discussed in
Sec.~\ref{sec:nucGFFs}. A nonmonotonic gluon $D_g^N(t)$
causes the traceless gluon pressure to 
have two nodes, which is different than the form of
the quark pressure distribution as found in Ref.~\cite{Burkert:2018bqq}. 
Future
lattice, experimental, and phenomenological extractions of
$D^N_g(t)$ and $D^N_q(t)$ will help to clarify
the picture.

From the nucleon density results, we find that the pressures in all frames and models
are consistent with the von Laue condition~\eqref{eq:vonLaue}; however, Eq.~\eqref{eq:mechstab} is only satisfied
for the tripole fit. Moreover, the $D$-term fit by the z-expansion is not strictly positive within uncertainty, and thus
we only present the 
mechanical 
radius of Eq.~\eqref{eq:mechrad} for the
tripole fit.
The mass radii definitions are shown in Appendix~\ref{sec:nucdensities},
and the corresponding numerical results of all radii are presented in Table~\ref{tab:allradii}.

\subsection{$\rho$ Meson}

Beyond the lowest-order energy, pressure, and shear force densities  
of the pion and the nucleon, the
structure of hadrons of spin $=1$ or higher depends on additional quadrupole densities. 
The BF3 densities and mass radii of the $\rho$ meson were derived in Refs.~\cite{Polyakov:2019lbq,Sun:2020wfo} and are listed in Appendix~\ref{sec:rhodensities}. We also derive expressions for the lowest-order BF2 and 
IMF distributions. 

\begin{figure*}[p!]
\centering
\subfloat[\centering  ]
{{\includegraphics[height=5.8cm,width=7.8cm]{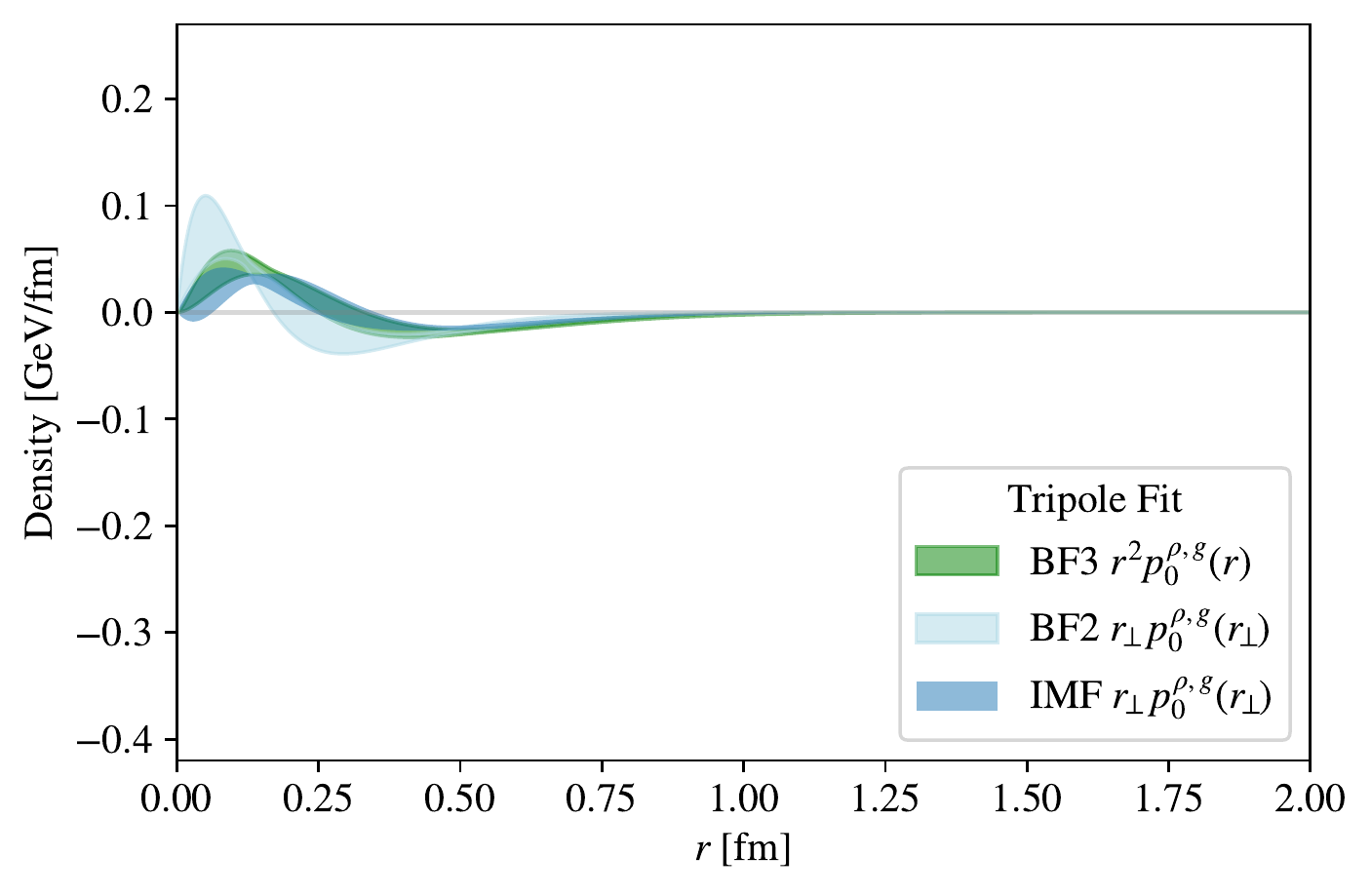} }}
\!
\subfloat[\centering  ]
{{\includegraphics[height=5.8cm,width=7.8cm]{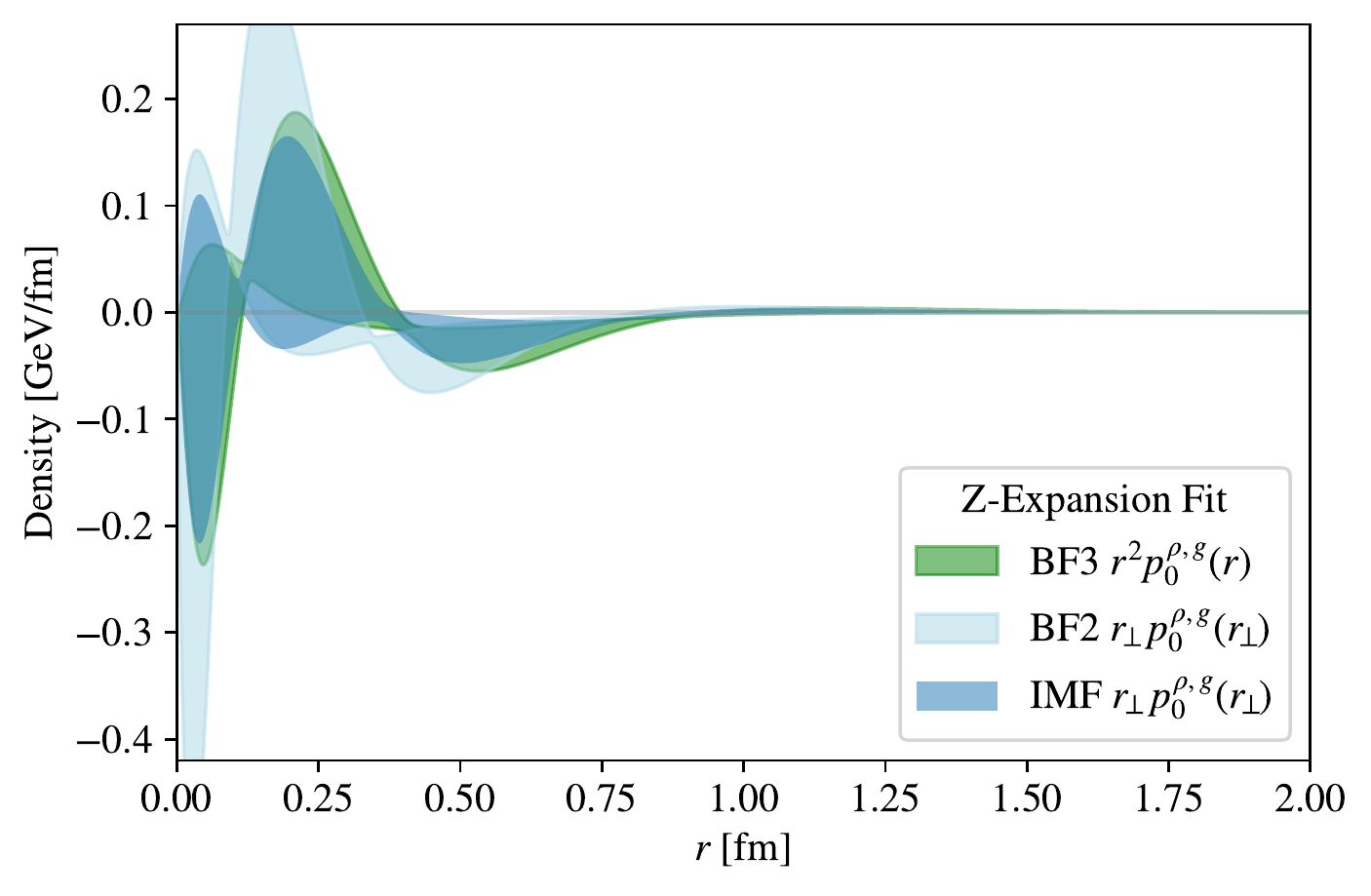} }}
\\
\subfloat[\centering  ]
{{\includegraphics[height=5.8cm,width=7.8cm]{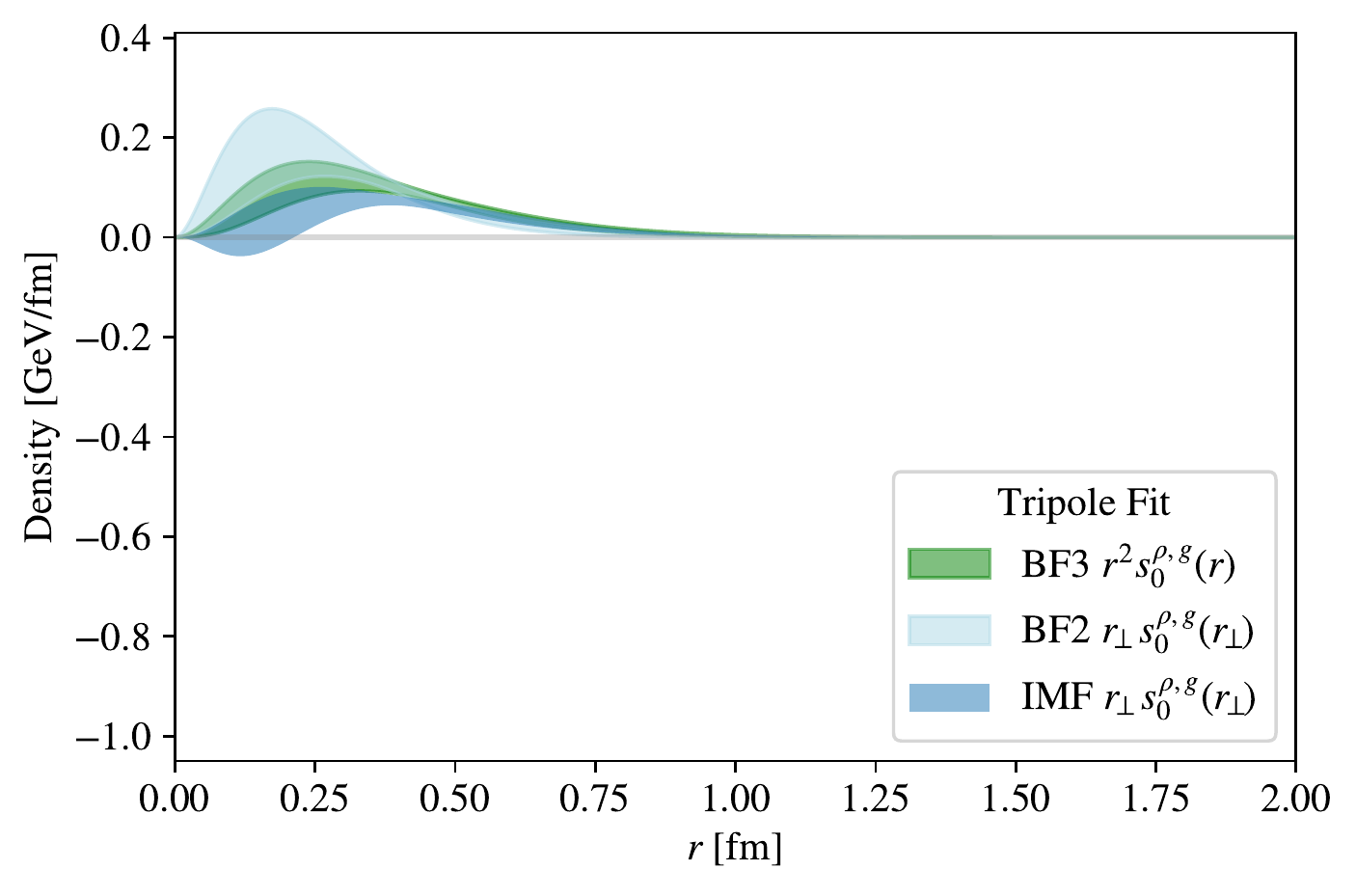} }}
\!
\subfloat[\centering  ]
{{\includegraphics[height=5.8cm,width=7.8cm]{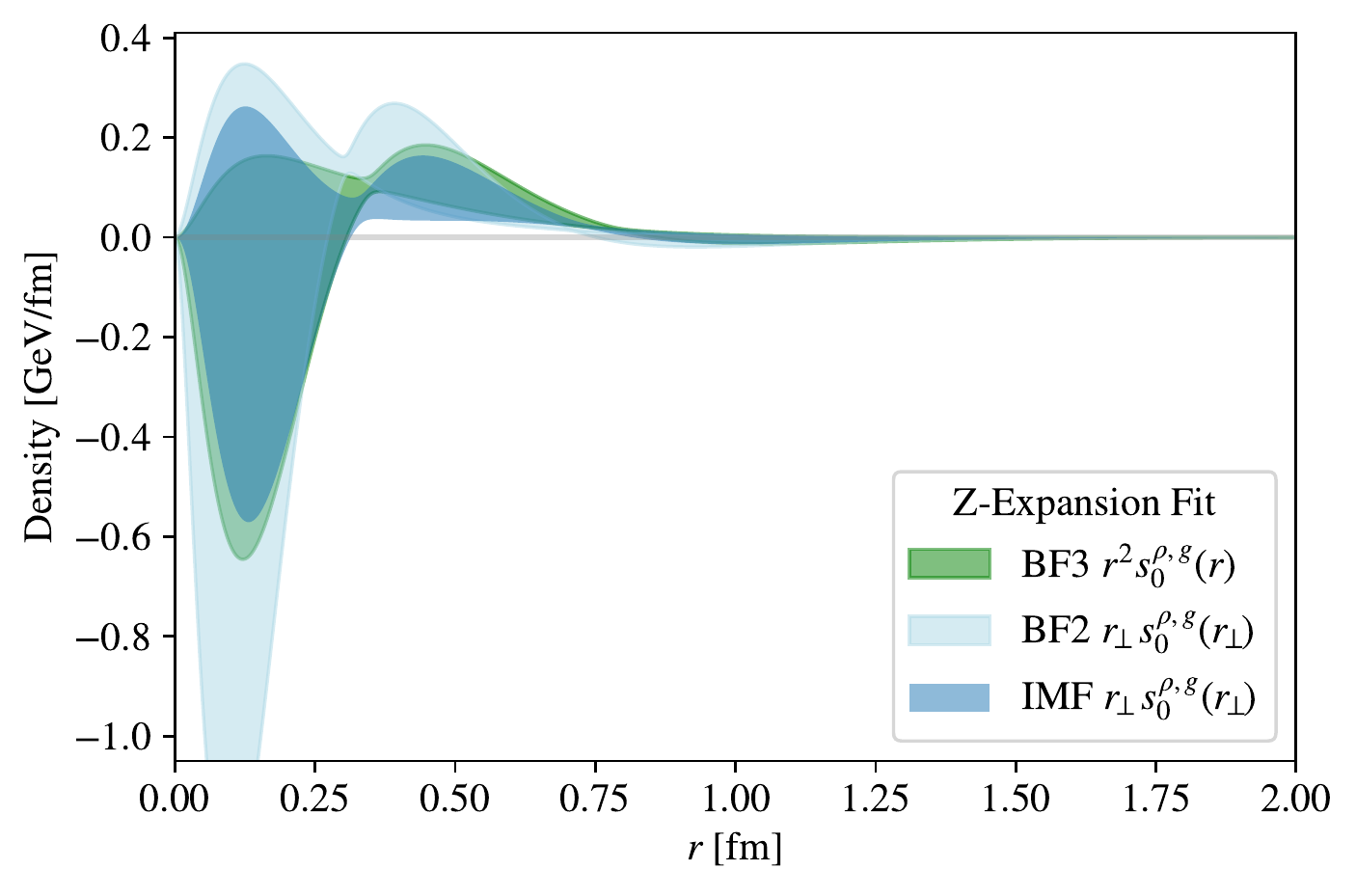} }}
\\
\subfloat[\centering  ]
{{\includegraphics[height=5.8cm,width=7.8cm]{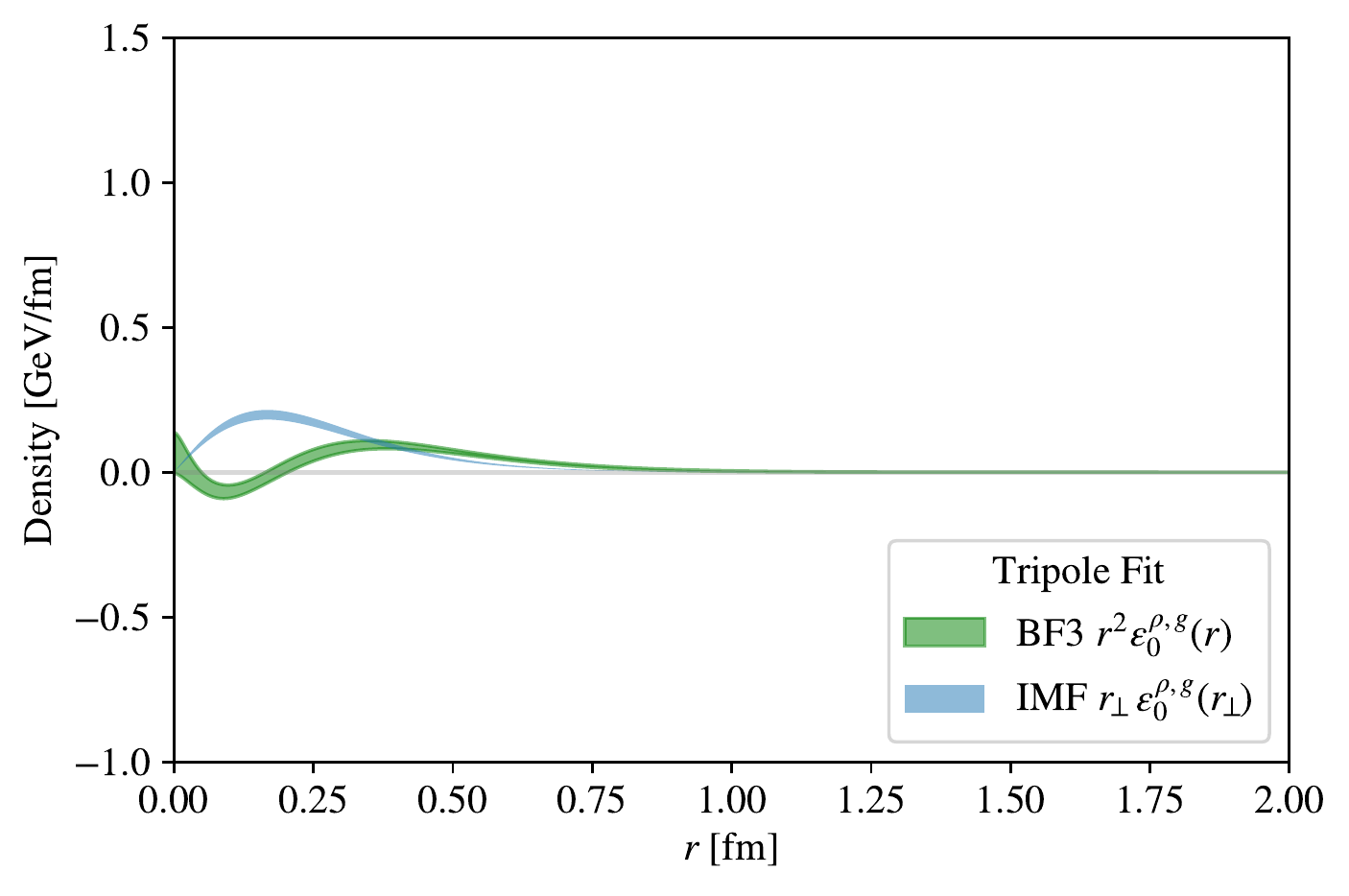} }}
\!
\subfloat[\centering  ]
{{\includegraphics[height=5.8cm,width=7.8cm]{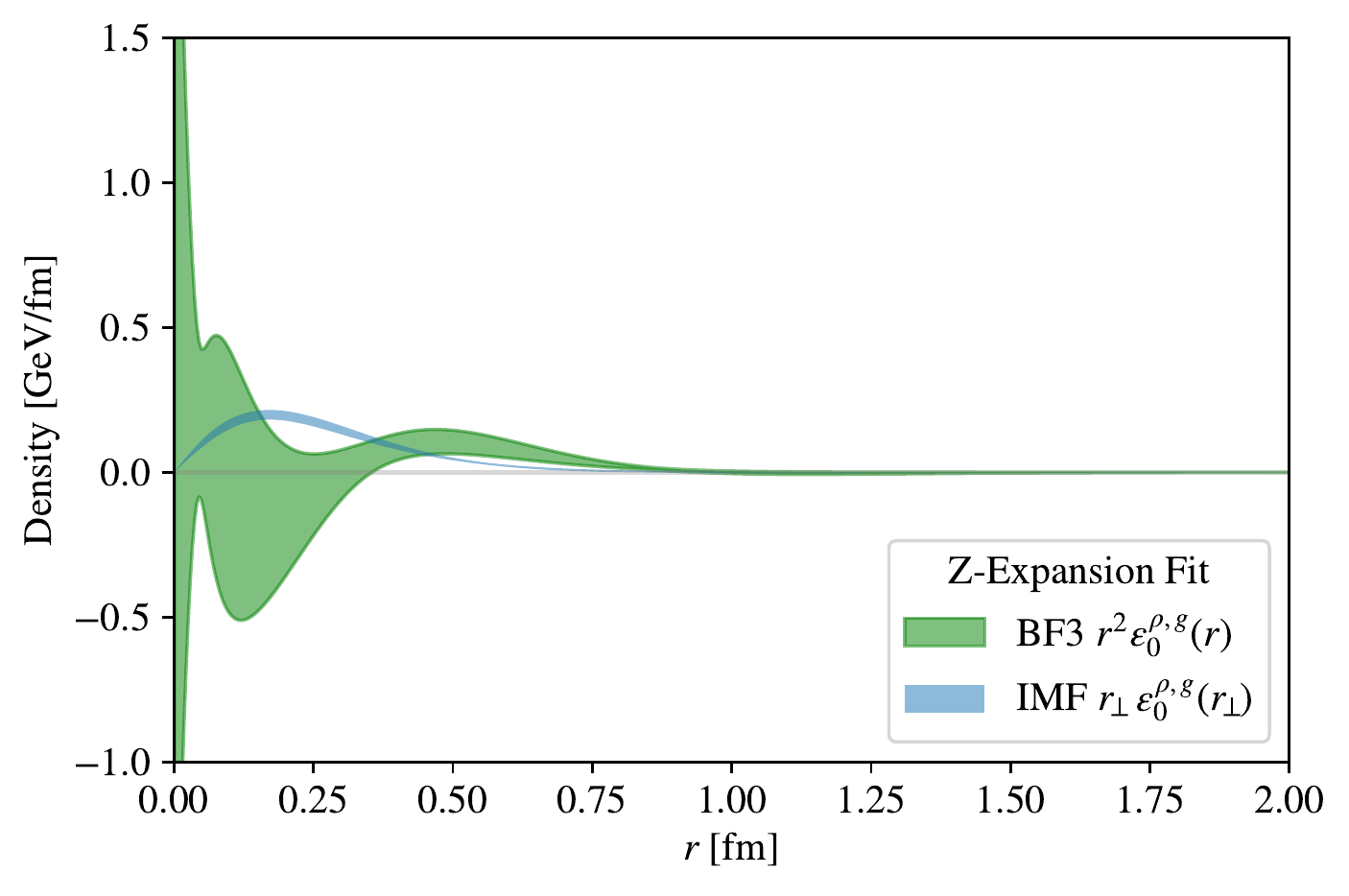} }}
\caption{Traceless (partial) gluon 
contributions to
the lowest-order pressure (a-b), shear force (c-d), and
energy (e-f) distributions of the $\rho$ meson
in the 3D BF, 2D BF, and in the IMF. 
The figures in the left (right) column
correspond to the tripole (z-expansion) fits, with fit
parameters in Table~\ref{tab:rho}. The BF2 contribution to the energy density is not constrained within the range of the figure axes.}
\label{fig:rhodens1}
\end{figure*}

\begin{figure*} [p!]
\centering
\subfloat[\centering  ]
{{\includegraphics[height=5.8cm,width=7.8cm]{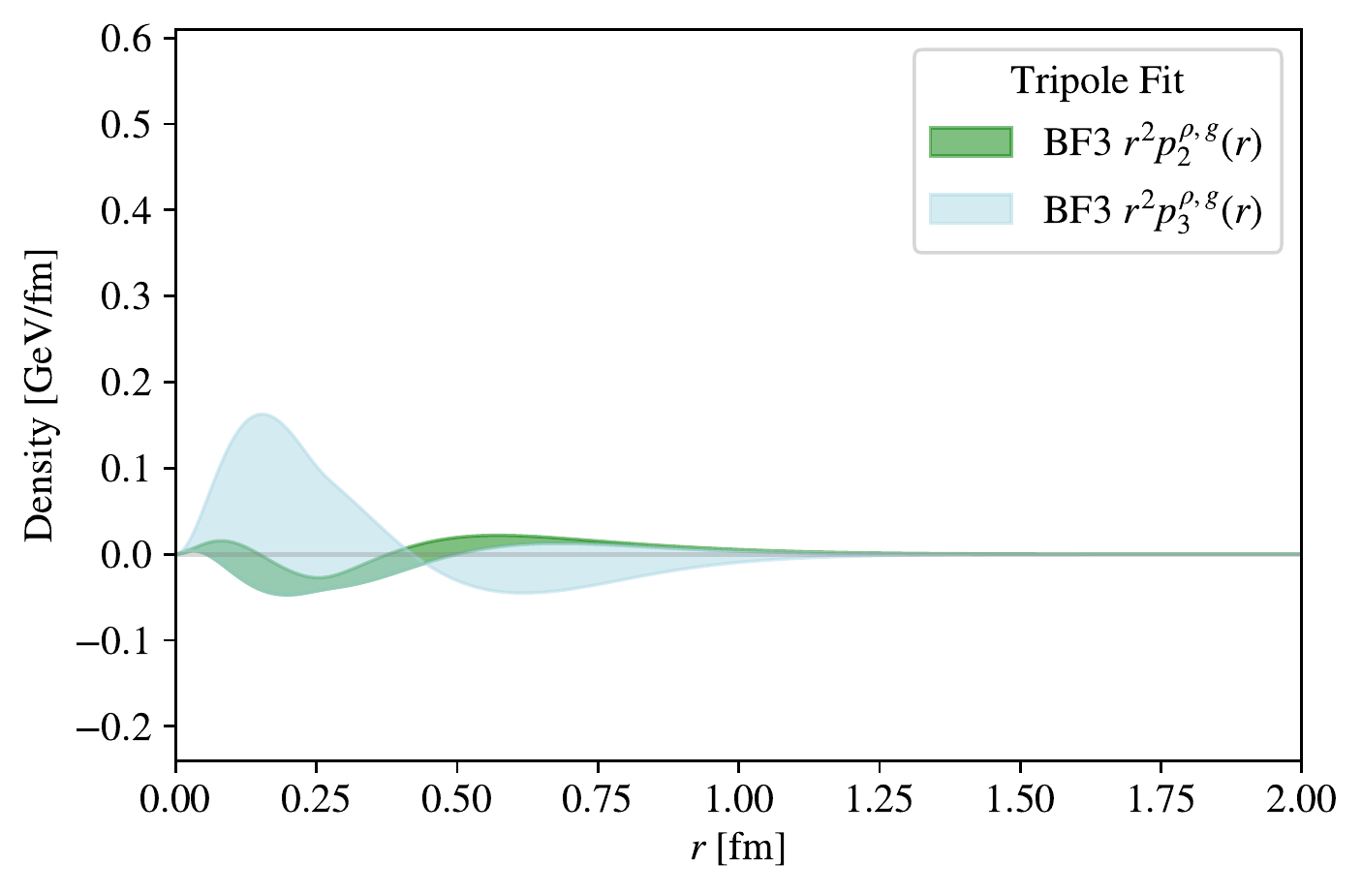} }}
\!
\subfloat[\centering  ]
{{\includegraphics[height=5.8cm,width=7.8cm]{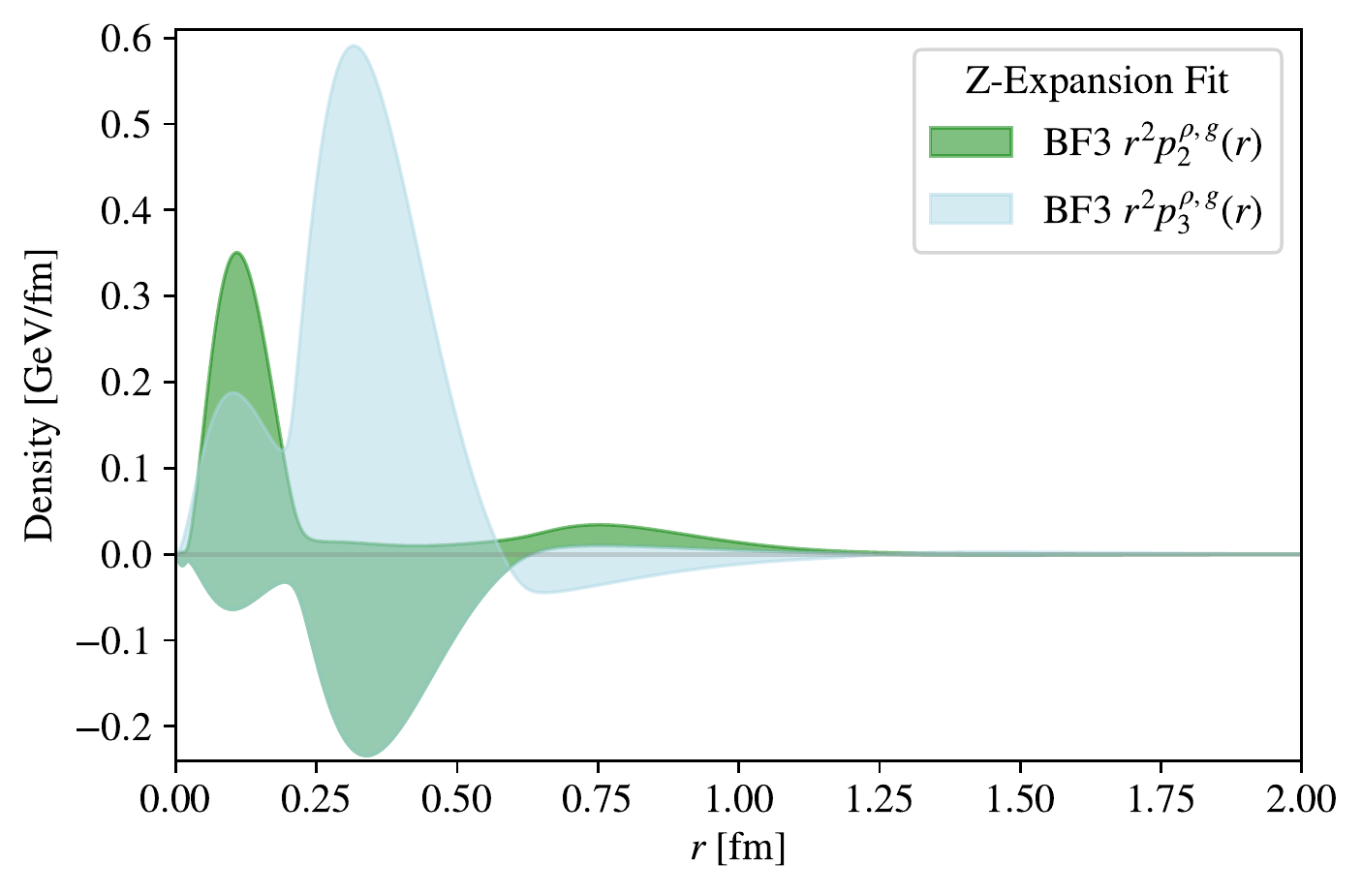} }}
\\
\subfloat[\centering  ]
{{\includegraphics[height=5.8cm,width=7.8cm]{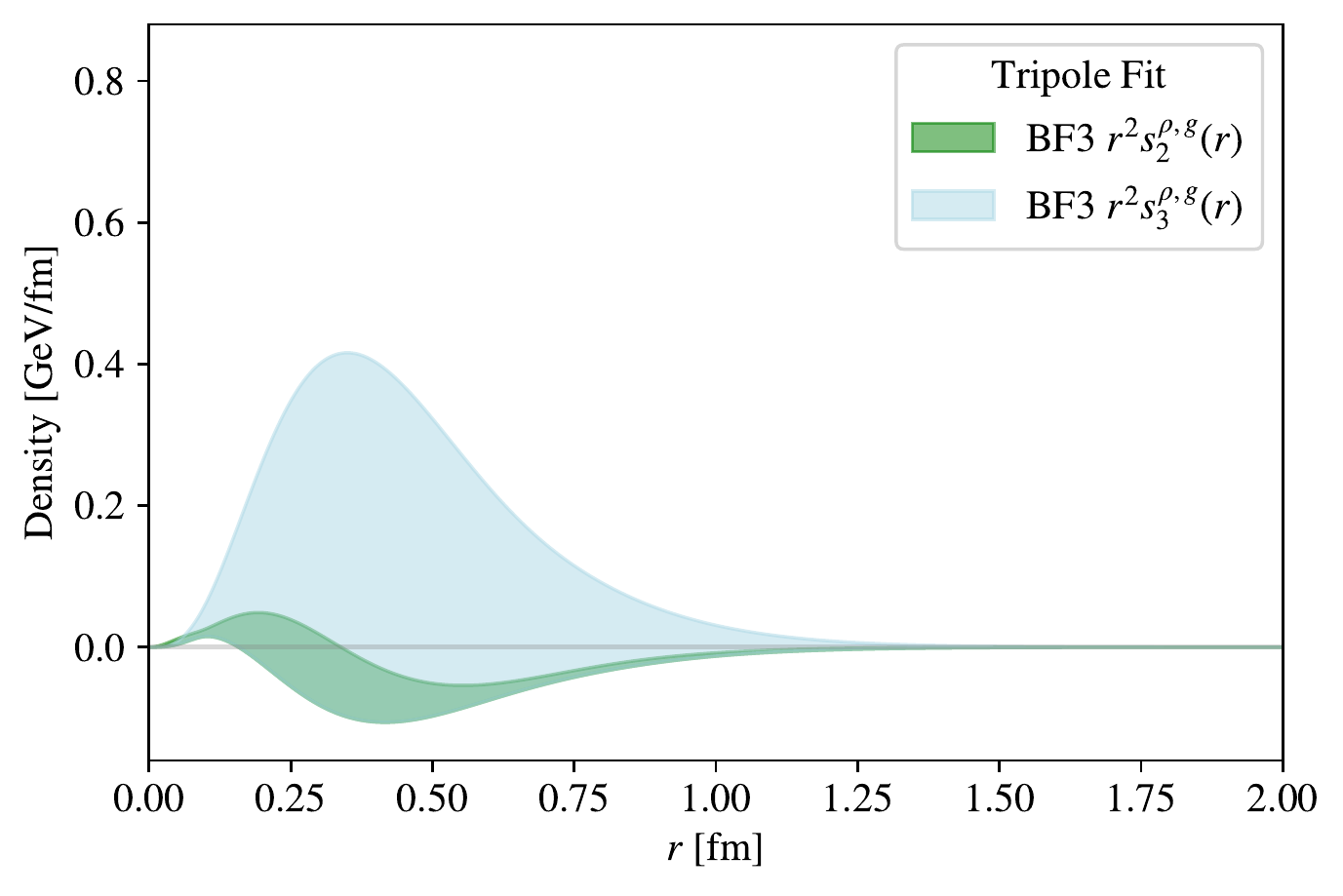} }}
\!
\subfloat[\centering  ]
{{\includegraphics[height=5.8cm,width=7.8cm]{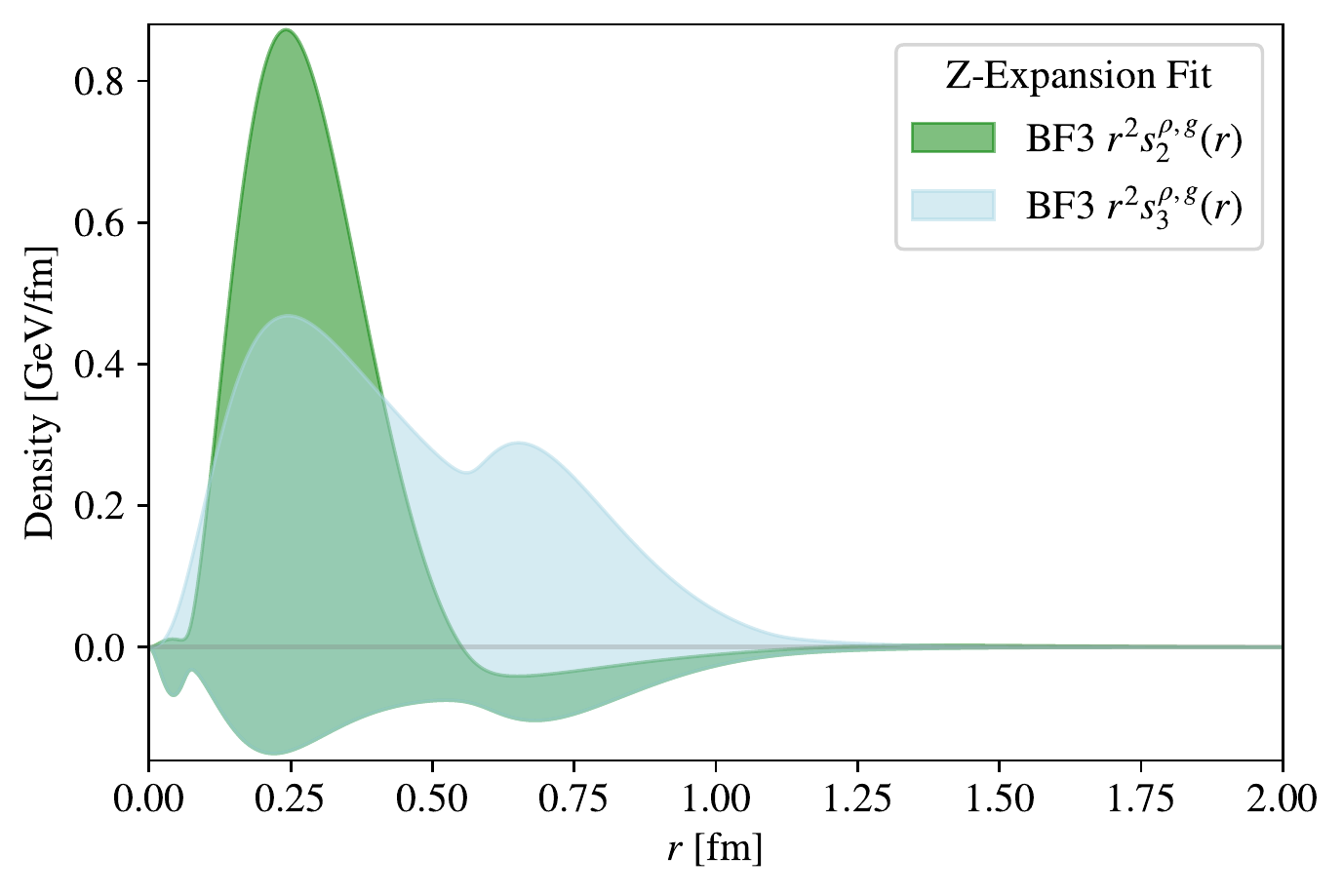} }}
\\
\subfloat[\centering  ]
{{\includegraphics[height=5.8cm,width=7.8cm]{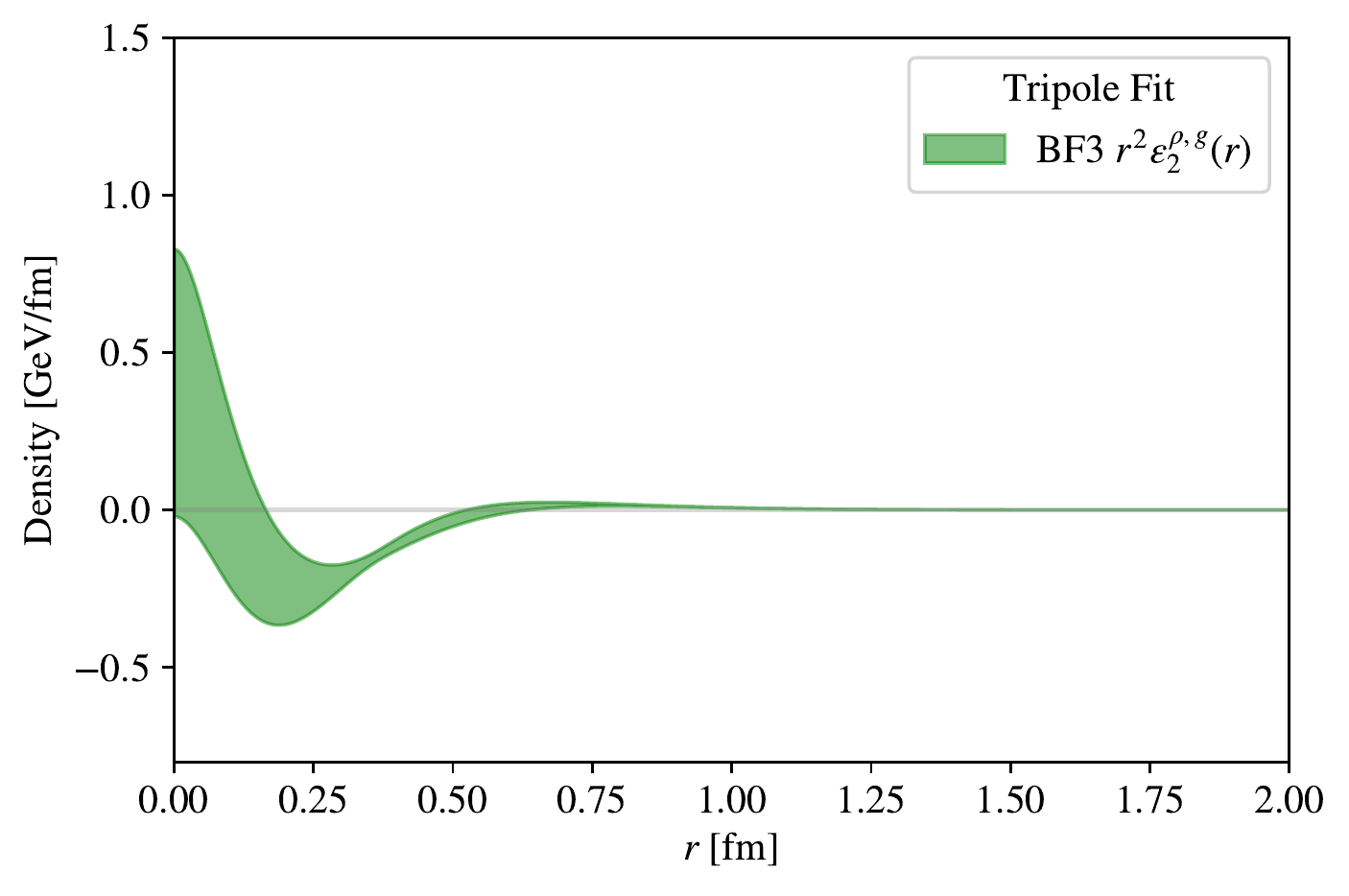} }}
\!
\subfloat[\centering  ]
{{\includegraphics[height=5.8cm,width=7.8cm]{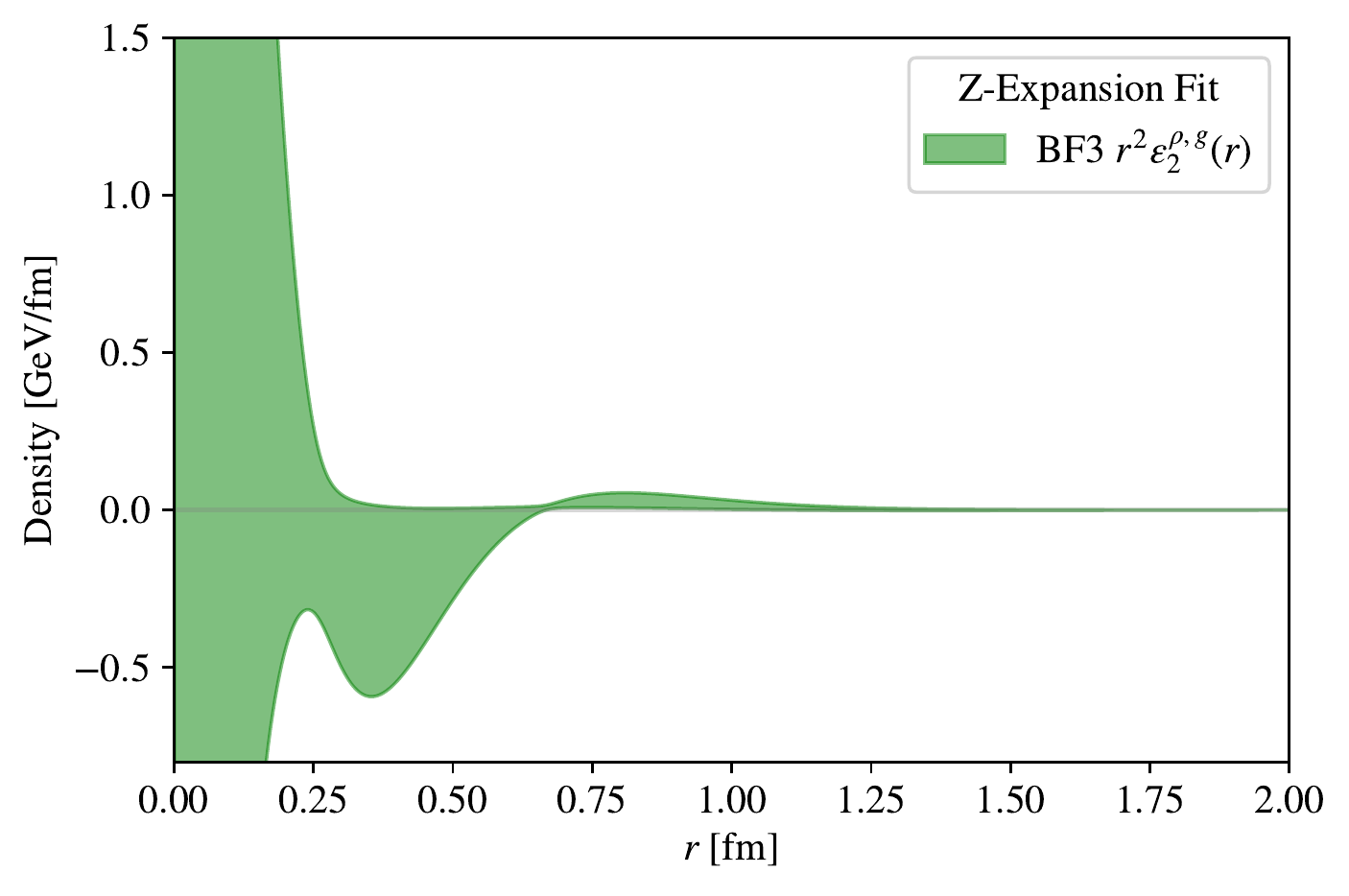} }}
\caption{Traceless (partial) gluon 
contributions to
the quadrupole order pressure (top), shear force (middle), and
energy (bottom) distributions of the $\rho$ meson
in the 3D BF. 
The figures in the left (right) column
correspond to the tripole (z-expansion) fits, with fit
parameters in Table~\ref{tab:rho}.}
\label{fig:rhodens2}
\end{figure*}

Our numerical results for the lowest-order (Fig.~\ref{fig:rhodens1}) and higher-order densities (Fig.~\ref{fig:rhodens2}) are partial
and exclude terms depending on $D_1^{\rho,g}(t)$, since its signal is not well modeled by the ans{\"a}tze considered here, as discussed
in Sec.~\ref{sec:rhoGFFs}. We note that almost all of the quadrupole densities are
poorly constrained, and therefore we only
consider the mechanical stability conditions
for the lowest-order densities.
In particular, Eq.~\eqref{eq:mechstab} only holds for the BF3 and BF2 from the tripole GFFs fits. The corresponding mechanical radii of Eq.~\eqref{eq:mechrad} and the mass radii are shown in Table~\ref{tab:allradii}.

\begin{figure*}[p!]
\centering
\subfloat[\centering  ]
{{\includegraphics[height=5.8cm,width=7.8cm]{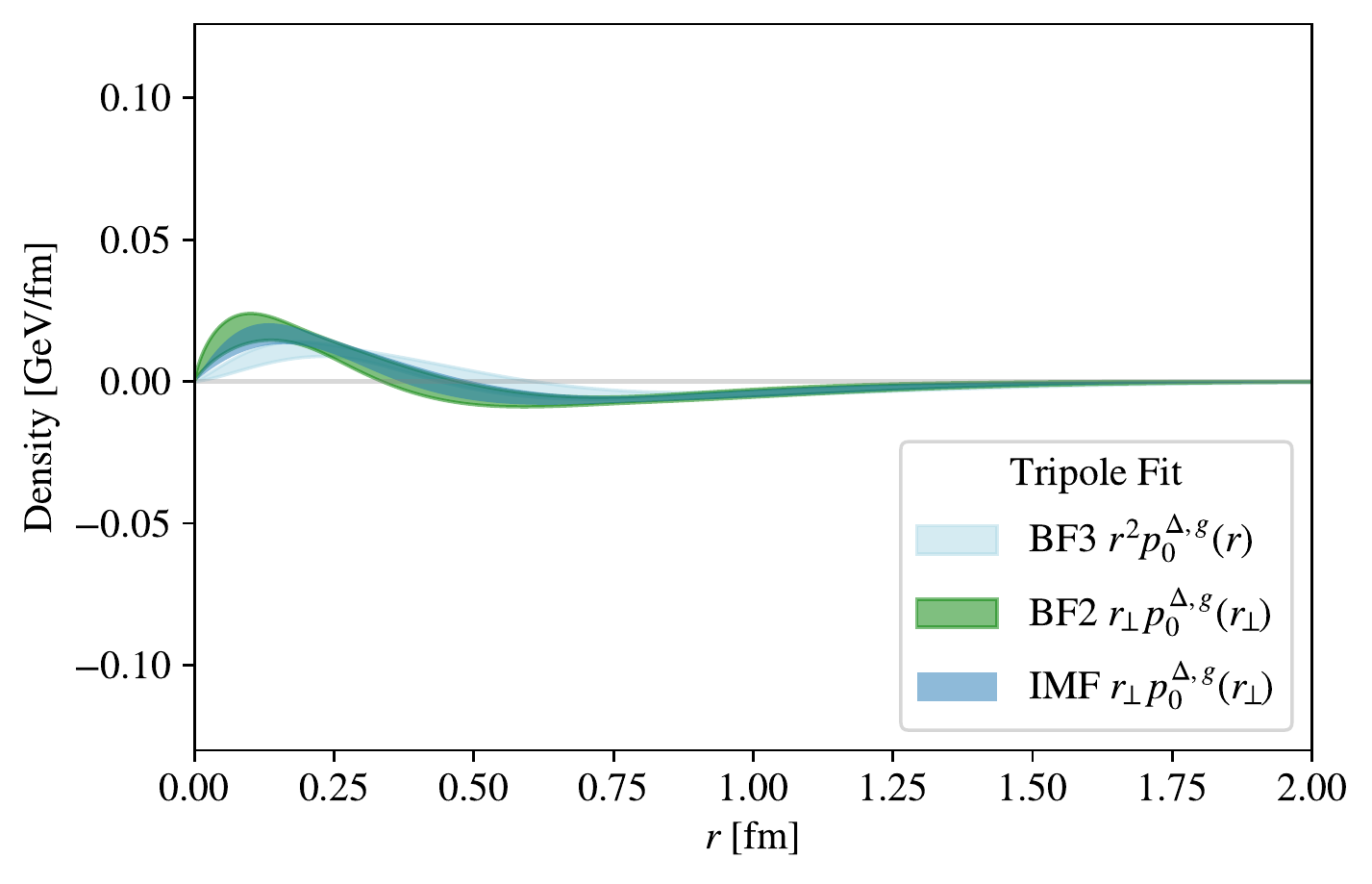} }}
\!
\subfloat[\centering  ]
{{\includegraphics[height=5.8cm,width=7.8cm]{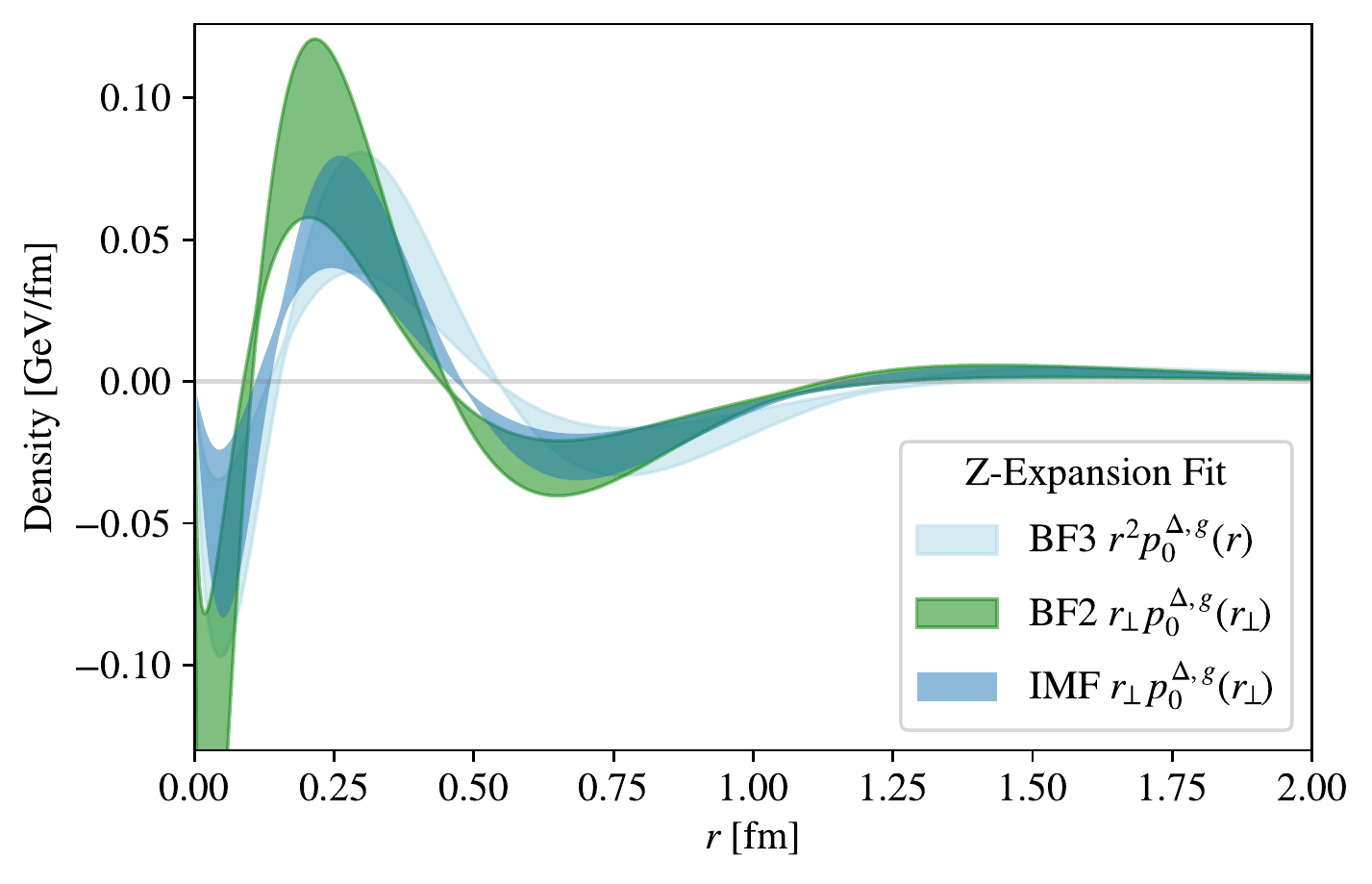} }}
\\
\subfloat[\centering  ]
{{\includegraphics[height=5.8cm,width=7.8cm]{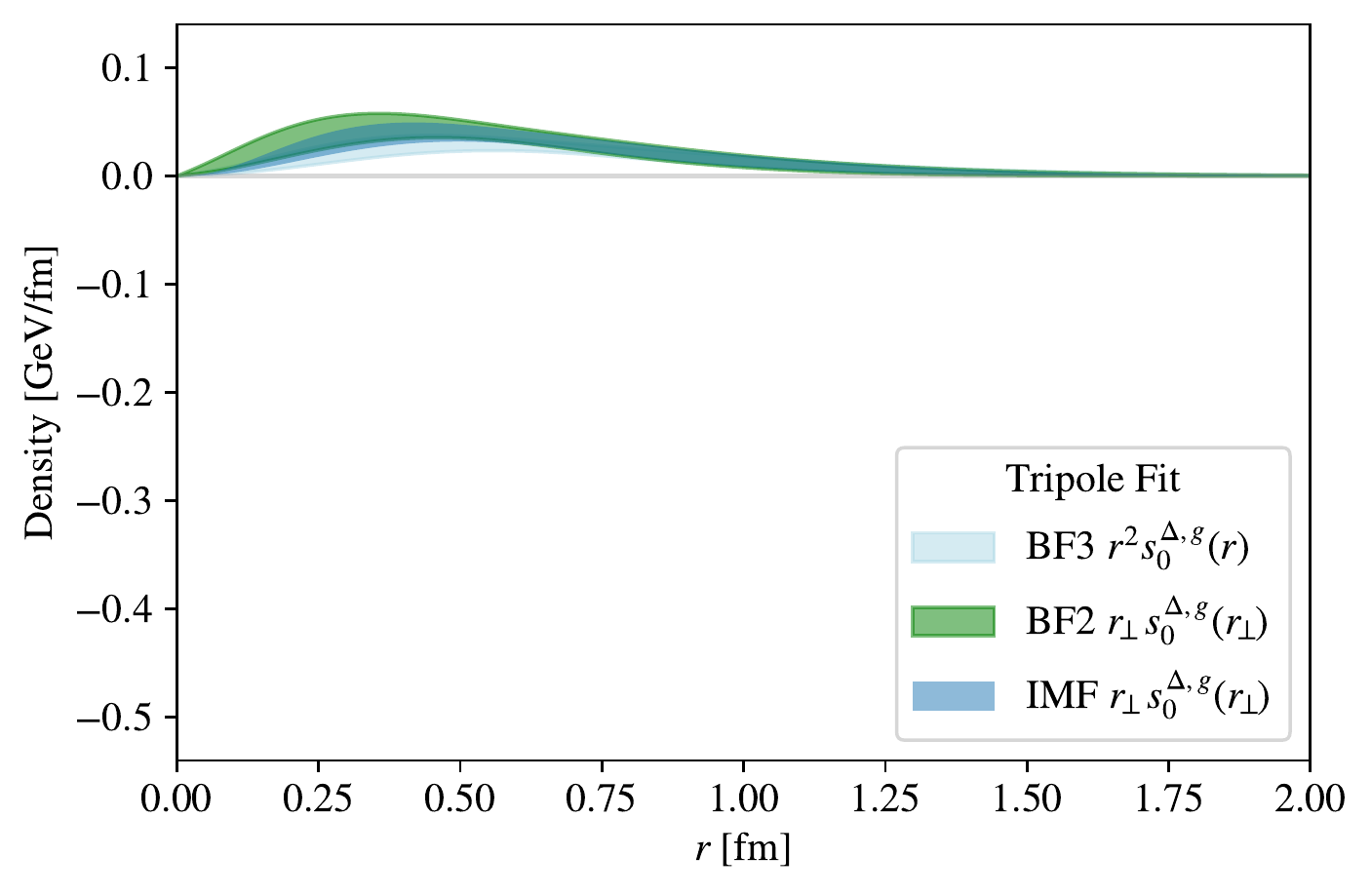} }}
\!
\subfloat[\centering  ]
{{\includegraphics[height=5.8cm,width=7.8cm]{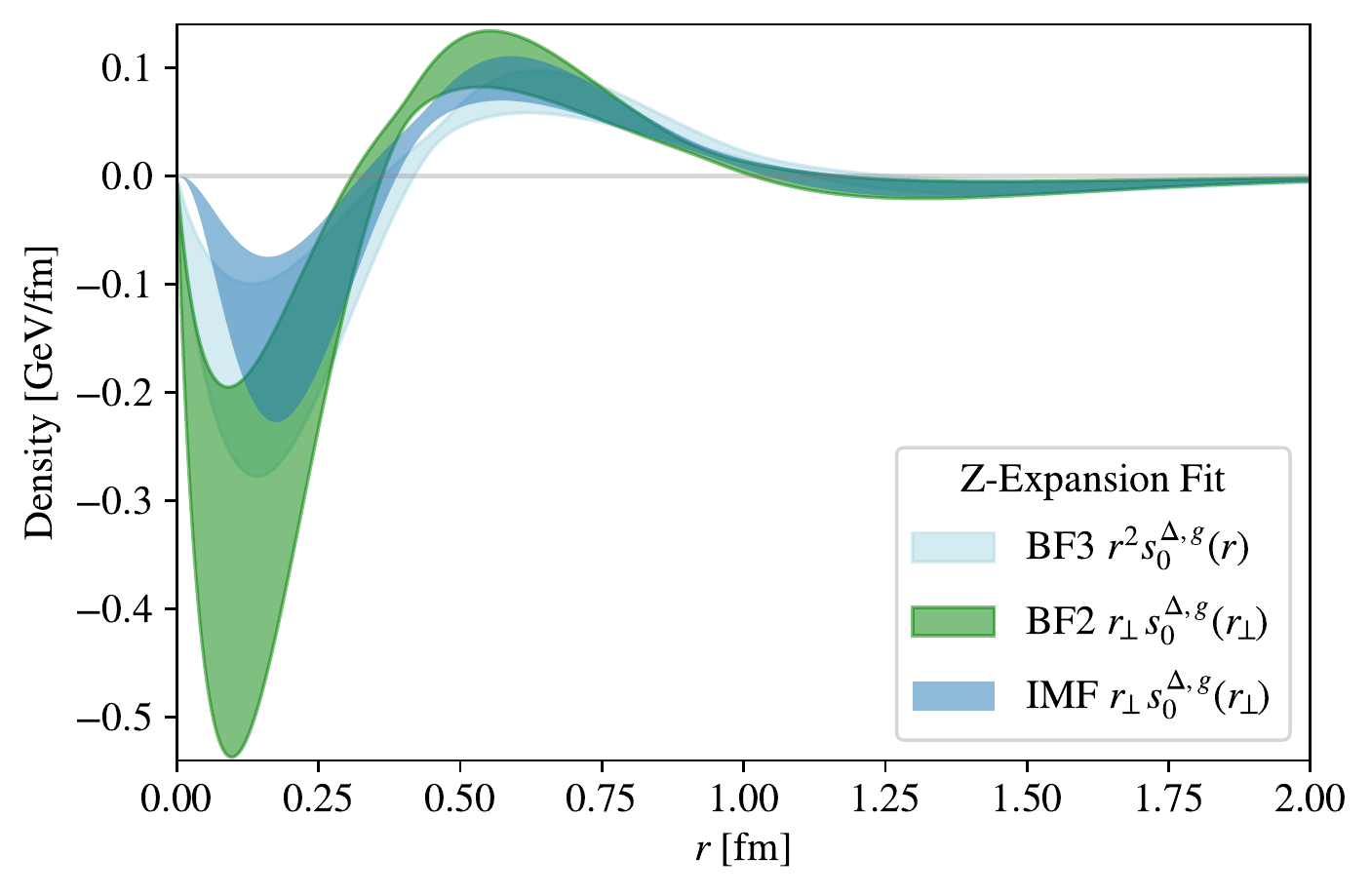} }}
\\
\subfloat[\centering  ]
{{\includegraphics[height=5.8cm,width=7.8cm]{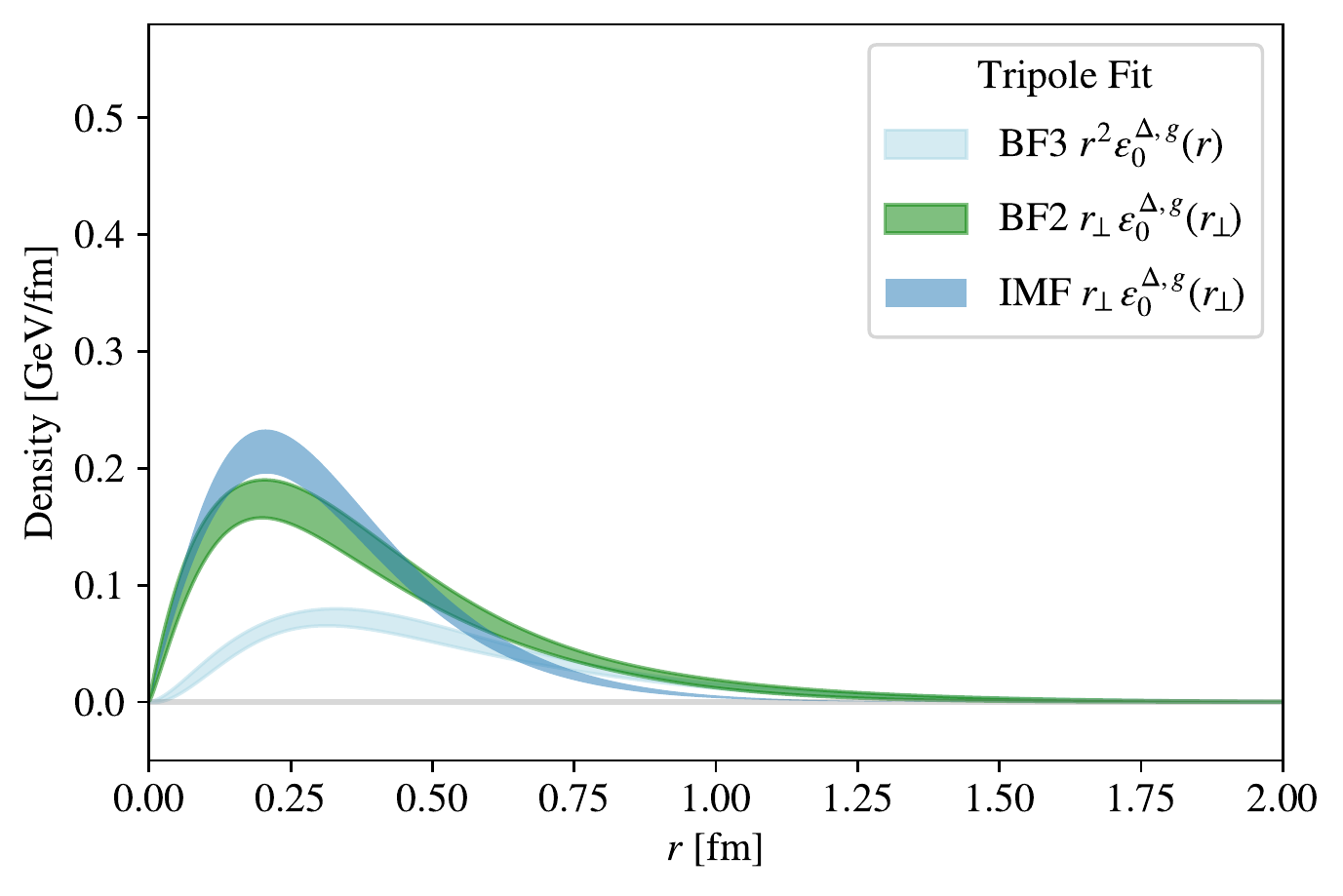} }}
\!
\subfloat[\centering  ]
{{\includegraphics[height=5.8cm,width=7.8cm]{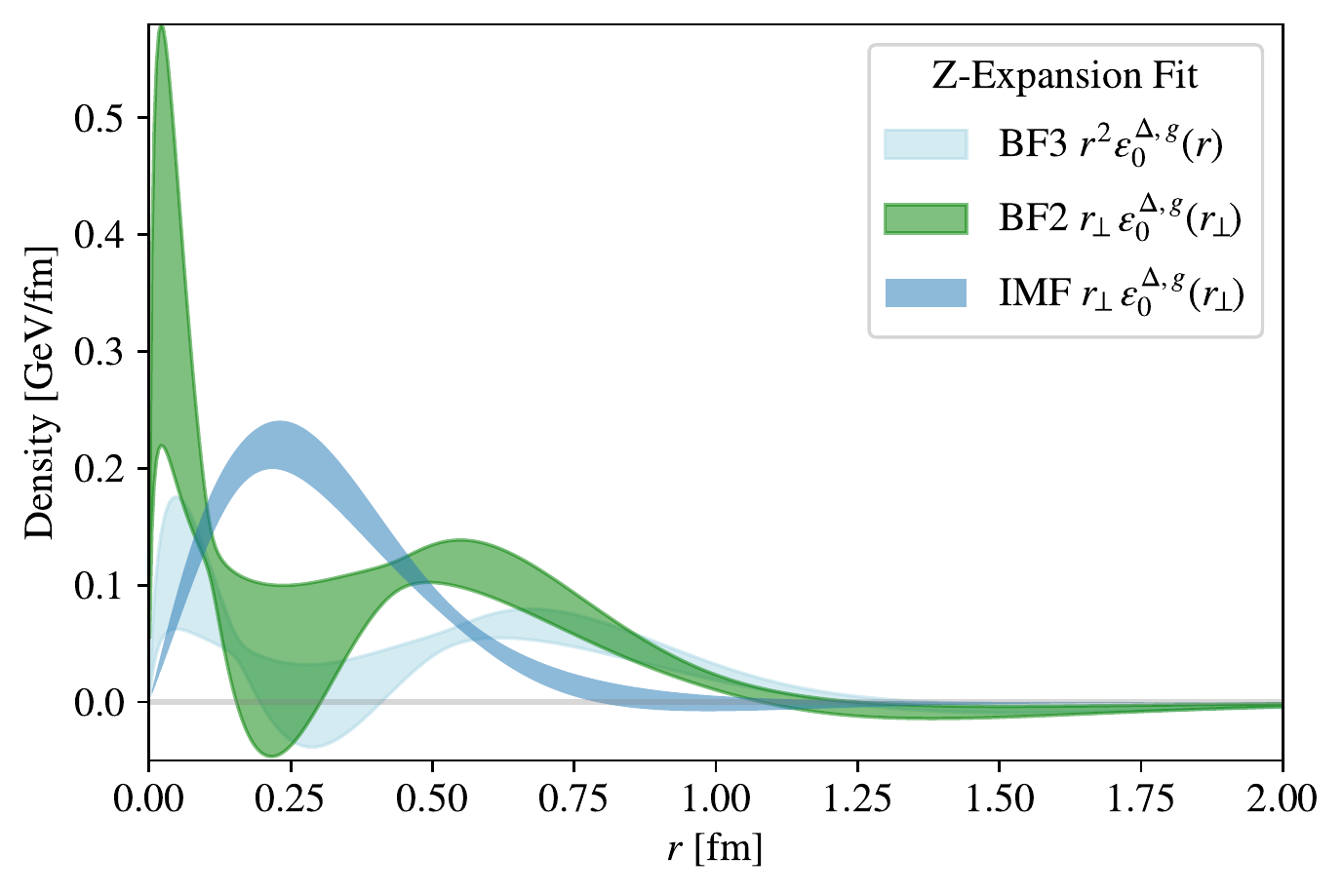} }}
\caption{Traceless (partial) gluon 
contributions to
the lowest-order pressure (a-b), shear force (c-d), and
energy (e-f) distributions of the $\Delta$ baryon
in the 3D BF, 2D BF, and in the IMF. 
The figures in the left (right) column
correspond to the tripole (z-expansion) fits, with fit
parameters in Table~\ref{tab:delta}.}
\label{fig:deltadens1}
\end{figure*}

\begin{figure*} [p!]
\centering
\subfloat[\centering  ]
{{\includegraphics[height=5.8cm,width=7.8cm]{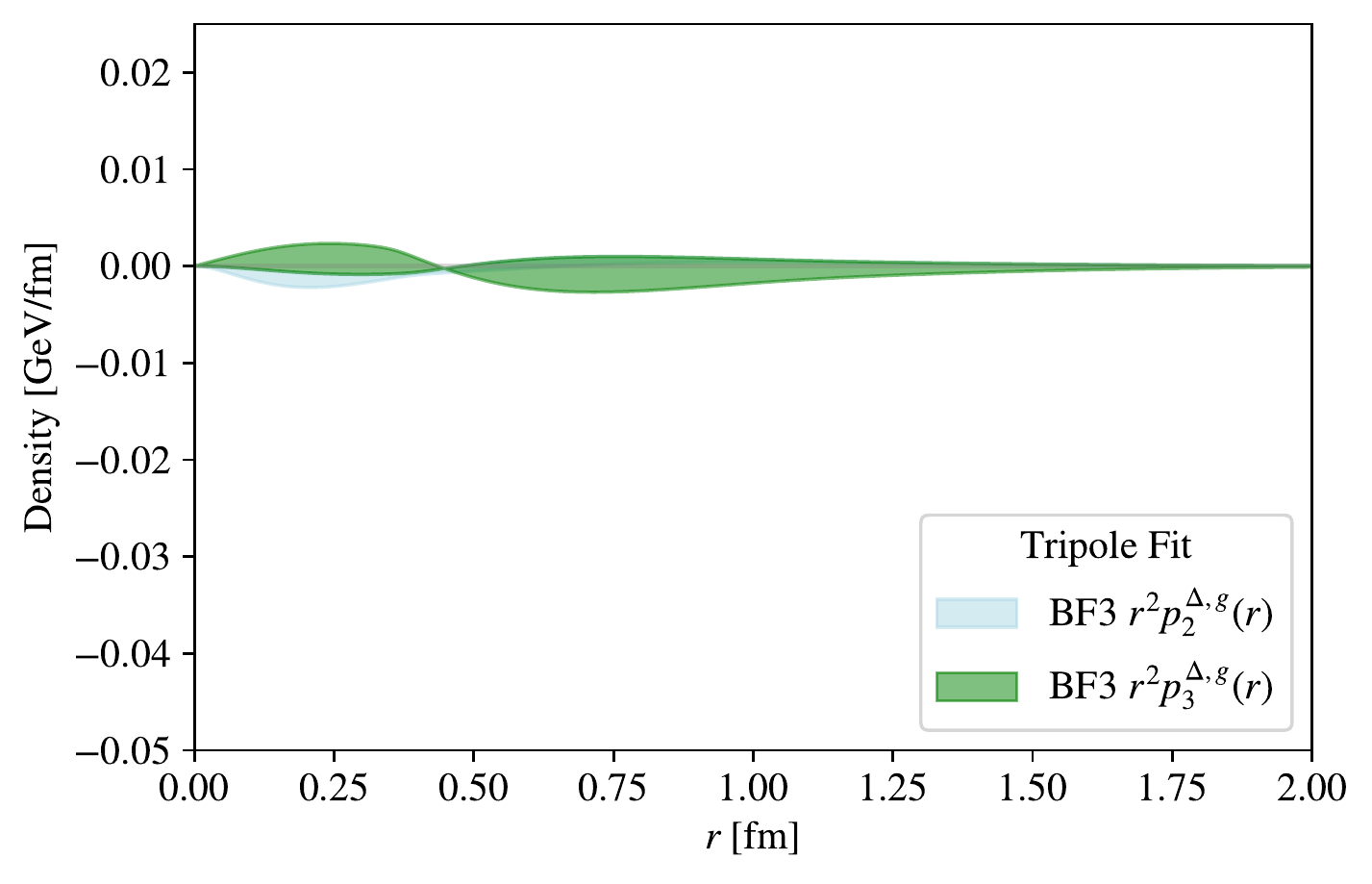} }}
\!
\subfloat[\centering  ]
{{\includegraphics[height=5.8cm,width=7.8cm]{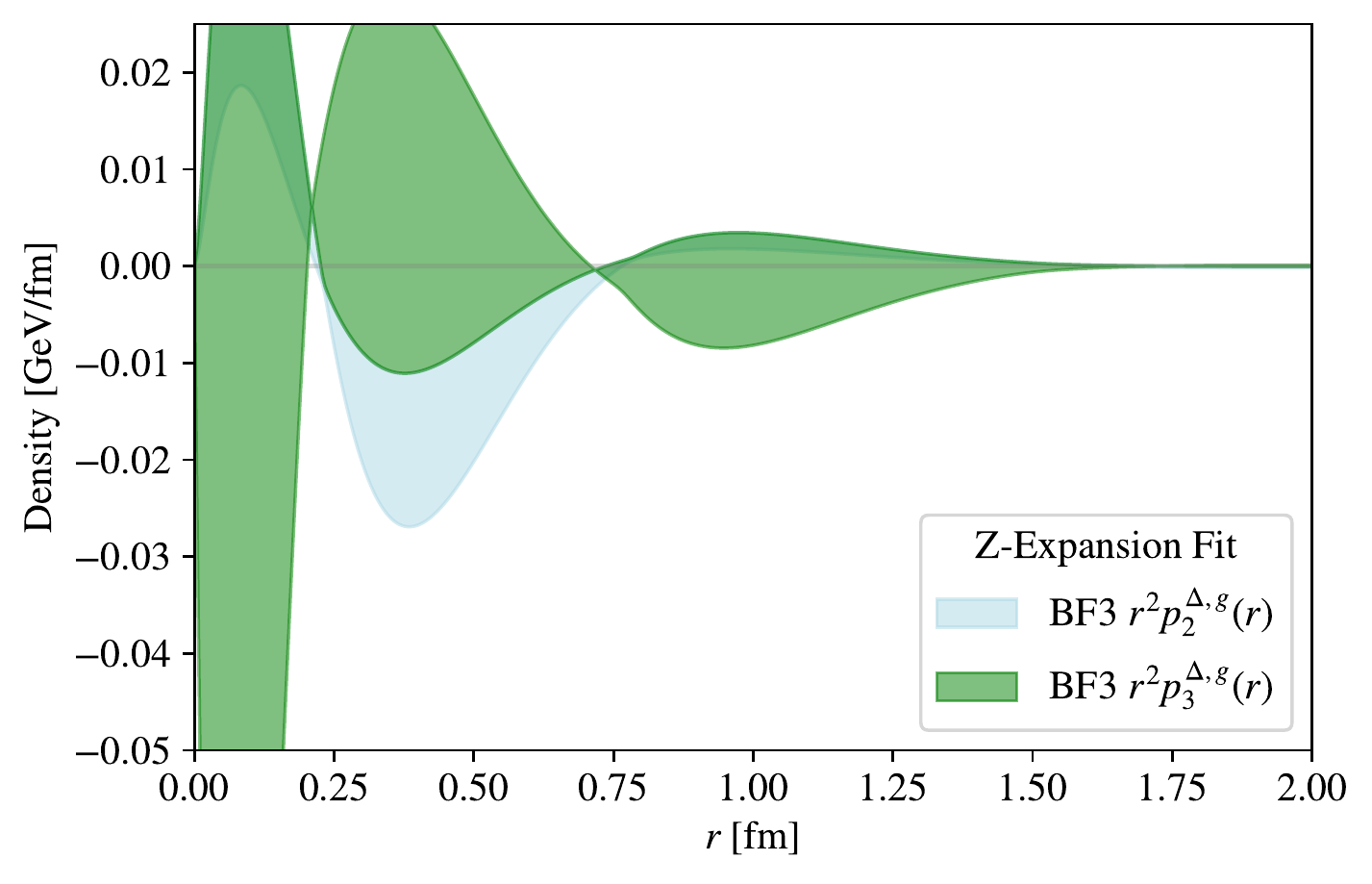} }}
\\
\subfloat[\centering  ]
{{\includegraphics[height=5.8cm,width=7.8cm]{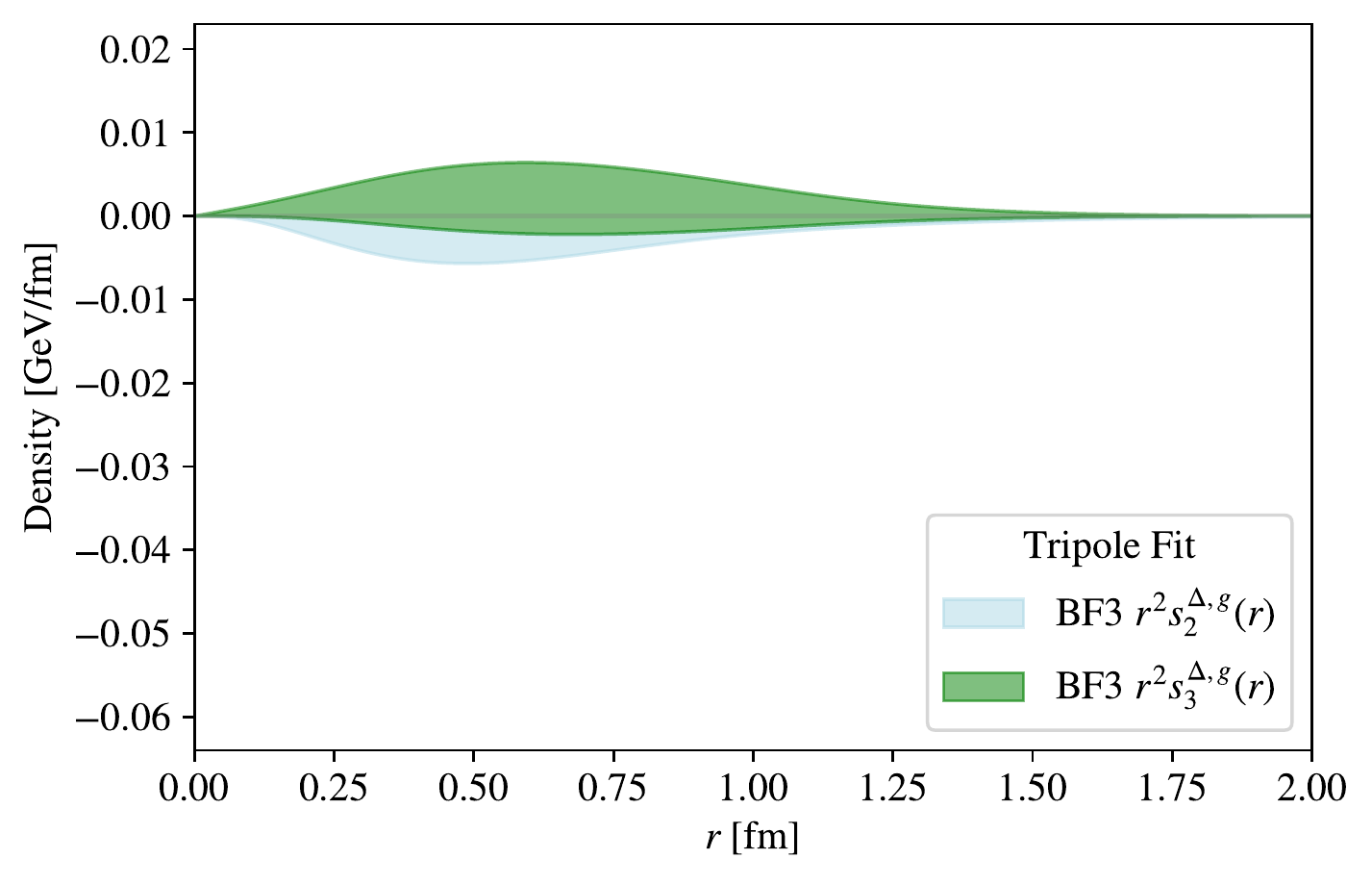} }}
\!
\subfloat[\centering  ]
{{\includegraphics[height=5.8cm,width=7.8cm]{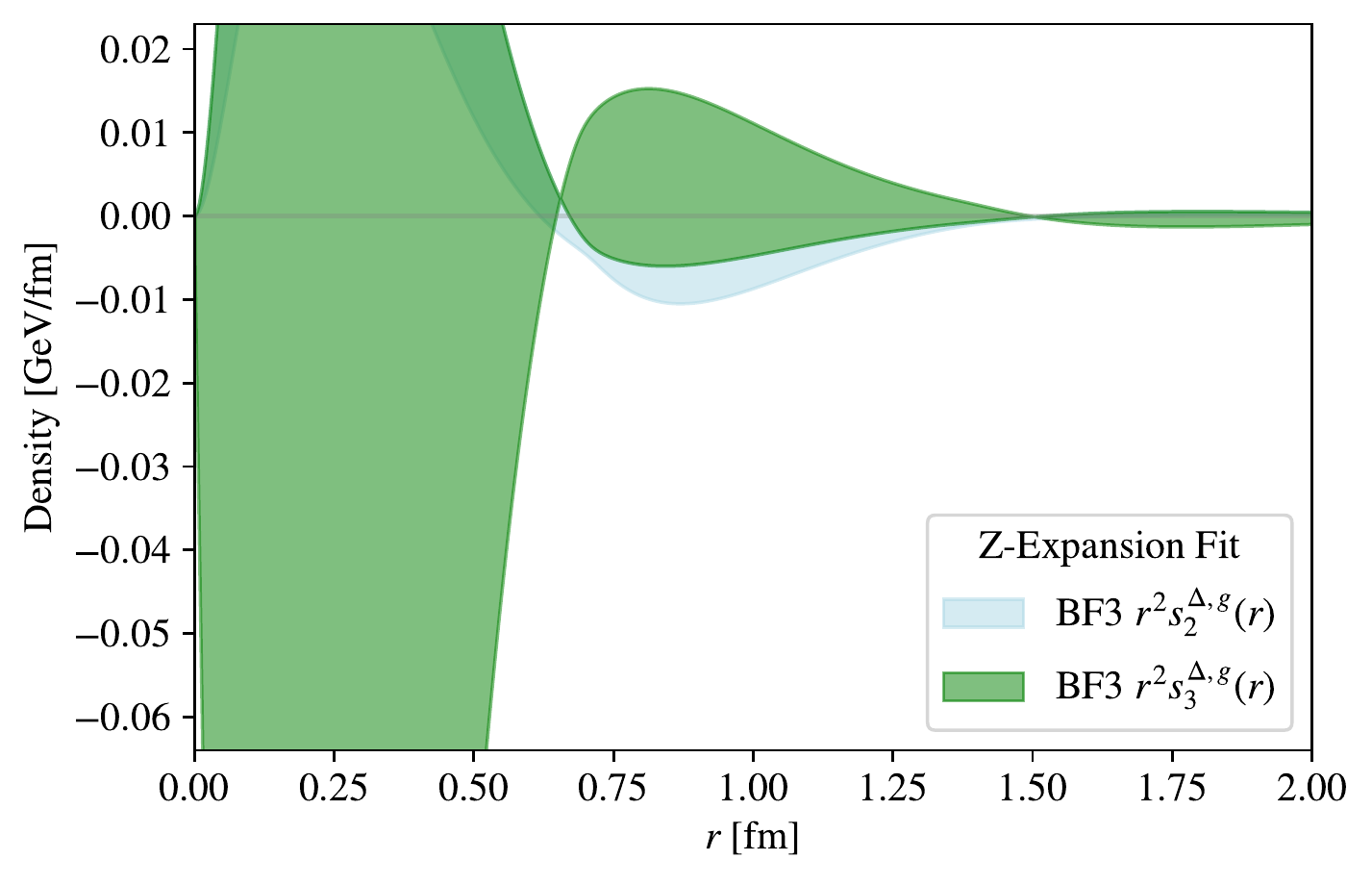} }}
\\
\subfloat[\centering  ]
{{\includegraphics[height=5.8cm,width=7.8cm]{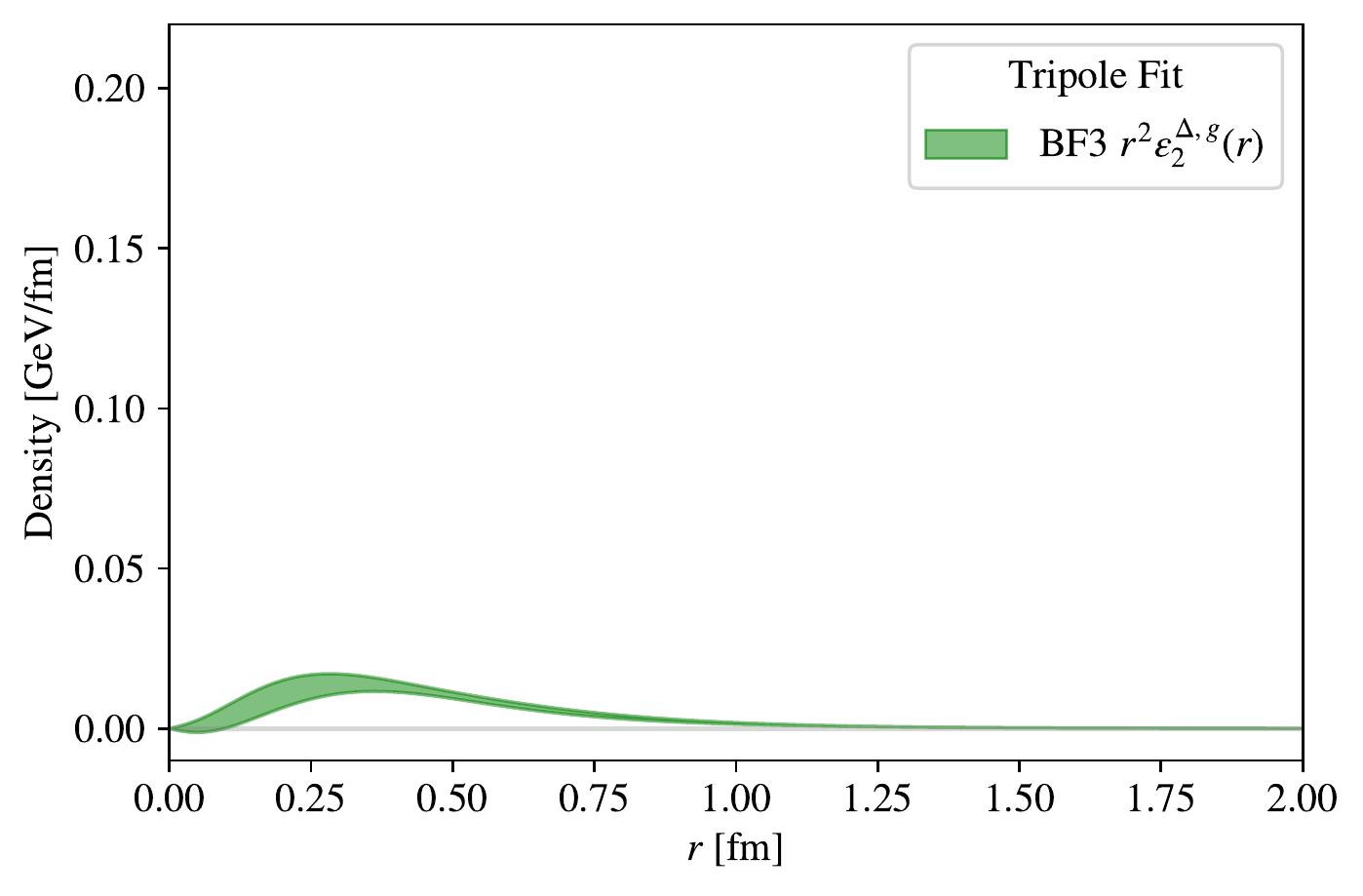} }}
\!
\subfloat[\centering  ]
{{\includegraphics[height=5.8cm,width=7.8cm]{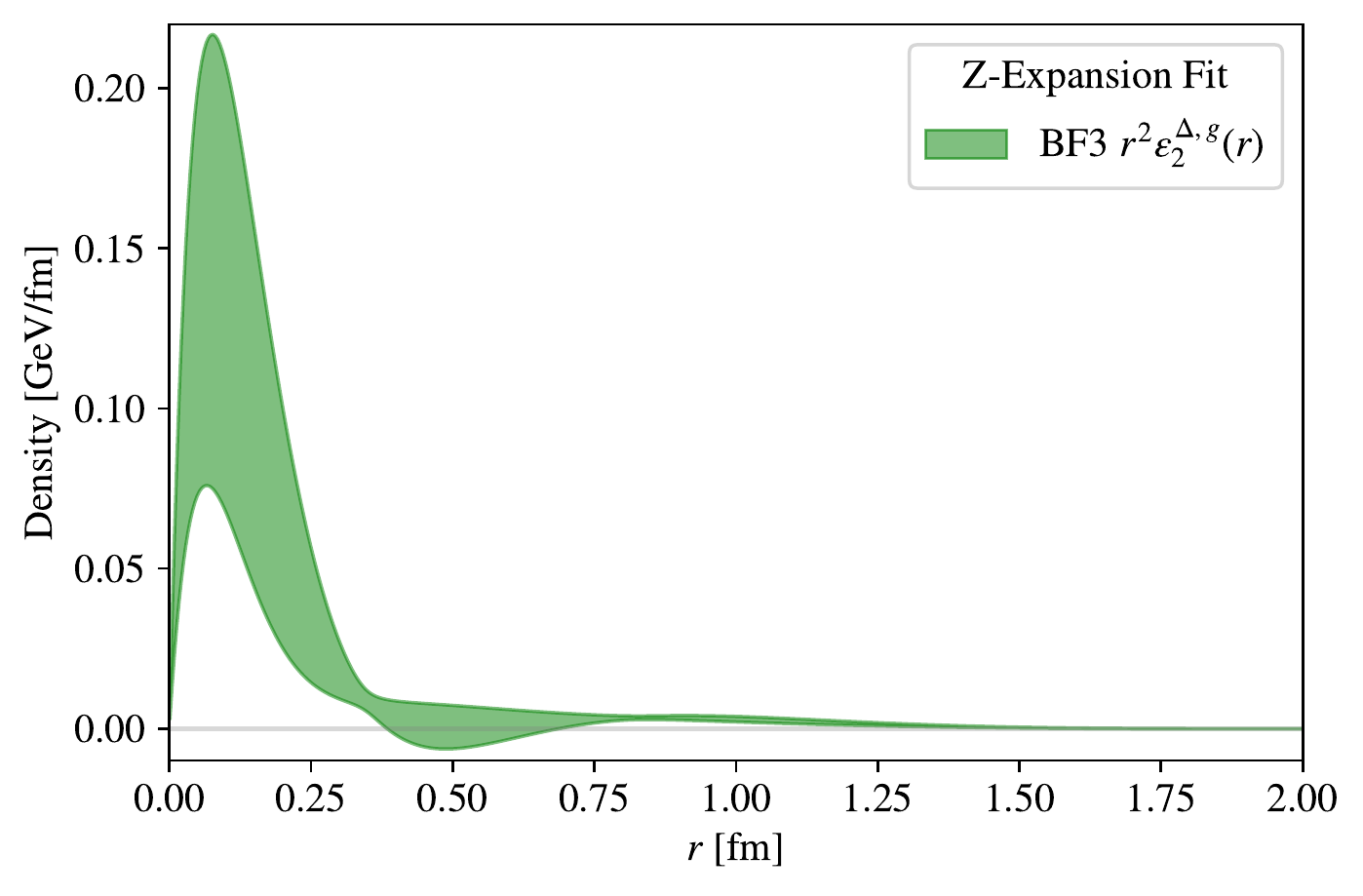} }}
\caption{Traceless (partial) gluon 
contributions to
the quadrupole order pressure (a-b), shear force (c-d), and
energy (e-f) distributions of the $\Delta$ baryon
in the 3D BF. 
The figures in the left (right) column
correspond to the tripole (z-expansion) fits, with fit
parameters in Table~\ref{tab:delta}.}
\label{fig:deltadens2}
\end{figure*}

\subsection{$\Delta$ Baryon}

The BF3 distributions for the $\Delta$ baryon were first presented in Ref.~\cite{Panteleeva:2020ejw} and extended in Ref.~\cite{Kim:2012ts}, and are listed in Appendix~\ref{sec:deltadensities},
along with our derived lowest-order BF2 and IMF contributions to the densities. The higher-order densities are not well constrained, and we limit our mechanical stability check to the lowest-order
densities only. We find that Eq.~\eqref{eq:mechstab} only holds for
the tripole fits to the GFFs.  Our estimates for the
accessible
gluon mechanical and mass radii are shown
in Table~\ref{tab:allradii}.
As discussed in Sec.~\ref{sec:DeltaGFFs}, we were
able to model only four of the $\Delta$ gluon GFFs with
the tripole and z-expansion fits, and therefore our results for the
gluon densities shown in Figs.~\ref{fig:deltadens1} and \ref{fig:deltadens2}
and the radii in Table~\ref{tab:allradii} are partial.

\section{SUMMARY AND CONCLUSION}
\label{sec:conclusion}

In this work, we present a lattice QCD extraction of the gluon GFFs of the pion, nucleon, $\rho$ meson, and $\Delta$ baryon at quark masses corresponding to a pion mass $m_{\pi} = 450(5)~\text{MeV}$ in the range $0 \leq -t < 2\;\text{GeV}^2$. All of the pion and nucleon GFFs, along with six of seven of the $\rho$ and four of eight of the $\Delta$ GFFs, are fit using the multipole and modified z-expansion models of Eqs.~\eqref{eq:multipole} and \eqref{eq:z-expansion}. Most of the GFF fits are consistent between the two models considered, with the most significant exception being the nucleon $D_g^N(t)$ which, while consistent within error, shows a qualitative difference in behavior between the (monotonic) tripole fit and (nonmonotonic) z-expansion fit. Further calculations
with different ensembles are needed in order to determine whether this nonmonotonicity is physical, or the result of poorly quantified systematic uncertainties or a statistical fluctuation in the data.
This inconsistency, however, brings to
attention the importance of considering
a variety of models when studying
the GFFs of hadrons
at different values of the
energy transfer.

\begin{figure}[h]
\centering
\subfloat[\centering  ]
{{\includegraphics[height=5.5cm,width=7.5cm]{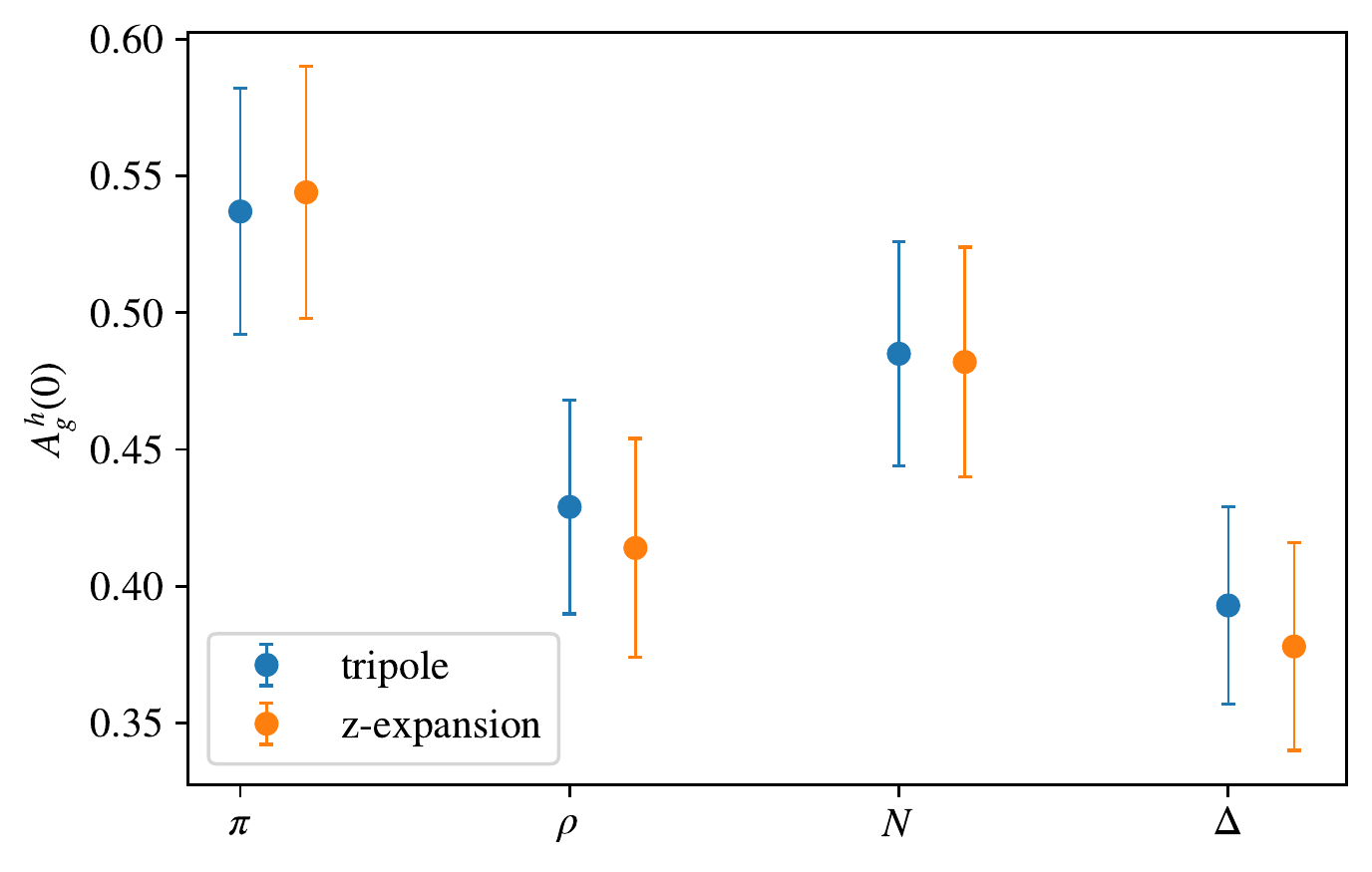} }}
\\
\subfloat[\centering  ]
{{\includegraphics[height=5.5cm,width=7.5cm]{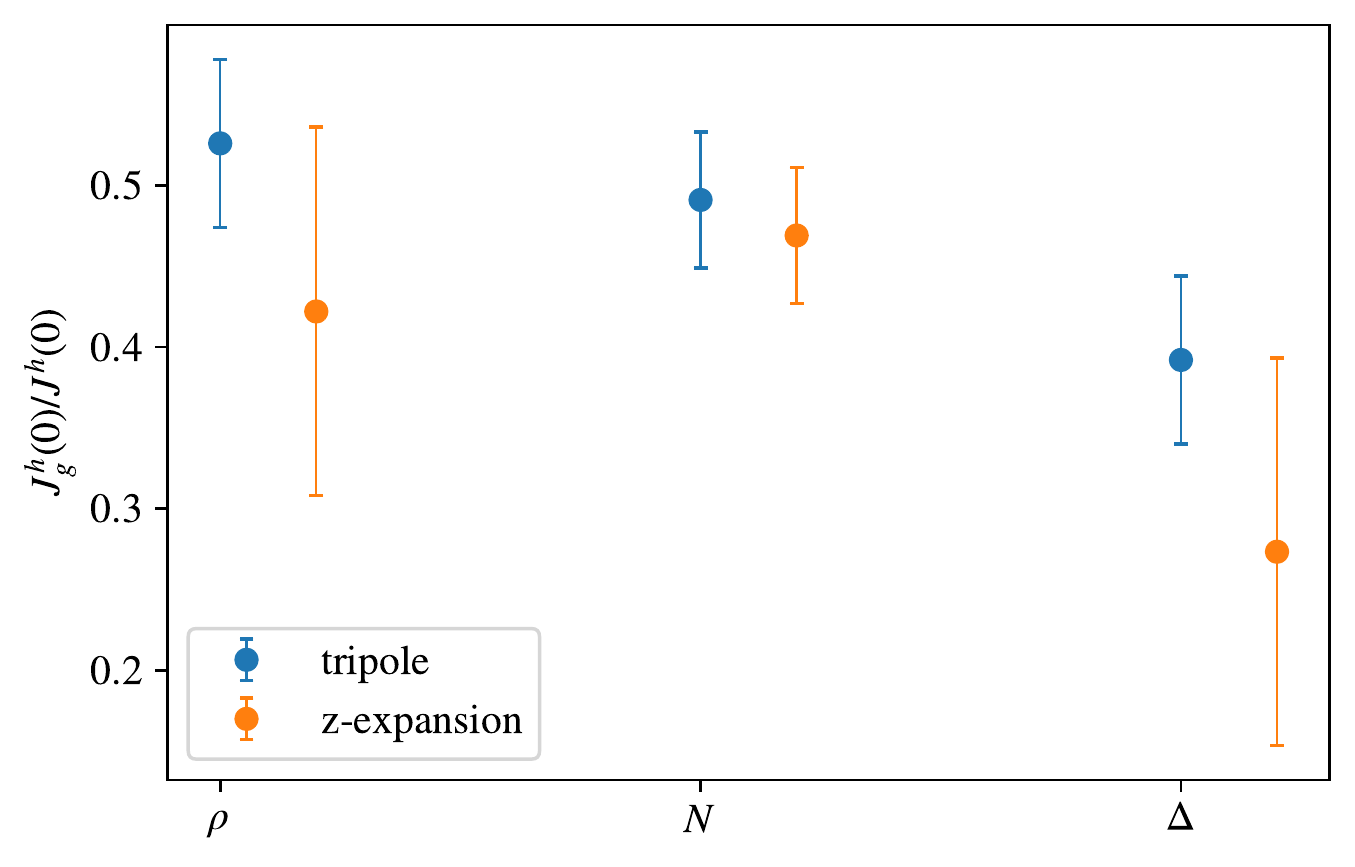} }}
\caption{Results for the gluon momentum (a) and spin (b) fractions of the pion, $\rho$ meson, nucleon, and $\Delta$ baryon as listed in Tables~\ref{tab:pionQuantities}, \ref{tab:nucQuantities}, \ref{tab:rhoQuantities}, and \ref{tab:deltaQuantities}}
\label{fig:comparison}
\end{figure}

\begin{figure}[h]
\centering
{\includegraphics[height=6.cm,width=8.cm]{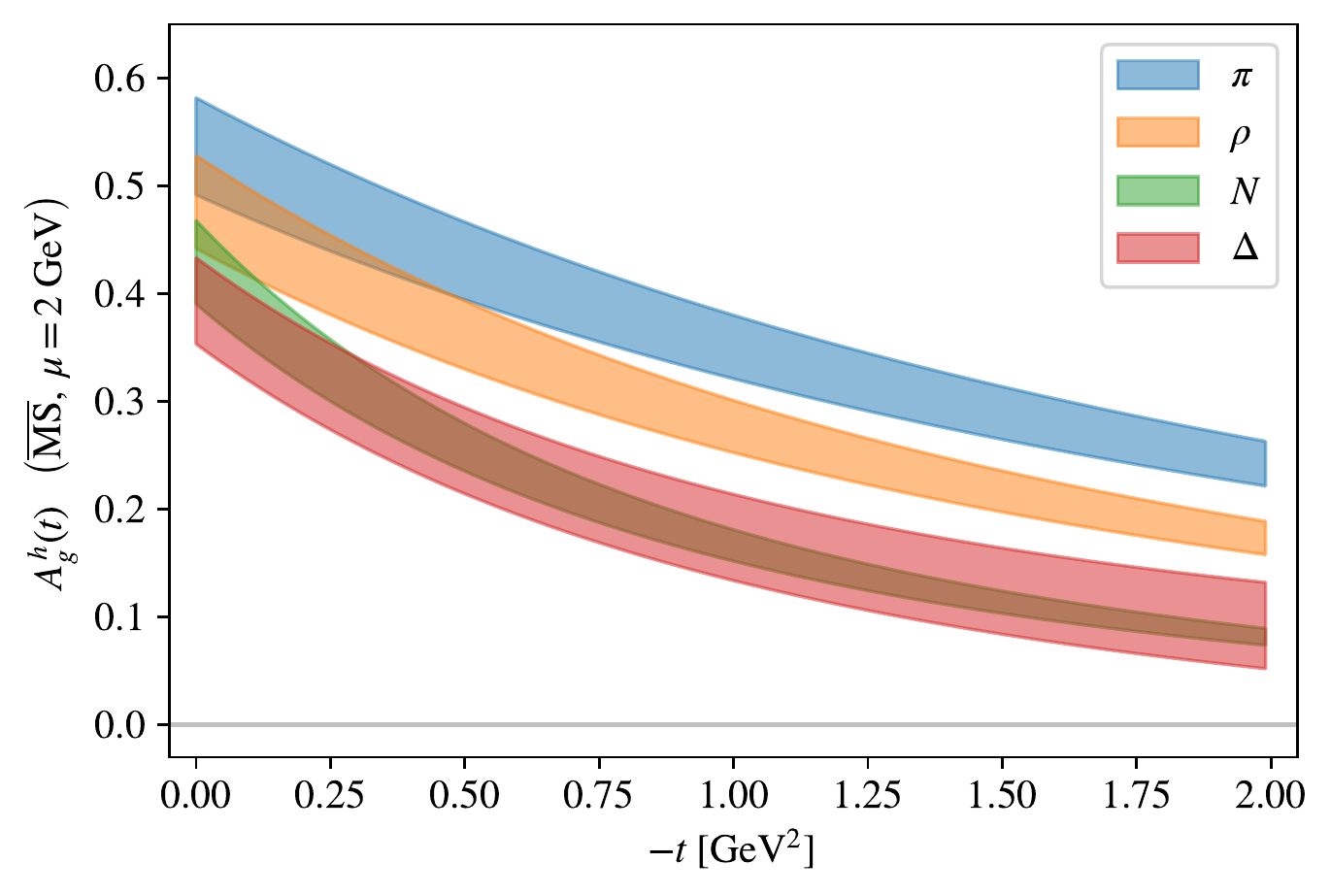} }
\caption{Tripole fits to the $A_g^h(t)$ form factor, corresponding in the forward limit
to the gluon contribution to the momentum fraction, for the four
hadrons, with fit parameters shown in Tables~\ref{tab:pion}, \ref{tab:nuc}, \ref{tab:rho}, and \ref{tab:delta}. The equivalent figure showing z-expansion fits is
indistinguishable.}
\label{fig:Acomparison}
\end{figure}

\begin{figure*}[h]
\centering
{\includegraphics[height=7.cm,width=16.7cm]{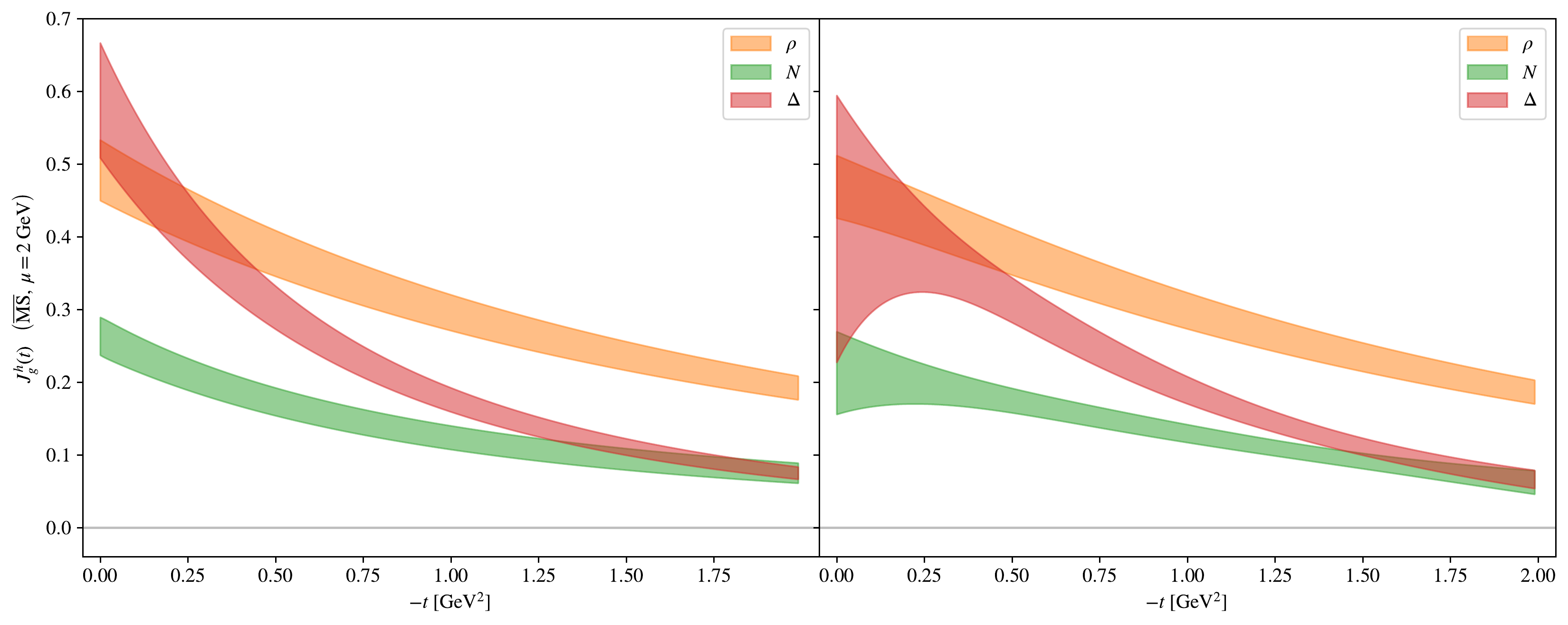} }
\caption{Tripole (z-expansion) fits in the left (right) panel to the $J_g^h(t)$ form factor, corresponding in the forward limit to the gluon contribution
to the angular momentum, for the three
hadrons, with fit parameters shown in Tables~\ref{tab:nuc}, \ref{tab:rho}, and~\ref{tab:delta}.}
\label{fig:Jcomparison}
\end{figure*}

\begin{figure*}[h]
\centering
{\includegraphics[height=7.cm,width=16.6cm]{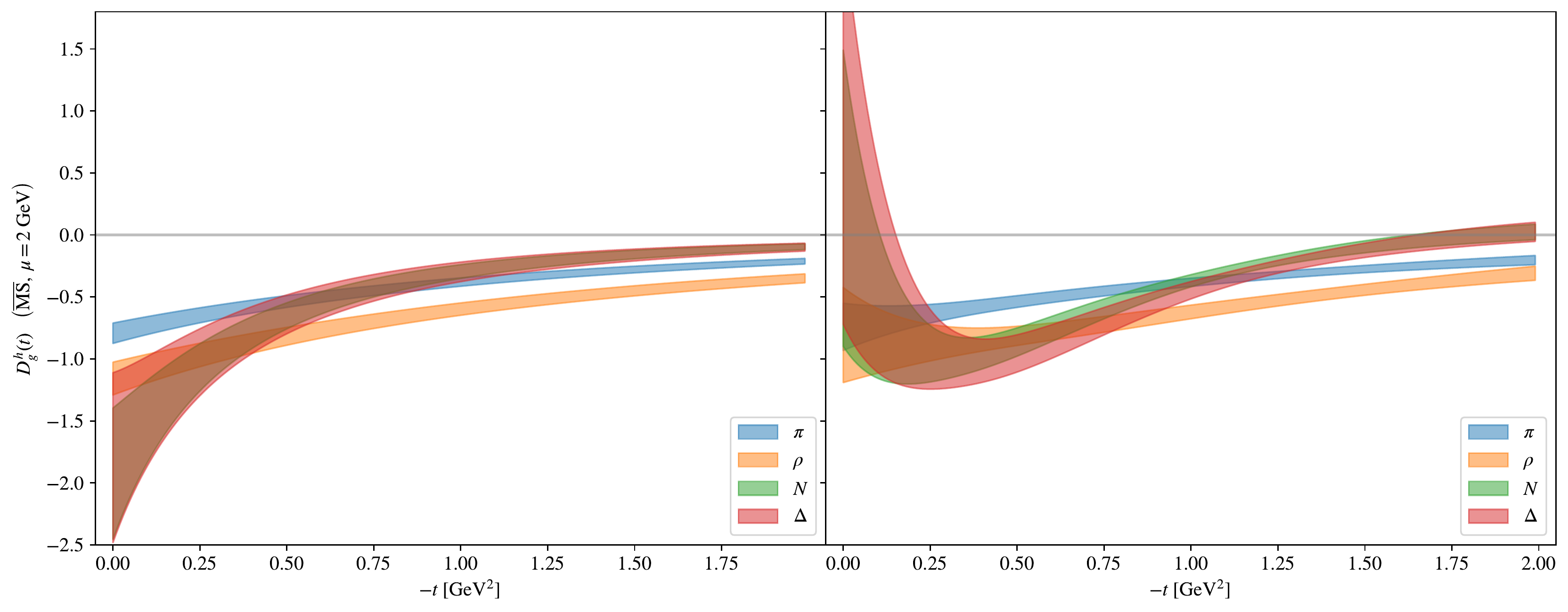} }
\caption{Tripole (z-expansion) fits in the left (right) panel to the $D_g^h(t)$ form factor for the four
hadrons, with fit parameters shown in Tables~\ref{tab:pion}, \ref{tab:nuc}, \ref{tab:rho} and \ref{tab:delta}.}
\label{fig:Dcomparison}
\end{figure*}

In Figs.~\ref{fig:Acomparison}, \ref{fig:Jcomparison}, and \ref{fig:Dcomparison} we provide
a summary of the gluon contributions
to the momentum fraction form 
factor $A^h(t)$, angular momentum 
form factor $J^h(t)$, and the $D^h(t)$ form factor. In all cases, the meson
GFFs fall off more slowly as functions of
$-t$ than the baryon GFFs. It will be important for future work to study them at higher magnitudes of energy transfer in order to fully quantify their behavior. In Fig.~\ref{fig:comparison}, we summarize
the forward limit results for 
the gluon momentum and spin fractions,
with the gluon spin fraction defined as the ratio $J^h_g(0)/J^h(0)$
of the gluon contribution to the forward-limit angular momentum $J^h_g(0)$ and the total
spins $J^h(0)$ of the corresponding hadrons. The gluon momentum
fraction is larger for the mesons and smaller
for the baryons, decreasing with increasing
hadron mass. 

Additionally, 
for each model of the $t$ dependence of the GFFs we compute the
gluon contributions to the hadron energy, pressure, and shear force spatial densities and mass radii, as well as their mechanical radii given stability conditions are satisfied.
We consider both the Breit frame and the infinite momentum frame of the densities and the radii.
The estimates made using each model
are consistent with each other, with the exception of a discrepancy in the nucleon gluon $D$-term and energy, pressure, and shear force densities 
between the two ans\"atze, that can be traced to the nonmonotonicity of the $D_g^N(t)$ z-expansion fit.\footnote{The IMF energy density does not depend on $D_g^N(t)$ and is consistent between the tripole and the modified z-expansion fit.}

For the aforementioned hadronic quantities that additionally depend on trace or antisymmetric GFFs, or on the $\rho$ and $\Delta$ GFFs that are not consistent with our fit models, we provide  partial contributions from only the GFFs that we have constrained. In order to better understand in what ways the gluons and quarks separately contribute to the gravitational structure of hadrons, it will be important to constrain all of the GFFs in future studies. Moreover, our renormalization procedure for the gluon EMT does not include mixing with quark operators. Due to the small magnitude of the mixing renormalization coefficient \cite{Alexandrou:2016ekb,Alexandrou:2020sml}, the effect is expected to contribute at the level of a few percent, which is negligible compared to the current statistical uncertainties of the calculation.

This study uses heavier-than-physical quark masses at a single lattice spacing and volume, and therefore our results are subject to unquantified systematic uncertainties that need to be addressed in future studies. 
For the pion
and nucleon, repeating the calculation of
the gluon GFFs using different ensembles is critical in order
to control the effect of systematic uncertainties
for comparisons with future
experimental data from $J/\psi$ and $\Upsilon$ production processes. 
For unstable hadrons like $\rho$ and $\Delta$, lattice QCD
methods are the only known way to access their gluon GFFs; studying them at lighter quark masses, where they are not stable, will require more computationally involved L{\"u}scher method analyses \cite{Baroni:2018iau, Briceno:2015tza,Luscher:1986pf,Luscher:1990ux}.

\begin{acknowledgements}
We thank Kyle Cranmer, Will Detmold, Bob Jaffe, William Jay, Andreas Kronfeld, Ethan Neil, and Maxim Polyakov for useful discussions. We also thank Cedric Lorc\'e for helpful comments on the calculation of densities in different frames, and June-Young Kim and Bao-Dong Sun for sharing their draft manuscript on the structure of spin-3/2 baryons.
This work is supported in part by the U.S. Department of Energy, Office of Science, Office of Nuclear Physics, under Grant Contract No.~DE-SC0011090. P.E.S. is additionally supported by the National Science Foundation under EAGER Grant No.~2035015, by the U.S. DOE Early Career Award DE-SC0021006, by a NEC research award, and by the Carl G and Shirley Sontheimer Research
Fund.
This research used resources of the National Energy Research Scientific Computing Center (NERSC), a U.S. Department of Energy Office of Science User Facility operated under Contract No. DE-AC02-05CH11231, as well as resources of the Argonne Leadership Computing Facility, which is a DOE Office of Science User Facility supported under Contract No.~DE-AC02-06CH11357, and the Extreme Science and Engineering Discovery Environment (XSEDE), which is supported by National Science Foundation Grant No.~ACI-1548562. Computations were carried out in part on facilities of the USQCD Collaboration, which are funded by the Office of Science of the U.S. Department of Energy. The authors thank Robert Edwards, Balint Joo, Kostas Orginos, and the NPLQCD Collaboration for generating the ensembles used in this study.
The Chroma~\cite{Edwards:2004sx}, QLua \cite{qlua},  QUDA \cite{Clark:2009wm,Babich:2011np,Clark:2016rdz}, QDP-JIT \cite{6877336}, and QPhiX~\cite{10.1007/978-3-319-46079-6_30} software libraries were used in this work.
Data analysis used NumPy~\cite{harris2020array}, SciPy~\cite{2020SciPy-NMeth}, pandas \cite{jeff_reback_2020_3715232,mckinney-proc-scipy-2010}, lsqfit~\cite{peter_lepage_2020_4037174}, and gvar~\cite{peter_lepage_2020_4290884}.
Figures were produced using matplotlib~\cite{Hunter:2007}, seaborn~\cite{Waskom2021}, and \textit{Mathematica}~\cite{Mathematica}.

\end{acknowledgements}

\onecolumngrid
\appendix

\section{NUMERICAL AND ANALYSIS DETAILS}
\label{sec:lattice-details}

\subsection{Symmetric traceless gluon EMT and renormalization}
\label{sec:lattice-gluon-EMT}

The symmetric traceless part of the EMT, $\hat{T}^{\mu\nu}$,
can be obtained via the Belinfante-Rosenfeld process\footnote{See e.g., the Appendix E of Ref.~\cite{Belitsky:2005qn} for a review.} \cite{Landau}
and can be decomposed into gluon and quark terms, $\hat{T}^{\mu\nu} =
\hat{T}_g^{\mu\nu} + \sum_{q} \hat{T}_{q}^{\mu\nu}$, where
\begin{align} \label{eq:belifante}
\begin{split}\;
\hat{T}_g^{\mu\nu} &= 2\; \mathrm{tr}\left[ F^{\mu\alpha}F_{\alpha}{}^{\nu} + \frac{1}{4}
g^{\mu\nu}F^{\alpha\beta}F_{\alpha\beta}\right] \\
\hat{T}_{q}^{\mu\nu} &= i\bar{\psi}_{q}\overleftrightarrow{\mathcal{D}}^
{\{\mu}\gamma^{\nu\}} \psi_{q} - 
i g^{\mu\nu}\bar{\psi}_q\overleftrightarrow{\cancel{\mathcal{D}}}\psi_{q} \;,
\end{split}
\end{align}
where $F^{\mu\nu}$ is the gluon
field-strength tensor of
QCD, $\mathcal{D}_{\mu}$ is the covariant
derivative, $\overleftrightarrow{\mathcal{D}}
^{\mu} = (\overrightarrow{\mathcal{D}}^{\mu}-
\overleftarrow{\mathcal{D}}^{\mu})/2$, $\gamma^{\mu}$ are the Dirac matrices, the repeated indices are contracted with the 
Minkowski space-time metric $g^{\mu\nu}$, and the trace is over the color indices.

The gluon field-strength tensor $F^{\mu\nu}$ can be defined up to finite lattice spacing corrections on a Euclidean spacetime lattice as
\begin{equation}
F_{\mu\nu}^E(x) = \frac{i}{8 g_0}(P_{\mu\nu}(x) - P^{\dagger}_{\mu\nu}(x)),
\end{equation}
where $g_0$ is the bare gauge coupling, the label $E$ denotes Euclidean spacetime, and $P_{\mu\nu}(x)$ is the clover term defined in terms of gauge links $U_\mu(x)$ as
\begin{equation}
\begin{split}
P_{\mu\nu}(x) &= U_{\mu}(x)U_{\nu}(x+\hat{\mu})U^{\dagger}_{\mu}(x+\hat{\nu})
U^{\dagger}_{\nu}(x) \\
&+ U_{\nu}(x)U_{\mu}^{\dagger}(x-\hat{\mu}-\hat{\nu})U_{\nu}^{\dagger}(x-\hat{\mu})
U_{\mu}(x-\hat{\mu}) \\
&+ U^{\dagger}_{\mu}(x-\hat{\mu})U^{\dagger}_{\nu}(x-\hat{\mu}-\hat{\nu})
U_{\mu}(x-\hat{\mu}-\hat{\nu})U_{\nu}(x-\hat{\nu}) \\
&+ U^{\dagger}_{\nu}(x-\hat{\nu})U_{\mu}(x-\hat{\nu})U_{\nu}(x-\hat{\nu}+\hat{\mu})
U^{\dagger}_{\mu}(x) ~.
\end{split}
\end{equation}
The momentum-projected traceless symmetric piece of the gluon EMT in Euclidean space can be defined as
\begin{equation}
\hat{T}_{\mu\nu}^{gE}(\vec{\Delta}, \tau_0) 
= \sum_{\vec{x}} e^{-i \vec{\Delta} \cdot \vec{x}} ~ \hat{T}_{\mu\nu}^{gE}(\vec{x}, \tau_0) 
= \sum_{\vec{x}} e^{-i \vec{\Delta} \cdot \vec{x}} 2\;\mathrm{tr}\left[ F_{\alpha\mu}^E (\vec{x}, \tau_0) F_{\alpha\nu}^E (\vec{x}, \tau_0) -\frac{1}{4}\delta_{\mu\nu}F^E_{\alpha\beta}F^E_{\alpha\beta}\right]\;,
\label{eqn:gluon-emt-opr-def}
\end{equation}
where the repeated indices are contracted with
the Euclidean metric $\delta_{\mu\nu}$ (i.e.~the
Kronecker delta in four dimensions) and the trace is over the
color indices.
In continuous spacetime, a traceless symmetric tensor transforms in the
$(1,1)$ representation of the Lorentz group. However, Lorentz symmetry
is reduced to hypercubic symmetry on the lattice and therefore lattice
operators transform in irreps of the symmetry
group $H(4)$. There are two choices of irreps that are safe from power-divergent mixing with lower-dimensional operators, namely $\tau_1^{(3)}$
and $\tau_3^{(6)}$ \cite{Gockeler:1996mu}. We thus project all operator measurements of $\hat{T}^{gE}_{\mu\nu}$ to particular bases for these two irreps,
\begin{equation}
\begin{gathered}
\hat{T}^{gE}_{\tau_{1,1}^{(3)}} = \frac{1}{2}(\hat{T}_{xx}^{gE} + \hat{T}_{yy}^{gE} - \hat{T}_{zz}^{gE} - \hat{T}_{tt}^{gE}),
\quad \hat{T}^{gE}_{\tau_{1,2}^{(3)}} =\frac{1}{\sqrt{2}}(\hat{T}_{zz}^{gE} - \hat{T}_{tt}^{gE}), \quad
\hat{T}^{gE}_{\tau_{1,2}^{(3)}} = \frac{1}{\sqrt{2}}(\hat{T}^{gE}_{xx}-\hat{T}^{gE}_{yy})\;, \\
\hat{T}^{gE}_{\tau_{3,i=\{1,...,6\}}^{(6)}} = \frac{1}{\sqrt{2}}(\hat{T}^{gE}_{\mu\nu}+\hat{T}^{gE}_{\nu\mu}),
\quad \mu\nu \in \{ xy, xz, xt, yz, yt, zt \}.
\end{gathered}
\label{eqn:hypercubic-irreps}
\end{equation}

We compute all operators in the $\tau_1^{(3)}$ and $\tau_3^{(6)}$ irreps for all lattice momenta satisfying $|\vec{\Delta}|^2 \le 18 (2 \pi / L)^2$.
To suppress gauge noise, we improve the operators by constructing them from gauge links that have been subjected to Wilson flow \cite{Luscher:2010iy,Narayanan:2006rf,Lohmayer:2012hs} to flow time $t/a^2=1$ (with integrator step size $\epsilon = 0.01$).
Different choices of flow time, as well as use of hypercubic smearing instead of Wilson flow, have been shown to give consistent results~\cite{Detmold:2016gpy,Detmold:2017oqb}.

At finite lattice spacing, the two irreps $\tau_1^{(3)}$ and $\tau_3^{(6)}$ renormalize differently and only coincide in the continuum limit, but all operators within each irrep share the same renormalization factor by symmetry.
Ref.~\cite{Shanahan:2018pib} carried out a nonperturbative RI-MOM calculation \cite{Martinelli:1994ty,Martinelli:1993dq} of the renormalization factors of these operators on a smaller-volume ensemble for the same parameters as those used in this work. With a one-loop perturbative matching to the $\overline{\text{MS}}$ scheme \cite{Yang:2016xsb}, this yielded the renormalization coefficients\footnote{Note that these values correspond to $1/g_0^2=\beta/4N_c$ with $\beta=6.1$, corresponding to the bare lattice operator definition used in this work. For the tadpole-improved L\"uscher-Weisz gauge action, this differs from the continuum normalization which is $\beta(1-2/5u_0^2)/2N_c$, where $u_0$ is the tadpole factor. The renormalized operator is independent of this choice.}
\begin{equation}\begin{aligned}
    Z_{\tau_1^{(3)}}^{\overline{\text{MS}}}(\mu=2\;\text{GeV}) &= 0.9(2) \\
    Z_{\tau_3^{(6)}}^{\overline{\text{MS}}}(\mu=2\;\text{GeV}) &= 0.78(7) ~ 
    \label{eqn:Z-factors}
\end{aligned}\end{equation}
which renormalize the lattice operators multiplicatively as $[\hat{T}^{gE}]^{\overline{\text{MS}}} = Z_{\mathcal{R}} [\hat{T}_{\mathcal{R}}^{gE}]^{\text{latt}}$ where $\mathcal{R}$ indexes the irrep.
The renormalization factors were computed for the same flowed definition as used in this calculation \cite{Shanahan:2018pib}.
The uncertainties on these quantities are dominantly systematic and, because they were computed on a different ensemble, uncorrelated with the rest of the data. We choose to model their distribution as uncorrelated Gaussians.
As in Ref.~\cite{Shanahan:2018pib}, we neglect mixing with the quark operators under renormalization, which is expected to contribute at the few-percent level \cite{Alexandrou:2016ekb,Alexandrou:2020sml}.

\subsection{Two-point correlation functions}
\label{sec:two-point-fns}

As detailed in the main text, we use a single lattice ensemble in this work; parameters for this ensemble are listed in Table~\ref{tab:ensemble}.
Using matching valence and sea quark actions, we compute two-point functions for varying numbers of light-quark sources on each configuration, using an average of 235 randomly chosen locations (240 for 80\% of the configurations, $\gtrsim$ 200 for 90\%).
As described below, our analysis accounts for differing numbers of sources by weighting configurations proportionately when drawing bootstrap ensembles.
For each source position, we invert from a smeared source (S) and construct propagators for both a point (P) and smeared sink (S), with matching source and sink smearing for the SS propagators, using APE smearing \cite{Falcioni:1984ei} with 35 steps of gauge-invariant Gaussian smearing with width $\rho = 4.7$.
From each propagator we construct two-point correlation functions for each hadron  using the interpolating operators
\begin{equation}\begin{aligned}
    &\chi^\pi(x) &&= \overline{\psi}_u(x) \gamma_5 \psi_d(x) \;, \\
    &\chi^N(x) &&= \left[ \psi_u(x) C \gamma_5 \psi_d(x) \right] \psi_u(x) \;, \\
    &\chi^\rho_\mu(x) &&= \overline{\psi}_u(x) \gamma_\mu \psi_d(x) \;, \\
    &\chi^\Delta_{\mu}(x) &&= \left[ \psi_u(x) C \gamma_\mu \psi_u(x) \right] \psi_u(x) \;,
\end{aligned}\end{equation}
where all gamma matrices are Euclidean, $C$ is the charge conjugation matrix, and
color and spinor indices are left implicit.

The interpolating operators overlap with the lowest-lying hadronic states as
\begin{equation}\begin{alignedat}{999}
    \bra{0} &\chi^\pi(x) &&\ket{\pi(\vec{p})} &&= Z^\pi_{\vec{p}}e^{i\vec{p}\cdot\vec{x}} \;,  \\
    \bra{0} &\chi^N(x) &&\ket{N(\vec{p},\sigma)} &&= Z^N_{\vec{p}}  u(\vec{p},\sigma)e^{i\vec{p}\cdot\vec{x}} \;, \\
    \bra{0} &\chi^\rho_a(x) &&\ket{\rho(\vec{p},\lambda)} &&= Z^\rho_{\vec{p}}  \epsilon_a(\vec{p},\lambda)e^{i\vec{p}\cdot\vec{x}} \;, \\
    \bra{0} &\chi^\Delta_a(x) &&\ket{\Delta(\vec{p},\xi)} &&= Z^\Delta_{\vec{p}}  u_{a}(\vec{p},\xi)e^{i\vec{p}\cdot\vec{x}} \;,
\end{alignedat}\end{equation}
where $Z_{\vec{p}}$ is an overlap factor, $\epsilon_a(\vec{p},\lambda)$ is a spin-1 polarization vector with $\lambda\in\{1,0,-1\}$ and in a spherical basis
 ${a\in\{+,-,0\}}$ such that
\begin{equation} \label{eq:basis}
    \epsilon_a(\vec{p},\lambda) = \begin{cases}
        \frac{1}{\sqrt{2}}[\epsilon_x(\vec{p},\lambda) \pm i \epsilon_y(\vec{p},\lambda)], & a = \pm \\
        \epsilon_z(\vec{p},\lambda), & a = 0 \;,
    \end{cases}
\end{equation}
$u(\vec{p},\sigma)$ with $\sigma \in \{\frac{1}{2},-\frac{1}{2}\}$ is a Dirac spinor, and $u_a(\vec{p},\xi)$ is a Rarita-Schwinger spin vector with $\xi \in\{\frac{3}{2},\frac{1}{2},-\frac{1}{2},-\frac{3}{2}\}$, 
written in the same spherical basis as $\epsilon_a(\vec{p},\lambda)$ in Eq.~\eqref{eq:basis}.

The momentum projected two-point correlation function of the pion can
be expressed as
\begin{equation} \label{eq:twopoint_pi}
C^{\pi,\text{2pt}}(\vec{p},t'; \vec{x}_0,t_0) =
\sum_{\vec{x}} e^{-i\vec{p} \cdot (\vec{x} - \vec{x}_0) }
 \vev{ \chi^\pi(\vec{x},t')  \chi^{\pi\dagger}(\vec{x}_0,t_0) }
\xrightarrow{t' \rightarrow \infty}
\frac{e^{-E^{\pi}_\vec{p} t'}}{2 E^{\pi}_\vec{p}} \tilde{Z}_{\vec{p}} Z_{\vec{p}} \;,
\end{equation}
where $E_{\vec{p}}^{\pi}$ is the energy of the lowest-lying state with momentum $\vec{p}$,
and $\tilde{Z}_{\vec{p}} \ne Z_{\vec{p}}$ when the source and sink are smeared differently.
The two-point correlation function of the nucleon for spin channel $\sigma \rightarrow \sigma'$ is
\begin{equation} \label{eq:twopoint_nuc}
C^{N,\text{2pt}}_{\sigma\sigma'}(\vec{p},t'; \vec{x}_0,t_0) =
\sum_{\vec{x}} e^{-i\vec{p} \cdot (\vec{x} - \vec{x}_0) }
\mathrm{tr} \big[\Gamma_{\sigma'\sigma}\vev{ \chi^N(\vec{x},t')  \bar{\chi}^N(\vec{x}_0,t_0) }\big]
\xrightarrow{t' \rightarrow \infty}
\frac{e^{-E^N_\vec{p} t'}}{2 E^N_\vec{p}} \tilde{Z}_{\vec{p}} Z_{\vec{p}}
\mathrm{tr} \big[\Gamma_{\sigma'\sigma}(\cancel{p}+m_N)\big] \;,
\end{equation}
where $\bar{\chi} = \chi^{\dagger}\gamma_t$, traces are
over Dirac indices, and $\Gamma_{\sigma'\sigma}$ is a $2\times2$ block matrix that
projects the four different spin channels of the nucleon \cite{greiner1990relativistic}, i.e.
\begin{equation}
\Gamma_{\sigma'\sigma} = 
\begin{pmatrix}
P_+(1+\gamma_x\gamma_y) & P_+\gamma_z(\gamma_x+i\gamma_y) \\
P_+\gamma_z(\gamma_x-i\gamma_y) & P_+(1-\gamma_x\gamma_y)
\end{pmatrix}
_{\sigma'\sigma} \;,
\end{equation}
where $P_+ \equiv \frac{1}{2}(1+\gamma_t)$ is a positive-energy projector. 
We use all four possible channels $\sigma,\sigma'\in\{+1/2,-1/2\}$, adding significant additional data over the analysis in Ref.~\cite{Shanahan:2018pib} where only the two spin-conserving channels were used. 
The two-point correlation function of the $\rho$ meson, in the spherical basis of one of the 9 spin channels $a\rightarrow a'$, can be expressed as
\begin{equation}
C^{\rho,\text{2pt}}_{aa'}(\vec{p},t'; \vec{x}_0,t_0) =
\sum_{\vec{x}} e^{-i\vec{p} \cdot (\vec{x} - \vec{x}_0) }
 \vev{ \chi_{a'}^{\rho}(\vec{x},t')  \chi_a^{\rho\dagger}(\vec{x}_0,t_0) }
\xrightarrow{t' \rightarrow \infty}
\frac{e^{-E^{\rho}_\vec{p} t'}}{2 E^{\rho}_\vec{p}} \tilde{Z}_{\vec{p}} Z_{\vec{p}}
\Lambda_{a'a}^{(\rho)}(\vec{p}) \;,
\end{equation}
where $\Lambda^{(\rho)}_{a'a}(\vec{p}) \equiv \sum_{\lambda}\epsilon_{a'}(\vec{p},
\lambda)\epsilon^*_a(\vec{p},\lambda)$ [cf.~Eq.~\eqref{eqn:lambda-rho-def}]. Finally, we compute the
two-point correlator of the $\Delta$ baryon 
for the 10 spin channels $\xi \rightarrow \xi'$ where $\xi \geq \xi'$,
\begin{equation}
\begin{split}
C^{\Delta,\text{2pt}}_{\xi \xi'}(\vec{p},t'; \vec{x}_0,t_0) &=
\sum_{\vec{x}} e^{-i\vec{p} \cdot (\vec{x} - \vec{x}_0) }
\mathrm{tr} \big[\mathcal{D}^{\xi}_{\sigma, a} \mathcal{D}^{\xi'}_{\sigma', a'}\Gamma_{\sigma'\sigma}\vev{ \chi_{a'}^{\Delta}(\vec{x},t')\bar{\chi}_a^{\Delta}(\vec{x}_0,t_0) }\big] \\
& \xrightarrow{t' \rightarrow \infty}
\frac{e^{-E^{\Delta}_\vec{p} t'}}{2 E^{\Delta}_\vec{p}} \tilde{Z}_{\vec{p}} Z_{\vec{p}} 
\mathrm{tr} \big[\mathcal{D}^{\xi}_{\sigma, a} \mathcal{D}^{\xi'}_{\sigma', a'}\Gamma_{\sigma'\sigma}\Lambda^{(\Delta)}_{a'a}(\vec{p})\big] \;,
\end{split}
\end{equation}
where repeated indices are summed over, $\Lambda^{(\Delta)}_{a'a}(\vec{p})\equiv\sum_\xi u_{a'}(\vec{p},\xi)\bar{u}_{a}(\vec{p},\xi)$ [cf.~Eq.~\eqref{eqn:lambda-delta-def}],
and the coefficients $\mathcal{D}^{\xi}_{\sigma, a}$ are defined
such that
\begin{equation}
\mathcal{D}_{1/2, +}^{3/2} = 
\mathcal{D}_{-1/2, -}^{-3/2} = 
\mathcal{D}_{1/2, 0}^{1/2} = 
\mathcal{D}_{-1/2, 0}^{-1/2} = 1 \;,
\end{equation}
and $\mathcal{D}^{\xi}_{\sigma, a} = 0$ for all other choices of $\{\xi,\sigma,a\}$.%
\footnote{For the $+3/2 \rightarrow -1/2$ channel we instead computed correlation functions corresponding to $\mathcal{D}^{3/2}_{1/2,+} = \mathcal{D}^{-1/2}_{1/2,-} = 1$.}

\begin{figure}
    \centering
    \includegraphics[width=0.55\textwidth]{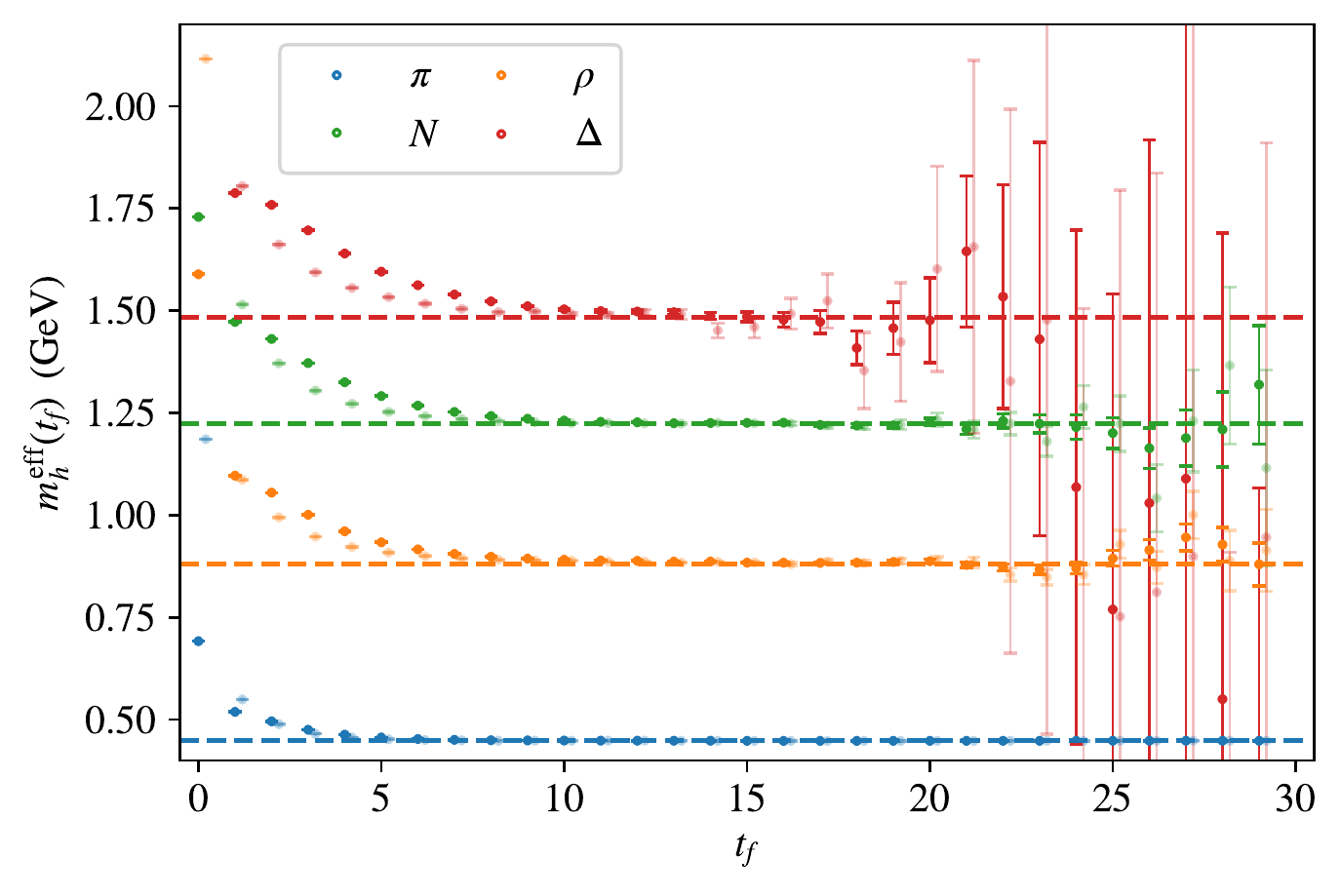}
    \caption{
        Effective mass functions for all four hadrons, with the results obtained using the SP (SS) correlation functions shown in solid (translucent) color. The dashed line indicates the numerical values used to compute kinematic coefficients, as listed in Table~\ref{tab:constraint-counts}.
    }
    \label{fig:meffs}
\end{figure}

We average over sources to obtain per-configuration measurements $C_{s s'}^{h,\text{2pt}}({\bf p},t_f=t'-t_0)$ for each hadron $h$, weighting this average by the number of sources on each configuration when forming bootstrap ensembles as discussed below. 
The effective mass for each hadron $h$ is defined as
\begin{equation}
\label{eq:meff}
m^{\text{eff}}_h(t_f) = \log \left( \frac{\sum_{s} C^{h,\text{2pt}}_{ss}(\vec {0},t_f)}{\sum_{s} C^{h,\text{2pt}}_{ss}(\vec{0},t_f+1)}\right)
\end{equation}
and constructed from the  spin-averaged (over diagonal spin channels for states with spin $\neq 0$) two-point
functions. The results for each hadron are shown in Fig.~\ref{fig:meffs}, along
with the numerical values that
we use for the hadron masses $m_h$
throughout this work, which are obtained via single-state correlated fits to Eq.~\eqref{eq:meff}, in regions
in which the excited-state contamination is smaller than the statistical uncertainties of the effective mass function. The numerical values are given in Table~\ref{tab:constraint-counts}.

\subsection{Three-point correlation functions and ratios}
\label{sec:three-pt-fns}

We construct hadronic three-point functions of the gluon EMT operator, which are defined as
\begin{align}
C_{\mathcal{R},i}^{\pi,\text{3pt}}(\vec{p},\vec{p}'; t',\tau; \vec{x_0},t_0) &=
\sum_{\vec{x},\vec{y}} e^{-i \vec{p}' \cdot (\vec{x} - \vec{x}_0) }  e^{ i \vec{\Delta} \cdot (\vec{y} - \vec{x}_0) } 
 \vev{
        \chi^\pi(\vec{x},t') 
         \hat{T}^{gE}_{\mathcal{R},i}(\vec{y},\tau + t_0) 
         \chi^{\pi\dagger} (\vec{x_0},t_0)
    }
\\
C_{\sigma\sigma'; \mathcal{R},i}^{N,\text{3pt}}(\vec{p},\vec{p}'; t',\tau; \vec{x_0},t_0) &=
\sum_{\vec{x},\vec{y}} e^{-i \vec{p}' \cdot (\vec{x} - \vec{x}_0) }  e^{ i \vec{\Delta} \cdot (\vec{y} - \vec{x}_0) } 
\text{tr} [ \Gamma^N_{\sigma'\sigma}  \vev{
        \chi^N(\vec{x},t') 
         \hat{T}^{gE}_{\mathcal{R},i}(\vec{y},\tau + t_0) ~
         \bar{\chi}^N (\vec{x_0},t_0)
        } ]
\\
C_{aa'; \mathcal{R},i}^{\rho,\text{3pt}}(\vec{p},\vec{p}'; t',\tau; \vec{x_0},t_0) &=
\sum_{\vec{x},\vec{y}} e^{-i \vec{p}' \cdot (\vec{x} - \vec{x}_0) }  e^{ i \vec{\Delta} \cdot (\vec{y} - \vec{x}_0) } ~
\vev{
        \chi_{a'}^\rho(\vec{x},t') 
         \hat{T}^{gE}_{\mathcal{R},i}(\vec{y},\tau + t_0)
         \chi_a^{\rho\dagger} (\vec{x_0},t_0)
    }
\\
C_{\xi\xi'; \mathcal{R},i}^{\Delta,\text{3pt}}(\vec{p},\vec{p}'; t',\tau; \vec{x_0},t_0) &=
\sum_{\vec{x},\vec{y}} e^{-i \vec{p}' \cdot (\vec{x} - \vec{x}_0) }  e^{ i \vec{\Delta} \cdot (\vec{y} - \vec{x}_0) } ~
\text{tr} [\mathcal{D}^{\xi}_{\sigma, a} \mathcal{D}^{\xi'}_{\sigma', a'} \Gamma_{\sigma'\sigma}^\Delta  \vev{
        \chi_{a'}^\Delta(\vec{x},t') 
         \hat{T}^{gE}_{\mathcal{R},i}(\vec{y},\tau + t_0) 
         \bar{\chi}_a^\Delta (\vec{x_0},t_0)
        } ]
\end{align}
where repeated indices are summed over, $(\vec{x}_0, t_0)$ and $(\vec{x}, t')$ are the source and sink positions, $(\vec{y},\tau)$ is the operator insertion position, $\vec{p}$ and $\vec{p}'$ are the three-momenta of the hadron at the source and sink, $\vec{\Delta} = \vec{p}' - \vec{p}$ is the three-momentum injected by the operator, and $\hat{T}^{gE}_{\mathcal{R},i}$ is a gluon EMT operator projected to irrep $\mathcal{R}$ and basis element $i$ as defined in Eq.~\eqref{eqn:hypercubic-irreps}.
These three-point functions are entirely disconnected and so may be computed by correlating two-point functions with measurements of the gluon EMT, i.e.~by computing
\begin{equation}
C_{ss'; \mathcal{R},i}^{h,\text{3pt}}(\vec{p},\vec{p}'; t',\tau; \vec{x_0},t_0) = 
e^{-i \vec{\Delta} \cdot \vec{x}_0 } ~
C^{h,\text{2pt}}_{ss'}(\vec{p}', t';\vec{x_0},t_0) ~
\hat{T}^{gE}_{\mathcal{R},i} (\vec{\Delta}, \tau + t_0) \;,
\end{equation} 
where $\hat{T}^{gE}_{\mathcal{R},i} (\vec{\Delta}, \tau)$ is the gluon EMT operator projected to momentum $\vec{\Delta}$ as in Eq.~\eqref{eqn:gluon-emt-opr-def}.
We average measurements of the three-point correlation functions over sources (translating appropriately) to obtain per-configuration measurements, denoted by $C_{ss'; \mathcal{R},i}^{h,\text{3pt}}(\vec{p},\vec{p}'; t_f=t'-t_0,\tau)$. Given per-configuration measurements of the two- and three-point functions, we draw 1000 bootstrap ensembles, weighting the probability of drawing each configuration by the number of sources measured on that configuration.
To improve signal-to-noise as discussed in Ref.~\cite{Detmold:2017oqb}, we perform a vacuum subtraction of each three-point correlation function 
\begin{equation}
    \vev{ \widetilde{C}_{ss'; \mathcal{R},i}^{h,\text{3pt}} (\vec{p}, \vec{p}'; t_f, \tau) } = \vev{C^{h,\text{3pt}}_{ss'; \mathcal{R},i} (\vec{p}, \vec{p}'; t_f, \tau) } - \vev{C^{h,\text{2pt}}_{ss'}(\vec{p}', t_f) } 
    \bigg\langle
    \frac{1}{N_{\text{src}}} \sum_{(\vec{x}_0, t_0)} e^{-i\vec{\Delta}\cdot \vec{x}_0} \hat{T}^{gE}_{\mathcal{R},i}(\vec{\Delta}, \tau + t_0) 
    \bigg\rangle
    \label{eqn:vac-sub-3pt}
\end{equation}
within each bootstrap ensemble, where $\vev{\ldots}$ indicates an ensemble average and the explicit sum is an average over sources.
We form ratios of two- and three-point functions to isolate the matrix elements of interest.
For all hadrons the appropriate ratio is the same, and is constructed as
\begin{equation} \label{eq:ratio}
R_{ss';\mathcal{R},i}(\vec{p},\vec{p}'; t_f,\tau)
= \frac{
    \vev{ \widetilde{C}^{\mathrm{3pt}}_{ss';\mathcal{R},i}(\vec{p},\vec{p}'; t_f,\tau) }
}{
    \vev{ C_{s's'}^{\mathrm{2pt}}(\vec{p}',t_f) }
}
\sqrt{\frac{
    \vev{ C^{\mathrm{2pt}}_{ss}(\vec{p},t_f-\tau) } ~
    \vev{ C^{\mathrm{2pt}}_{s's'}(\vec{p}',t_f) } ~
    \vev{ C^{\mathrm{2pt}}_{s's'}(\vec{p}',\tau) }
}{
    \vev{ C^{\mathrm{2pt}}_{s's'}(\vec{p}',t_f-\tau) } ~
    \vev{ C^{\mathrm{2pt}}_{ss}(\vec{p},t_f) } ~
    \vev{ C^{\mathrm{2pt}}_{ss}(\vec{p},\tau) }
}}
\end{equation}
within each bootstrap ensemble.
We have suppressed dependence on the source and sink smearing, but we carry out this computation separately using correlation functions constructed from SS- and SP-smeared propagators, yielding separate SS and SP measurements of each ratio.

\subsection{Coefficients, binning, and ratio fits}
\label{sec:coeffs-binning-and-ratio-fits}

The ratio in Eq.~\eqref{eq:ratio} is chosen such that the leading-order $t_f,\tau$ dependence and the overlap factors between the hadronic ground state and
the interpolating operator cancel. Thus, for sufficiently large separation between the source, sink, and operator insertion times, the ratio asymptotically
approaches a value proportional
to the matrix element $\bra{h(p',s')}\hat{T}_{\mu\nu}^g
\ket{h(p,s)}$ with exponentially suppressed excited state contamination.
Specifically, for the four states of interest $h \in \{ \pi,N,\rho,\Delta \}$, the computed ratios are related to the matrix elements of interest as
\begin{alignat}{999}
R_{\mathcal{R},i}^{(\pi)} (\vec{p},\vec{p}'; t_f,\tau)
&& \, \xrightarrow{t_f\gg\tau\gg0} \,
&&R^{(\pi)}_{\mathcal{R},i}(P^\mu, \Delta^\mu) &\equiv
\frac{1}{2\sqrt{E^{\pi}_{\vec{p}}E^{\pi}_{\vec{p}'}}}
\mathcal{O}_{\mathcal{R},i}^{E(\pi)} \;,
\label{eqn:pion-ratio}
\\
R_{\sigma\sigma';\mathcal{R},i}^{(N)} (\vec{p},\vec{p}'; t_f,\tau) 
&& \, \xrightarrow{t_f\gg\tau\gg0} \,
&&R^{(N)}_{\sigma\sigma';\mathcal{R},i}(P^\mu, \Delta^\mu) &\equiv 
\frac{\text{tr} \left[
    \Gamma_{\sigma'\sigma}^N ~
    (\cancel{p}\,'+m_N) ~
    \mathcal{O}_{\mathcal{R},i}^{E(N)} ~
    (\cancel{p}+m_N)
\right]}{
    4 \sqrt{E_{\vec{p}}^N E_{\vec{p}'}^N (E_{\vec{p}}^N+m_N)(E_{\vec{p}'}^N +m_N)} \;,
} \label{eqn:nucleon-ratio}
\\
R_{aa';\mathcal{R},i}^{(\rho)} (\vec{p},\vec{p}'; t_f,\tau) 
&& \, \xrightarrow{t_f\gg\tau\gg0} \,
&&R^{(\rho)}_{aa';\mathcal{R},i}(P^\mu, \Delta^\mu) &\equiv
\frac{
    \Lambda^{(\rho)}_{a' \alpha'}(\vec{p}') ~
    \mathcal{O}_{\mathcal{R},i}^{E(\rho) \alpha \alpha'} ~
    \Lambda^{(\rho)}_{\alpha a}(\vec{p})
}{
2 \sqrt{ 
    E^{\rho}_{\vec{p}} E^{\rho}_{\vec{p}'}
    ~ \Lambda^{(\rho)}_{a a}(\vec{p}) ~ \Lambda^{(\rho)}_{a' a'}(\vec{p}') \;,
}} \label{eqn:rho-ratio}
\\
R_{\xi\xi';\mathcal{R},i}^{(\Delta)} (\vec{p},\vec{p}'; t_f,\tau) 
&& \, \xrightarrow{t_f\gg\tau\gg0} \,
&&R^{(\Delta)}_{\xi\xi';\mathcal{R},i}(P^\mu, \Delta^\mu) &\equiv
\frac{ \text{tr} \left[
    \mathcal{D}^{\xi}_{\sigma,a}
    \mathcal{D}^{\xi'}_{\sigma',a'}
    \Gamma^\Delta_{\sigma'\sigma} ~
    \Lambda^{(\Delta)}_{a' \alpha'}(p') ~
    \mathcal{O}_{\mathcal{R},i}^{E(\Delta) \alpha\alpha'} ~
    \Lambda^{(\Delta)}_{\alpha a}(p)
\right]}{ 2 \sqrt{
    E^\Delta_{\vec{p}} E^\Delta_{\vec{p}'} ~
    \text{tr}[
    \mathcal{D}^{\xi}_{\kappa,b}
    \mathcal{D}^{\xi}_{\kappa',b'}
    \Gamma^\Delta_{\kappa'\kappa} ~ \Lambda^{(\Delta)}_{b' b}(p') ] ~
    \text{tr}[
    \mathcal{D}^{\xi'}_{\lambda,c}
    \mathcal{D}^{\xi'}_{\lambda',c'}
    \Gamma^\Delta_{\lambda'\lambda} ~ \Lambda^{(\Delta)}_{c'c}(p) ]
}} \;, \label{eqn:delta-ratio}
\end{alignat}
where $P = (p+p')/2$, $\Delta=p'-p$, the
repeated Lorentz indices $\alpha$ and $\alpha'$ are contracted with the Minkowski metric, and other repeated indices  are summed over besides the external spin indices $\sigma, \sigma'$ in Eq.~\eqref{eqn:nucleon-ratio}, $a,a'$ in Eq.~\eqref{eqn:rho-ratio}, and $\xi,\xi'$ in Eq.~\eqref{eqn:delta-ratio}.

The matrix elements $\mathcal{O}^{E(h)}_{\mathcal{R},i}$ are constructed as the Euclidean analogs of the decompositions of $\mathcal{O}^{(h)}_{\mu\nu} = \mathcal{O}^{M(h)}_{\mu\nu}$ into GFFs in Eqs.~\eqref{eq:pionME}, \eqref{eq:protonME}, \eqref{eq:rhoME}, and \eqref{eq:deltaME}, projected to the hypercubic irrep bases $\mathcal{R},i$ defined in Eq.~\eqref{eqn:hypercubic-irreps}.
The free Lorentz indices on $\mathcal{O}^{M(h)}_{\mu\nu}$ are Euclideanized using
 the Euclidean-to-Minkowski matching relation
\begin{equation}\begin{gathered}
    \left[x^{M}\right]^{\mu} = i^{\delta_{\mu t}} \left[x^{E}\right]^{\mu}, \quad
    \left[\partial^{M}\right]^{\mu} = (-i)^{\delta_{\mu t}} \left[\partial^{E}\right]^{\mu} \;,
\end{gathered}\end{equation}
where $i^{\delta_{\mu t}}$ generates a factor of $i$ on the temporal component.
It follows directly from $F_{\mu\nu} \propto [\mathcal{D}_\mu, \mathcal{D}_\nu]$ that the Euclidean and Minkowski matrix elements of the gluon EMT are related as
\begin{equation}
\mathcal{O}^{M(h)}_{\mu\nu} = - i^{\delta_{\mu t}} i^{\delta_{\nu t}} \mathcal{O}^{E(h)}_{\mu \nu}.
\end{equation}

Each ratio is associated with a different set of momenta $\Delta^\mu$ and $P^\mu$, operator basis element $\mathcal{R},i$, and spin channel $s \rightarrow s'$, all of which define a set of kinematic coefficients $K^{h,j}_{ss';\mathcal{R},i}(P^\mu, \Delta^\mu)$ for the bare GFFs for irrep $\mathcal{R}$ in the decomposition
\begin{equation}
    R^{(h)}_{ss';\mathcal{R},i}(P^\mu, \Delta^\mu) = \sum_j K^{h,j}_{ss';\mathcal{R},i}(P^\mu, \Delta^\mu) ~ G^{h,j}_{\mathcal{R}}(t) .
\end{equation}
The GFFs are real, but the kinematic coefficients and ratio measurements are generically complex, so the real and imaginary parts of each ratio measurement provide independent constraints on the GFFs; we thus treat each part as a separate real-valued ratio associated with real coefficients.
We discard any ratio for which all kinematic coefficients are zero.
Energies appearing in the expressions for the kinematic coefficients of each hadron and $t=\Delta^2$ are set using the dispersion relation $E_\vec{p}^h = \sqrt{m_h^2+\vec{p}^2}$.
Although the kinematic coefficients and values of $t=\Delta^2$ associated with each ratio are functions of the hadron mass $m_h$ and lattice spacing $a$ and so in principle are only known up to some uncertainty (correlated with the ratios), these errors are subdominant, so we neglect them and evaluate the coefficients using $a = 0.1167~\mathrm{fm}$ and the numerical values of $m_h$ listed in Table~\ref{tab:constraint-counts}, obtained from single-state fits to the effective mass as shown in Fig.~\ref{fig:meffs}.

\begin{figure*} 
\centering
\subfloat[\centering $t$ binning for $\pi$]{{
    \includegraphics{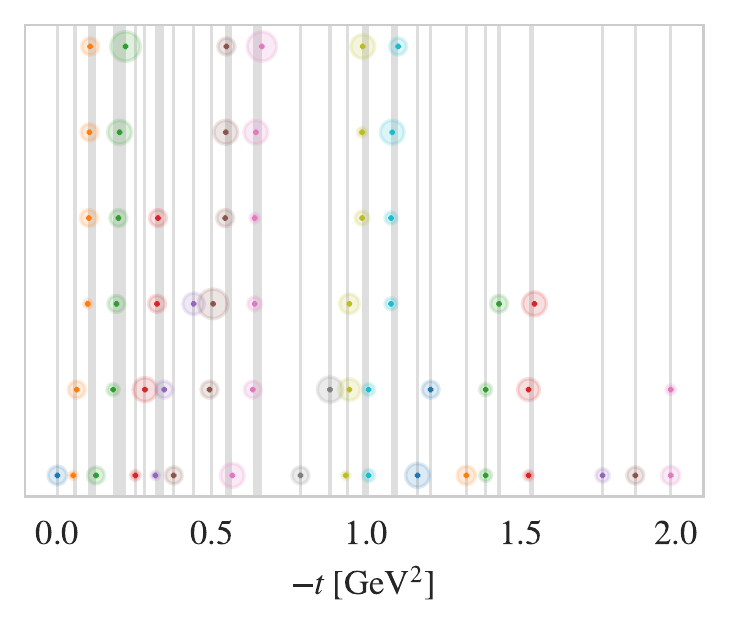}
}}
\!
\subfloat[\centering $t$ binning for $N$]{{
    \includegraphics{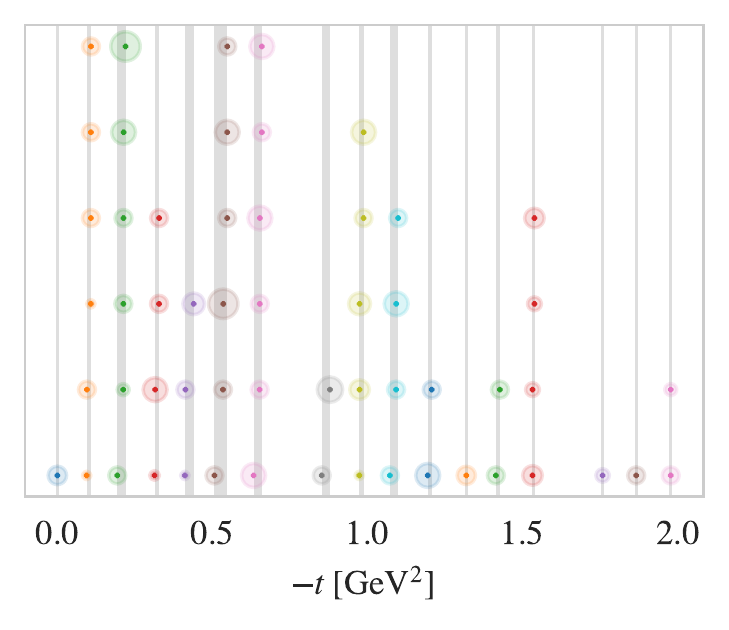}
}}
\\
\subfloat[\centering $t$ binning for $\rho$]{{
    \includegraphics{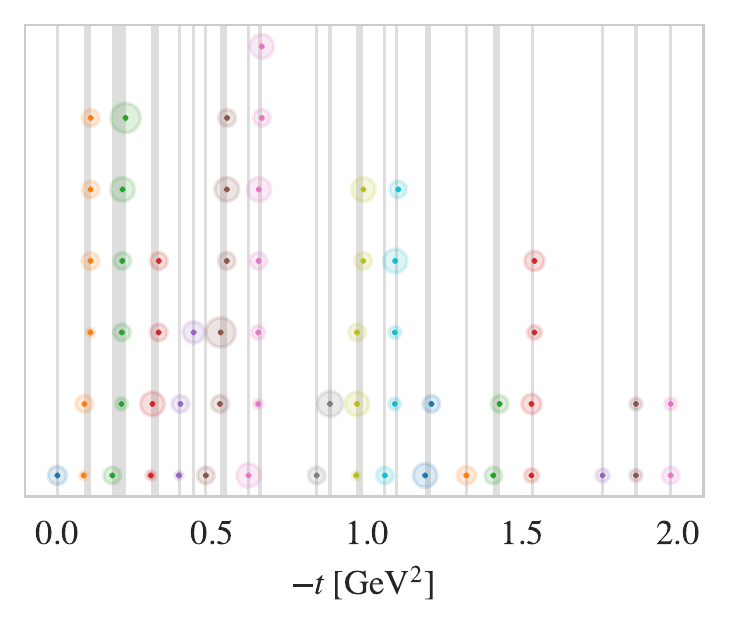}
}}
\!
\subfloat[\centering $t$ binning for $\Delta$]{{
    \includegraphics{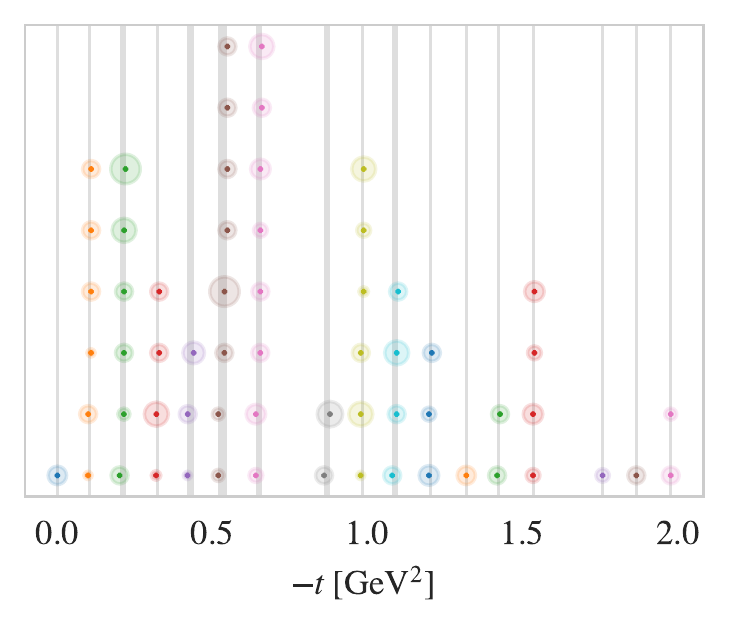}
}}
\caption{
    For each of the states $\pi$, $N$, $\rho$, and $\Delta$, the figure shows how different combinations of the discretized momenta $p$, $p'$, and $\Delta = p'-p$ used in this calculation are associated with discrete $t$ bins as described in the text.
    Each marker represents a collection of different momentum combinations that result in the same
    value of $-t$.
    Colors correspond to different values of $|\vec{\Delta}|^2$.
    The area of each marker is proportional to the number of associated momentum combinations.
    Gray bands indicate each $t$ bin, and markers are associated with the band that contains their central point.
    Table~\ref{tab:constraint-counts} lists the number of bins for each state.
}
\label{fig:t-bins}
\end{figure*}

We associate each ratio with a ``$t$ bin'' so that we can estimate model-independent values of the GFFs at discrete values of $t$. Bins are defined by grouping together any two ratios associated with values of $t$ that differ by less than $0.03~\mathrm{GeV}^2$, with no additional restriction on the maximum width of each bin.\footnote{This binning algorithm is identical to the one used in Ref.~\cite{Shanahan:2018pib}.}
We define the value of $t$ for each bin as the average over $t$ for all ratios in the bin.
Figure~\ref{fig:t-bins} illustrates the resulting associations for each hadron.
There is a one-to-one correspondence
between $t$-bins and values of $|\vec{\Delta}|^2$ in the case of the baryons, but not
in that of the mesons, which is due to the smaller 
masses of $\rho$ and $\pi$ compared to $N$ and $\Delta$.
Within each $t$-bin, we average any ratios associated with kinematic coefficients related by an overall sign within each bootstrap draw, with each ratio multiplied by the appropriate sign.
We do not combine ratios from different irreps, as they are renormalized differently, and we continue to keep SS and SP ratios separate.
This additional averaging helps to compensate for gauge noise, providing clearer signals for subsequent fitting.
The resulting averaged ratios $\overline{R}_{\mathcal{R} t c}(t_f, \tau)$ in momentum bin $t$ for irrep $\mathcal{R}$ are no longer associated with specific momenta, irrep basis elements, spin channels, or real/imaginary parts, and are instead associated simply with some particular set of kinematic coefficients indexed by $c$.
The resulting reduction in data volume is significant, as tabulated in Table~\ref{tab:constraint-counts}.

\begin{table}
\begin{ruledtabular}
\begin{tabular}{cccccc}
State & $a m_h$ & \# spin channels & \# $R_{ss';\mathcal{R},i}(P^\mu, \Delta^\mu)$ & \# $t$-bins & \# $\overline{R}_{\mathcal{R}tc}$
\\ \hline
$\pi$ & 0.266 & 1 & 24086 & 26 & 672
\\ 
$N$  & 0.724 & 4 & 175244 & 17 & 1940
\\ 
$\rho$  & 0.534 & 9 & 385182 & 22 & 8084
\\ 
$\Delta$  & 0.878 & 10 & 453868 & 17 & 17839
\end{tabular}
\end{ruledtabular}
\caption{For each hadron: the mass used to
calculate the kinematic coefficients of the GFFs in lattice units, the number of spin channels incorporated, the number of squared momentum transfer bins ($t$-bins), and the number
of ratios before and after combining ratios with kinematic coefficients
related by an overall sign.
}
\label{tab:constraint-counts}
\end{table}

\begin{figure*} [h]
\captionsetup[subfloat]{captionskip=-3pt}
\centering
\subfloat[\centering $h=\pi$, $-t=0\;\text{GeV}^2$]
{{\includegraphics[height=5.2cm,width=7.5cm]{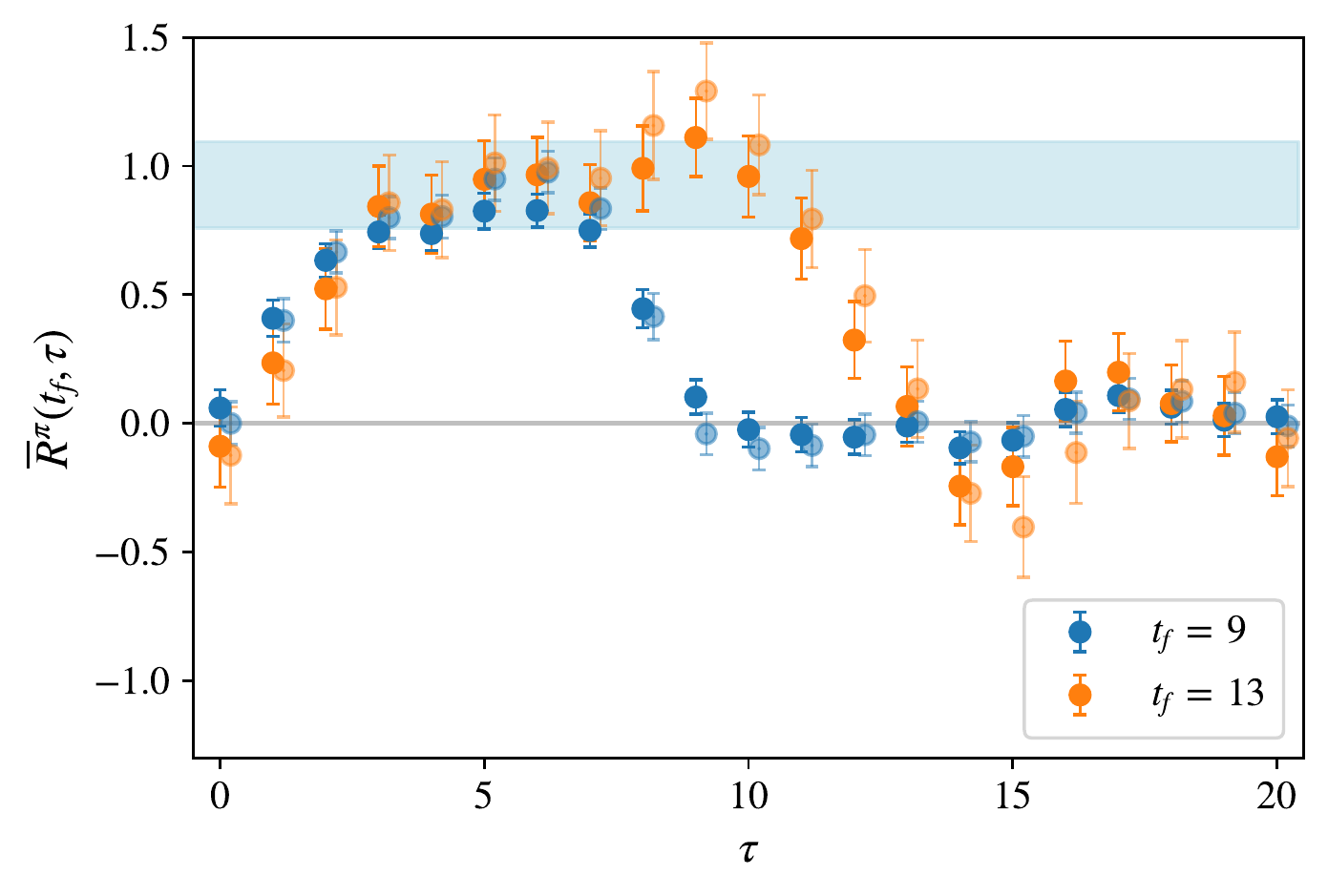} }}
\!
\subfloat[\centering $h=\pi$, $-t=2\;\text{GeV}^2$]
{{\includegraphics[height=5.2cm,width=7.5cm]{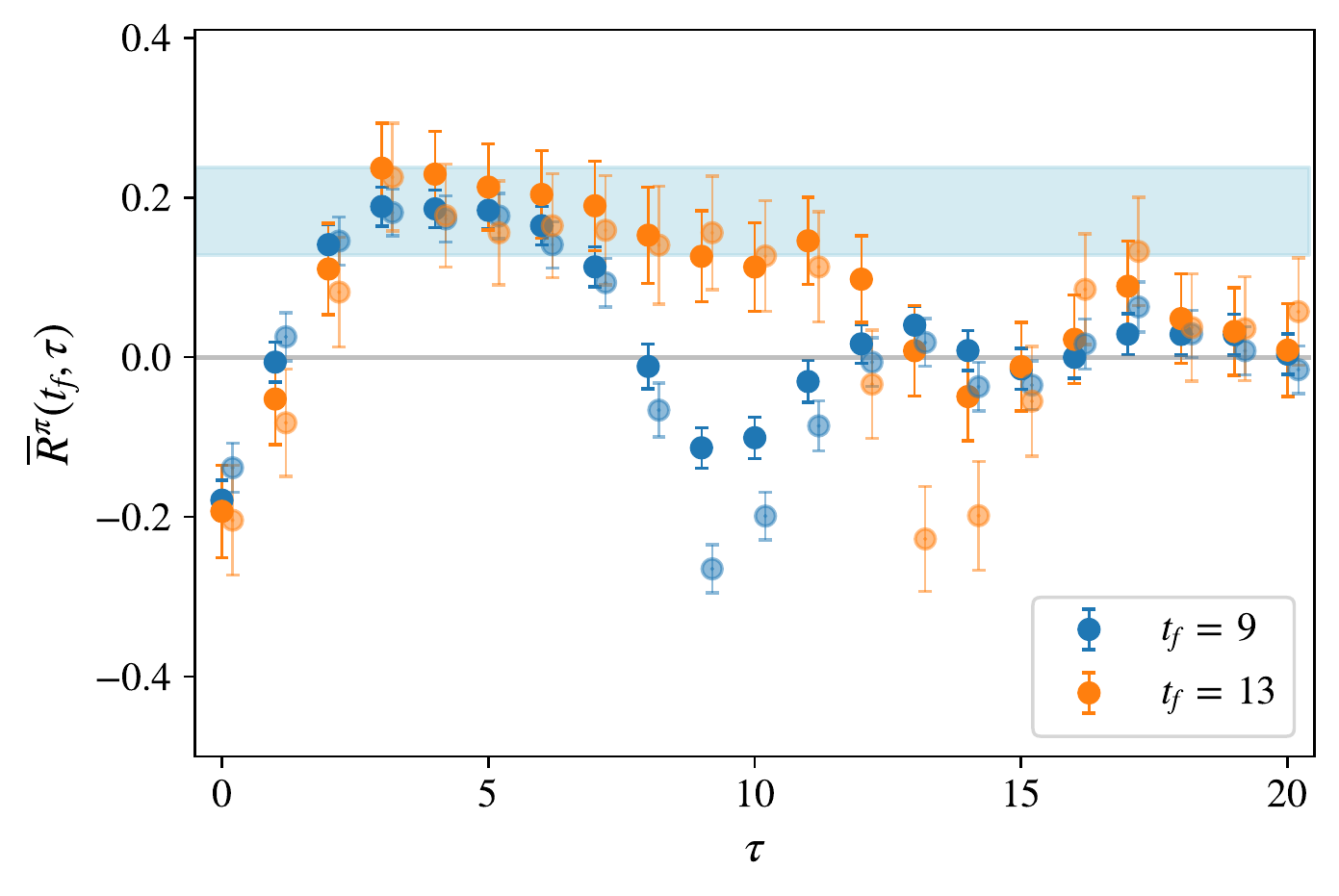} }}
\\[-2ex]
\subfloat[\centering $h=N$, $-t=0\;\text{GeV}^2$]
{{\includegraphics[height=5.2cm,width=7.5cm]{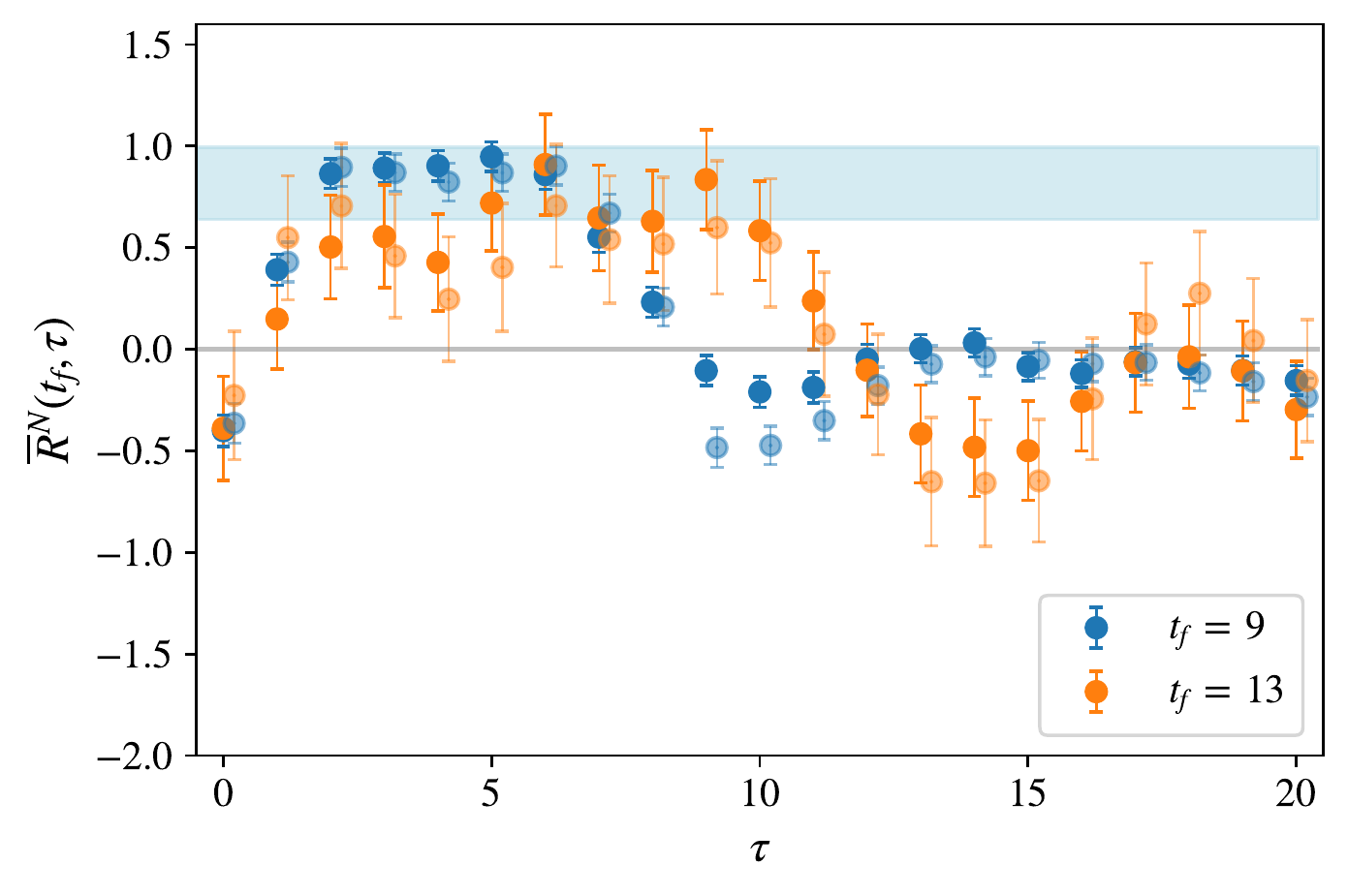} }}
\!
\subfloat[\centering $h=N$, $-t=2\;\text{GeV}^2$]
{{\includegraphics[height=5.2cm,width=7.5cm]{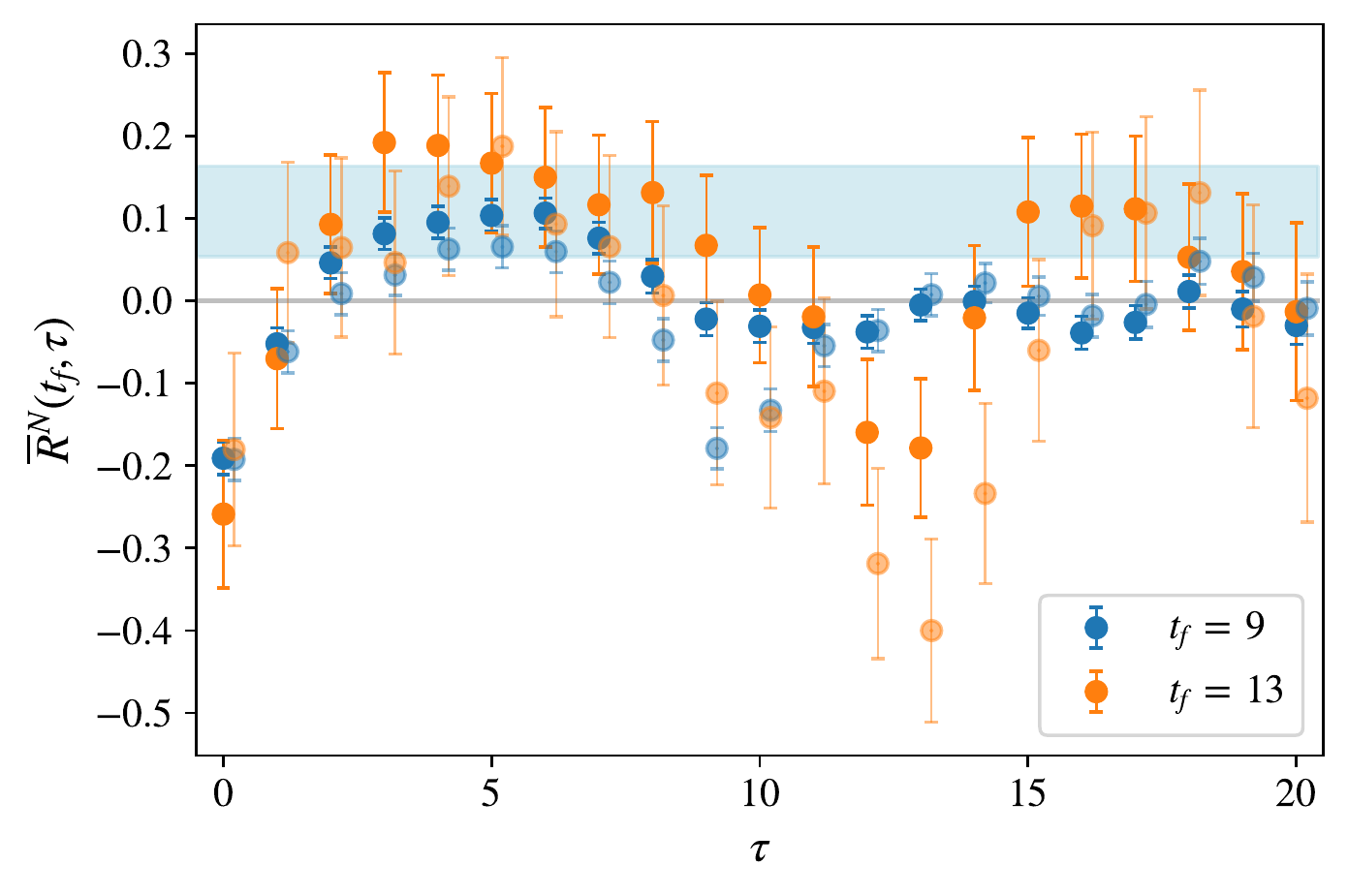} }}
\\[-2ex]
\subfloat[\centering $h=\rho$, $-t=0\;\text{GeV}^2$]
{{\includegraphics[height=5.2cm,width=7.5cm]{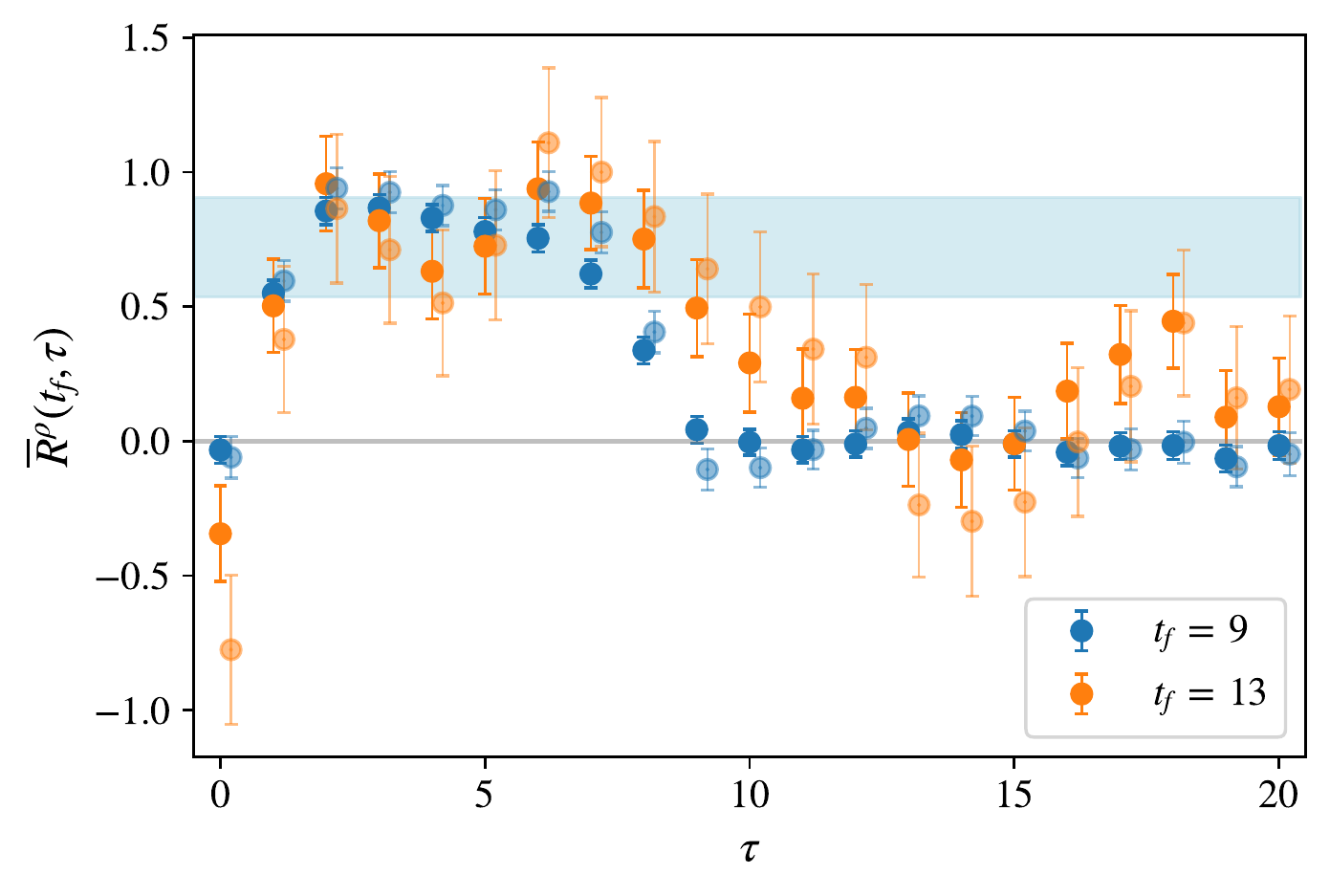} }}
\!
\subfloat[\centering $h=\rho$, $-t=2\;\text{GeV}^2$]
{{\includegraphics[height=5.2cm,width=7.5cm]{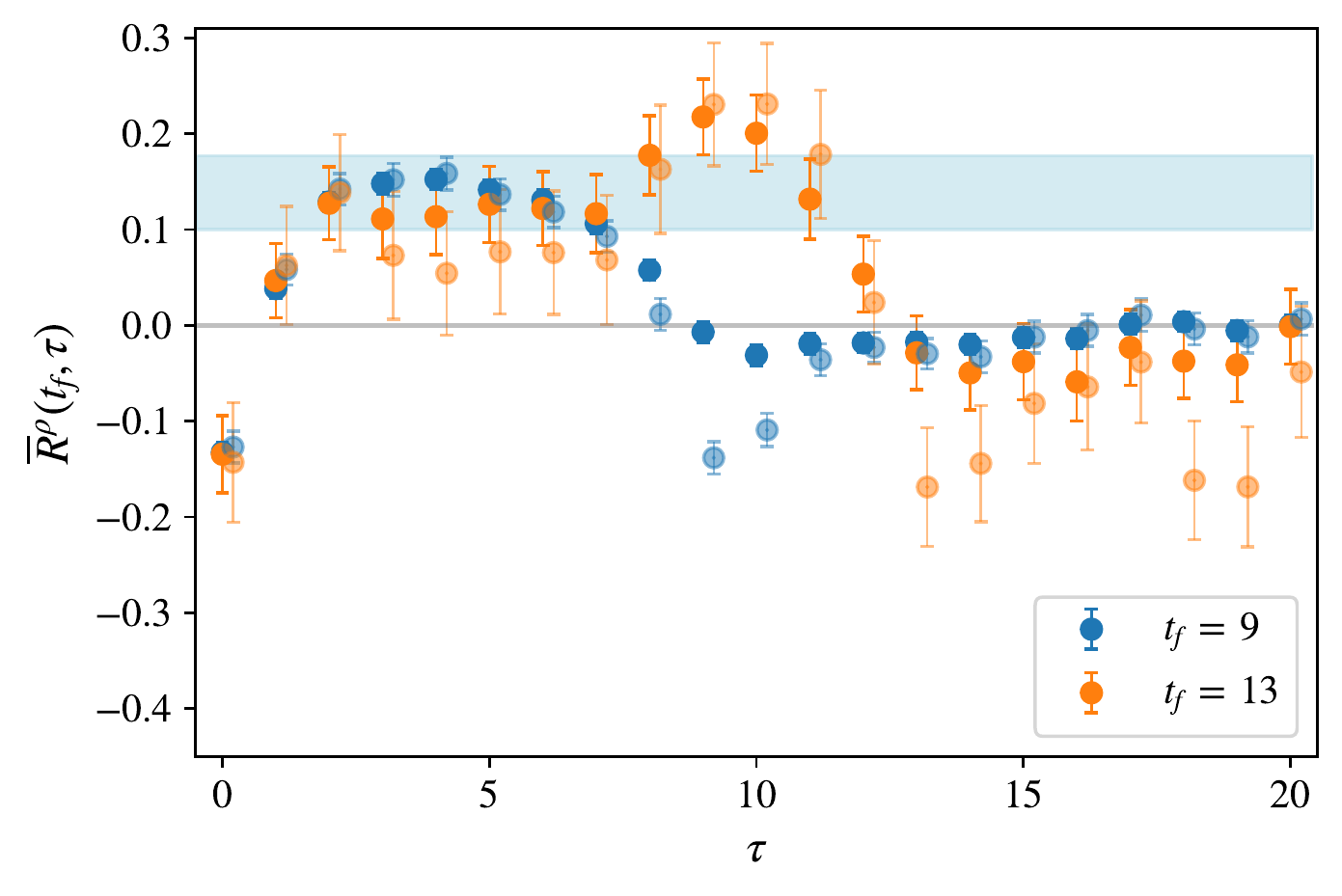} }}
\\[-2ex]
\subfloat[\centering $h=\Delta$, $-t=0\;\text{GeV}^2$]
{{\includegraphics[height=5.2cm,width=7.5cm]{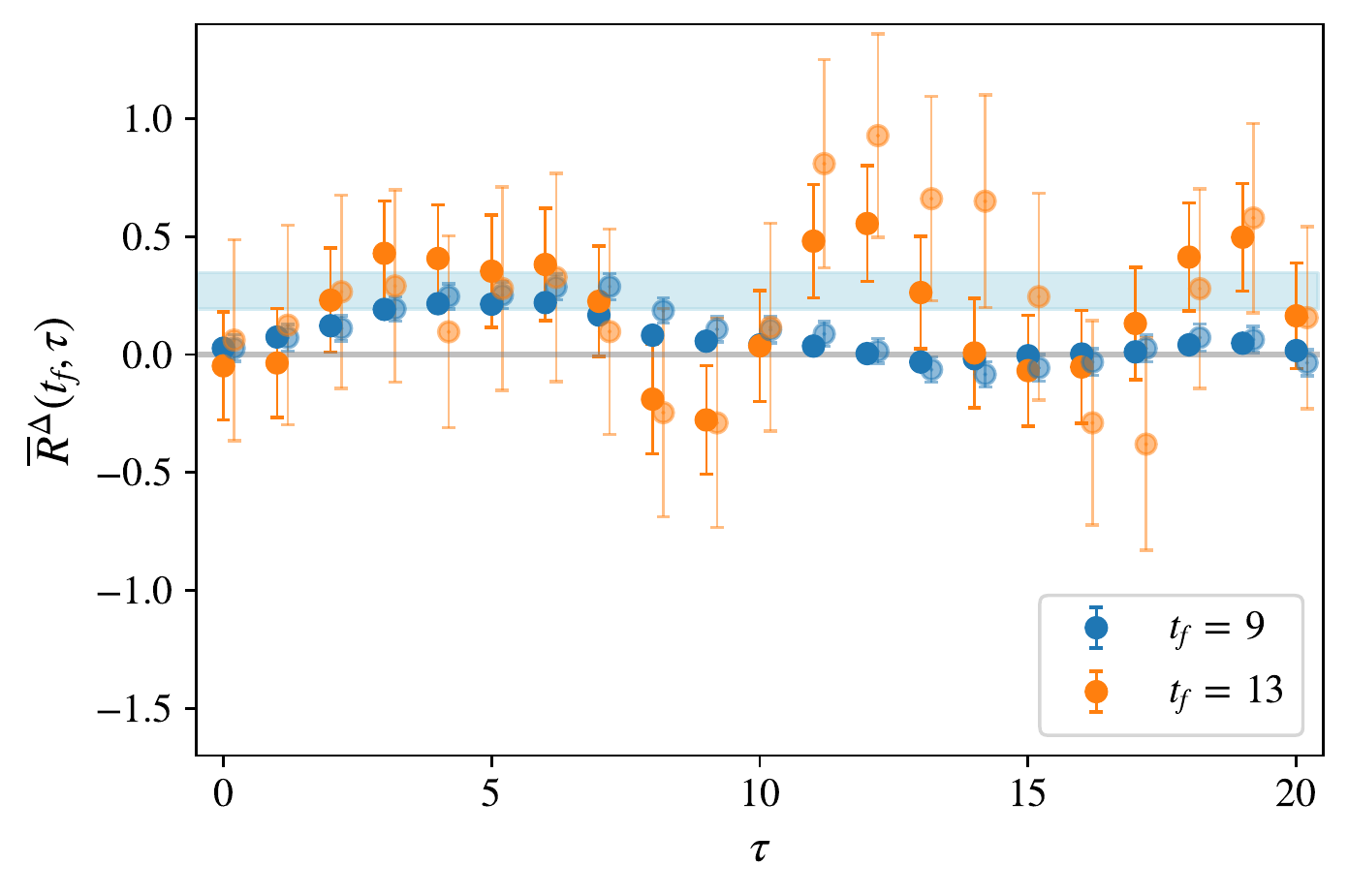} }}
\!
\subfloat[\centering $h=\Delta$, $-t=2\;\text{GeV}^2$]
{{\includegraphics[height=5.2cm,width=7.5cm]{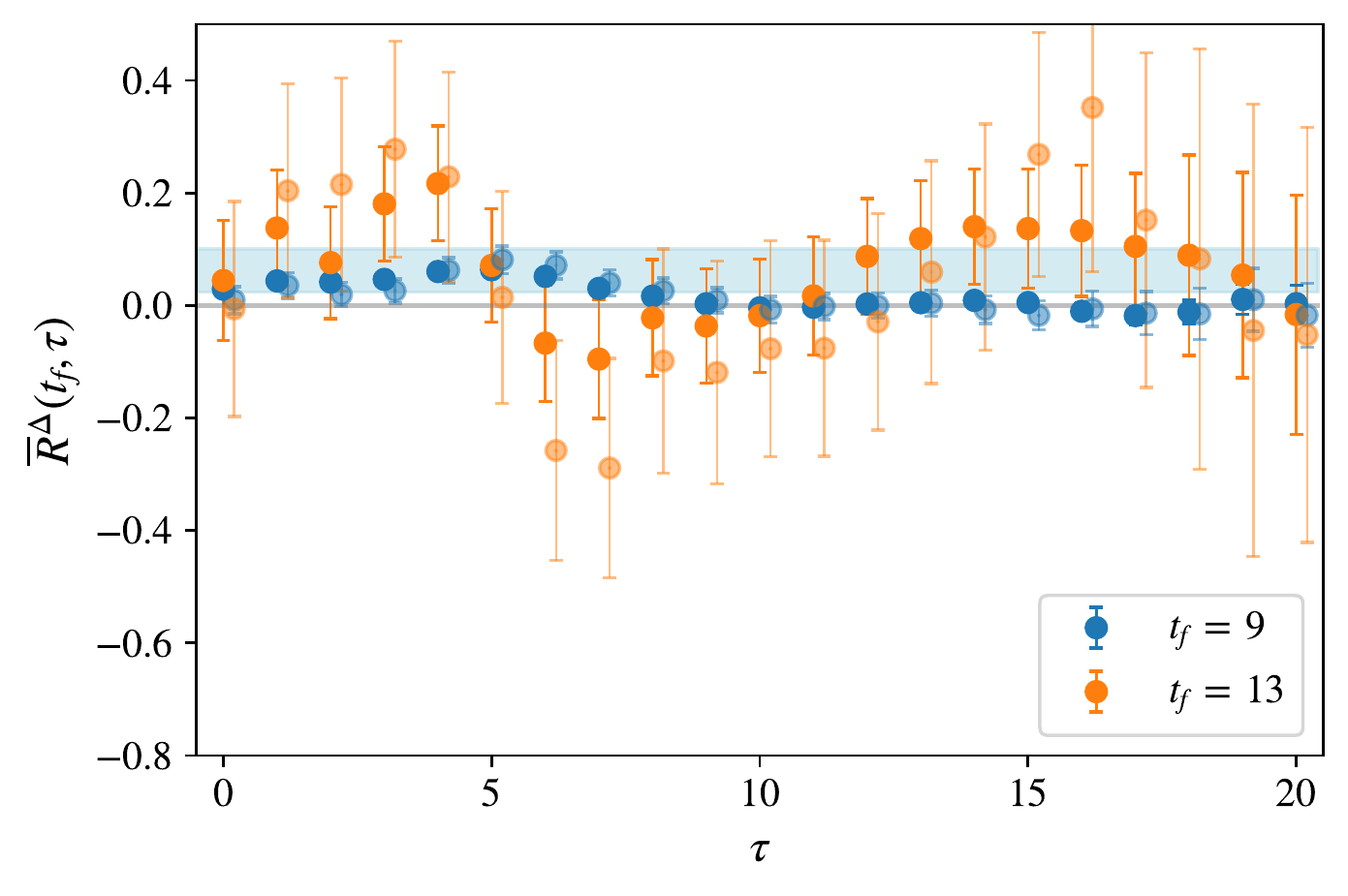} }}
\caption{Examples of averaged ratios $\bar{R}^{(h)}_{\mathcal{R}tc}$ as a function of operator insertion time $\tau$, with one binned ratio at two different sink times per figure, along with the $p$-value-averaged fit bands. In the left (right) column are examples of ratios at squared momentum transfer $-t = 0\;\text{GeV}^2$ ($-t = 2\;\text{GeV}^2 $). The solid (translucent) points correspond to results computed with SP (SS) smeared propagators.
}
\label{fig:bumps}
\end{figure*}

\begin{figure*} [h]
\captionsetup[subfloat]{captionskip=-3pt}
\centering
\subfloat[\centering $h=\pi$, $-t=0\;\text{GeV}^2$]
{{\includegraphics[height=5.2cm,width=7.5cm]{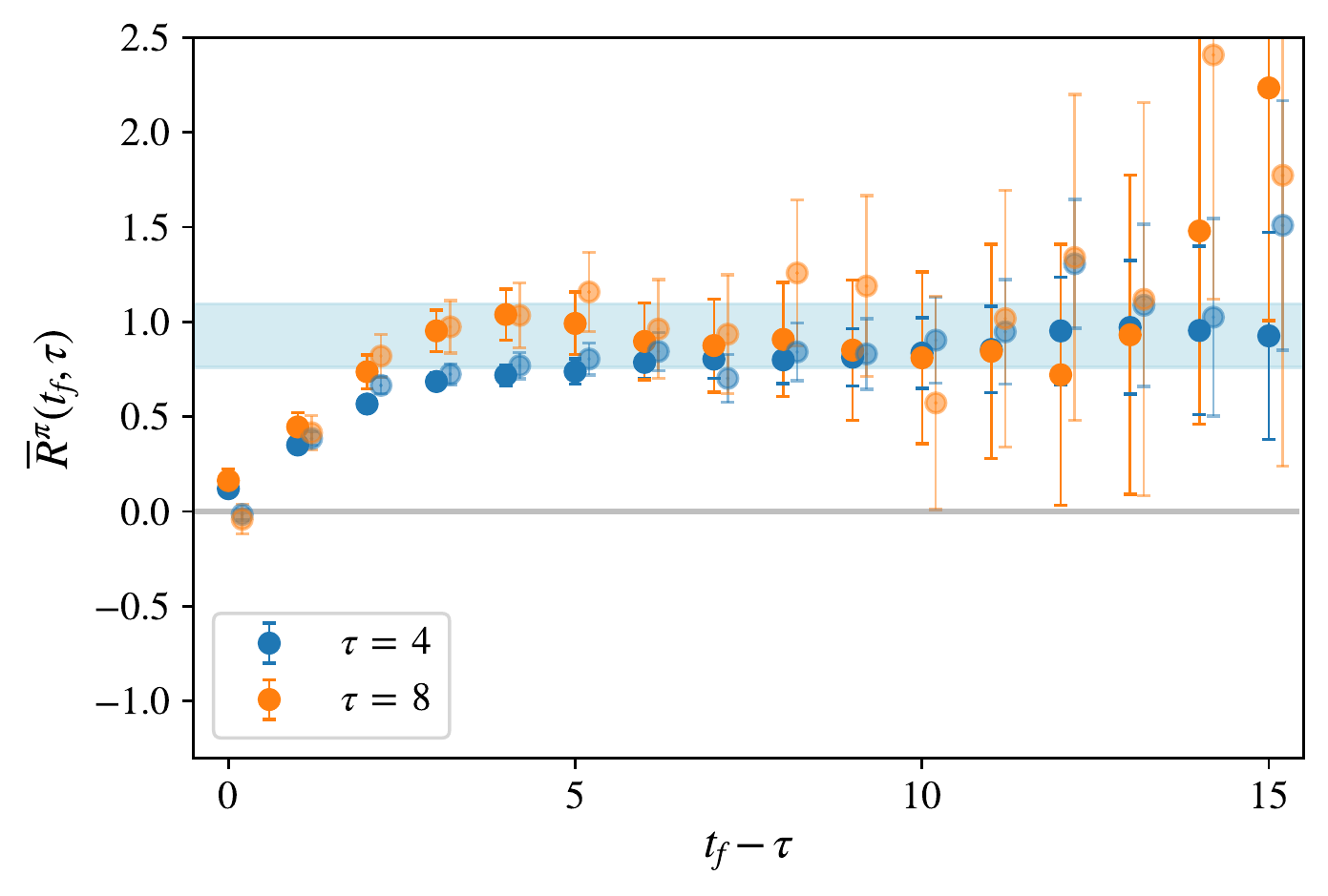} }}
\!
\subfloat[\centering $h=\pi$, $-t=2\;\text{GeV}^2$]
{{\includegraphics[height=5.2cm,width=7.5cm]{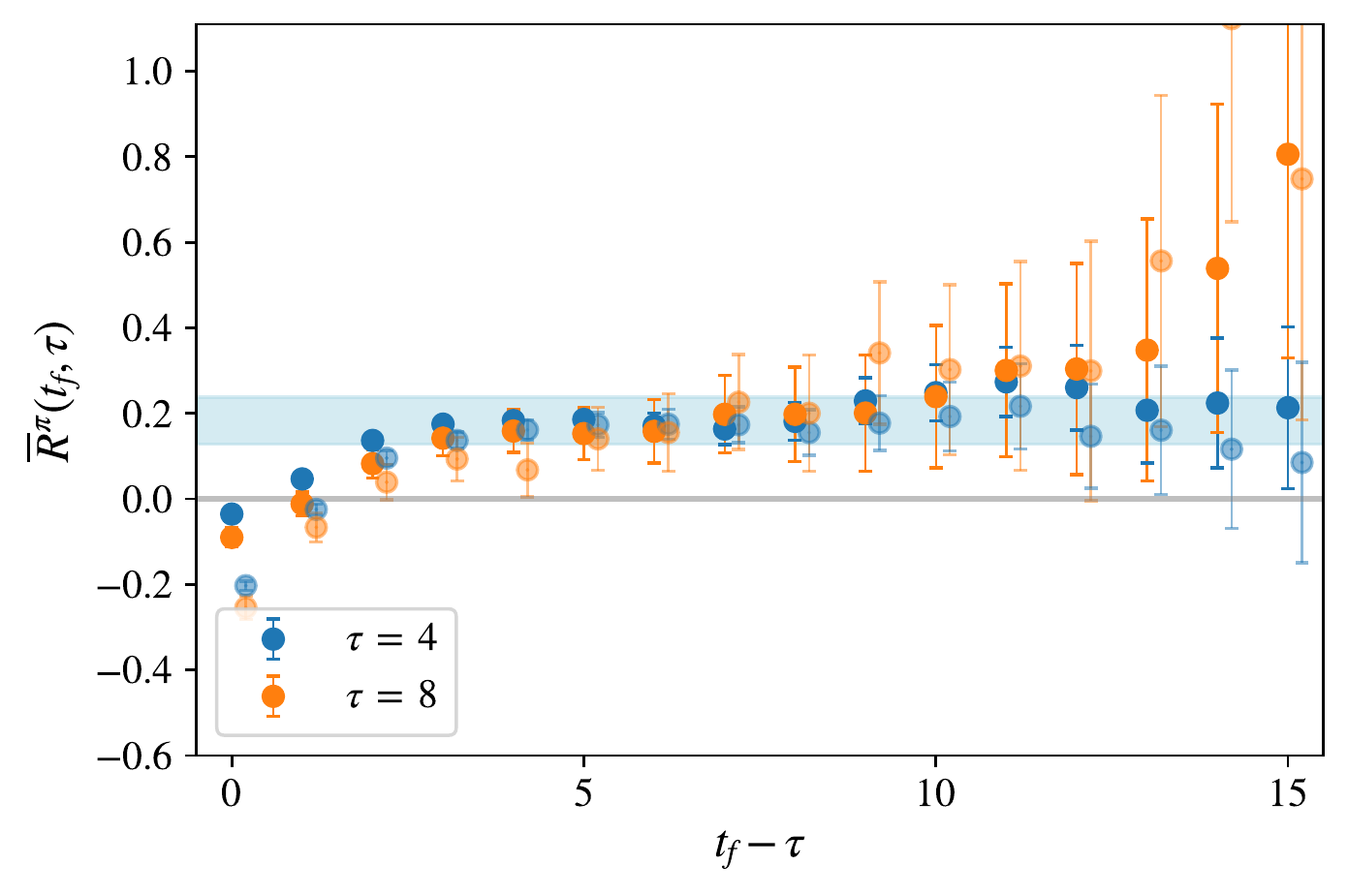} }}
\\[-2ex]
\subfloat[\centering $h=N$, $-t=0\;\text{GeV}^2$]
{{\includegraphics[height=5.2cm,width=7.5cm]{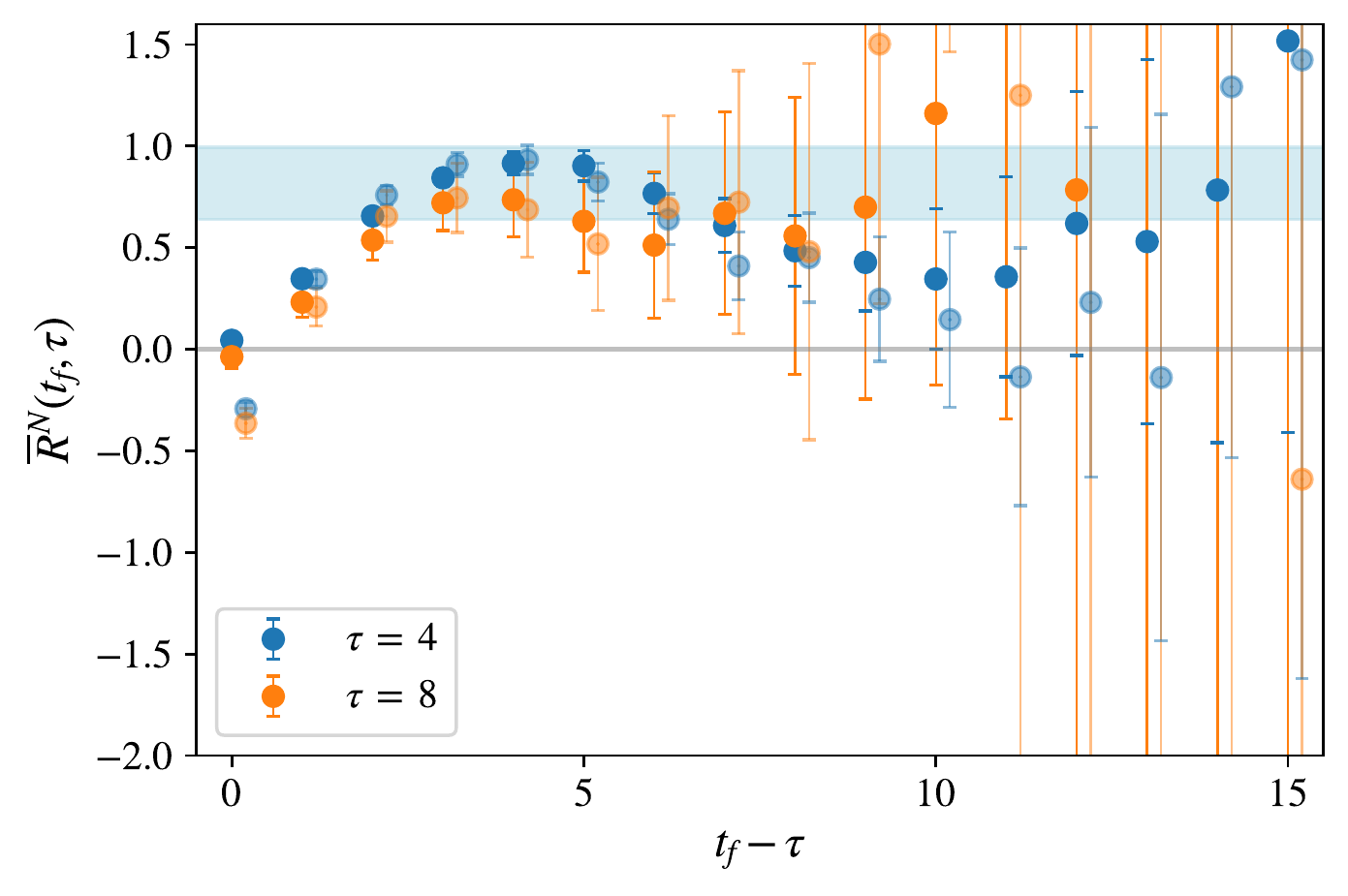} }}
\!
\subfloat[\centering $h=N$, $-t=2\;\text{GeV}^2$]
{{\includegraphics[height=5.2cm,width=7.5cm]{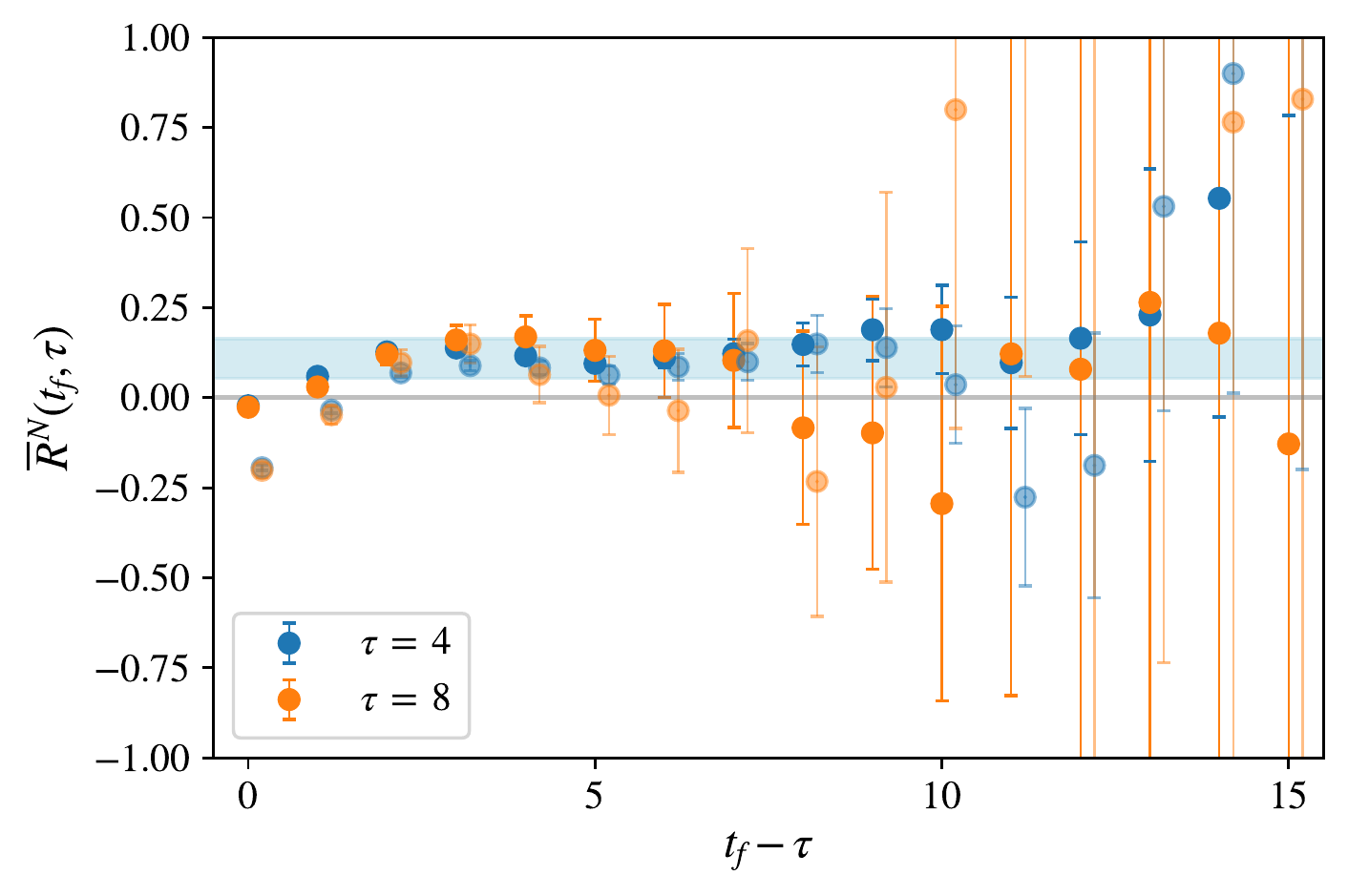} }}
\\[-2ex]
\subfloat[\centering $h=\rho$, $-t=0\;\text{GeV}^2$]
{{\includegraphics[height=5.2cm,width=7.5cm]{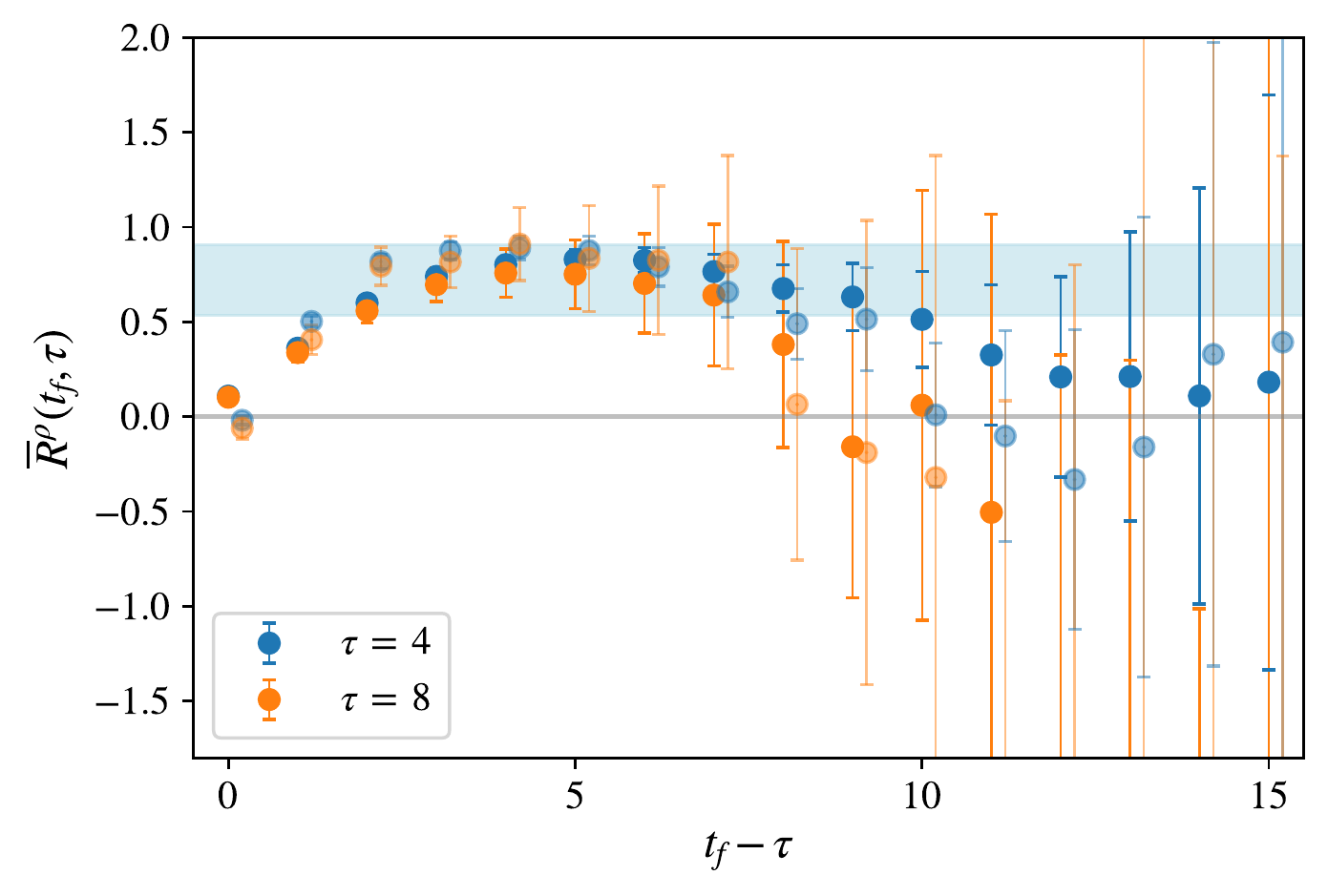} }}
\!
\subfloat[\centering $h=\rho$, $-t=2\;\text{GeV}^2$]
{{\includegraphics[height=5.2cm,width=7.5cm]{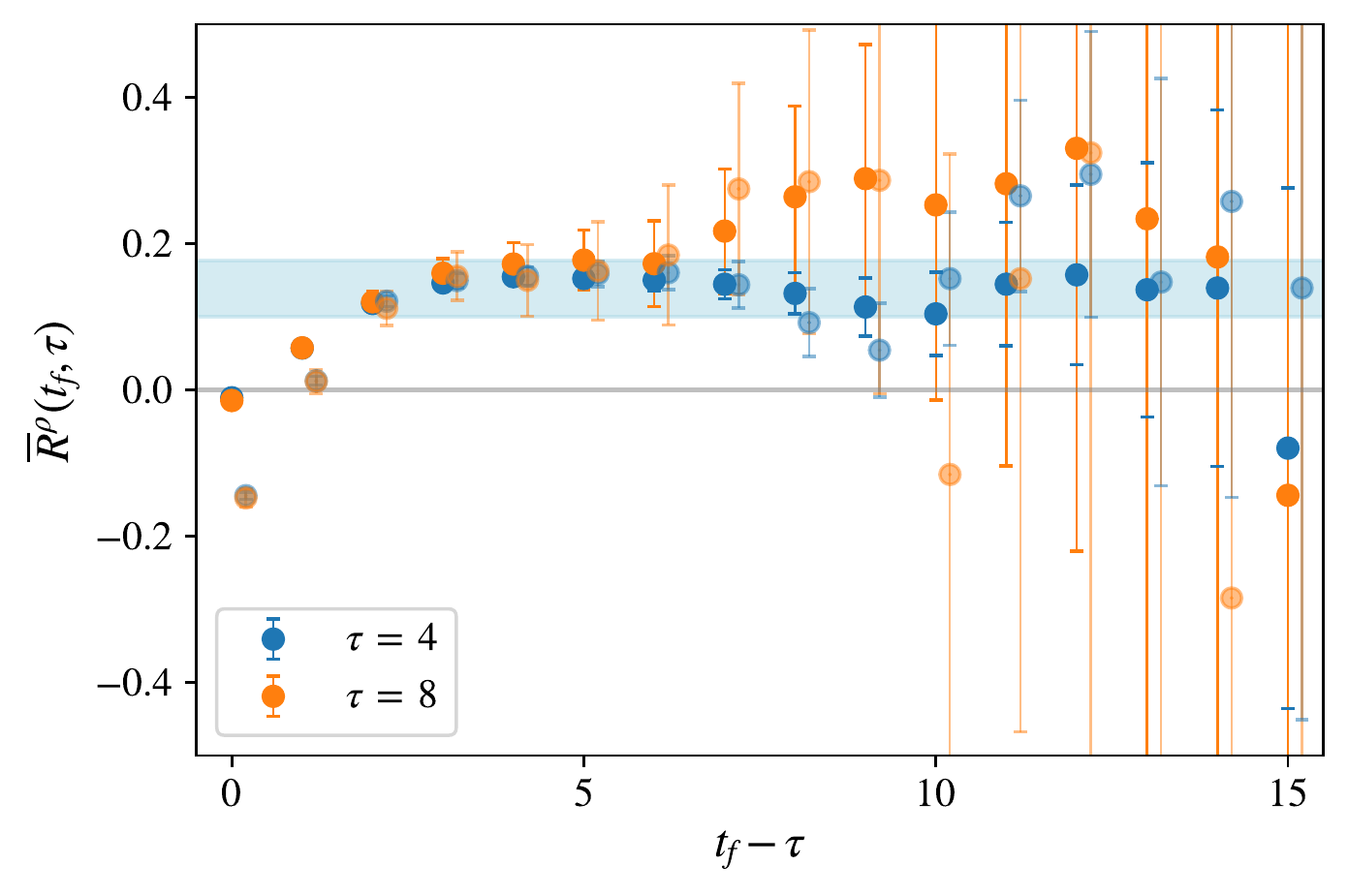} }}
\\[-2ex]
\subfloat[\centering $h=\Delta$, $-t=0\;\text{GeV}^2$]
{{\includegraphics[height=5.2cm,width=7.5cm]{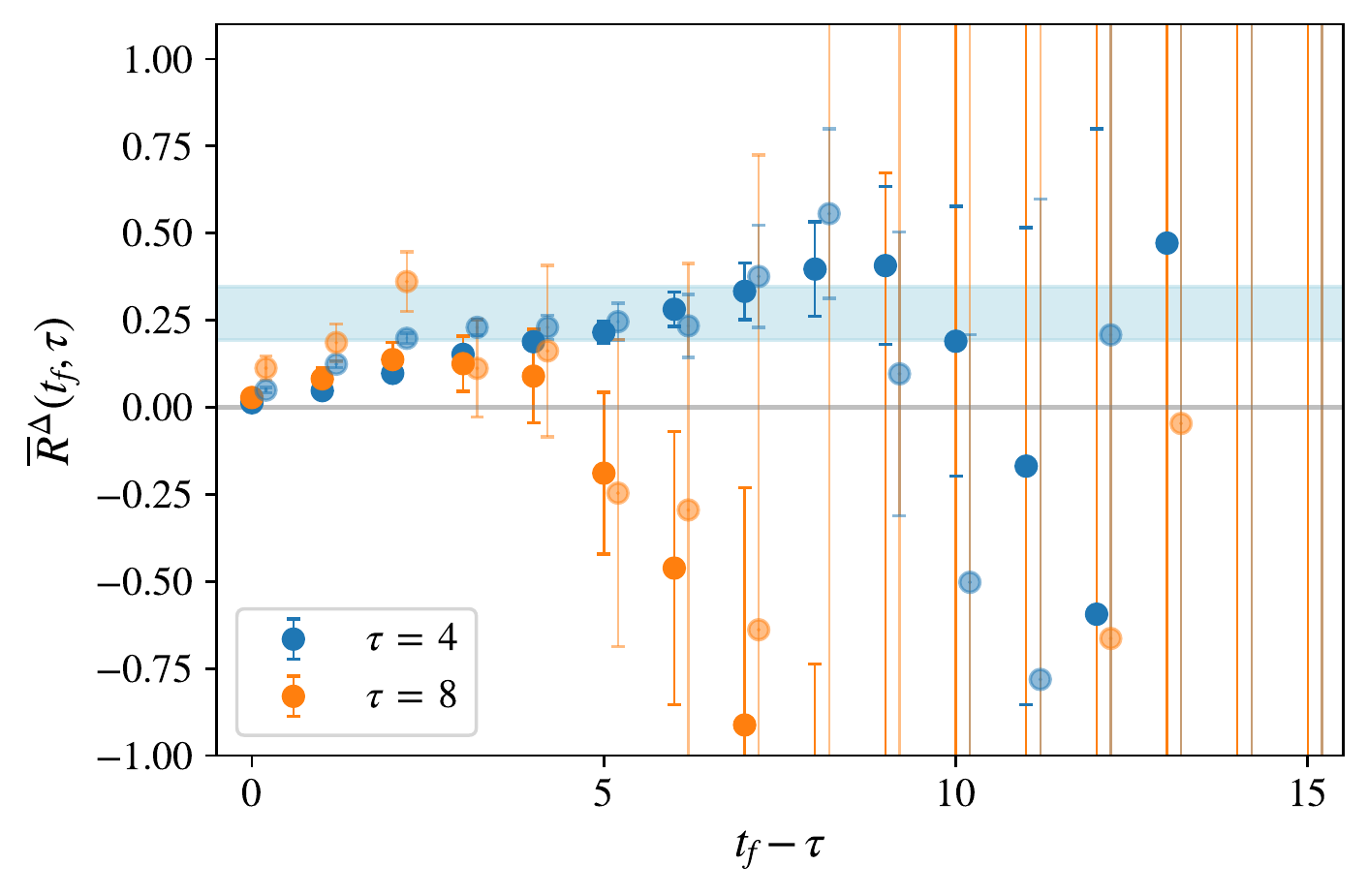} }}
\!
\subfloat[\centering $h=\Delta$, $-t=2\;\text{GeV}^2$]
{{\includegraphics[height=5.2cm,width=7.5cm]{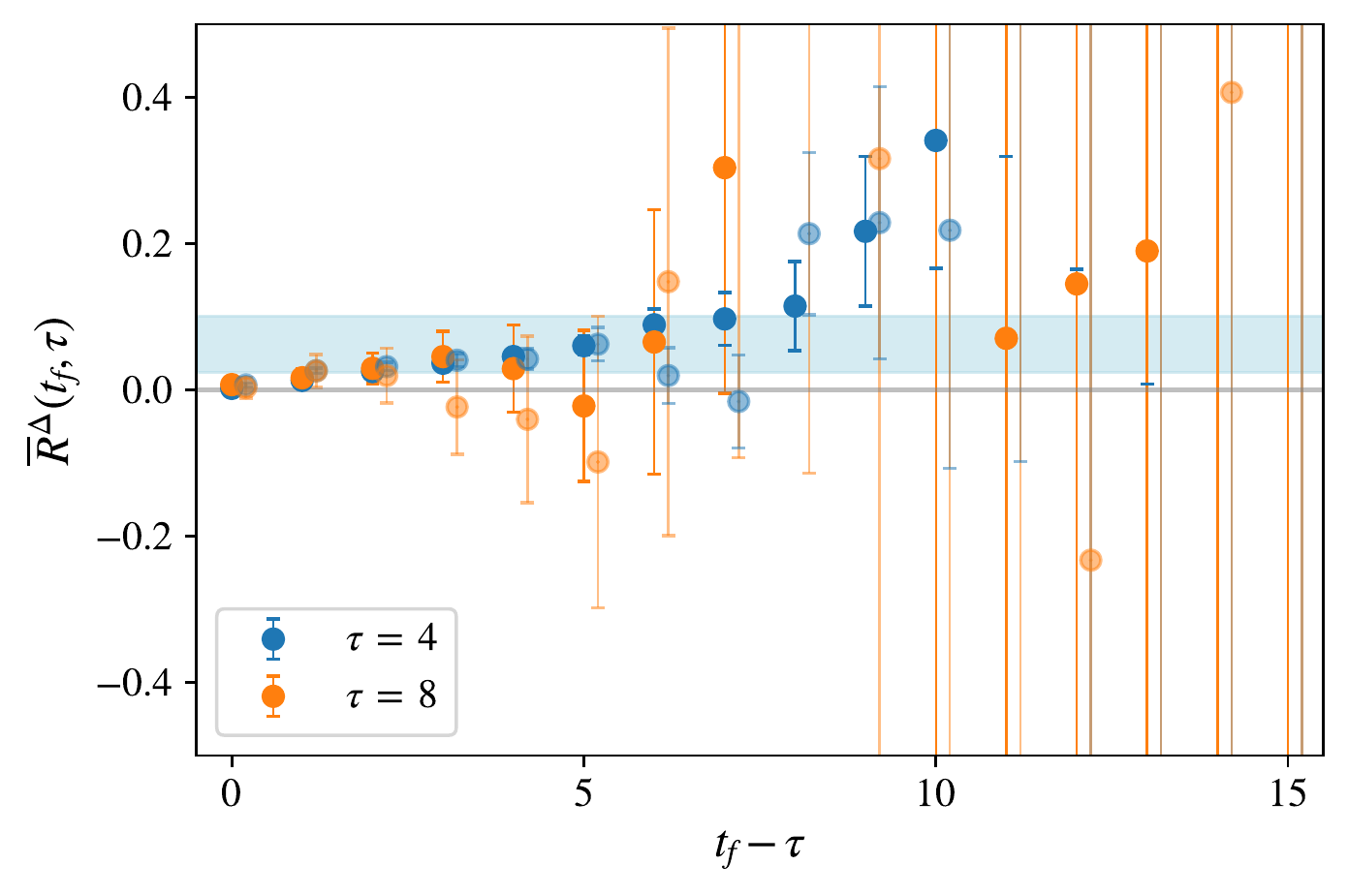} }}
\caption{Examples of averaged ratios $\bar{R}^{(h)}_{\mathcal{R}tc}$ as a function of sink-operator separation $t_f - \tau$, with one binned ratio at two different operator insertion times $\tau$ per figure, along with the $p$-value-averaged fit bands. 
In the left (right) column are examples of ratios at squared momentum transfer $-t = 0\;\text{GeV}^2$ ($-t = 2\;\text{GeV}^2 $). The solid (translucent) points correspond to results computed with SP (SS) smeared propagators.
}
\label{fig:taubumps}
\end{figure*}

To extract the $t_f \gg \tau \gg 0$ asymptotic values of the ratios $\overline{R}_{\mathcal{R}tc}(t_f, \tau)$, which we denote by $\overline{R}_{\mathcal{R}tc}$ with no argument, we perform correlated $\chi^2$ fits of a constant to each ratio for every triangular connected region in the $(t_f,\tau)$ plane that satisfies $t_f < 25$, $\tau > 4$, and $t_f - \tau > 4$. 
The minimum cuts on $\tau$ and $t_f - \tau$ guarantee a transfer matrix exists between the source and operator insertion, and the insertion and sink.
$t_f \approx 8$ is the approximate time after which the effective masses are consistent with a single state for all hadrons, momenta, and smearings, and the upper bound $t_f \approx 25$ removes the bulk of the noise-dominated region.
To combine the separate SS and SP ratios, we simultaneously fit the same region in each to a single value of $\overline{R}_{\mathcal{R}tc}$. We combine the results of fits to different regions of $(t_f, \tau)$ using a scheme inspired by Bayesian model averaging~\cite{Jay:2020jkz}.
Denoting by $r_m$ the values of $\overline{R}$ found by fits to each region $m$, we associate each fit with a weight~\cite{Rinaldi:2019thf}
\begin{equation}
    w_m \propto p_m (\delta r_m^{\text{stat}})^{-2}
\end{equation}
where for the fit to region $m$, $\delta r_m^{\text{stat}}$ is the statistical error found by the fit and $p_m = \text{Prob}(\chi^2_{N_{\text{d.o.f.}}} < \chi^2_m) = 1 - \mathrm{CDF}_{\chi^2|N_{\text{d.o.f.}}}(\chi^2_m)$ is the $p$-value of the fit.
Normalizing the weights such that $\sum_m w_m = 1$, we obtain the mean value as ${\hat{R} = \sum_m w_m r_m}$ and the total variance $(\delta \hat{R})^2$ as the sum of statistical and systematic contributions defined as \cite{Jay:2020jkz}
\begin{equation}
(\delta \hat{R}_{\text{stat}})^2 = \sum_m w_m (\delta r_m^{\text{stat}})^2
\quad \text{and} \quad
(\delta \hat{R}_{\text{syst}})^2 = \sum_m w_m (r_m-\hat{R})^2.
\end{equation}
We find that typically $\delta \hat{R}_{\text{stat}} \approx \delta \hat{R}_{\text{syst}}$.
In practice, we perform ``central-value fits'' to the median of $\overline{R}(t_f,\tau)$ over bootstraps, from which we compute a set of weights $w^*_m$ and averaged error $\delta \hat{R}^*$.
For subsequent error propagation, we compute bootstrapped fit results by averaging over fits within bootstraps $b$ using the central-value weights $w^*_m$, then rescaling to obtain a set of results whose spread reproduces $\delta \hat{R}^*$.
In detail:
we fit all $\overline{R}_b(t_f,\tau)$ to obtain $r_{bm}$ for only the subset of highest-weight regions making up $99\%$ of the total weight, which reduces the computational cost by excluding the bulk of fit regions.
We then average to obtain $\hat{R}_b = \sum_m w^*_m r_{bm}$, with $w^*_m$ suitably re-normalized to account for the exclusion of low-weight fits.
The spread in $\hat{R}_b$ obtained in this way only reproduces $\delta \hat{R}_{\text{stat}}^*$, so we rescale each set of $\hat{R}_b$ around their mean by $\delta \hat{R}^*/\delta \hat{R}_{\text{stat}}^*$.
Note that we use the same covariance matrix for both the central-value fits and bootstrap fits, computed over $\overline{R}_b(t_f, \tau)$ using an outlier-robust $\pm 1 \sigma$ percentile definition of the error.\footnote{Based on the percentile method for confidence intervals \cite{efron1994introduction,efron1986bootstrap} and as implemented in the \texttt{gvar} package \cite{peter_lepage_2020_4290884}, this procedure computes the error for each dimension as the maximum of the differences between the median and the percentiles corresponding to $\pm 1 \sigma$ in a Gaussian distribution, then rescales the (Pearson) correlation matrix by these errors to construct the covariance matrix.}
As shown in Figs.~\ref{fig:bumps} and \ref{fig:taubumps}, the ratios typically exhibit plateaus in $t_f$, suggesting that excited-state contamination will not significantly affect the results. To check this, we perform a simplified version of this analysis for the pion, nucleon, and rho using a two-state ansatz and only bootstraps from the highest-weight fits; the resulting GFFs are consistent within uncertainties in all cases. The precision of the ratio data for the delta baryon does not admit two-state fits.
% We have not attempted fits with extra terms parametrizing subleading time dependence due to excited states; however, as shown in Figures~\ref{fig:bumps} and \ref{fig:taubumps}, the ratios typically exhibit plateaus in $t_f$, so we do not expect this choice to bias our results.

The results of this fitting and averaging procedure are generically robust against varying the lower bounds on fit regions, but increasing the upper bound on $t_f$ results in sudden catastrophic increases in error and destabilization of central values.
This effect can be traced back to fits to pure-noise regions which are excluded by the $t_f$ cut.
These fits are apparently good, as measured by their $\chi^2/\text{d.o.f.}$ or $p$-values, but the loss of Gaussianity in noise regions (the onset of which occurs at $t_f \approx 25$ in the nucleon two-point correlator as diagnosed using both cumulant expansions \cite{Wagman:2016bam} and Shapiro-Wilk testing \cite{shapiro1965analysis}) renders these metrics of fit quality meaningless.
Noisy regions must thus be excluded using a $t_f$ cut to prevent them from dominating the averages.
We chose to use the \textit{ad hoc} weight definition described above because we found it to be practically more robust against this effect (due to the inverse variance factor) than the better-motivated AIC weighting of Ref.~\cite{Jay:2020jkz}.

In the analysis described above, the choice to rescale the bootstraps around their means amounts to an assumption that systematic errors due to the choice of fit range have the same correlation structure as the statistical errors.
This is different from the typical assumption of uncorrelated systematics \cite{Shanahan:2018pib}, but both are strong assumptions.
To check that this choice does not bias our results, we applied the subsequent analysis to the nucleon data with all correlations between ratios either artificially scaled down by overall factors or completely neglected, as well as using best fits or fits to a fiducial $(t_f,\tau)$ region rather than averaging, and found no systematic shift in the results.
Further work is needed to more gracefully reconcile frequentist resampling techniques with Bayesian model averaging methods and avoid the need for such \textit{ad hoc} constructions.
For further analysis, we take the median over (rescaled) bootstraps for the central value of each $\overline{R}_{\mathcal{R}tc}$ and construct their covariance matrix using the outlier-robust estimator noted above.
Parametrizing the fit results as central values and a covariance matrix amounts to modeling their distribution as a multivariate Gaussian. We check this assumption by examining the bootstrap distribution of fit results, and find that histograms of marginal distributions are either consistent with or contained in their Gaussian approximations.
We have also checked that bootstrapping through the further analysis detailed below produces marginal distributions consistent with or narrower than the ones presented in the main text, which are obtained using linear error propagation from this Gaussian model.

\subsection{Constraint fitting}
\label{sec:constraint-fitting}

To compactify notation, throughout this section we use 1 for $\tau_1^{(3)}$ and 2 for $\tau_1^{(6)}$ whenever an irrep label appears in a subscript, and switch to vector notation for the kinematic coefficients and GFFs, i.e.~$K_j, G_j \Leftrightarrow \vec{K},\vec{G}$.

The procedure described in the previous section yields a set of measurements which constrain the bare GFFs of each irrep $\mathcal{R} \in \{ \tau_1^{(3)}, \tau_3^{(6)} \}$ separately as
\begin{equation}
    \vec{K}_{\mathcal{R}tc} \cdot \vec{G}_{\mathcal{R} t} = \overline{R}_{\mathcal{R}tc} \;,
    \label{eqn:gff-constraint}
\end{equation}
where $\vec{K}$ and $\vec{G}$ are $N_h$-element vectors over the set of different GFFs, $t$ indexes the discrete $t$-bin, and $c$ indexes the different combined ratios with shared kinematic factors as described in Sec.~\ref{sec:coeffs-binning-and-ratio-fits}.
Extracting the renormalized GFFs from these constraints, as well as subsequent model fitting of the GFFs, requires careful treatment to avoid the d'Agostini bias \cite{DAgostini:1993arp}.
This bias is an effect caused by violation of implicit Gaussianity assumptions in correlated $\chi^2$ fitting by non-Gaussianity arising from multiplication by the renormalization factors.
To circumvent it, we use a Bayesian version of the ``penalty trick''~\cite{DAgostini:1993arp}, performing combined fits of data from both irreps to estimate the bare GFFs $\vec{G}_{1t}$ and update the renormalization factors $Z_1, Z_2 \rightarrow Z_1', Z_2'$.
The updated renormalization may be applied immediately to obtain the renormalized GFFs $\vec{G}_{t} = Z_1' \vec{G}_{1t}$, or deferred until after subsequent model fitting to again circumvent the bias as discussed below.
We defer detailed discussion of the bias and the derivation of the fitting procedure presented here to Sec.~\ref{sec:dagostini}.

Our procedure estimates a Gaussian approximation of the posterior distribution
\begin{equation}
p \left( \vec{G}_{1t}, Z_1', Z_2' \bigg| \overline{R}_1, \overline{R}_2 \right) 
= \frac{1}{p\left( \overline{R}_1, \overline{R}_2 \right)} L\left( \overline{R}_1, \overline{R}_2 \bigg| \vec{G}_{1t}, \frac{Z_1'}{Z_2'} \right) p\left( Z_1', Z_2' \right) \;,
\label{eqn:constraint-fit-posterior}
\end{equation}
where $\vec{G}_{1t}$ are the $t$-bin-dependent bare GFFs for irrep $\tau_1^{(3)}$, $Z_1'$ and $Z_2'$ are the updated renormalization factors which are shared across all $t$ bins, $\overline{R}_1$ and $\overline{R}_2$ represent the full set of ratio fit results $\overline{R}_{\mathcal{R}tc}$,
the factor $p(\overline{R}_1, \overline{R}_2 )$ is the usual uninteresting data normalization factor in Bayes's theorem, 
the prior $p\left( Z_1', Z_2' \right)$ is the multivariate Gaussian defined by Eq.~\eqref{eqn:Z-factors},
and the likelihood $L$ is multivariate Gaussian,
\begin{equation}\begin{aligned}
    L\left( \overline{R}_1, \overline{R}_2 \bigg| \vec{G}_{1t}, \frac{Z_1'}{Z_2'} \right)
    &\propto
    \exp \left[
        \sum_{\mathcal{R}tc \mathcal{R'}t'c'}
        \vec{\Delta}_{\mathcal{R}c}(\vec{G}_{1t}, Z_1'/Z_2')^T  ~
        \Sigma^{-1}_{\mathcal{R}tc, \mathcal{R}'t'c'}  ~
        \vec{\Delta}_{\mathcal{R'}c'}(\vec{G}_{1t'}, Z_1'/Z_2')
    \right]
    \\
    \vec{\Delta}_{\mathcal{R}c}(\vec{G}_{1t}, Z_1'/Z_2') &= 
    \begin{cases}
        \phantom{(Z_1'/Z_2')} 
        \hat{R}_{\mathcal{R}tc} -
        \vec{K}_{\mathcal{R} t c} \cdot \vec{G}_{1t}, & \mathcal{R} = \tau_1^{(3)} \\
        \hat{R}_{\mathcal{R}tc} -
        (Z_1'/Z_2') \vec{K}_{\mathcal{R} t c} \cdot \vec{g}_{1t}, & \mathcal{R} = \tau_3^{(6)} \\
    \end{cases}
    \;,
\end{aligned}
\end{equation}
defined in terms of the measured means $\hat{R}_{\mathcal{R}tc}$ and covariance matrix $\Sigma^{-1}_{\mathcal{R}tc, \mathcal{R}'t'c'}$ of the ratio fit results $\overline{R}_{\mathcal{R}tc}$.
Equation~\eqref{eqn:constraint-fit-posterior} should be read as one overall distribution for all $t$ bins and not a set of separate equations for each bin.
The data only constrain the ratio of the $Z'$ factors and not their overall magnitude, which corresponds to a flat direction in the likelihood function that is only regulated in the posterior by $p(Z_1',Z_2')$.
Note that we have left implicit the uniform prior over $\vec{G}_{1t}$ to emphasize that, although our analysis is phrased in Bayesian language, it involves no informative priors.

We estimate the parameters of the posterior distribution using two stages of fitting.
In the first stage, we introduce a separate ratio $(Z_1'/Z_2')_t$ for each $t$ bin, defining an extended version of the likelihood
which we approximate as a Gaussian distribution around the maximum likelihood parameters $\vec{G}_{1t}^*$ and $(Z_1'/Z_2')_t^*$.
We obtain these parameters by fitting each $t$ bin separately, using linear error propagation to obtain covariances between the parameters (both within and between $t$ bins).
The posterior of interest can then be written in terms of this extended likelihood function as 
\begin{equation}
\int \left[ \prod_t d(Z_1'/Z_2')_t ~ \delta( (Z_1'/Z_2')_t - (Z_1'/Z_2') ) \right] ~
L\left( R_{1 t}, R_{2 t} \bigg| \vec{G}_{1 t}, (Z_1'/Z_2')_t \right) 
p\left( Z_1', Z_2' \right)
\end{equation}
which, after evaluating the $\delta$ functions, provides a new merit function which we can re-fit (i.e. minimize and expand about) to estimate the parameters of the Gaussian posterior. This second stage of fitting incorporates the measured distribution of $Z$ factors [Eq.~\eqref{eqn:Z-factors}] and the constraint that the ratio $Z_1'/Z_2'$ is the same for all $t$ bins.
We again estimate the covariances of this distribution using linear error propagation.

For the pion and nucleon, $\chi^2/\text{d.o.f.} \approx 1$ and $p > 0.1$ for all first-stage fits to individual $t$ bins; for most $t$ bins, $p \approx 1$. 
The second-stage fits are of similarly high quality. However, for the $\rho$ and $\Delta$, we observe that $p \ll 1$ in fits to $t$ bins with more than $\approx 600$ constraints.
We trace the source of this effect to finite-statistics limitations, which we circumvent by combining constraints.
When more than 600 constraints are present in a $t$ bin, we apply a ``pair binning'' procedure to that bin to reduce the number of constraints before fitting.
To choose which pairs of constraints are binned together in a way that heuristically minimizes loss of orthogonality in the set of constraints, we use a greedy algorithm which repeatedly associates the two unpaired constraints $\vec{K}$ and $\vec{K}'$ with the least angle $\cos^{-1}( \vec{K} \cdot \vec{K}' / |\vec{K}| |\vec{K}'| )$ between them until all constraints are paired (with possibly one left unpaired, which is retained).
Paired constraints are combined by taking weighted averages at the per-bootstrap level, using weights proportional to the number of ratio measurements averaged into each constraint.
For the $\rho$, no $t$ bin requires more than one application of this procedure, while for the $\Delta$, some bins require two applications.
After applying this procedure, first-stage fits to the pair-binned constraints for the $\rho$ and $\Delta$ satisfy $p > 0.1$ for all bins, with $p \approx 1$ for most; second-stage fits are also of high quality.
This procedure could have instead been applied to combine the {$(t_f, \tau)$-dependent} ratios before fitting, and less naive clustering algorithms than the one used here may allow more effective use of the data; we did not explore either direction in this study, but they are interesting topics for future work.
To check that pair binning does not bias the results, we instead discard random subsets of the data to equivalently reduce the number of constraints and find consistent but noisier results.
The statistical limitations addressed by pair binning may be artificial and due to the limited number ($B=1000$) of bootstraps, as we observe similar failures in the fits for the pion and nucleon when using $B=200$ that are resolved when using more;\footnote{The simple solution of drawing more bootstraps is not guaranteed to solve this problem: regardless of the number of bootstrap draws $B$ taken of an $N$-sample dataset, one needs $\sim N^2$ independent samples to estimate an $N \times N$ covariance matrix \cite{Michael:1993yj} and $N \sim \mathcal{O}(10^3)$ for this study, insufficient for the larger $t$ bins. It is also logistically prohibitive as, before sign-averaging, the ratios occupy $\mathcal{O}(10s)$ of TBs of storage with $B=1000$, and storage as well as the computational cost of fitting the ratios scales linearly in the number of bootstraps.} however, we find that our results do not depend significantly on the number of bootstraps $B$, after pair binning or discarding constraints to ensure all first-stage fits are of good quality.

The renormalized GFFs are distributed as the product under the posterior distribution of the bare GFFs $\vec{G}_{1t}$ and renormalization factor $Z_1'$ for irrep $\tau_1^{(3)}$.
Starting from the Gaussian approximation of the posterior computed using the procedure described above, we obtain the uncertainties of the renormalized GFFs presented in the main text using linear error propagation; we find this approximation to be consistent with the spreads in $Z_1' \vec{G}_{1t}$ computed over samples drawn from the Gaussian posterior.
However, as discussed in Appendix~\ref{sec:dagostini}, the renormalized GFFs are sufficiently non-Gaussian that subsequent fits of the models of Eqs.~\eqref{eq:multipole} and \eqref{eq:z-expansion} to them would again be victim to the d'Agostini bias.
We instead fit the models to the bare GFFs $\vec{G}_{1t}$, then renormalize afterwards by multiplying with $Z_1'$.
Due to the structure of the model functions, the factor of $Z_1'$ may be absorbed into the parameters $\alpha$ and $\alpha_k$, defining the renormalized fit parameters presented throughout this work.

Reference~\cite{Shanahan:2018pib} instead circumvented the d'Agostini bias by neglecting additional correlations between renormalized constraints induced by common factors of $Z$.
The results obtained using the method presented here are consistent with the ones from that study, but with narrower and more correlated uncertainties on the GFF estimates and wider ones on the model fits and densities.
For all GFFs, the results of this sampling procedure are consistent within error with the results of the procedure used here.
Employing the sampling procedure while accounting for correlations induced by shared $Z$ factors would require performing the entire analysis for each sample from the $Z$ distribution, including the expensive density estimations, which would require significant additional computational effort.

As mentioned throughout the discussion above, starting from the model of the ratio distribution as Gaussian, we use linear error propagation to propagate uncertainty through the rest of the analysis and obtain the presented results, amounting to repeatedly approximating intermediate distributions as Gaussian.
Other than the checks of these approximations described above, we have also checked that bootstrapping through the entire analysis, as well as just through the first stage of fitting and using the bootstrap results to construct a covariance matrix before the second stage, produces marginal distributions of GFFs consistent with or contained within the marginal distributions obtained with linear error propagation.

\subsection{Gaussianity and the d'Agostini bias}
\label{sec:dagostini}

In this section we discuss the d'Agostini bias, identify where the non-Gaussianities that trigger it arise in our analysis, introduce and discuss the penalty trick fitting procedure in a Bayesian framework, and motivate and derive the modified version described in Sec.~\ref{sec:constraint-fitting}.

In its simplest form, the d'Agostini bias occurs when performing a correlated $\chi^2$ fit of some linear model [or a nonlinear model whose form accommodates arbitrary rescaling, like the model ansatz\"e Eqs.~\eqref{eq:multipole} and \eqref{eq:z-expansion}] to some data which has been multiplied by an overall normalization factor with a large relative uncertainty; the result is different than what is obtained by first fitting then normalizing after, and thus obviously incorrect.
This occurs because the $\chi^2$ fitting procedure takes the covariance matrix of the data as input, and thus implicitly truncates the data distribution to Gaussian; products of Gaussian-distributed variables are not Gaussian distributed, and the bias occurs when this truncation yields a poor approximation of the true product distribution.
While resampling through a fit allows for treatment of non-Gaussianity in distributions of fit parameters due to nonlinear model functions, it cannot correct for the d'Agostini bias, which occurs because the fit assumes an inaccurate representation of the data.

\begin{figure*}[b]
\centering
\subfloat[\centering ]{{
    \includegraphics[width=1.55in]{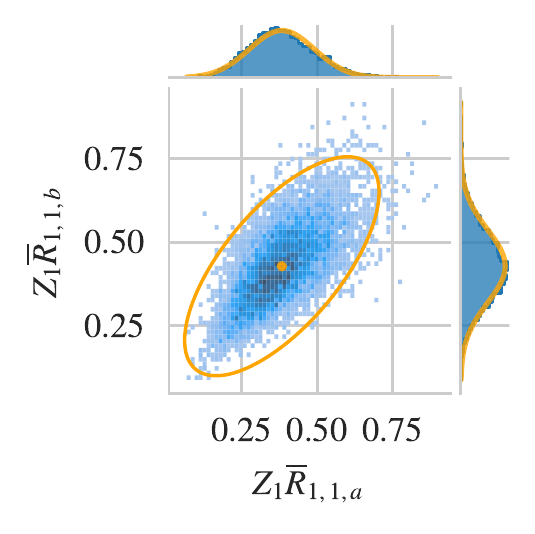} 
    \includegraphics[width=1.55in]{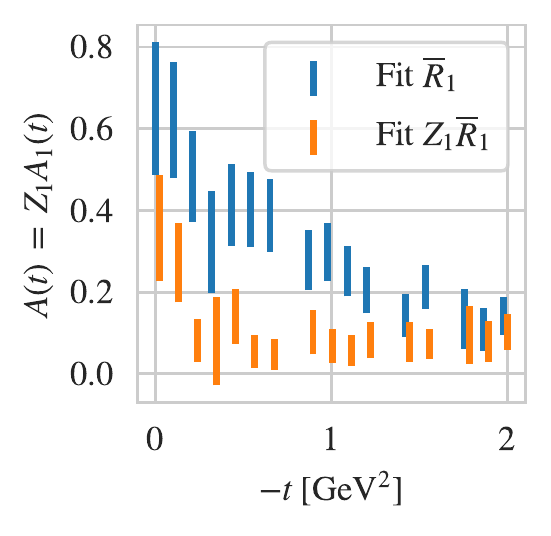}
    \label{fig:dagostini-nongauss-examples:ratios}
    \label{fig:dagostini-bias-examples:g1Fit}
}}
\!
\subfloat[\centering ]{{
    \includegraphics[width=1.55in]{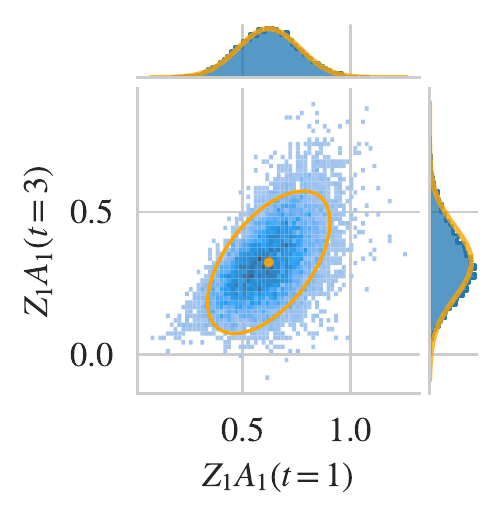}
    \includegraphics[width=1.55in]{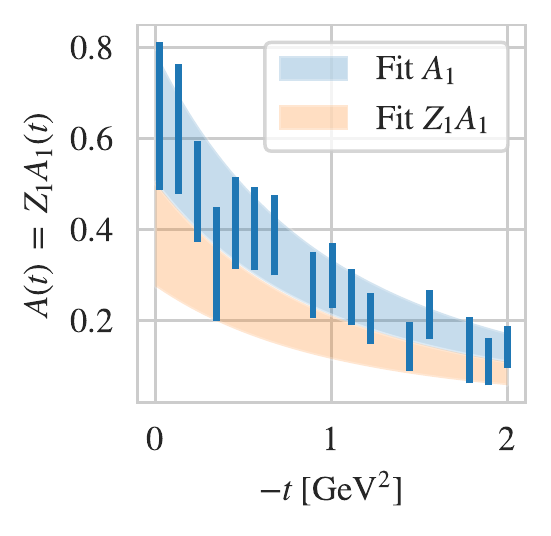}
    \label{fig:dagostini-nongauss-examples:gff-g1fits}
    \label{fig:dagostini-bias-examples:tripole-g1}
}}
\\
\subfloat[\centering  ]{{
    \includegraphics[width=1.55in]{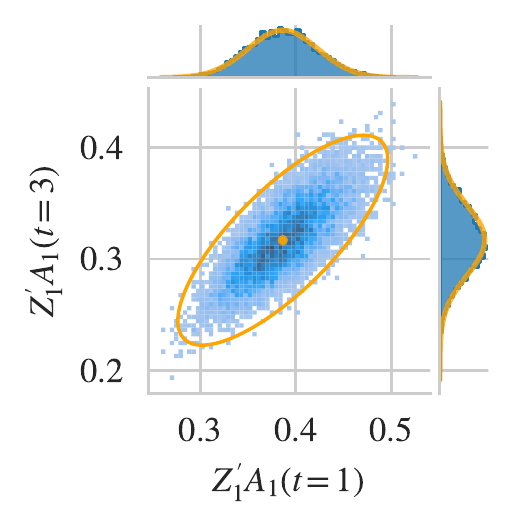}
    \includegraphics[width=1.55in]{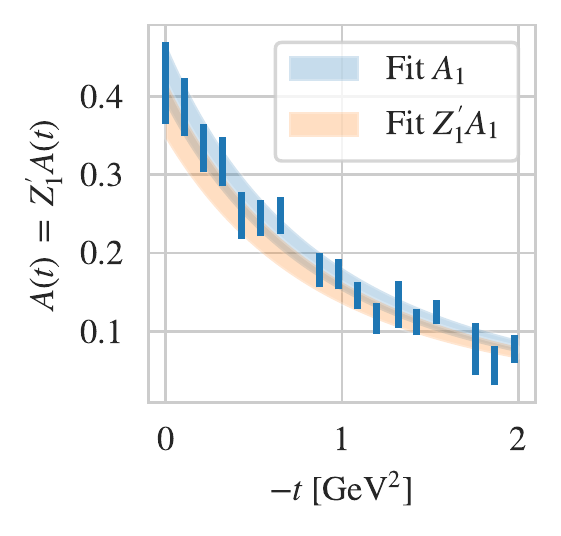} 
    \label{fig:dagostini-nongauss-examples:gff-gRfits}
    \label{fig:dagostini-bias-examples:tripole-gR}
}}
\caption{
Examples of non-Gaussianities in various distributions in the analysis of the nucleon data (left panels), and the resulting effects of d'Agostini bias (right panels).
\textbf{Note:} these plots are an illustration of the bias and are not the final results of our calculation.
In left panels, the orange features show the Gaussian approximations to these distributions obtained with linear error propagation. In the joint histograms, the ellipses denote the $3 \sigma$ contour and the dot denotes the mean.
In right panels, blue bands show results obtained by fitting a bare quantity first then renormalizing afterwards to circumvent the bias, whereas orange bands are biased fits to data renormalized before fitting.
{\protect\subref{fig:dagostini-nongauss-examples:ratios}} For irrep $\tau_1^{(3)}$, example of a joint distribution of two different renormalized ratio fit results $Z_1 \overline{R}_{\mathcal{R}tc}$ in the $t=1$ bin for the nucleon, and the renormalized GFF $A(t) = Z_1 A_1(t)$ obtained by fitting irrep $\tau_1^{(3)}$ constraints.
{\protect\subref{fig:dagostini-nongauss-examples:gff-g1fits}} For renormalized GFFs obtained by fitting the bare constraints from irrep $\tau_1^{(3)}$ only to obtain $A_1$ then renormalizing afterwards, example of a joint distribution of the renormalized GFF $A(t) = Z_1 A_1(t)$ in two different $t$ bins, and fits of a tripole model to these GFFs.
{\protect\subref{fig:dagostini-nongauss-examples:gff-gRfits}} For renormalized GFFs obtained using the fitting procedure described in Sec.~\ref{sec:constraint-fitting} incorporating data from both irreps, joint distribution of the same renormalized GFFs as in \protect\subref{fig:dagostini-nongauss-examples:gff-gRfits}, and fits of a tripole model to these GFFs.
}
\label{fig:dagostini-nongauss-examples}
\label{fig:dagostini-bias-examples}
\end{figure*}

Given our multivariate Gaussian models of the distributions of the bare ratios and renormalization factors, the bare GFFs are Gaussian but the renormalized ratios and GFFs are not.
The bare ratios are Gaussian by assumption and constrain the bare GFFs linearly per Eq.~\eqref{eqn:gff-constraint}, so the bare GFFs inherit the Gaussianity of the ratios.
However, the renormalized ratios $Z_{\mathcal{R}} R_{\mathcal{R}tc}$ are non-Gaussian, as shown in Fig.~\ref{fig:dagostini-nongauss-examples:ratios} and discussed in the caption.
It follows that the renormalized GFFs, which are linearly constrained by the renormalized ratios, are also non-Gaussian, intrinsically and independently of how we extract them, as shown in Figs.~\ref{fig:dagostini-nongauss-examples:gff-g1fits} and \ref{fig:dagostini-nongauss-examples:gff-gRfits}.
These non-Gaussianities trigger the d'Agostini bias both when fitting ratios to extract GFFs, as demonstrated in Fig~\ref{fig:dagostini-bias-examples:g1Fit}, as well as subsequently when fitting the GFFs to model functions, as shown in Figs.~\ref{fig:dagostini-bias-examples:tripole-g1} and \ref{fig:dagostini-bias-examples:tripole-gR}.
The fitting procedure described in Sec.~\ref{sec:constraint-fitting} circumvents the bias in the former case using the penalty trick, and in the latter case by extracting the Gaussian-distributed bare GFFs for one irrep and allowing the problematic multiplication by a $Z$ factor to be deferred until after fitting models to the bare GFFs.
Note that while the histograms of marginal distributions shown in Fig.~\ref{fig:dagostini-bias-examples} naively appear close enough to Gaussian to justify approximation as Gaussian, inspection of the joint histograms reveals the asymmetry of the distribution that leads to the bias.

The penalty trick is a common prescription for circumventing the d'Agostini bias \cite{DAgostini:1993arp}.
Our choice to phrase the fitting problem as an estimation of a posterior distribution (as described in Sec.~\ref{sec:constraint-fitting}), with the measured distribution of the renormalization factors entering as a prior to be updated, amounts to a Bayesian reframing of this technique.
Generally, for a fit of a model function $\vec{f}(\vec{\theta})$ to some data $\vec{y}$ times a normalization factor $Z$, where $\vec{y} \sim \mathcal{N}(\hat{\vec{y}}, \Sigma_{\vec{y}})$ and $Z \sim \mathcal{N}(\hat{Z}, \sigma_Z^2)$ are Gaussian, but a Gaussian $\mathcal{N}(\widehat{\vec{y} Z}, \Sigma_{\vec{y}Z})$ is a poor approximation of the distribution of the product $\vec{y} Z$, the penalty trick prescribes the replacement
\begin{equation}
    \begin{split}
        & \vec{\Delta} = \vec{f}(\vec{\theta}) - \widehat{\vec{y} Z} \\
        & \chi^2(\theta) = \Delta^T \Sigma_{\vec{y}Z}^{-1} \Delta
    \end{split}
    \quad \xrightarrow{\text{Penalty trick}} \quad
    \begin{split} 
        & \vec{\Delta}' = \vec{f}(\vec{\theta})/z - \hat{\vec{y}} \\
        & \chi^2(\theta,z) = 
         \vec{\Delta}'^T \Sigma_{\vec{y}}^{-1} \vec{\Delta}'
        + (z - \hat{Z})^2/\sigma_{Z}^2
    \end{split}
\label{eqn:penalty-trick}
\end{equation}
allowing a fit using the original covariance matrix $\Sigma_{\vec{y}}$, assumed to be a good description of the data.
This comes at the cost of replacing the fixed normalization $Z$ with an additional nuisance parameter $z$ which is constrained to be consistent with the provided $Z$ and discarded after fitting.
In the limit $\sigma_Z \rightarrow 0$ the two fit procedures are equivalent.

While usually motivated as an \textit{ad hoc} frequentist procedure, the penalty trick can be more naturally understood in a Bayesian context, wherein it is structurally equivalent to updating a prior for $Z$ with the data then marginalizing over it, assuming a Gaussian posterior.
The right-hand side of Eq.~\eqref{eqn:penalty-trick} can be interpreted as a log-likelihood and log-prior for the data and $z$, defining a posterior distribution via Bayes's theorem as
\begin{equation}\begin{gathered}
    p(\theta,z|\vec{y}) = L(\vec{y}|\theta,z) ~ p(z) ~ p(\theta) ~ / ~ p(\vec{y}) \\
    -2 \log L(\vec{y}|\theta,z) = \vec{\Delta}'^T \Sigma_{\vec{y}}^{-1} \vec{\Delta}' \\
    -2 \log p(z) = (z - \hat{Z})^2/\sigma_{Z}^2
\end{gathered}\end{equation}
where $p(\vec{y})$ is the data normalization and $p(\theta)$ is a trivial factor of the uniform distribution added as a prior for the fit parameters.
Fitting the penalty trick $\chi^2$ to obtain the best-fit $\theta^*$ and $z^*$ and fit parameter covariance matrix $\Sigma^*_{\theta,z}$ corresponds to approximating the posterior distribution as Gaussian, i.e.
\begin{equation}\begin{gathered}
    p(\theta,z| \vec{y}) \propto \exp[-\frac{1}{2} \chi^2(\theta,z)]
    \approx
    \exp[ -\frac{1}{2} \vec{\Delta}_{\theta z}^T \Sigma^*_{\theta,z} \vec{\Delta}_{\theta z} ]
    \\
    \vec{\Delta}_{\theta z}^T \equiv \begin{bmatrix}
        \theta - \theta*, &
        z - z*
    \end{bmatrix}
\end{gathered}\end{equation}
suppressing normalization factors.
Discarding $z$ after fitting corresponds to marginalizing over $z$ in the posterior, as marginalizing over a dimension of a multivariate Gaussian is equivalent to dropping it.
The generalization to the case of multiple different normalization factors for different subsets of the data is straightforward: the prior $p(z)$ becomes multidimensional, and now
\begin{equation}
    \vec{\Delta}'_i = \vec{f}(\vec{\theta}) / z_i - \vec{\hat{y}}_i
\end{equation}
where $i$ indexes different subsets of the data.

We modify the penalty trick procedure to estimate the bare GFFs $\vec{G}_{1t}$ (corresponding to $\theta/z_1$) instead of the non-Gaussian renormalized GFFs $\vec{G}_{t}$ (corresponding to $\theta$).
The modification singles out one particular normalization as special, multiplying it onto the model function $\vec{f}$ so that the data are modeled as
\begin{equation}
    \vec{\Delta}'_i = \begin{cases}
        \phantom{\frac{z_1}{z_i}} \vec{f}(\theta') - \vec{\hat{y}}_1, & i=1 \\
        \frac{z_1}{z_i} \vec{f}(\theta') - \vec{\hat{y}}_i, & i \ne 1
    \end{cases}.
\end{equation}
If the model function $\vec{f}$ is linear in the parameters (e.g.~$\vec{K} \cdot \vec{G}$ is linear in the GFFs $\vec{G}$), then this procedure extracts $\theta' = \theta/z_1$ (corresponding to  $\vec{G}_{1t}$) rather than $\theta$ (corresponding to $\vec{G}_t$).
In this modified form one still (trivially) marginalizes over all $z_i$ for $i \ne 1$, but $z_1$ must be retained to examine $\theta = z_1 \theta'$ (corresponding to renormalizing the bare GFFs as $\vec{G}_R = Z_1' \vec{G}_1$).

While fitting procedures exist for treating the d'Agostini bias 
other than the penalty trick \cite{Ball:2009qv}, a model function and a data distribution define a distribution of model parameters (e.g.~GFFs) independent of the choice of bias-circumventing fitting procedure.
The Bayesian framework makes clear that the renormalized GFFs extracted by this procedure may themselves be non-Gaussian, such that subsequent fits are also vulnerable to the bias.
This will hold independent of the fitting procedure used.

\section{DENSITY DEFINITIONS}
\label{sec:density-defs}

\allowdisplaybreaks

This section lists the expressions for the energy, pressure, and shear force distributions in the 3D Breit frame (BF3), 2D Breit frame (BF2), and infinite momentum frame (IMF) used to generate the results of Sec.~\ref{sec:densities}.
To simplify the expressions below, we define bracket notation for the relevant integrals,
\begin{equation}
\begin{aligned}
[\mathcal{I}]_{\text{BF2}}(r) &= \int \frac{d^2\Delta_{\mybot}}{(2\pi)^2}
e^{-i\vec{\Delta_{\mybot}}\cdot\vec{r}} \mathcal{I}(t)\biggr\rvert_{\vec{P}=0} 
&&= \int \frac{d|\Delta_{\mybot}||\Delta_{\mybot}|}{2\pi}J_0\left(|\Delta_{\mybot}|r
\right)
\mathcal{I}(t)\biggr\rvert_{\vec{P}=0} 
\\
[\mathcal{I}]_{\text{BF3}}(r) &= \int \frac{d^3\Delta}{(2\pi)^3}
e^{-i\vec{\Delta}\cdot\vec{r}}\mathcal{I}(t)\biggr\rvert_{\vec{P}=0} 
&&= \int \frac{d|\Delta| |\Delta|}{2\pi^2 r}\text{sin}\left(|\Delta| r\right)
\mathcal{I}(t)\biggr\rvert_{\vec{P}=0}
\\
[\mathcal{I}]_{\text{IMF}}(r) &= \int \frac{d^2\Delta_{\mybot}}{(2\pi)^2}
e^{-i\vec{\Delta_{\mybot}}\cdot\vec{r}}\mathcal{I}(t)\biggr\rvert_{\vec{P}\cdot \vec{\Delta}=0}^{P_z
\rightarrow \infty} 
&&= \int \frac{d|\Delta_{\mybot}||\Delta_{\mybot}|}{2\pi}J_0\left(|\Delta_{\mybot}|r
\right) \mathcal{I}(t) \biggr\rvert_{\vec{P}\cdot
\vec{\Delta}=0}^{P_z=0}
\end{aligned}
\label{eqn:density-brackets}
\end{equation}
where $\mathcal{I}$ is a generic integrand and $J_0$ is a Bessel function of the first kind.

We compute the presented densities, defined by Eq.~\eqref{eqn:density-brackets} and the expressions below, using numerical integration.
The analysis of Ref.~\cite{Shanahan:2018pib} propagated uncertainty on model parameters into the densities by sampling from the multivariate Gaussian distribution of the model parameters, evaluating the integrals for each draw.
The large number of densities considered in the present study make this approach impractical.
We instead used linearized error propagation: by differentiating under the integral sign with respect to model parameters $\vec{\theta}$, we obtain the Jacobian $J(r;\vec{\theta})_i = \partial I(r;\vec{\theta}) / \partial \theta_i$, where $I(r;\vec{\theta})$ is an integral evaluated to obtain a density at radius $r$, as a matrix of integrals that can each be evaluated numerically.
The covariance matrix for the $r$-dependent density is then obtained as $\mathrm{Cov}[\rho(r), \rho(r')] = \sum_{ij} J(r;\vec{\theta})_i \mathrm{Cov}[\theta_i, \theta_j] J(r';\vec{\theta})_j$
where $\mathrm{Cov}[\theta_i, \theta_j]$ is the covariance matrix of the parameters of the model integrated to obtain the density.

The model functions are linear in some parameters ($\alpha$ for the multipole and $\alpha_k$ for the modified z-expansion) but not others (multipole masses), so this approach is approximate.
However, for all densities for the nucleon and pion, as well as for the monopole densities for the $\rho$ meson, we found consistent results for all $r$ by computing integrals for samples from the distribution of renormalized model parameters.
For tripole models of the nucleon GFFs in the 3D Breit frame, we also checked our numerically integrated density results against ones derived from the closed-form solution
\begin{equation}
    \int \frac{d^3 \Delta}{(2 \pi)^3} 
    e^{-i \vec{\Delta} \cdot \vec{r} }
    \frac{\alpha}{\left(1 + \frac{\vec{\Delta}^2}{\Lambda^2} \right)^3}
    = \alpha (1 + \Lambda r) \frac{\Lambda^3}{32 \pi} e^{-\Lambda r}
    \label{eqn:tripole-closed-form}
\end{equation}
using linear error propagation from the tripole model parameters $\alpha$ and $\Lambda$, and found indistinguishable results.

The uncertainties on the densities presented in the main text are derived from the distribution of renormalized model parameters, obtained by combining the uncertainties of the bare $\alpha$ and $\alpha_k$ parameters and fitted values of $Z_1'$ using linear error propagation as described in Sec.~\ref{sec:constraint-fitting}.
We found consistent results by computing densities from bare model parameters, using linear error propagation to obtain correlations between $Z_1'$ and the resulting bare densities, then applying the renormalization factor $Z_1'$ and propagating uncertainities either linearly or by drawing correlated samples of $Z_1'$ and the bare densities and multiplying within samples.

The mass mean square radii are  defined identically for all hadrons as
\begin{align}
\braket{r^2_i}^{h,\text{mass}}_{\text{BF3}} &=
\frac{\int d^3r \; r^2 \varepsilon^{h,i}_{\text{BF3}}(r)}{\int d^3r \; \varepsilon^{h,i}_{\text{BF3}}(r) } \\
\braket{r^2_i}^{h,\text{mass}}_{\text{BF2/IMF}} &=
\frac{\int d^2r_{\mybot} \; r^2_{\mybot} \varepsilon^{h,i}_{\text{BF2/IMF}}(r_{\mybot})}{\int d^2r_{\mybot} \; \varepsilon^{h,i}_{\text{BF2/IMF}}(r_{\mybot})} \;,
\end{align}
while the mechanical mean square radii are defined as
\begin{align} \label{eq:mechrad}
\braket{r^2_i}^{h,\text{mech}}_{\text{BF3}} &=
\frac{\int d^3r \; r^2 (p^{h,i}_{\text{BF3}}(r) +
\frac{2}{3}s^{h,i}_{\text{BF3}}(r))}{\int d^3r \; (p^{h,i}_{\text{BF3}}(r) +
\frac{2}{3}s^{h,i}_{\text{BF3}}(r))} \\
\braket{r^2_i}^{h,\text{mech}}_{\text{BF2/IMF}} &=
\frac{\int d^2r_{\mybot} \; r_{\mybot}^2 (p^{h,i}_{\text{BF2/IMF}}(r_{\mybot}) +
\frac{1}{2}s^{h,i}_{\text{BF2/IMF}}(r_{\mybot}))}{\int d^2r_{\mybot} \; (p^{h,i}_{\text{BF2/IMF}}(r_{\mybot}) +
\frac{1}{2}s^{h,i}_{\text{BF2/IMF}}(r_{\mybot}))} \; .
\end{align}
The mechanical radius results presented throughout this work are computed by numerically approximating the integrals with the trapezoidal rule, evaluated at 500 values of the integrands evenly spaced in $0\leq r \leq 2~\mathrm{fm}$.
We obtain error estimates using linear error propagation from the values of $p(r)$ and $s(r)$ at each $r$, computed as described above, and corresponding to the results presented in Table~\ref{tab:allradii}.
To check discretization errors, we instead use simple Riemann sums and obtain results which are consistent within uncertainty.
To check the error induced by truncating the range of integration from $[0, \infty]$ to $[0, 2~\mathrm{fm}]$, we derive the exact expression for $[0, \infty]$ in the tripole case as $\braket{r^2_i}^{N,\text{mech}}_{\text{BF3}} = 12/\Lambda^2$ from Eq.~\eqref{eqn:tripole-closed-form} and find it yields results consistent within uncertainty.
The shear and pressure densities for other models, frames, and hadrons are comparably small by $r = 2~\mathrm{fm}$, so we expect this quality of approximation to hold generally.

The subsection below lists the various densities computed for each hadron. In all expressions for the densities and radii, we use the definitions
$\partial^2 = \frac{1}{r^2}\frac{d}{dr}r^2\frac{d}{dr}$ and
$\partial^2_{\mybot} = \frac{1}{r_{\mybot}}\frac{d}{dr_{\mybot}}
r_{\mybot}\frac{d}{dr_{\mybot}}$,
and the $\approx$ symbol in the IMF definitions when suppressing higher-order terms in $\mathcal{O}(P_z)$.
Moreover, the symbol $X$ is defined such that
\begin{equation}
X = \begin{cases} 6 \quad \text{for BF3} \\ 8\quad \text{for BF2/IMF}\end{cases}\;.
\end{equation} 

\subsection{Pion}
\label{sec:piondensities}

Below we list expressions for the BF energy ($\epsilon$), pressure ($p$), and shear force ($s$) densities, and the mass radii of the
pion \cite{Polyakov:2018zvc,Freese:2019bhb}, as well as the contributions to the IMF densities and radii at lowest order in
$\mathcal{O}(1/P_z)$. The IMF densities are derived
by considering the
matrix elements $\gamma \bra{\pi(\vec{p}')}T^{00}_g\ket{\pi(\vec{p})}$ and $ \bra{\pi(\vec{p}')}T^{ij}_g\ket{\pi(\vec{p})}/\gamma$, where
$\gamma$ is the relativistic boost factor.
\begin{align} \label{eq:piondensities}
\varepsilon^{\pi,i}_{\text{BF3(2)}}(r_{(\mybot)}) &= m_{\pi}^2\left[\frac{1}
{\sqrt{m_{\pi}^2 -t/4}}
\left(A^{\pi}_i(t)+\bar{c}^{\pi}_i(t)-
\frac{t}{4m_{\pi}^2}(A^{\pi}_i(t)+D^{\pi}_i(t))\right)\right]_{\text{BF3(2)}} \;, \\
p^{\pi,i}_{\text{BF3(2)}}(r_{(\mybot)}) &= \frac{1}{X }\partial^2_{(\mybot)}\left[\frac{1}{\sqrt{m_{\pi}^2 -t/4}}D^{\pi}_i(t)\right]_{\text{BF3(2)}}-m_{\pi}^2\left[
\frac{1}{\sqrt{m_{\pi}^2 -t/4}}\bar{c}^{\pi}_i(t)\right]_{\text{BF3(2)}} \;,\\
s^{\pi,i}_{\text{BF3(2)}}(r_{(\mybot)}) &= -\frac{1}{4}r_{(\mybot)}\frac{d}{dr_{(\mybot)}}
\frac{1}{r_{(\mybot)}}\frac{d}{dr_{(\mybot)}}\left[\frac{1}{\sqrt{m_{\pi}^2 -t/4}}D^{\pi}_i(t)\right]_{\text{BF3(2)}}\;, \\
\varepsilon^{\pi,i}_{\text{IMF}}(r_{\mybot}) &\approx m[A^{\pi}_i(t)]_{\text{IMF}}\;, \\
p^i_{\text{IMF}}(r_{\mybot}) &\approx \frac{1}{8m_{\pi}}\frac{1}{r_{\mybot}}\frac{d}{dr_{\mybot}}
r_{\mybot}\frac{d}{dr_{\mybot}}\left[D^{\pi}_i(t)\right]_{\text{IMF}}
-m_{\pi}\left[\bar{c}^{\pi}_i(t)\right]_{\text{IMF}}\;,\\
s^i_{\text{IMF}}(r_{\mybot}) &\approx -\frac{1}{4m_{\pi}}r_{\mybot}\frac{d}{dr_{\mybot}}
\frac{1}{r_{\mybot}}\frac{d}{dr_{\mybot}}
\left[D^{\pi}_i(t)\right]_{\text{IMF}}\;,
\end{align}
\begin{align} \label{eq:pionradii}
\braket{r^2_i}_{\text{BF3}}^{\pi,\text{mass}} &= 
 \lim_{\vec{\Delta}\rightarrow 0}
-\frac{1}{\sqrt{m_{\pi}^2 -t/4}}\nabla^2_{\Delta}\left[\frac{1}{2\sqrt{m_{\pi}^2 -t/4}}\bra{\pi(p')}
T_i^{00}\ket{\pi(p)}\rvert_{\Delta_0=0}\right] \nonumber \\
&= 6\frac{dA^{\pi}_i(t)}{dt}\rvert_{t=0}-\frac{3}{4m_{\pi}^2}(A^{\pi}_i(0)
+2D^{\pi}_i(0))\;, \\
\braket{r^2_i}_{\text{BF2}}^{\pi,\text{mass}} &= 4\frac{dA^{\pi}_i(t)}{dt}\rvert_{t=0} -\frac{1}{2m_{\pi}^2}(A^{\pi}_i(0)+2D^{\pi}_i(0))
= \frac{2}{3} \braket{r^2_i}_{\text{BF3}}^{\pi,\text{mass}} \;, \\
\braket{r^2_i}_{\text{IMF}}^{\pi,\text{mass}} &=\frac{\int dr_{\mybot} r_{\mybot}^2  \varepsilon^{\pi,i}_{\text{IMF}}(r)} 
{\int dr_{\mybot} \varepsilon^{\pi,i}_{\text{IMF}}(r)}
= 4\frac{dA^{\pi}_i(t)}{dt}\rvert_{t=0} \;.
\end{align}

\subsection{Nucleon}
\label{sec:nucdensities}

The BF \cite{Polyakov:2018zvc} and lowest-order
IMF \cite{Lorce:2018egm} densities and mass radii of the
nucleon can be expressed as

\begin{align}  \label{eq:nucdensities}
\varepsilon^{N,i}_{\text{BF3(2)}}(r_{(\mybot)}) &=
m_N \left[A^N_i(t) - \frac{t}{4m_N^2}(D^N_i(t) - B^N_i(t)) + \bar{c}^N_i(t)\right]_
{\text{BF3(2)}} \;,\\
\varepsilon^{N,i}_{\text{IMF}}(r_{\mybot}) &= m_N [A^N_i(t)]_{\text{IMF}} \;,\\
p^{N,i}_{\text{BF3(2)/(IMF)}}(r_{(\mybot)}) &= \frac{1}{X m_N}\partial^2_{(\mybot)}[D^N_i(t)]
_{\text{BF3(2)/(IMF)}} - m[\bar{c}^N_i(t)]_{\text{BF3(2)/(IMF)}} \;,\\
s^{N,i}_{\text{BF3(2/IMF)}}(r_{(\mybot)}) &= -\frac{1}{4m_N}r_{(\mybot)}\frac{d}{dr_{(\mybot)}}\frac{1}{r_{(\mybot)}}\frac{d}{dr_{(\mybot)}}
[D^N_i(t)]_{\text{BF3(2/IMF)}} \;,
\end{align}
\begin{align} \label{eq:nucradii}
\braket{r^2_i}_{\text{BF3}}^{N,\text{mass}} &= 
 6\frac{dA^N_i(t)}{dt}\rvert_{t=0}+\frac{3}{2m_N^2}(B^N_i(0)-D^N_i(0)) \;,\\
\braket{r^2_i}_{\text{BF2}}^{N,\text{mass}} &= \frac{2}{3} \braket{r^2_i}_{\text{BF3}}^{N,\text{mass}}\;, \\
\braket{r^2_i}_{\text{IMF}}^{N,\text{mass}} &= 4\frac{dA^N_i(t)}{dt}\rvert_{t=0}\;.
\end{align}

\subsection{$\rho$ meson}
\label{sec:rhodensities}

The BF3 densities of the $\rho$ meson were derived in Refs.~\cite{Sun:2020wfo,Polyakov:2019lbq} and can be expressed as
\begin{align} \label{eq:rhodensitiesBF3}
\varepsilon_{0,\text{BF3}}^{\rho,i}(r) &= m_{\rho}^2\left[\frac{1}{\sqrt{m_{\rho}^2 -t/4}}\left(A^{\rho,i}_0(t) + \frac{1}{4}
\bar{f}^{\rho,i}(t)-\frac{1}{2}\bar{c}^{\rho,i}_0(t) 
\vphantom{\frac{t^3}{192m_{\rho}^6}}
\right.\right. \nonumber \\ & 
+\frac{t}{12m_{\rho}^2}[-5A^{\rho,i}_0(t) + 3D^{\rho,i}_0(t)+4J^{\rho,i}(t)-2E^{\rho,i}(t)
+A^{\rho,i}_1(t)+\frac{1}{2}\bar{f}^{\rho,i}(t)+\bar{c}^{\rho,i}_0(t)+\frac{1}{2}
\bar{c}^{\rho,i}_1(t)] 
\nonumber\\&
-\frac{t^2}{24m_{\rho}^4}[-A^{\rho,i}_0(t)+D^{\rho,i}_0(t)+2J^{\rho,i}(t)
-2E^{\rho,i}(t)+A^{\rho,i}_1(t)+\frac{1}{2}D^{\rho,i}_1(t)+\frac{1}{4}\bar{c}^{\rho,i}_1]
\nonumber\\&\left.\left.
+\frac{t^3}{192m_{\rho}^6}[A^{\rho,i}_1(t)+D^{\rho,i}_1(t)]\right)\right]_{\text{BF3}}\;, \\
\varepsilon_{2,\text{BF3}}^{\rho,i}(r) &= - \frac{r}{2}\frac{d}{dr}\frac{1}{r}\frac{d}{dr}
\left[\frac{1}{\sqrt{m_{\rho}^2 -t/4}}\left(-A^{\rho,i}_0(t)+2J^{\rho,i}(t)-E^{\rho,i}(t)+\frac{1}{2}A^{\rho,i}_1(t) + 
\frac{1}{4}\bar{f}^{\rho,i}(t)+\frac{1}{2}\bar{c}_0^{\rho,i}(t)
+\frac{1}{4}\bar{c}_1^{\rho,i}(t) 
\vphantom{\frac{t}{4m_{\rho}^2}}
\right.\right. \nonumber\\
&-\frac{t}{4m_{\rho}^2}[-A^{\rho,i}_0(t)+D^{\rho,i}_0(t)
+2J^{\rho,i}(t)-2E^{\rho,i}(t)+A^{\rho,i}_1(t)
+\frac{1}{2}D^{\rho,i}_1(t)+\frac{1}{4}\bar{c}^{\rho,i}_1(t)] \nonumber \\
&\left.\left.+\frac{t^2}{32m_{\rho}^4}[A^{\rho,i}_1(t)+D^{\rho,i}_1(t)]\right)\right]
_{\text{BF3}} \;,
\end{align}
\begin{align}
p_{0,\text{BF3}}^{\rho,i}(r)  &= \frac{1}{6}\partial^2\left[\frac{1}{\sqrt{m_{\rho}^2 -t/4}}\left(
-D^{\rho,i}_0(t)+\frac{4}{3}E^{\rho,i}(t)+\frac{t}{12m_{\rho}^2}[2D^{\rho,i}_0(t)
-2E^{\rho,i}(t)+D^{\rho,i}_1(t)]-\frac{t^2}{48m_{\rho}^4}D^{\rho,i}_1(t)
\right)\right]_{\text{BF3}} \;,\\ 
p_{2,\text{BF3}}^{\rho,i}(r) & = \frac{1}{6}\partial^2\left[\frac{-E^{\rho,i}(t)}{\sqrt{m_{\rho}^2 -t/4}}\right]_{\text{BF3}} \nonumber \\
&+\frac{1}{6 m_{\rho}}\partial^2 \left(\frac{d}{dr}\frac{d}
{dr}-\frac{2}{r}\frac{d}{dr}\right)
\left[\frac{1}{\sqrt{m_{\rho}^2 -t/4}} \left(\frac{1}{2}D^{\rho,i}_0(t)-\frac{1}{2}E^{\rho,i}(t)+
\frac{1}{4}D^{\rho,i}_1(t) - \frac{t}{16m_{\rho}^2}D^{\rho,i}_1(t)\right)\right]_{\text{BF3}} \;,\\
p_{3,\text{BF3}}^{\rho,i}(r)& = -\frac{2}{6 m_{\rho}}\partial^2\left(\frac{d}{dr}\frac{d}
{dr}-\frac{3}{r}\frac{d}{dr}\right) \nonumber\\
&\left[\frac{1}{\sqrt{m_{\rho}^2 -t/4}}\left(
\frac{1}{2}D^{\rho,i}_0(t)-\frac{1}{2}E^{\rho,i}(t)+\frac{1}{4}D^{\rho,i}_1(t) - \frac{t}{16m_{\rho}^2}D^{\rho,i}_1(t)
\right)\right]_{\text{BF3}} \;,\\
s_{0,\text{BF3}}^{\rho,i}(t) &= -\frac{1}{4}r\frac{d}{dr}\frac{1}{r}\frac{d}{dr} \nonumber\\
&\left[\frac{1}{\sqrt{m_{\rho}^2 -t/4}}\left(
-D^{\rho,i}_0(t)+\frac{4}{3}E^{\rho,i}(t)+\frac{t}{12m_{\rho}^2}[2D^{\rho,i}_0(t)
-2E^{\rho,i}(t)+D^{\rho,i}_1(t)]-\frac{t^2}{48m_{\rho}^4}D^{\rho,i}_1(t)
\right)\right]_{\text{BF3}} \;,\\
s_{2,\text{BF3}}^{\rho,i}(r) &= -\frac{1}{4}r\frac{d}{dr}\frac{1}{r}\frac{d}{dr}
\left[\frac{-E^{\rho,i}(t)}{\sqrt{m_{\rho}^2 -t/4}}\right]_{\text{BF3}} \nonumber\\
&-\frac{1}{4m_{\rho}}r\frac{d}{dr}\frac{1}{r}\frac{d}{dr}\left(\frac{d}{dr}\frac{d}
{dr}-\frac{2}{r}\frac{d}{dr}\right) 
\nonumber\\&
\left[\frac{1}{\sqrt{m_{\rho}^2 -t/4}}\left(\frac{1}{2}D^{\rho,i}_0(t)-\frac{1}{2}E^{\rho,i}(t)+\frac{1}{4}D^{\rho,i}_1(t) - \frac{t}{16m_{\rho}^2}D^{\rho,i}_1(t)\right)\right]_{\text{BF3}}\;, \\
s_{3,\text{BF3}}^{\rho,i}(r) &= \frac{1}{2m_{\rho}}r\frac{d}{dr}\frac{1}{r}\frac{d}{dr}\left(\frac{d}{dr}\frac{d}
{dr}-\frac{3}{r}\frac{d}{dr}\right) 
\nonumber\\&
\left[\frac{1}{\sqrt{m_{\rho}^2 -t/4}}\left(
\frac{1}{2}D^{\rho,i}_0(t)-\frac{1}{2}E^{\rho,i}(t)+\frac{1}{4}D^{\rho,i}_1(t) - \frac{t}{16m_{\rho}^2}D^{\rho,i}_1(t)
\right)\right]_{\text{BF3}} \;.
\end{align}
Using the same methods but restricting the analysis to a two-dimensional plane, we obtain the following expressions for the BF2
leading-order contributions to the
EMT monopole densities:
\begin{align} \label{eq:rhodensitiesBF2}
\varepsilon_{0,\text{BF2}}^{\rho,i}(r_{\mybot}) &= m_{\rho}^2\left[\frac{1}{\sqrt{m_{\rho}^2 -t/4}}\left(A^{\rho,i}_0(t)+\frac{1}{4}\bar{f}^{\rho,i}(t)-\frac{1}{2}\bar{c}^{\rho,i}_0(t) 
\vphantom{\frac{t}{8m_{\rho}^2}}
\right.\right. \nonumber \\   
 &+\frac{t}{8m_{\rho}^2}[-4A^{\rho,i}_0(t) + 2D^{\rho,i}_0(t)+4J^{\rho,i}(t)-2E^{\rho,i}(t)
+A^{\rho,i}_1(t)+\frac{1}{2}\bar{f}^{\rho,i}(t)+\bar{c}^{\rho,i}_0(t)+\frac{1}{2}\bar{c}^{\rho,i}_1(t)]  
\nonumber \\&+
\frac{t^2}{16m_{\rho}^4}[A^{\rho,i}_0(t)-D^{\rho,i}_0(t)-2J^{\rho,i}(t)
+2E^{\rho,i}(t)-A^{\rho,i}_1(t)-\frac{1}{2}D^{\rho,i}_1(t)-\frac{1}{4}\bar{c}^{\rho,i}_1(t)]
\nonumber \\&\left.\left.
+\frac{t^3}{128m_{\rho}^6}[A^{\rho,i}_1(t)+D^{\rho,i}_1(t)]\right)\right]_{\text{BF2}} \;,\\
s^{\rho,i}_{0,\text{BF2}}(r_{\mybot}) &= -\frac{1}{4} r_{\mybot}\frac{d}{dr_{\mybot}}\frac{1}
{r_{\mybot}}\frac{d}{dr_{\mybot}}
\nonumber \\&
\left[\frac{1}{\sqrt{m_{\rho}^2 -t/4}}\left(-D_0^{\rho,i}(t)+2E^{\rho,i}(t)
+\frac{t}{4m_{\rho}^2}(D_0^{\rho,i}(t)+\frac{1}{2}D_1^{\rho,i}(t)-E^{\rho,i}(t))-\frac{t^2}{32m_{\rho}^4}D_1^{\rho,i}(t)\right)\right]_{\text{BF2}} \;,\\
p^{\rho,i}_{0,\text{BF2}}(r_{\mybot}) &=
\frac{1}{8}\partial^2_{\mybot}\left[\frac{1}{\sqrt{m_{\rho}^2 -t/4}}\left(-D^{\rho,i}_0(t)+
2E^{\rho,i}(t)+
\frac{t}{4m_{\rho}^2}(D_0^{\rho,i}(t)+\frac{1}{2}D_1^{\rho,i}(t)-E^{\rho,i}(t))-\frac{t^2}{32m_{\rho}^4}D_1^{\rho,i}(t)\right)\right]_{\text{BF2}} \;.
\end{align}
By considering the
matrix elements $\gamma \bra{\rho(p',s')}T^{00}_g\ket{\rho(p,s)}$ and $ \bra{\rho(p',s')}T^{ij}_g\ket{\rho(p,s)}/\gamma$, we obtain the lowest-order contributions
to the monopole densities in the IMF as
\begin{align}
\label{eq:rhodensitiesIMF}
\varepsilon^{\rho,i}_{0,\text{IMF}}(r_{\mybot}) &
\approx m_{\rho}[A^{\rho,i}_0(t)]_{\text{IMF}} \;,\\
s^{\rho,i}_{0,\text{IMF}}(r_{\mybot}) &\approx -\frac{1}{4m_{\rho}}r_{\mybot}\frac{d}{dr_{\mybot}}\frac{1}
{r_{\mybot}}\frac{d}{dr_{\mybot}}[-D_0^{\rho,i}(t)+2E^{\rho,i}(t)]_{\text{IMF}}\;, \\
p^{\rho,i}_{0,\text{IMF}}(r_{\mybot}) &\approx 
\frac{1}{8m_{\rho}}\partial^2_{\mybot}[-D^{\rho,i}_0(t)+
2E^{\rho,i}(t)]_{\text{IMF}} \,.
\end{align}
The corresponding conserved mass radii are
\begin{align} \label{eq:rhoradii}
\braket{r^2_i}_{\text{BF3}}^{\rho,\text{mass}}
&= 6\frac{dA^{\rho,i}_0(t)}{dt}\rvert_{t=0}+\frac{1}{m_{\rho}^2}(-\frac{7}{4}
A^{\rho,i}_0(0)+\frac{1}{2}A^{\rho,i}_1(0)+\frac{3}{2}D^{\rho,i}_0(t)+2J^{\rho,i}(0)
-E^{\rho,i}(0)) \;,\\
\braket{r^2_i}_{\text{BF2}}^{\rho,\text{mass}}
&= 4 \frac{dA^{\rho,i}_0(t)}{dt}\rvert_{t=0} +\frac{1}{2 m_{\rho}^2}(-4A^{\rho,i}_0(0)+A^{\rho,i}_1(0)+2D^{\rho,i}_0(0)-2E^{\rho,i}(0)+4J^{\rho,i}(0)) \;,\\
\braket{r^2_i}_{\text{IMF}}^{\rho,\text{mass}} 
&= 4\frac{dA^{\rho,i}_0(t)}{dt}\rvert_{t=0} \,.
\end{align}
Note that the IMF energy density corresponds to a different component of the EMT than the Drell-Yan frame (DYF) energy, as discussed in Ref.~\cite{Lorce:2018egm} for the case of
the nucleon, and therefore the
IMF mass radius is different
than the DYF radius found
in Ref.~\cite{Freese:2019bhb}.

\subsection{$\Delta$ baryon}
\label{sec:deltadensities}

The BF3 densities of the $\Delta$ baryon were derived in
Ref.~\cite{Kim:2020lrs} and are
\begin{align} \label{eq:deltadensities}
\varepsilon_{0,\text{BF3}}^{\Delta,i}(r) &= m_{\Delta}\bigg[F^{\Delta,i}_{10}(t)+F^{\Delta,i}_{30}(t)\nonumber \\
& +\frac{t}{6m_{\Delta}^2}\left(-\frac{5}{2}F^{\Delta,i}_{10}(t)-F^{\Delta,i}_{11}(t)
-\frac{3}{2}F^{\Delta,i}_{20}(t)+4F^{\Delta,i}_{50}(t)+3F^{\Delta,i}_{40}(t)-F^{\Delta,i}_{30}(t)
-F^{\Delta,i}_{31}(t)-F^{\Delta,i}_{60}(t)\right)\nonumber \\
& +\frac{t^2}{12m_{\Delta}^4}\left(\frac{1}{2}F^{\Delta,i}_{10}(t)+F^{\Delta,i}_{11}(t)
+\frac{1}{2}F^{\Delta,i}_{20}(t)+\frac{1}{2}F^{\Delta,i}_{21}(t)-4F^{\Delta,i}_{50}(t)
-F^{\Delta,i}_{40}(t)-F^{\Delta,i}_{41}(t)+\frac{1}{2}F^{\Delta,i}_{31}(t)\right)\nonumber \\
&+\frac{t^3}{48m_{\Delta}^6}\left(-\frac{1}{2}F^{\Delta,i}_{11}(t)-\frac{1}{2}
F^{\Delta,i}_{21}(t)+F^{\Delta,i}_{41}(t)\right) \bigg]_{\text{BF3}} \\
\varepsilon_{2,\text{BF3}}^{\Delta,i}(r) &= -\frac{1}{m_{\Delta}}
r_{(\mybot)}\frac{d}{dr}\frac{1}{r}\frac{d}{dr}\bigg[-\frac{1}{6}\left(F^{\Delta,i}_{10}(t)+F^{\Delta,i}_{11}(t)-4F^{\Delta,i}_{50}(t)
+F^{\Delta,i}_{30}(t)+F^{\Delta,i}_{31}(t)+F^{\Delta,i}_{60}(t)\right) \nonumber \\
&+ \frac{t}{12m_{\Delta}^2}\left(\frac{1}{2}F^{\Delta,i}_{10}(t)+F^{\Delta,i}_{11}(t)+\frac{1}{2}
F^{\Delta,i}_{20}(t)+\frac{1}{2}F^{\Delta,i}_{21}(t)-4F^{\Delta,i}_{50}(t)
-F^{\Delta,i}_{40}(t)-F^{\Delta,i}_{41}(t)+\frac{1}{2}F^{\Delta,i}_{31}(t)\right) \nonumber\\
&+ \frac{t^2}{48m_{\Delta}^4}\left(-\frac{1}{2}F^{\Delta,i}_{11}(t)
-\frac{1}{2}F^{\Delta,i}_{21}(t)+F^{\Delta,i}_{41}(t)\right)\bigg]_{\text{BF3}} \\
p_{0,\text{BF3}}^{\Delta,i}(r) &= \frac{1}{6 m_{\Delta}}\partial^2\bigg[
F^{\Delta,i}_{20}(t)-\frac{16}{3}F^{\Delta,i}_{50}(t) - \frac{t}{6m_{\Delta}^2}\left(F^{\Delta,i}_{20}(t)+
F^{\Delta,i}_{21}(t)-4 F^{\Delta,i}_{50}(t)\right) + \frac{t^2}{24 m_{\Delta}^4}F^{\Delta,i}_{21}(t)\bigg]
_{\text{BF3}} \\
p_{2,\text{BF3}}^{\Delta,i}(r) &=
\frac{1}{6 m_{\Delta}}\partial^2\bigg[\frac{4}{3}F^{\Delta,i}_{50}(t)\bigg]_
{\text{BF3}} \nonumber \\
+&\frac{1}{6 m_{\Delta}^2}\partial^2 \left(\frac{d}{dr}\frac{d}
{dr}-\frac{2}{r}\frac{d}{dr}\right)\bigg[\frac{1}{6}\left(-F^{\Delta,i}_{20}(t)
-F^{\Delta,i}_{21}(t)+4F^{\Delta,i}_{50}(t)\right)+\frac{t}{24m_{\Delta}^2}F^{\Delta,i}_{21}(t)\bigg]_
{\text{BF3}} \\
p_{3,\text{BF3}}^{\Delta,i}(r) &=
-\frac{2}{6 m_{\Delta}^2}\partial^2\left(\frac{d}{dr}\frac{d}
{dr}-\frac{3}{r}\frac{d}{dr}\right)\bigg[\frac{1}{6}\left(-F^{\Delta,i}_{20}(t)
-F^{\Delta,i}_{21}(t)+4F^{\Delta,i}_{50}(t)\right)+\frac{t}{24m_{\Delta}^2}F^{\Delta,i}_{21}(t)\bigg]_
{\text{BF3}} \\
s_{0,\text{BF3}}^{\Delta,i}(r) &= -\frac{1}{4m_{\Delta}}r
\frac{d}{dr}\frac{1}{r}\frac{d}{dr} \\
&\bigg[F^{\Delta,i}_{20}(t)-\frac{16}{3}F^{\Delta,i}_{50}(t) - \frac{t}{6m_{\Delta}^2}\left(F^{\Delta,i}_{20}(t)+
F^{\Delta,i}_{21}(t)-4 F^{\Delta,i}_{50}(t)\right) + \frac{t^2}{24 m_{\Delta}^4}F^{\Delta,i}_{21}(t)\bigg]
_{\text{BF3}} \\
s_{2,\text{BF3}}^{\Delta,i}(r) &= -\frac{1}{4m_{\Delta}}r
\frac{d}{dr}\frac{1}{r}\frac{d}{dr}
\bigg[\frac{4}{3}F^{\Delta,i}_{50}(t)\bigg]_
{\text{BF3}} \nonumber\\
&-\frac{1}{4m_{\Delta}^2}r\frac{d}{dr}\frac{1}{r}\frac{d}{dr}\left(\frac{d}{dr}\frac{d}
{dr}-\frac{2}{r}\frac{d}{dr}\right) 
\bigg[\frac{1}{6}\left(-F^{\Delta,i}_{20}(t)
-F^{\Delta,i}_{21}(t)+4F^{\Delta,i}_{50}(t)\right)+\frac{t}{24m_{\Delta}^2}F^{\Delta,i}_{21}(t)\bigg]_
{\text{BF3}} \\
s_{3,\text{BF3}}^{\Delta,i}(r) &= \frac{1}{2m_{\Delta}^2}r\frac{d}{dr}\frac{1}{r}\frac{d}{dr}\left(\frac{d}{dr}\frac{d}
{dr}-\frac{3}{r}\frac{d}{dr}\right) 
\bigg[\frac{1}{6}\left(-F^{\Delta,i}_{20}(t)
-F^{\Delta,i}_{21}(t)+4F^{\Delta,i}_{50}(t)\right)+\frac{t}{24m_{\Delta}^2}F^{\Delta,i}_{21}(t)\bigg]_
{\text{BF3}} \;.
\end{align}
We obtain the following expressions for the BF2 and IMF 
leading-order contributions to the
EMT monopole densities:
\begin{align} \label{eq:deltadensitiesIMF}
\varepsilon_{0,\text{BF2}}^{\Delta,i}(r_{\mybot}) &= m_{\Delta}\bigg[F^{\Delta,i}_{10}(t) +F^{\Delta,i}_{30}(t)\nonumber \\
& +\frac{t}{4m_{\Delta}^2}\left(-2F^{\Delta,i}_{10}(t)-F^{\Delta,i}_{11}(t)
-F^{\Delta,i}_{20}(t)+4F^{\Delta,i}_{50}(t)+2F^{\Delta,i}_{40}(t)
-F^{\Delta,i}_{60}(t)-F^{\Delta,i}_{30}(t)-F^{\Delta,i}_{31}(t)\right)\nonumber \\
& +\frac{t^2}{16m_{\Delta}^4}\left(F^{\Delta,i}_{10}(t)+2F^{\Delta,i}_{11}(t)
+F^{\Delta,i}_{20}(t)+F^{\Delta,i}_{21}(t)-8F^{\Delta,i}_{50}(t)
-2F^{\Delta,i}_{40}(t)-2F^{\Delta,i}_{41}(t)+F^{\Delta,i}_{31}(t)\right)\nonumber \\
&+\frac{t^3}{64m_{\Delta}^6}\left(-F^{\Delta,i}_{11}(t)-
F^{\Delta,i}_{21}(t)+2F^{\Delta,i}_{41}(t)\right) \bigg]_{\text{BF2}} \\
p_{0,\text{BF2}}^{\Delta,i}(r_{\mybot}) &= \frac{1}{8 m_{\Delta}}\partial_{\mybot}^2\bigg[
F^{\Delta,i}_{20}(t)-8F^{\Delta,i}_{50}(t) - \frac{t}{4m_{\Delta}^2}\left(F^{\Delta,i}_{20}(t)+
F^{\Delta,i}_{21}(t)-4 F^{\Delta,i}_{50}(t)\right) + \frac{t^2}{16 m_{\Delta}^4}F^{\Delta,i}_{21}(t)\bigg]
_{\text{BF2}} \\
s_{0,\text{BF2}}^{\Delta,i}(r_{\mybot}) &= -\frac{1}{4m_{\Delta}}r_{\mybot}
\frac{d}{dr_{\mybot}}\frac{1}{r_{\mybot}}\frac{d}{dr_{\mybot}} \\
&\bigg[F^{\Delta,i}_{20}(t)-8F^{\Delta,i}_{50}(t) - \frac{t}{4m_{\Delta}^2}\left(F^{\Delta,i}_{20}(t)+
F^{\Delta,i}_{21}(t)-4 F^{\Delta,i}_{50}(t)\right) + \frac{t^2}{16 m_{\Delta}^4}F^{\Delta,i}_{21}(t)\bigg]
_{\text{BF2}} \\
\varepsilon^{\Delta,i}_{0,\text{IMF}}(r_{\mybot}) &\approx
m_{\Delta}[F^{\Delta,i}_{10}(t)]_{\text{IMF}} \\
s^{\Delta,i}_{0,\text{IMF}}(r_{\mybot}) &\approx
-\frac{1}{4m_{\Delta}}r_{\mybot}\frac{d}{dr_{\mybot}}\frac{1}
{r_{\mybot}}\frac{d}{dr_{\mybot}}[F^{\Delta,i}_{20}(t)-8F^{\Delta,i}_{50}(t)
]_{\text{IMF}} \\
p^{\Delta,i}_{0,\text{IMF}}(r_{\mybot}) &\approx
\frac{1}{8m_{\Delta}}\partial^2_{\mybot}[F^{\Delta,i}_{20}(t)-8F^{\Delta,i}_{50}(t)]_{\text{IMF}} \, .
\end{align}
The corresponding conserved mass radii formulas are
\begin{align} \label{eq:deltaradii}
\braket{r^2_i}_{\text{BF3}}^{\Delta,\text{mass}}
&= 6\frac{dF^{\Delta,i}_{10}(t)}{dt}\rvert_{t=0}+\frac{1}{m_{\Delta}^2}\left
(-\frac{5}{4}
F^{\Delta,i}_{10}(0)-F^{\Delta,i}_{11}(0)-\frac{3}{2}F^{\Delta,i}_{20}(0)+
4F^{\Delta,i}_{50}(0)+4F^{\Delta,i}_{40}\right)\\
\braket{r^2_i}_{\text{BF2}}^{\text{mass}}
&= 4\frac{dF^{\Delta,i}_{10}(t)}{dt}\rvert_{t=0} +\frac{1}{m_{\Delta}^2}(-2F^{\Delta,i}_{10}(0)-F^{\Delta,i}_{11}(0)-F^{\Delta,i}_{20}(0)+2F^{\Delta,i}_{40}(0)+4F^{\Delta,i}_{50}(0)) \\
\braket{r^2_i}_{\text{IMF}}^{\Delta,\text{mass}} 
&= 4\frac{dF^{\Delta,i}_{10}(t)}{dt}\rvert_{t=0} \, .
\end{align}

\twocolumngrid
\bibliography{main}

\end{document}